\documentclass[11pt,english]{article} 
\usepackage[scale=0.7]{geometry}

\usepackage[utf8x]{inputenc}
\usepackage[T1]{fontenc}
\usepackage{ucs}

\usepackage{xcolor,enumitem}

\usepackage{subcaption}

\usepackage{graphicx,interval}
\usepackage{mathtools,amsfonts,amssymb}
\usepackage{mathrsfs} 

\definecolor{vertfonce}{rgb}{0.20, 0.46, 0.25}
\definecolor{rougefonce}{rgb}{0.64, 0.09, 0.20}

\usepackage{babel}
\usepackage[breaklinks=true,
            colorlinks=true,
            linkcolor=rougefonce,
            citecolor=vertfonce]{hyperref}
						
\usepackage{float}						
\usepackage{tikz}

\newcommand{\RM}{\mathbb{R}}
\newcommand{\ZM}{\mathbb{Z}}

\newcommand{\NM}{\mathbb{N}}
\newcommand{\CM}{\mathbb{C}}
\newcommand{\T}{\mathbb{T}}
\newcommand{\h}{\hbar}
\newcommand{\deriv}[2]{\frac{\partial #1}{\partial #2}}
\newcommand{\dd}[1]{\ensuremath{\operatorname{d}\!{#1}}}
\newcommand{\Cinf}{C^\infty}
\newcommand{\formel}[1]{[\![#1]\!]}
\newcommand{\snorm}[1]{\|#1\|}

\newcommand{\pscal}[2]{\langle #1,#2\rangle}

\newcommand{\abs}[1]{\left|#1\right|}

\newcommand{\phy}{\varphi}
\newcommand{\ham}[1]{\mathcal{X}_{#1}}
\newcommand{\cqfd}{\hfill $\square$\par\vspace{1ex}}
\renewcommand{\O}{\mathscr{O}}
\newtheorem{theo}{Theorem}[section]
\newtheorem{defi}[theo]{Definition}
\newtheorem{prop}[theo]{Proposition}
\newtheorem{lemm}[theo]{Lemma}
\newtheorem{coro}[theo]{Corollary}
\newtheorem{ex}[theo]{Example}
\newenvironment{rema}
{\par\vspace{1ex}\refstepcounter{theo}
\noindent\textbf{Remark~\thetheo} }
{~\hfill\mbox{$\triangle$}\par\vspace{1ex}}
\newcommand{\demons}[1][$\!\!$]{\noindent\textbf{Proof\ }\textsl{#1}\textbf{.}~}
\newenvironment{demo}[1][$\!\!$]
{\demons[#1]\ }
{\cqfd}
\newenvironment{algo}
{\par\vspace{1ex}\refstepcounter{theo}
\noindent\textbf{Algorithm~\thetheo} }
{~\hfill\mbox{$\triangle$}\par\vspace{1ex}}

\DeclarePairedDelimiter\floor{\lfloor}{\rfloor}

\newcommand\blfootnote[1]{
  \begingroup
  \renewcommand\thefootnote{}\footnote{#1}
  \addtocounter{footnote}{-1}
  \endgroup
}

\title{The inverse spectral problem for quantum semitoric systems}

\author{Yohann \textsc{Le Floch}\footnote{Univ Strasbourg, CNRS, IRMA
    - UMR 7501, F-67084 Strasbourg, France} \and \textsc{V\~u Ng\d{o}c}
  San\footnote{Univ Rennes, CNRS, IRMAR - UMR 6625, F-35000 Rennes,
    France}}

\date{}

\begin{document}
\maketitle

\begin{abstract}
  Given a quantum semitoric system composed of pseudodifferential
  operators, Berezin-Toeplitz operators, or a combination of both, we
  obtain explicit formulas for recovering, from the semiclassical
  asymptotics of the joint spectrum, all symplectic invariants of the
  underlying classical semitoric system.

  Our formulas are based on the possibility to obtain good quantum
  numbers for joint eigenvalues from the bare data of the joint
  spectrum. In the spectral region corresponding to regular values of
  the momentum map, the algorithms developed by Dauge, Hall and the
  second author~\cite{san-dauge-hall-rotation} produce such
  labellings. In our proof, it was crucial to extend these algorithms
  to the boundary of the spectrum, which led to the new notion of
  asymptotic half-lattices, and to globalize the resulting labellings.

  Using the construction given by Pelayo and the second author
  in~\cite{san-alvaro-II}, our results prove that semitoric systems
  are completely spectrally determined in an algorithmic way~: from
  the joint spectrum of a quantum semitoric system one can construct a
  representative of the isomorphism class of the underlying classical
  system. In particular, this recovers the uniqueness result
  obtained by Pelayo and the authors
  in~\cite{san-alvaro-yohann,san-alvaro-yohann-erratum}, and completes
  it with the explicit computation of all invariants, including the
  twisting index.

  In the cases of the spin-oscillator and the coupled angular momenta,
  we implement the algorithms and illustrate numerically the
  computation of the invariants from the joint spectrum.
\end{abstract}

\blfootnote{\,\,\emph{2020 Mathematics Subject Classification.}
  81S10, 81Q20, 58J50, 37J35, 58J40, 53D50, 53D20.\\
  \indent\indent \emph{Key words and phrases.} Semitoric integrable
  systems. Inverse spectral theory. Quantum mechanics. Semiclassical
  analysis. Pseudodifferential operators. Berezin-Toeplitz
  operators. Symplectic invariants. Focus-focus singularity. Joint
  spectrum. Asymptotic lattice. Good labelling. Quantum
  numbers. Lattice structure detection. Singular Lagrangian
  fibration.}

\clearpage

\tableofcontents

\section{Introduction}

The goal of this paper is to answer the question ``can one hear a
semitoric system?'', which belongs to a long lineage of inverse
spectral problems popularized by Kac in his famous article
\cite{kac}. As often, the aim is to recover a classical geometry, up
to isomorphism, from the data of a discrete set obtained as a quantum
spectrum.

Semitoric systems form a class of completely integrable Hamiltonian
systems with two degrees of freedom.  Their introduction as a
mathematical object, more than 15 years ago, was motivated both by
symplectic geometry~\cite{symington-four} and quantum
physics~\cite{san-polytope}. Indeed, they play a fundamental role in
explaining stable couplings between two particles, through the
celebrated Jaynes-Cummings model and its
variants~\cite{jaynes-cummings}. For instance, an atom, seen as a
multi-spin system, trapped in a potential cavity, is a semitoric
system of great importance in entanglement experiments and quantum
computing (constructing and controlling quantum dots), as explained in
the colloquium paper by
Raimond-Brune-Haroche~\cite{raimond-brune-haroche01}.  Semitoric
systems can describe numerous models, from a photon in an optical
cavity to a symmetric molecule near a relative equilibrium, and have
been widely used in quantum chemistry and spectrocopy,
see~\cite{sadovski-zhilinski, joyeux-sado-tenny} and references
therein. The precise structure of the \emph{quantum spectrum} of
semitoric systems, in particular its ``non-linear'' behaviour with
respect to the harmonic oscillator ladder, has been used as a proof of
the true quantum mechanical nature of matter-light
interaction~\cite{fink08}, and it was suggested that this spectral
feature should also impact the dynamical control of quantum
dots. Recently, the spectral structure of a seemingly different model
(Rydberg-dressed atoms) was used to propose an ``experimental
isomorphism'' with the classical Jaynes-Cummings sytem~\cite{lee17}.

On the mathematical side, semitoric systems have been extensively
studied in the last 15 years (see for instance the review
\cite{alonso-hohloch19}), and the intriguing connections between the
spectra of quantum semitoric systems and the symplectic geometry of
the underlying classical systems have been a driving force in the
development of the theory. Thus, naturally, when a complete set of
``numerical'' symplectic invariants of classical simple
semitoric systems was discovered~\cite{san-alvaro-I,san-alvaro-II},
the question was raised of whether these invariants were
\emph{spectrally determined}. This was stated in~\cite[Conjecture
9.1]{san-alvaro-survey}, further advertised in several papers as the
\emph{inverse spectral conjecture for semitoric systems} (see for
instance~\cite{san-daniele}, \cite[Section
7.2]{bolsinov-matveev-miranda-tabachnikov} or the recent
surveys~\cite{alonso-hohloch19,pelayo2020symplectic}, and the
references therein), and investigated in particular
in~\cite{san-alvaro-taylor, san-alvaro-yohann,
  san-alvaro-yohann-erratum,charles-pelayo-vungoc, san-alvaro-leonid}.
The aim of the current article is to give explicit formulas and
algorithms to obtain all the invariants from the spectrum. This
provides a complete proof of the aforementioned conjecture.

Before giving detailed definitions in the next sections, let us simply
mention, in this introduction, that a semitoric system on a
4-dimensional symplectic manifold $(M,\omega)$ is a pair of commuting
Hamiltonians $F=(J,H)$ on $M$, where $J$ is the momentum map of an
effective $S^1$-action, and $F:M\to\RM^2$, viewed as a singular
Lagrangian fibration, has singularities of a certain Morse-Bott type,
with compact, connected fibers. These systems are of course very
natural from the physical viewpoint where $S^1$-symmetry is ubiquitous
and can be seen mathematically as a surprisingly far-reaching
generalization of the \emph{toric systems} studied by Atiyah,
Guillemin-Sternberg,
Delzant~\cite{atiyah-convex,guillemin-sternberg,delzant} and many
others. In this paper we will restrict ourselves to the generic
  case of simple semitoric systems (see
  Definition~\ref{defi:simple}). Then, the \emph{symplectic
  invariants} of $F$, which completely characterize
$(M,\omega,F)$~\cite{san-alvaro-I,san-alvaro-II}, are a sequence of
numbers and combinatorial objects that describe the associated
singular integral affine structure; and these invariants can be
expressed as five objects, with some mutual relations:

\begin{enumerate}
\item A rational, convex \emph{polygonal set} $\Delta\subset\RM^2$.
\item A discrete set of distinguished \emph{points} $c_j\in\Delta$,
  representing the isolated critical points (the focus-focus singularities) of $F$.
\item Each point $c_j$ is decorated with the following:
  \begin{enumerate}
  \item \label{item:height} a real number representing a symplectic
    volume (usually called the \emph{height invariant});
  \item \label{item:twist} an integer $k\in\ZM$ called the
    \emph{twisting number};
  \item \label{item:taylor} a formal \emph{Taylor series} in two
    indeterminates.
  \end{enumerate}
\end{enumerate}
 
A quantum semitoric system is a pair of commuting selfadjoint
operators (quantum Hamiltonians), depending on the semiclassical
parameter $\hbar$, whose joint principal symbol is $F$ as above,
acting on a Hilbert space quantizing the symplectic manifold
$(M,\omega)$. It defines a \emph{joint spectrum}, which is a set of
points in $\RM^2$, and the natural inverse spectral problem is to
recover the classical system $(M,\omega,F)$, up to symplectic
equivalence, from the raw data of this point set as $\hbar \to
0$. This question naturally originates from quantum spectroscopy,
where it is crucial to recover the nature of molecules through the
observation of their spectrum; it is still a very active area of
research, with many approaches and algorithms for detection; see for
instance~\cite{roucou-dhont-al17}. In this paper we shall adopt a
semiclassical viewpoint, which takes advantage of symplectic
invariants in phase space and was already advocated
in~\cite{heller81}.

One of the first results concerning the inverse spectral conjecture
was to solve the particular case of \emph{toric}
systems~\cite{charles-pelayo-vungoc,san-alvaro-leonid}, where only the
first invariant (the polygon) subsists. This was crucially based on
Delzant's theorem~\cite{delzant} (since the polygon is obtained
explicitly as the semiclassical limit of the joint spectrum), and on
the properties of Berezin-Toeplitz quantization, or more general
quantizations for \cite{san-alvaro-leonid}. Naturally, the techniques
were not transferable to more general cases, for which the main
challenge is the treatment of focus-focus singularities, which do not
appear in toric systems. Remark that, even when $M$ is fixed and when
there is only one focus-focus singularity, the moduli space of
semitoric systems on $M$ immediately becomes infinite
dimensional. In~\cite{san-alvaro-taylor}, it was proven that the last
invariant, the Taylor series, was spectrally determined, in the sense
of a \emph{uniqueness} statement: if two systems have the same quantum
spectrum, then their Taylor series must coincide. Finally, the best
result (prior to the present work) was obtained by Pelayo and the
authors in~\cite{san-alvaro-yohann,san-alvaro-yohann-erratum} and is
also a uniqueness statement: based on the former cited article, it was
proven in~\cite[Theorem A]{san-alvaro-yohann} that two quantum
semitoric systems having the same joint spectrum (modulo $\O(\h^2)$)
must share the same invariants, with possibly the exception of the
twisting index. As a consequence~\cite[Theorem
B']{san-alvaro-yohann-erratum}, two Jaynes-Cummings systems (semitoric
systems with only one critical fiber) with the same quantum spectrum
and the same twisting number must be symplectically
isomorphic. However, it was not known whether the twisting number
could be determined from the spectrum, or not.

\medskip In this paper, we focus on the constructive part of the
semitoric inverse spectral conjecture. Our main result can be
informally stated as follows.

\begin{theo}[Theorem~\ref{theo:main}]
  From the joint spectrum (modulo $\O(\h^2)$) in a vertical strip of
  bounded width $S\subset \RM^2$ of a quantum semitoric system, one
  can compute, in an algorithmic way, all symplectic
  invariants of the underlying classical semitoric system restricted to
  $F^{-1}(S)$.
\end{theo}

Using the construction given by Pelayo and the second author
in~\cite{san-alvaro-II}, this completes the proof the inverse spectral
conjecture for general semitoric systems (with an arbitrary number of
focus-focus singularities).
  
\begin{coro}
  If two quantum semitoric systems have the same spectrum, then their
  underlying classical systems are symplectically isomorphic.
\end{coro}

This corollary completes the result previously obtained by Pelayo and
the authors in~\cite{san-alvaro-yohann,san-alvaro-yohann-erratum}.
The word ``quantum'' in these statements refers to both
$\h$-pseudodifferential and Berezin-Toeplitz quantizations, which
respectively appear in the quantization of cotangent bundles and
compact symplectic manifolds. To the authors' knowledge, this is the
first algorithmic inverse spectral result that holds for a large class
of quantum integrable systems on a possibly compact phase space, with
possibly non-toric dynamics. Recall that in the specific case of
compact toric systems, the result was proven by recovering the
associated Delzant
polytope~\cite{charles-pelayo-vungoc,san-alvaro-leonid}.

The uniqueness result
of~\cite{san-alvaro-yohann,san-alvaro-yohann-erratum} implies that,
within the class of Jaynes-Cummings systems, the symplectic invariants
other than the twisting index are implicitly determined by the
asymptotics of the joint spectrum. In view of this, and the above
discussion, the main achievements of the present work are to consider
the whole class of semitoric systems, and within this class:
\begin{enumerate}
\item to constructively recover the \emph{twisting number}
  associated with each focus-focus critical value
  (Theorem~\ref{theo:recover-sigma_1});
\item to constructively recover the full \emph{Taylor series
    invariant} associated with each focus-focus critical value
  (Theorem \ref{theo:full_taylor});
\item to find a global procedure to construct the \emph{polygon
    invariant} from the spectral data (Theorem
  \ref{theo:poly_twisting});
\item to obtain an explicit formula that gives the \emph{height
    invariant} from the joint spectrum (Proposition
  \ref{prop:height-invariant}).
\end{enumerate}
It is known from the classification of semitoric systems that the
first and third item are not independent, since changing the polygon
invariant implies a global shift of all the twisting numbers; hence
the procedures to obtain them from the joint spectrum are
intricate. In proving the second item, we additionally recover for the
first time the full infinite jet of the Eliasson diffeomorphism, which
brings the system near a focus-focus singularity into a Morse-Bott
normal form and is known to be an invariant of the map $F$, see for
instance~\cite[Definition 4.37]{san-daniele}. In fact, the Taylor
series and the Eliasson diffeomorphism are not specific to semitoric
systems: they are invariants of a singular Lagrangian fibration near a
focus-focus fiber, and our techniques allow to compute them explicitly
from the joint spectrum of any quantum integrable system possessing
such a singularity.

For the sake of completeness, we have also included the proof of some
Bohr-Sommerfeld rules that were missing in the literature, in
particular for Berezin-Toeplitz operators in the case of a
transversally elliptic singularity. However, it is worth noticing that
our strategy does not necessitate the more delicate uniform
description of the joint spectrum in a neighborhood of a focus-focus
singularity, which has been proved for $\h$-pseudodifferential
operators \cite{san-focus} but is still conjectural for
Berezin-Toeplitz operators (see also~\cite{babelon-spin}). This can be
circumvented by taking two consecutive limits, one as $\h \to 0$ for a
given regular value $c$, then one as $c$ goes to the focus-focus
value.

Recently, a renewal of interest on semitoric invariants was triggered
by their explicit (algebraic and numerical) computations in a large
number of important
examples~\cite{san-alvaro-spin,alvaro-yohann,ADH-spin,ADH-angular}. Thus,
we also wanted to take advantage of this to test our results on
several cases, by implementing numerical algorithms along the proof of
the theoretical results. This also means that in most of the proofs,
we put some emphasis on practical formulas, errors and convergence
rates.

Our proof is a combination of microlocal analysis, asymptotic
analysis, symplectic geometry, but also, and crucially, combinatorial
and algorithmic techniques borrowed from the recent
work~\cite{san-dauge-hall-rotation}. That work, motivated by detecting
the \emph{rotation number} on the joint spectrum of a quantum
integrable system, introduced general tools for dealing with so-called
\emph{asymptotic lattices} of eigenvalues; these tools turned out to
be essential in our approach. Indeed, contrary to the usual cases of
inverse spectral theory where the spectral data is a sequence of real
(and hence ordered) eigenvalues, here we have to deal with joint
spectra of commuting operators, which are two-dimensional point
clouds, moving with the semiclassical parameter. Thus, the first step
in all our results is to consistently define good quantum numbers for
such joint eigenvalues. Coming up with these quantum numbers is
already non trivial near a regular value of the underlying momentum
map where these eigenvalues are a deformation of the standard lattice,
see \cite{san-dauge-hall-rotation}. Here, it will be crucial not only
to develop a local-to-global theory of quantum numbers (because the
presence of focus-focus singularities is known to obstruct the
existence of global labellings), but also to obtain good labels near
transversally elliptic singular values as well; in this case the joint
spectrum is a deformation of the intersection of the standard lattice
with a half-plane, which leads us to introduce the notion of
\emph{asymptotic half-lattice}.

In the aforementioned article, the emphasis was put on
$\h$-pseudodifferential operators. In our case however, it is very
important to also consider Berezin-Toeplitz operators, since many
relevant examples of semitoric systems are defined on compact
symplectic manifolds. Throughout this manuscript, we will make sure
that the results that we use hold in both contexts.

Some general ideas of proof were already present
in~\cite{san-alvaro-first-steps}. We were able to make some of them
concrete. However, in that paper, the difficulty for finding good
labellings for the joint spectrum was overlooked, and no strategy was
given for the twisting index, since at that time the relationship
between that invariant and the Taylor series was not understood.

The structure of the article is as follows.
\begin{itemize}
\item In Section \ref{sec:semitoric_invariants}, we recall the
  essential properties of semitoric systems and present their symplectic
  invariants in a new way which is more adapted to the inverse problem.
\item In Section \ref{sec:quantum_semitoric}, we introduce a notion of
  semiclassical operators which allows us to deal with
  $\h$-pseudodifferential operators and Berezin-Toeplitz operators
  simultaneously. Then we define quantum semitoric systems and their
  joint spectra, and state our main result; Sections
  \ref{sec:lattices} to \ref{sect:BScrit} are devoted to its proof.
\item In Section \ref{sec:lattices}, we review asymptotic lattices and
  their labellings, define and study asymptotic half-lattices, and
  show how to construct global labellings for unions of asymptotic
  lattices and half-lattices.
\item In Section \ref{sec:twisting}, we explain how to compute the
  twisting number and the semitoric polygon from the joint spectrum,
  effectively recovering the twisting index invariant.
\item In Section \ref{sec:taylor_series}, we give a procedure to
  obtain explicitly the height invariant, the full Taylor series
  invariant and the full infinite jet of the Eliasson diffeomorphism
  from the spectral data.
\item In Section \ref{sect:BScrit}, we give a proof of the
  Bohr-Sommerfeld rules near an elliptic-transverse critical value of
  an integrable system, valid both for $\h$-pseudodifferential
  operators and Berezin-Toeplitz operators.
\item In Section \ref{sect:coupled_spins}, we illustrate
    numerically the various formulas giving the symplectic invariants
    on the example of coupled angular momenta, which takes place on a
    compact manifold.
\item In the Appendix, we briefly review $\h$-pseudodifferential and
  Berezin-Toeplitz operators, emphasizing the non-compact cases
  required for our analysis.
\end{itemize}

Section \ref{sect:coupled_spins} is not the only place where we
  discuss examples. We also give numerical results for the
  Jaynes-Cummings system, which is a semitoric system on a non-compact
  manifold that consists in coupling a classical spin and a harmonic
  oscillator; this system is defined in
  Examples~\ref{ex:spin-osc_class} and~\ref{ex:spin-osc_quant}, and is
  used to illustrate the different definitions throughout the whole
  manuscript.

\begin{rema}
  The presentation of the classifying space of semitoric systems by
  means of the five invariants mentioned above has proven quite useful
  in the development of the inverse theory, allowing various studies
  to focus on a particular item; but the separation between the five
  invariants is somewhat arbitrary. For instance, the last three
  invariants \ref{item:height}, \ref{item:twist},
  and~\ref{item:taylor}, could be naturally combined into a single
  Taylor series (see Section~\ref{sec:tayl-seri-invar}). This was
  already partly observed in~\cite{jaume-thesis};
  in~\cite{palmer-pelayo-tang} the authors even prefer to pack all
  invariants into a single object. However, it makes sense to single
  out the last one (the Taylor series invariant), as it is the
  complete \emph{semi-global} invariant for a neighborhood of the
  critical fiber associated with $c_j$, not only for semitoric
  systems, but also for any integrable system with a focus-focus leaf
  carrying only one focus-focus critical point~\cite{san-semi-global}
  (this was used in~\cite{san-alvaro-taylor} to show that the joint
  spectrum near the singular value determines the semi-global
  classification). On the contrary, the height invariant and the
  twisting number characterize the \emph{global} location of the fiber
  within the whole system.

  Actually, while the main goal of our work is to solve the inverse
  problem for semitoric systems, it is interesting to notice that a
  large part of our analysis, which concerns the Taylor series
  invariant and the Eliasson diffeomorphism, is local in action
  variables and hence not specific to semitoric systems.
\end{rema}

\begin{rema}
  As mentioned above, the present article goes beyond the uniqueness
  statement of the inverse problem by proposing a constructive
  approach. Therefore, the methods in play are necessarily quite
  different from those used in the previous inverse spectral
  results~\cite{san-alvaro-taylor} and~\cite{san-alvaro-yohann}. As a
  consequence, in addition to completely solving the inverse spectral
  conjecture, our analysis also provides a new proof of the main
  results of these articles.
\end{rema}

\begin{rema}\label{rema:generalizations}
  After their original definition and classification, several natural
  generalizations of semitoric systems have been
  proposed~\cite{san-pelayo-ratiu-affine, HolSabSepSym18,
    palmer-pelayo-tang, wacheux-local-preprint}. It would be
  interesting to investigate the inverse problem for the generalized
  classes, and in particular in the case of multiple pinches in the
  focus-focus fibers~\cite{pelayo-tang,palmer-pelayo-tang}, because in
  this case a negative answer seems plausible.
\end{rema}

\begin{rema}
  There are many interesting connections between semiclassical inverse
  spectral theory of quantum integrable systems, as presented here,
  and other inverse spectral problems in geometric analysis and PDEs;
  on this matter, we refer the reader to the existing literature; see
  for instance~\cite{san-qmath14} and references therein for a quick
  and recent survey. Let us simply recall two salient aspects.

  The first one concerns the inverse spectral theory of the Riemannian
  Laplacian, certainly the most well-known of all inverse spectral
  problems; see the survey~\cite{datchev-hezari-survey13}. In that
  case, semiclassical asymptotics are clearly present through the
  high-energy limit, and the consequences of $S^1$-invariant geometry
  (surfaces of revolution) have been derived in important cases,
  see~\cite{zelditch-revolution, dryden-macedo-senadias16}. In order
  to completely fill the gap between these types of systems and the
  semitoric framework, one would need to lift the properness
  assumption on the $S^1$-momentum map $J$, which is one of the
  generalizations alluded to in the previous
  remark~\ref{rema:generalizations}.

  The second one concerns the generalization of inverse problems from
  Schrödinger operators to general Hamiltonians,
  see~\cite{iantchenko-sjostrand-zworski}; in that paper, a ``Taylor
  series'' plays an important role, and comes from a Birkhoff normal
  form. This is related (although in an indirect fashion,
  see~\cite{dullin-pendulum}) to our Taylor series and the Eliasson
  diffeomorphism discussed in Section~\ref{sec:tayl-seri-invar}. The
  use of such formal series in inverse problems was already crucial in
  Zelditch's milestone paper~\cite{zelditch-inverse-II}. Under a toric
  hypothesis, this formal series can
  disappear~\cite{dryden-guillemin-senadias12}, giving way to more
  geometric invariants like Delzant polytopes. In our semitoric case,
  we need to combine both worlds: Birkhoff-type invariants \emph{and}
  toric-type invariants.
\end{rema}

\paragraph{Acknowledgements.} We thank \'Alvaro Pelayo for
  interesting remarks which clarified the introduction. We also thank
  an anonymous referee for helpful suggestions which improved the
  exposition.

\section{Symplectic invariants of semitoric systems}
\label{sec:semitoric_invariants}

In this section, after recalling the definition of semitoric
  systems and describing their singularities, we give a new formula
to compute the twisting number in relation with the Eliasson
diffeomorphism and the linear part of the Taylor series invariant (see
Lemma \ref{lemm:def_sigma1_zero} and Proposition
\ref{prop:relation_inv}), which will be crucial for our analysis of
the joint spectrum. We also describe all the other symplectic
invariants of semitoric integrable systems introduced in
\cite{san-alvaro-I,san-alvaro-II} (see also \cite{san-daniele} for a
more recent account), and we illustrate them with the so-called
  spin-oscillator system, introduced in Example
  \ref{ex:spin-osc_class}.

\subsection{Symplectic preliminaries}
\label{sec:symp_prel}

We endow $\RM^4$ with canonical coordinates $(x_1,x_2,\xi_1,\xi_2)$
and the standard symplectic form
$\omega_0 = d\xi_1 \wedge dx_1 + d\xi_2 \wedge dx_2$. If $(M,\omega)$
is a four-dimensional symplectic manifold and $m \in M$, there always
exist local Darboux coordinates $(x_1,x_2,\xi_1,\xi_2)$ centered at
$m$ in which $\omega = \omega_0$. If
$f \in \Cinf(M;\RM)$, we define the Hamiltonian vector
field $\mathcal{X}_f$ as the unique vector field such that
$\dd{f} + \omega(\mathcal{X}_f, \cdot) = 0$. The Poisson bracket of two
functions $f, g \in \Cinf(M;\RM)$ is defined as
$\{f,g\} = \omega(\mathcal{X}_f, \mathcal{X}_g)$.

A \emph{Liouville integrable system} on the four-dimensional
symplectic manifold $(M,\omega)$ is the data of two functions
$J, H \in \Cinf(M;\RM)$ such that $\{J,H\} = 0$ and
$\ham{J}, \ham{H}$ are almost everywhere linearly independent. In this
article, we will use the terminology ``integrable system'' for
``Liouville integrable system''. The map $F = (J,H): M \to \RM^2$ is
called the momentum map of the system. A point $m \in M$ where the
above linear independence condition holds (which is equivalent to the
linear independence of $\dd{J}$ and $\dd{H}$) is called a regular point
of $F$; otherwise, $m$ is called a critical point of $F$. A point
$c \in \RM^2$ is called a regular value of $F$ if $F^{-1}(c)$ contains
only regular points, and a critical value of $F$ otherwise.

Let $(M,\omega, F= (J,H))$ be an integrable system on a
four-dimensional manifold, and let $c \in \RM^2$ be a regular value of
the momentum map $F$. The action-angle theorem \cite{mineur} (see also
\cite{duistermaat}) states that if $F^{-1}(c)$ is compact and
connected, then there exist a local diffeomorphism
$G_0: (\RM^2,0) \to (\RM^2,c)$ and a local symplectomorphism $\phi$
from a neighborhood of $F^{-1}(c)$ in $M$ to a neighborhood of
the zero section in $T^* \mathbb T^2$ with coordinates
$(\theta_1, \theta_2, I_1, I_2)$ and symplectic form
$\dd{I_1} \wedge \dd{\theta_1} + \dd{I_2} \wedge \dd{\theta_2}$ such
that $F \circ \phi^{-1} = G_0(I_1,I_2)$. Our convention is
$\mathbb{T}^2 = \RM^2 \slash (2\pi \ZM)^2$, so that the angles
$\theta_i$ belong to $\RM \slash 2\pi \ZM$.

It is standard to call $I_1, I_2$ \emph{action variables}; in what
follows, we call $G_0^{-1}$ an \emph{action diffeomorphism}. These are
not unique; if $A \in \mathrm{GL}(2,\ZM)$ and $\kappa \in \RM^2$, and
if we let
\begin{equation}
  \begin{pmatrix} L_1 \\ L_2 \end{pmatrix} = A \begin{pmatrix} I_1 \\ I_2 \end{pmatrix} + \kappa,
  \label{equ:change-action}
\end{equation}
then $(L_1, L_2)$ is another set of action variables near $m$, and
every pair of action variables is obtained in this fashion. We will
mainly be interested in the case where $G_0^{-1}$ is an \emph{oriented
  action diffeomorphism}, \emph{i.e.} satisfying $\det({\rm d} G_0(0)) > 0$;
in this case the above statements remain true with
$A \in \mathrm{SL}(2,\ZM)$.

Action diffeomorphisms define a natural integral affine structure on
the set of regular values of $F$; recall that an integral affine
manifold of dimension $d$ is a smooth manifold with an atlas whose
transition maps are of the form $A \cdot + b$ where
$A \in \mathrm{GL}(d,\ZM)$ and $b \in \RM^d$.

\subsection{Semitoric systems}
\label{sec:semitoric-systems}

There exists a notion of non-degenerate critical point of an
integrable system which we will not describe here, see \cite[Section
1.8]{bolsinov-fomenko}. A consequence of this definition is the
following symplectic analogue of the Morse lemma, which we state here
only in dimension four:

\begin{theo}[Eliasson normal form~\cite{eliasson-these}]
  \label{theo:eliasson}
  Let $(M,\omega,F = (J,H))$ be an integrable system on a
  four-dimensional manifold and let $m \in M$ be a non-degenerate
  critical point of $F$. Then there exist local symplectic coordinates
  $(x,\xi) = (x_1,x_2,\xi_1,\xi_2)$ on an open neighborhood
  $U \subset M$ of $m$ and $Q = (q_1, q_2): U \to \RM^2$ whose
  components $q_i$ belong to the following list:
  \begin{itemize}
  \item $q_i(x,\xi) = \frac{1}{2}(x_i^2 + \xi_i^2)$ (elliptic),
  \item $q_i(x,\xi) = x_i \xi_i$ (hyperbolic),
  \item $q_i(x,\xi) = \xi_i$ (regular),
  \item $q_1(x,\xi) = x_1 \xi_2 - x_2 \xi_1$,
    $q_2(x,\xi) = x_1 \xi_1 + x_2 \xi_2$ (focus-focus),
  \end{itemize}
  such that $m$ corresponds to $(x,\xi) = (0,0)$ and
  $\{J,q_i\} = 0 = \{H,q_i\}$ for every $i \in \{1,2\}$. Furthermore,
  if none of these components is hyperbolic, there exists a local
  diffeomorphism $g: (\RM^2,0) \to (\RM^2,F(m))$ such that for every
  $(x,\xi) \in U$, $F(x,\xi) = (g \circ Q)(x,\xi)$.
\end{theo}

Strictly speaking, a complete proof of this theorem was published only
for analytic Hamiltonians~\cite{vey}, and for $\Cinf$ Hamiltonians in
several cases: the fully elliptic case in any
dimension~\cite{dufour-molino,eliasson}, the focus-focus case in
dimension $4$~\cite{san-wacheux,chaperon-focus}, the general
(hyperbolic and elliptic) case in dimension 2~\cite{colin-vey}. Based
on this theorem, the extension to partial action-angle coordinates
corresponding to the regular components $\xi_i$, or in the presence of
additional compact group action, was proven in~\cite{miranda-zung}.

A semitoric system $(M,\omega,F=(J,H))$ is the data of a connected
four-dimensional symplectic manifold $(M,\omega)$ and smooth functions
$J,H: M \to \RM$ such that
\begin{enumerate}
\item $(J,H)$ is a Liouville integrable system,
\item $J$ generates an effective Hamiltonian $S^1$-action,
\item $J$ is proper,
\item $F$ has only non-degenerate singularities with no hyperbolic
  components.
\end{enumerate}

Consequently, a semitoric system only displays singularities of
elliptic-elliptic, elliptic-regular (commonly called
elliptic-transverse) and focus-focus type.

\begin{rema}
  The properness of $J$ implies that of the momentum map $F$; while
  the properness of $F$ is crucial throughout the analysis, that of
  $J$ itself can be seen as a technical condition, enabling the use of
  Morse theory. It implies in particular that the fibres of $F$ and
  $J$ are connected, see~\cite{san-polytope}. However, in order to
  include classical examples from mechanics, like the spherical
  pendulum, which live on cotangent bundles, it would be important to
  relax this assumption. First steps in this direction were made
  in~\cite{san-pelayo-ratiu-connectivity,san-pelayo-ratiu-affine};
  in~\cite{san-dauge-hall-rotation}, the properness of $J$ was not
  assumed. In this work, however, we shall keep this assumption,
  because in most places we rely on the classification of semitoric
  systems, which only exists for proper $J$.
\end{rema}

\begin{ex}\label{ex:spin-osc_class} The spin-oscillator system
    (also known as the classical Jaynes-Cummings system
    \cite{jaynes-cummings}) is obtained by coupling a harmonic
    oscillator and a classical spin. Concretely, we consider the
    symplectic manifold
    $(\RM^2 \times S^2 , \omega = \omega_0 \oplus \omega_{S^2})$, with
    coordinates $(u,v,x,y,z)$, where $\omega_{S^2}$ and $\omega_0$ are
    the standard symplectic forms on $S^2$ and $\RM^2$, respectively,
    and the momentum map
  \[
    F = (J,H), \qquad J = \frac{1}{2}(u^2 + v^2) + z, \quad H =
    \frac{1}{2}(ux + vy). \] This is the momentum map of a semitoric
  integrable system, with one focus-focus singularity
  $m = (0,0,1,0,0)$, so that $F(m) = (1,0)$. The image of $F$ can be
  seen in \cite[Section 4]{san-alvaro-spin}, see also Figure
  \ref{fig:jsp_spinosc}.
\end{ex}

\begin{defi}
  \label{defi:simple}
  A semitoric system is
  called \emph{simple} if each level set of $J$ contains at most one
  focus-focus point.
\end{defi}
Throughout the rest of the article, we will always assume that
semitoric systems are simple.

\begin{defi}[\cite{san-alvaro-I}]
  Two semitoric systems $(M,\omega,F)$ and $(M',\omega',F')$ are
  \emph{isomorphic} if there exist a symplectomorphism
  $\phy:(M,\omega) \to (M',\omega')$ and a smooth map
  $g(x,y)=(x,f(x,y))$, with $\partial_y f>0$, such that
  \[
    F'\circ \phy = g\circ F.
  \]
\end{defi}
The main result of~\cite{san-alvaro-I} is to exhibit a list of
concrete invariants such that two semitoric systems that possess the
same set of invariants are isomorphic. Then~\cite{san-alvaro-II} shows
how to construct a semitoric system given an arbitrary choice of
invariants.  Let us now introduce these invariants more precisely,
and illustrate this presentation with the spin-oscillator system
  (Example \ref{ex:spin-osc_class}).

Note that the symplectic invariants of the spin-oscillator were
  computed in \cite{ADH-spin} (using a convention that differs from
  the one we use here; this discrepancy has been fixed in
  \cite{jaume-thesis}). In fact, these works extend results from
  \cite{san-alvaro-spin}, in which the polygonal invariant, the height
  invariant and the linear coefficients of the Taylor series invariant
  were computed (with yet another convention).

Let $(M, \omega, F = (J,H))$ be a (simple) semitoric system. Its first
symplectic invariant is the number $m_f \in \NM$ of focus-focus
singularities. If $m_f = 0$, i.e. if the system is of \emph{toric
  type}, the only remaining invariant is the semitoric polygon.

\subsection{Semitoric polygons}
\label{sec:semitoric-polygons}

We first assume that $m_f \geq 1$ and denote by
$(x_1, y_1), \ldots, (x_{m_f}, y_{m_f})$ the images of the focus-focus
singularities by $F$, numbered in such a way that
$x_1 < \ldots < x_{m_f}$. Let $B_{\mathrm{reg}}$ be the set of regular
values of $F$. The polygonal invariant is given as an equivalence
class of convex polygonal sets; each representative in this class can
be constructed after making a choice of initial action diffeomorphism
and cut directions $\vec{\epsilon} \in \{-1,1\}^{m_f}$. (In the
non-compact case, these polygons are not bounded in general, and the
terms \emph{convex polygon} will have the meaning given in
\cite[Definition 5.19]{san-daniele}; in particular a convex polygon
has a discrete set of vertices.)

More precisely, let
$\vec{\epsilon} = (\epsilon_1, \ldots, \epsilon_{m_f}) \in
\{-1,1\}^{m_f}$ and, for $i \in \{1, \ldots, m_f\}$, let
$\ell_i^{\epsilon_i} = \{ (x_i, y) \ | \ \epsilon_i y \geq \epsilon_i
y_i \}$ be the vertical half-line starting at $(x_i, y_i)$ and going
upwards if $\epsilon_i = 1$ and downwards if $\epsilon_i =
-1$. Finally, let
$\ell^{\vec{\epsilon}} = \cup_{i=1}^{m_f} \ell_i^{\epsilon_i}$. By
\cite[Theorem 3.8]{san-polytope}, there exists a homeomorphism
$\Phi_{\vec{\epsilon}}: F(M) \to \Phi_{\vec{\epsilon}}(F(M)) \subset
\RM^2$ whose restriction to $F(M) \setminus \ell^{\vec{\epsilon}}$ is
a diffeomorphism into its image, of the form
\begin{equation}
  \Phi_{\vec{\epsilon}}(x,y) = \left(x,\Phi_{\vec{\epsilon}}^{(2)}(x,y)\right),
  \qquad \frac{\partial \Phi_{\vec{\epsilon}}^{(2)}}{\partial y} > 0,
  \label{equ:cartographic}
\end{equation}
whose image $\Delta_{\vec{\epsilon}} = \Phi_{\vec{\epsilon}}(F(M))$ is
a convex polygon, which sends the integral affine structure of
$B_{\mathrm{reg}} \setminus \ell^{\vec{\epsilon}}$ given by
action-angle coordinates to the standard integral affine structure on
$\RM^2$, and extends to a smooth multivalued map from
$B_{\mathrm{reg}} $ to $\RM^2$ such that for any
$i \in \{1, \ldots, m_f\}$ and for any
$c \in \ell_i^{\epsilon_i} \setminus \{(x_i,y_i)\}$,
\begin{equation}
  \lim_{\substack{(x,y) \to c \\ x < x_i}} \dd\Phi_{\vec{\epsilon}} =
  T \lim_{\substack{(x,y) \to c \\ x > x_i}} \dd\Phi_{\vec{\epsilon}},
  \qquad T = \begin{pmatrix} 1 & 0 \\ 1 & 1 \end{pmatrix}.
  \label{equ:matriceT}
\end{equation}
Following \cite{san-daniele}, such a homeomorphism
$\Phi_{\vec{\epsilon}}$ is called a \emph{cartographic
  homeomorphism}. For a given $\vec{\epsilon}$, it is unique modulo
left composition by an element of the subgroup $\mathcal{T}$ of
$GL(2,\ZM) \ltimes \RM^2$ consisting of the composition of $T^k$ for
some $k \in \ZM$ and a vertical translation. Indeed, a cartographic
homeomorphism is constructed from action variables above
$B_{\mathrm{reg}}$, and this degree of freedom corresponds to the
choice of initial action variables of the form $(J,L)$.

One can formalize the action of changing cut directions as
follows. For $x_0 \in \RM$ and $n \in \NM$, let
$t_{x_0}^n: \RM^2 \to \RM^2$ be the map defined as the identity on
$\{ x \leq x_0\}$ and as $T^n$ (relative to any choice of origin on
the line $\{x = x_0\}$) on $\{ x \geq x_0 \}$. For
$\vec{x} = (x_1, \ldots, x_s) \in \RM^s$ and
$\vec{n} = (n_1, \ldots, n_s) \in \NM^s$, let
$t_{\vec{n},\vec{x}} = t_{x_1}^{n_1} \circ \ldots \circ
t_{x_s}^{n_s}$. Let
$\vec{\epsilon}, \vec{\epsilon'} \in \{-1,1\}^{m_f}$ and let
$\Delta_{\vec{\epsilon}}$,
$\Delta_{\vec{\epsilon'} \star \vec{\epsilon}}$ be the two polygons
constructed as above with the same initial set of action variables and
the two choices of cut directions $\vec{\epsilon}$ and
$\vec{\epsilon'} \star \vec{\epsilon} = \left( \epsilon_1' \epsilon_1,
  \ldots, \epsilon_{m_f}' \epsilon_{m_f} \right)$. Then one may check
that
\begin{equation}
\Delta_{\vec{\epsilon'} \star \vec{\epsilon}} =
t_{\vec{u},\vec{x}}(\Delta_{\vec{\epsilon}}), \qquad \vec{u} =
\left( \frac{\epsilon_1 - \epsilon_1 \epsilon_1'}{2}, \ldots,
  \frac{\epsilon_{m_f} - \epsilon_{m_f} \epsilon_{m_f}'}{2} \right),
\quad \vec{x} = (x_1, \ldots, x_{m_f})\,,
\label{equ:epsilon_etoile}
\end{equation}
see Figure~\ref{fig:some_reps_poly_spinosc}. The polygonal invariant
is the orbit of any of the convex polygons $\Delta_{\vec{\epsilon}}$
constructed as above under the action of
$\mathcal{T} \times \{-1,1\}^{m_f}$. We will denote by
$(\Delta_{\vec{\epsilon}}, \Phi_{\vec{\epsilon}})$ the representative
of this invariant constructed using $\Phi_{\vec{\epsilon}}$.

Finally, if $m_f = 0$, this construction is still valid but there is
no $\vec{\epsilon}$, no cut direction and the invariant is the orbit
of any of the polygons under the action of $\mathcal{T}$, see
\cite[Section 5.2.2]{san-daniele} for more details.

\begin{ex}  In Figure \ref{fig:some_reps_poly_spinosc} below, we show a few representatives of the polygonal invariant of the spin-oscillator system defined in Example \ref{ex:spin-osc_class}.
\begin{figure}[H]
    \begin{center}
      \begin{tikzpicture}
        \path[fill, color=gray!40] (-2, -1) -- (0, 1) -- (2, 1) -- (2,
        -1) -- (-2,-1) -- cycle;

        \draw[line width = 1pt] (-2, -1) -- (0, 1);
        \draw[line width = 1pt] (-2, -1) -- (2, -1);
        \draw[line width = 1pt] (0, 1) -- (2, 1);

        \draw [dashed] (0,0) -- (0,1); \draw (0,0) node[] {$\times$};

        \draw (-2,-1) node[below left] {$(-1,-1)$}; \draw (0,1)
        node[above] {$(1,1)$};
        
        \begin{scope}[xshift=7cm]
                \path[fill, color=gray!40] (-2, -1) -- (0,-1) -- (2, 1) -- (0,1) -- (-2,-1) -- cycle;

        \draw[line width = 1pt] (-2, -1) -- (0, 1); 
        
        \draw[line width = 1pt] (-2, -1) -- (0, -1);
        
        \draw[line width = 1pt] (0, -1) -- (2, 1);

        \draw [dashed] (0,0) -- (0,-1); \draw (0,0) node[] {$\times$};

        \draw (-2,-1) node[below] {$(-1,-1)$}; 
        \draw (0,-1) node[below] {$(1,-1)$};
        \end{scope}
        
        \begin{scope}[yshift=-2cm]
         
         \path[fill, color=gray!40] (-2, -1) -- (0, -1) -- (2, -3) -- (0,
        -3) -- (-2,-1) -- cycle;

        \draw[line width = 1pt] (-2, -1) -- (0, -1); 
        \draw[line width = 1pt] (-2, -1) -- (0, -3);
         \draw[line width = 1pt] (0, -1) -- (2, -3);

        \draw [dashed] (0,-2) -- (0,-1); 
        \draw (0,-2) node[] {$\times$};

        \draw (-2,-1) node[above left] {$(-1,-1)$}; 
        \draw (0,-1) node[above right] {$(1,-1)$};
        \end{scope}
        
         \begin{scope}[xshift=7cm,yshift=-2cm]
         
         \path[fill, color=gray!40] (-2, -1) -- (2, -1) -- (2, -3) -- (0,-3) -- (-2,-1) -- cycle;

        \draw[line width = 1pt] (-2, -1) -- (2, -1); 
        \draw[line width = 1pt] (-2, -1) -- (0, -3);
         \draw[line width = 1pt] (0, -3) -- (2, -3);

        \draw [dashed] (0,-2) -- (0,-3); 
        \draw (0,-2) node[] {$\times$};

        \draw (-2,-1) node[above] {$(-1,-1)$}; 
        \draw (0,-3) node[below] {$(1,-3)$};
        \end{scope}
        
      \end{tikzpicture}
    \end{center}
    \caption{\small A few representatives of the polygonal 
    invariant for the spin-oscillator system. The polygons in the
    second row are obtained from those in the first row by
    applying the global transformation $T^{-1}$ defined
    in~\eqref{equ:matriceT} followed by the vertical translation
    by {\tiny $\protect\begin{pmatrix} 0 \\ -1 \protect\end{pmatrix}$}.  The polygons
    in the second column are obtained from those
    in the first column by changing the cut direction from upwards
    to downwards, see~\eqref{equ:epsilon_etoile}.  }
    \label{fig:some_reps_poly_spinosc}
  \end{figure}
\end{ex}

\subsection{Twisting number and twisting index}
\label{sec:twist-numb-twist}

In this section, we express the definition of the twisting numbers and
twisting index for a (simple) semitoric integrable system
$(M,\omega,F = (J,H))$ on a four-dimensional manifold, in terms of a
geometric object that we call the radial curve
(Definition~\ref{defi:radial-curve}). The construction we use here is
somewhat different from the initial definition in~\cite{san-alvaro-I},
and more adapted to the inverse problem.

We first introduce the twisting number, an integer associated with
each focus-focus singularity $m_0 \in M$ of $F$.  In order to simplify
notation, we may let $F(m_0)=0$.  By assumption, $m_0$ is the only
singularity of the connected critical fiber
$\Lambda_0:=F^{-1}(0)$. Let $\Omega_0\subset M$ be a saturated
neighborhood of $\Lambda_0$; then one can show that
$F(\Omega_0)\subset \RM^2$ is a neighborhood of the origin. Let
$B\subset F(\Omega_0)$ be a small ball centered at the origin, such
that $B\setminus\{0\}$ consists of regular values of $F_{| \Omega_0}$.

Let $U\subset B \cap \{ (x,y) \in \RM^2 \ | \ x > 0 \}$ be a simply connected open set. Let us
choose oriented action coordinates in $F^{-1}(U)$ of the form
$I=(J, L)$. Recall from~\eqref{equ:change-action} (and the fact that
the first component, $J$, must be preserved) that $L$ is not unique,
but any other choice can only be of the form $L' = L + n J + c$, for
some $n \in\ZM$ and $c \in \RM$. Nevertheless, there are two natural
ways of selecting $L$. One comes from the global geometry of the
momentum map, and consists in choosing the affine coordinates used in
the construction of the semitoric polygon
(Section~\ref{sec:semitoric-polygons}): $I$ must coincide with
$\Phi_{\vec\epsilon}\circ F$, where $\Phi_{\vec\epsilon}$ is the
cartographic map~\eqref{equ:cartographic} (so this depends on the
choice of a representative $\Delta_{\vec\epsilon}$ of the polygon
invariant). We will use the notation $L_{\Phi_{\vec{\epsilon}}}$ for
this global choice. Then, there is a local choice $L_{\textup{priv}}$,
which is dictated by the singular behaviour on $\Lambda_0$, and which
was called `privileged action variable' in~\cite[Definition
5.7]{san-alvaro-I}. 

\begin{defi}
\label{defi:twisting_number}
The integer $p \in \ZM$ such that
${\rm d}L_{\Phi_{\vec{\epsilon}}} = {\rm d}L_{\textup{priv}} + p \
{\rm d}J$ is called the \emph{twisting number} corresponding to the
global choice of $L_{\Phi_{\vec{\epsilon}}}$.
\end{defi}

\begin{figure}[H]
\begin{center}
\includegraphics[scale=0.6]{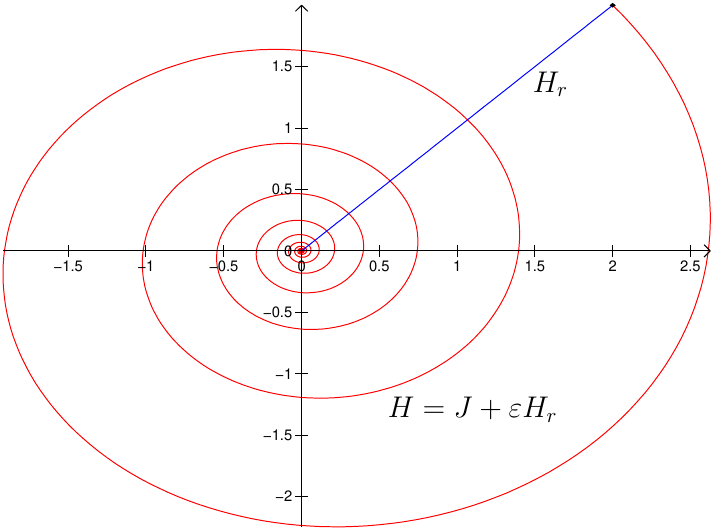}
\end{center}
\caption{Projection of the Hamiltonian flows of $H_r$ (in blue) and $H = J + \varepsilon H_r$ (in red) in the $(x_1,x_2)$-plane, for $\varepsilon = 0.01$. In other words, in this example $f_r(x,y) = (x,\frac{x-y}{\varepsilon})$. The projection of the flow of $H$ is the map $(t,x_1,x_2) \mapsto e^{(i+\varepsilon)t} (x_1 + i x_2)$, while the projection of the flow of $H_r$ is the map $(t,x_1,x_2) \mapsto e^t (x_1 + i x_2)$, see for instance \cite[Section 6.2]{san-focus}.}
\label{fig:radial}
\end{figure}

Let us recall the definition of the privileged action variable
$L_{\textup{priv}}$. Let $\Omega$ be a sufficiently small neighborhood
of $m_0$ in $M$. In $\Omega$, the fiber $\Lambda_0\cap \Omega$ is the
union of two surfaces intersecting transversally at $m_0$. For a
generic Hamiltonian of the form $f(J,H)$, these surfaces respectively
constitute the local stable and unstable manifolds for the flow of the
associated Hamiltonian vector field, and on each of them, the
trajectories tending to the fixed point $m_0$ are of `focus' type,
\emph{i.e.} they are spirals that wind infinitely many times around
$m_0$. However, Theorem~\ref{theo:eliasson} implies that there is a
precise choice $f_r$ of $f$, which is unique up to sign, addition of a
constant, and addition of a flat function, such that, in some local
Darboux coordinates $(x_1,x_2,\xi_1,\xi_2)$,
\[
  f_r(J,H) = x_1\xi_1 + x_2 \xi_2.
\]
(For the uniqueness, see~\cite[Lemma 4.1]{san-semi-global}). Let us
call such $H_r := f_r(J,H)$ the ``radial'' Hamiltonian, because its
trajectories inside $\Lambda_0\cap\Omega$ are line segments tending to
the origin in the above Darboux coordinates, and hence have
intrinsically a zero winding number around $m_0$, see Figure \ref{fig:radial}.  

Imposing $f_r(0)=0$ and $(J,H_r)$ to be oriented with respect to
$(J,H)$, meaning that $\partial_H f_r >0$, the function $f_r$ becomes
unique, modulo addition of a flat function at $0\in\RM^2$. The map
$q:=(J,H_r):M\to\RM^2$ is called the quadratic, or Eliasson momentum
map, because, modulo the aforementioned uniqueness, it must coincide
with the quadratic map $Q$ expressed in Eliasson's coordinates
(Theorem~\ref{theo:eliasson}). Note that $q=g^{-1}\circ F$ with
$g^{-1}(x,y) = (x,f_r(x,y))$. The function $f_r$ itself will be called
the Eliasson function.

\begin{ex} For the spin-oscillator system of Example
    \ref{ex:spin-osc_class}, we can infer the Taylor expansion of
    Eliasson's function $f_r$ from \cite[Lemma 4.1]{ADH-spin}; we
    illustrate this by obtaining this expansion up to cubic terms. The
    formula contained in that lemma says that if
    $(\xi_1, \xi_2) = (x_1, f_r(x_1, x_2))$ then
  \[
    x_2 = \frac{1}{2} \xi_2 + \frac{1}{16} \xi_1 \xi_2 + \O(3);
  \]
  here $\O(3)$ means cubic or higher order terms, to simplify
  notation. This means that
  \[
    \begin{split} \xi_2 & = f_r\left(\xi_1, \frac{1}{2} \xi_2 +
        \frac{1}{16} \xi_1 \xi_2
        + \O(3)\right) \\
      & \begin{aligned} = f_r(0) + \partial_x f_r(0) \xi_1 +
        \frac{\partial_y f_r(0)}{2} \xi_2 + \frac{\partial^2_x
          f_r(0)}{2} \xi_1^2 + \left( \frac{\partial_y f_r(0)}{16} +
          \frac{\partial_x \partial_y f_r(0)}{2} \right) \xi_1 \xi_2
        \\ + \frac{\partial^2_y f_r(0)}{8} \xi_2^2 + \O(3).
      \end{aligned}
    \end{split}
  \]
   Hence we can identify the coefficients to find
  \begin{equation}
    \partial_x f_r(0) = 0, \quad \partial_y f_r(0) = 2, \quad
    \partial^2_x f_r(0) = 0, \quad \partial_x \partial_y f_r(0) =
    -\frac{1}{4}, \quad \partial^2_y f_r(0) = 0.
    \label{eq:fr_elia_spinosc}
  \end{equation}
  This reasoning could be used to compute higher order derivatives of
  $f_r$.
\end{ex}

Assume now that $U$ is contained in the open set $F(\Omega)$.  Any
vector field $\mathcal{X}$ in $\Omega\cap F^{-1}(U)$ that is tangent
to the leaves of $F$ (\emph{i.e.} included in the kernel of $\dd{F}$)
decomposes in a unique way as
\begin{equation}
  \label{equ:tau1-tau2}
  \mathcal{X} = \tilde\tau_1 \ham{J} + \tilde\tau_2 \ham{H_r},
\end{equation}
where $\tilde\tau_1,\tilde\tau_2$ are smooth functions on the local
leaf space, \emph{i.e.} $\tilde\tau_j = F^*\tau_j$, where $\tau_j$ is
smooth on $U$. Let $L:M\to\RM$ be such that $(J,L)$ is a set of action
coordinates. Since action diffeomorphisms form a flat sheaf, they
admit a unique extension in any simply connected open subset of the
set of regular values of $F$. In particular we can extend $L$ inside
$F(\Omega)\setminus\ell$, where $\ell$ is the upward vertical
half-line from the origin. We may apply the
decomposition~\eqref{equ:tau1-tau2} to the Hamiltonian vector field
$\ham{L}$ to get smooth functions $\tau_1, \tau_2$ on
$F(\Omega)\setminus\ell$.

\begin{prop}[{\cite{san-semi-global},\cite[Lemma 4.46]{san-daniele}}]
  \label{prop:sigma_smooth}
  Let $\log$ be the determination of the complex logarithm obtained by
  choosing arguments in $(-\frac{3\pi}{2}, \frac{\pi}{2}]$. The
  functions
  \[
    \begin{cases} \sigma_1: c \mapsto \tau_1(c) +
      \frac{1}{2\pi} \Im(\log(c_1 + i f_r(c_1,c_2))), \\[2mm]
      \sigma_2: c \mapsto \tau_2(c) + \frac{1}{2\pi} \Re(\log(c_1 + i
      f_r(c_1,c_2))) \end{cases}
  \]
  extend smoothly at $c = (0,0)$.
\end{prop}

It follows that $\tau_1$ is multivalued and that $\tau_2$ exhibits a logarithmic singularity. In order to deal with this, it will be convenient to study $\sigma_1$ and $\sigma_2$ along a special curve that we describe below.

\begin{defi}
  \label{defi:radial-curve}
  The image by $F$ of the zero-set of $H_r$ in $\Omega$ is a local
  curve $\gamma_r$ given by the equation $f_r(x,y)=0$, which we call
  the \emph{radial curve}.
\end{defi}

From the implicit function theorem $\gamma_r$ is, locally near the origin, a
graph parameterized by $x$, say the graph of $\varphi: \RM \to \RM$.
Let $\Gamma$ be the intersection of $F(\Omega)$ with an open vertical
half-plane whose boundary contains the origin.  Shrinking $\Omega$ if
necessary, we may assume that $\Gamma$ contains a branch of $\gamma_r$
accumulating at the origin. In what follows, we will always choose
the half-plane defining $\Gamma$ to be the right half-plane.

\begin{lemm}
  \label{lemm:def_sigma1_zero}
  The function $\nu_1: x \mapsto \tau_1(x,\varphi(x))$, defined for
  $x > 0$ (so that $(x,\varphi(x)) \in \Gamma$), extends smoothly at
  $x = 0$, and $\lim_{x \to 0^+} \nu_1(x) = \sigma_1(0)$.
\end{lemm}

\begin{demo}
  We know from Proposition \ref{prop:sigma_smooth} that the function
  $\sigma_1$ extends smoothly at the origin. But for $x > 0$
  \[
    \sigma_1(x, \varphi(x)) = \nu_1(x) + \frac{1}{2\pi} \Im(\log(x + i
    f_r(x,\varphi(x)))) = \nu_1(x) + \frac{1}{2\pi} \Im(\log \ x) =
    \nu_1(x)
  \]
  and $\varphi(0) = 0$, so $\nu_1$ extends smoothly at $x = 0$, with
  value $\sigma_1(0)$ at this point.
\end{demo}

\begin{rema}
  \label{rema:gamma}
  Observe that the choice of $\Gamma$ is indeed important: for
  $x < 0$,
  \[
    \nu_1(x) = \sigma_1(x,\varphi(x)) - \frac{1}{2} \underset{x \to
      0}{\longrightarrow} \sigma_1(0) - \frac{1}{2},
  \]
  so choosing the left half-plane for $\Gamma$ would have shifted the
  above limit by a factor $\frac{1}{2}$.
\end{rema}

Because $f_r$ is unique up to a flat function, $\sigma_1(0)$ does not
depend on any choice made but $L$. As remarked earlier, any other $L'$ on $F^{-1}(\Gamma)$
is of the form $L'=L - n J + c$ for some integer $n$ and some constant
$c \in \RM$, leading to $\tau_1' = \tau_1 - n$ and hence
$\sigma_1'(0) = \sigma_1(0) - n$.  By definition, we call $L'$ a
\emph{privileged action variable} and we denote it by
$L_{\textup{priv}}$, when
\[
  \sigma_1'(0) \in \interval[open right]01
\]
and in this case we write $\sigma_1^{\textup{p}}$ instead of
$\sigma_1'$, to emphasize the fact that we are working with this
privileged choice. Notice that a privileged action variable is defined
only up to an additive constant $c$; one may fix its value if needed,
see Equation \eqref{equ:S} and the discussion below.

Summing up, given any fixed choice $L$ of action variable, defining
$(\tilde\tau_1, \tilde\tau_2) = (F^* \tau_1, F^* \tau_2)$ by
\begin{equation}
  \label{equ:tau-prime}
  \ham{L} = \tilde\tau_1 \ham{J} + \tilde\tau_2 \ham{H_r},
\end{equation}
and letting $\sigma_1(0)$ be the limit of $\tau_1$ at the origin along
the radial curve $\gamma_r$, we have $L_{\textup{priv}} = L - n J$ where $n$
is the integer part of $\sigma_1(0)$; remark how this formula confirms
that $L_{\textup{priv}}$ does not depend on the choice of $L$ (while
$n$ does).

\begin{defi}
  \label{defi:priv_poly}
  If we let $\vec{\epsilon} = (1, \ldots, 1)$ and choose
  $\Phi_{\vec{\epsilon}}$ so that the twisting number of the
  focus-focus point $m_0$ vanishes, then $\Phi_{\vec{\epsilon}}$ is
  called the \emph{privileged cartographic map at $m_0$}, and the
  corresponding polygon
  $\Delta_\textup{priv}^{m_0}:=\Phi_{\vec{\epsilon}}(F(M))$ is called
  the \emph{privileged polygon at $m_0$} for this semitoric system.
\end{defi}

\begin{rema}
  As explained earlier, this privileged polygon is defined up to a
  vertical translation (because the privileged action at $m_0$ is
  defined up to addition of a constant). Although one should keep this
  in mind, for simplicity we will often talk about the privileged
  polygon.
\end{rema}

\begin{ex}  We consider the spin-oscillator system of Example
    \ref{ex:spin-osc_class}. A representative of the polygonal
    invariant corresponding to $\epsilon = +1$ and with vanishing
    twisting number (in other words, the privileged polygon of the
    system), is represented in Figure \ref{fig:polygon_spinosc}.

  \begin{figure}[H]
    \begin{center}
      \begin{tikzpicture}

        \path[fill, color=gray!40] (-2, -1) -- (0, 1) -- (3, 1) -- (3,
        -1) -- (-2,-1) -- cycle;

        \draw[line width = 1pt] (-2, -1) -- (0, 1); \draw[line width =
        1pt] (-2, -1) -- (3, -1); \draw[line width = 1pt] (0, 1) --
        (3, 1);

        \draw [dashed] (0,0) -- (0,1); \draw (0,0) node[] {$\times$};

        \draw (-2,-1) node[below left] {$(-1,-1)$}; \draw (0,1)
        node[above] {$(1,1)$};
      \end{tikzpicture}
    \end{center}
    \caption{\small The privileged polygon for the spin-oscillator
      system.}
    \label{fig:polygon_spinosc}
  \end{figure}
\end{ex}

We may now recall the definition of the twisting index, which is a
global invariant taking into account all twisting numbers and the
choice of a semitoric polygon. Using the notation of
Section~\ref{sec:semitoric-polygons}, given a choice of cartographic
map $\Phi_{\vec\epsilon}$, we have $m_f$ twisting numbers
$p_1,\dots,p_{m_f}$. Each of them, individually, may be set to zero
using its associated privileged cartographic map, but in general one
cannot set all these numbers to zero simultaneously. The twisting
index is precisely the equivalence class of the tuple
$(p_1,\dots,p_{m_f})$ modulo the choice of a cartographic map;
see~\cite[Definition 5.9]{san-alvaro-I} and \cite[Remark 3.6]{AHP}.

\begin{rema}\label{rema:taylor2}
  Contrary to the twisting index, the twisting number $p$, being
  defined as an integer part, is sensitive to perturbations when
    $\sigma_1(0)$ is an integer. This is the case for the
    spin-oscillator system of Example \ref{ex:spin-osc_class}, see
    Equation~\eqref{eq:taylor_series_spinosc} together with
    Proposition \ref{prop:relation_inv}. This may lead to uncertainty
    when recovering $p$ from the joint spectrum, see Figure
  \ref{fig:sigma_spinosc_kvar} for an illustration of this fact. When
  a reference Hamiltonian $L$ is given, a better symplectic invariant
  of the triple $(J,H,L)$ is the coefficient $\sigma_1(0)$ itself.
  Its fractional part $\sigma_1^{\textup{p}}(0) = \sigma_1(0) - p$,
  which is independent of $L$, is the ``second Taylor series
  invariant'' of the foliation induced by $F=(J,H)$, as defined
  in~\cite{san-semi-global}: see Section~\ref{sec:tayl-seri-invar}
  below.
\end{rema}
\begin{rema}
  Given a triple $(J,H,L)$ as in Remark~\ref{rema:taylor2}, it follows
  directly from~\eqref{equ:tau-prime} that $-\tau_1$ is the
  \emph{rotation number} of the radial Hamiltonian $H_r$ computed in
  the action variables $(J,L)$. From the point of view of the toric
  action induced by $(J,L)$, it can be further interpreted as
  follows. Let $\mu:=(J,L) : M \to \RM^2$; it is a toric momentum map,
  defined on the saturated open set $\Omega':=F^{-1}(U)$, with the
  notation of the beginning of this section. It defines an isomorphism
  between the space of symplectic vector fields $X_{\beta}$ in
  $\Omega'$ that are tangent to the $F$-foliation, and closed
  one-forms $\beta$ on the affine space $\mu(\Omega')\subset \RM^2$,
  via the formula
  \[
    \iota_{X_{\beta}}\omega = - \mu^* \beta.
  \]
  Taking $\beta = -\tau_1 \dd{j} + \dd{\ell}$, where $(j,\ell)$ are
  the canonical affine coordinates in $\RM^2$, we see from~\eqref{equ:tau-prime} that
  $\iota_{\tilde{\tau}_2 \ham{H_r}}\omega = - \mu^* \beta$; hence
  $\beta$ gives the \emph{direction of the radial vector field
    $\ham{H_r}$}. Therefore, the tangent to $\gamma_r$, expressed in
  the coordinates $(j,\ell)$, is $\ker \beta$, \emph{i.e.} the line
  spanned by the vector $(1,\tau_1)$.
\end{rema}

Now let us relate $\tau_1$ with the original momentum map
$F=(J,H)$. Define the functions
$(\tilde a_1, \tilde a_2) = (F^* a_1, F^* a_2)$ in $\Gamma$ by
\begin{equation}
  \label{equ:a}
  \ham{L} = \tilde a_1 \ham{J} + \tilde a_2 \ham{H},
\end{equation}
and let
\begin{equation}
  s:=-\partial_x f_r / \partial_y f_r\,;
  \label{equ:petit_s}
\end{equation}
the latter is the ``slope'' of the tangent to the level sets of
$f_r$. In particular $s(0)$ is the slope of the tangent to $\gamma_r$
at the origin. Equating~\eqref{equ:tau-prime} with~\eqref{equ:a}, we
get
\begin{equation}
  \label{equ:a1_a2}
  \begin{cases}
    a_1 = \tau_1 + \tau_2 \partial_x f_r\\
    a_2 = \tau_2\partial_y f_r,
  \end{cases}
\end{equation}
which gives, in $\Gamma$,
\begin{equation}
  \label{equ:alpha}
  \tau_1 = a_1 + s a_2.
\end{equation}
Recall that $a_1,a_2,\tau_1,\tau_2$ are all ill-defined (and really
singular) at the origin, while $s$ is smooth in a neighborhood of $0$.

\begin{rema}
  In the papers \cite{san-alvaro-spin,alvaro-yohann}, the notation was
  slightly different and the matrix
  \[
    B = \begin{pmatrix} 1 & 0 \\ b_1 & b_2 \end{pmatrix}
  \]
  such that $q = \mathrm{Hess}(B \circ F)$ was considered. We claim
  that
  \[
    s(0) = -\frac{b_1}{b_2}.
  \]
  Indeed, on the one hand we have that
  $B \circ F = (q_1, b_1 q_1 + b_2 H)$, which yields
  \[
    (q_1, q_2) = (q_1, b_1 q_1 + b_2 \mathrm{Hess}(H)).
  \]
  On the other hand, $(q_1, q_2) = g \circ F = (J, f_r(J,H))$, hence
  $ (q_1, q_2) = (q_1, f_r(q_1,H))$. So we obtain that
  \[
    \dd q_2 = \partial_x f_r(q_1,H) \dd q_1 + \partial_y f_r(q_1,H)
    \dd H,
  \]
  and since $\dd q_1$ and $\dd q_2$ vanish at the origin, we finally
  get
  \[
    q_2 = \partial_x f_r(q_1,H) q_1 + \partial_y f_r(q_1,H)
    \mathrm{Hess}(H)
  \]
  plus a term that vanishes at the origin, so we identify
  $b_1 = \partial_x f_r(0)$ and $b_2 = \partial_y f_r(0)$.
\end{rema}

\subsection{The Taylor series invariant}
\label{sec:tayl-seri-invar}

The Taylor series invariant is not specific to semitoric systems. It
is the classifying invariant of any singular Lagrangian fibration
around a focus-focus fiber~\cite{san-semi-global}, and has been used
for instance in \cite{symington-blowdowns} to study rational
blowdowns. However, in this article we specialize its definition to
the semitoric case. (This is mainly a matter of simplifying notation,
since a neighborhood of a focus-focus fiber is always isomorphic, in a
natural sense, to a semitoric system.)

We keep the same notation as Section~\ref{sec:twist-numb-twist}. In
particular we fix a focus-focus point $m_0$, $(J,L)$ are action
variables in $F^{-1}(U)$, and $U$ is a small simply connected open set
close to the critical value $F(m_0)= 0\in\RM^2$. We can write
$L=\tilde L\circ q$, where $q=(J,H_r)$ and $\tilde L = \tilde L(X,Y)$
is smooth. From~\eqref{equ:tau-prime} we have
\[
  \tilde \tau_1 = \deriv{\tilde L}{X}\circ q, \quad \tilde \tau_2 =
  \deriv{\tilde L}{Y} \circ q.
\]
Thus, it follows from Proposition~\ref{prop:sigma_smooth} that the
function
\begin{equation}
  S(X,Y) := \tilde L(X,Y) + \Im(w\log w - w),
  \label{equ:S}
\end{equation}
where $w:=X+iY$, extends to a smooth function $S$ in a neighborhood of
$0$, with $g^* \dd S = \sigma_1 \dd c_1 + \sigma_2 \dd c_2$. We denote
the Taylor series of $S$ at the origin by
\[
  S^\infty = \sum_{\ell,m \geq 0} S_{\ell,m} X^{\ell} Y^m.
\]
The main result of~\cite{san-semi-global} is that the equivalence
class of $S^\infty$ in the quotient
$\frac{\RM\formel{X,Y}}{\RM\oplus\ZM X}$ is a complete symplectic
invariant for the singular foliation defined by $F$, in a neighborhood
of $\Lambda_0=F^{-1}(0)$.  The first terms $[S_{1,0}]\in\RM/\ZM$ and
$S_{0,1}\in\RM$ are called the linear invariants (of this Taylor
series). 

\begin{ex} The Taylor series invariant of the spin-oscillator
    system (see Example \ref{ex:spin-osc_class} and the computations
    in~\cite{san-alvaro-spin,ADH-spin}) starts as
  \[   S^\infty = \frac{5 \ln 2}{2 \pi} \ Y + \frac{1}{8\pi} \ XY + \O(3). \]
 In other words,
  \begin{equation}
    [S_{1,0}] = 0, \qquad S_{0,1} = \frac{5 \ln 2}{2\pi}, \qquad S_{2,0}
    = 0, \qquad S_{1,1} = \frac{1}{8\pi}, \qquad S_{0,2} = 0.
  \label{eq:taylor_series_spinosc}\end{equation}
\end{ex}

Let $\Phi_{\vec{\varepsilon}}$ be a cartographic map, and let $p$ be the twisting number associated with $\Phi_{\vec{\varepsilon}}$ and $m_0$, see Definition \ref{defi:twisting_number}; then 
\[ \dd L_{\Phi_{\vec{\varepsilon}}} = \dd L_{\text{priv}} + p \dd J.  \]
Moreover, let $n \in \ZM$ be such that $\dd L = \dd L_{\text{priv}} + n \dd J$; by definition (see the discussion after Remark \ref{rema:gamma}) $\sigma_1(0) = \sigma_1^{\text{p}}(0) + n$. Hence if $L_{\Phi_{\vec{\varepsilon}}} = L$, then $n = p$. This gives the following.

\begin{prop} 
\label{prop:relation_inv}
The linear invariants of the Taylor series and the quantities $\sigma_1, \sigma_2$ introduced in Section~\ref{sec:twist-numb-twist} are related by
\[
  [S_{1,0}] = \sigma_1(0) \mod \ZM, \qquad S_{0,1} = \sigma_2(0),
\]
and more precisely
\begin{equation*}
  S_{1,0} = \sigma_1^\textup{p}(0) + p,
\end{equation*}
with $\sigma_1^\textup{p}(0)\in\interval[open right]01$, and $p\in\ZM$
is the twisting number associated with $m_0$ for the choice of a cartographic map $\Phi_{\vec{\varepsilon}}$ such that $L_{\Phi_{\vec{\varepsilon}}} = L$ (see Definition \ref{defi:twisting_number}).
\end{prop}

The link between the twisting index and the Taylor series was
presented independently in~\cite{palmer-pelayo-tang}.

\paragraph{The height invariant.} The constant term $S_{0,0}$ is
irrelevant as far as the semi-global classification near $\Lambda_0$
is concerned. However, once the global picture is taken into account,
there is a way to get a meaningful value $S_{0,0}>0$. Since $L$ is
defined up to a constant, we may decide that $L=0$ where $H$ reaches
its minimal value on the compact set $J^{-1}(0)$. We see
from~\eqref{equ:S} that, if $X=0$ is fixed, and $Y\to 0$, the function
$\tilde L(X,Y)$ must tend to $S_{0,0}$. With this convention,
$S_{0,0}$ is precisely the height invariant defined
in~\cite[Definition 5.2]{san-alvaro-I}.

\begin{ex} For the spin-oscillator system of Example
    \ref{ex:spin-osc_class}, the height invariant was computed
    in~\cite{san-alvaro-spin,ADH-spin}: we have
    \begin{equation} S_{0,0} = 1. \label{eq:height_invariant_spinosc}
    \end{equation}
\end{ex}

\begin{rema}
  If we relax the orientation-preserving hypothesis for the image in
  $\RM^2$ of the joint momentum maps, and also the orientation of the
  $S^1$-action (\emph{i.e.} allowing to replace $J$ by $-J$), then we
  have an interesting finite group acting on all invariants, and in
  particular on the Taylor series. This was studied
  in~\cite{san-daniele}.
\end{rema}

\begin{rema}
  The reader should be aware that there are slight differences in
  convention and notation in the literature (regarding for instance
  the sign of the standard symplectic form on $\RM^4$, the respective
  parts played by $q_1$ and $q_2$, the complex structure on $\RM^4$,
  the choice of $\tau_1 \in \RM \slash \ZM$ or
  $\tau_1 \in \RM \slash 2\pi\ZM$, etc.), resulting in possible
  differences in the value of the Taylor series invariant: difference
  by a multiplicative factor $2\pi$, change in sign, shift of
  $S_{1,0}$ by $\pm \frac{1}{4}$ (or $\pm \frac{\pi}{2}$ when working
  modulo $2\pi\ZM$), etc. See \cite[Remark 6.2]{pelayo-tang} or
  \cite[Remark 4.11]{jaume-thesis}. Here we have mostly adopted the
  notation and convention from \cite{san-daniele}; the thesis
  \cite{jaume-thesis} is also an extremely reliable source for the
  computation of the symplectic invariants with comparable convention
  (up to normalization by $2\pi$).
\end{rema}

\section{Quantum semitoric systems}
\label{sec:quantum_semitoric}

What are the quantum analogues of semitoric systems? Of course, the
``old'' Jaynes-Cummings model from quantum optics was already a
quantum semitoric system, and so were the models studied
in~\cite{sadovski-zhilinski}. The mathematical formulation of
``quantized'' semitoric systems is hence very natural, and follows the
physics intuition, see~\cite{san-alvaro-first-steps}: a quantum
semitoric system is a pair of commuting operators which should be
semiclassical quantizations of the components of the momentum map of a
semitoric system. Here we need to make all these statements very
precise. The type of quantization that we use will depend on the
underlying phase space. Throughout this article, we will consider the
following three situations:
\begin{enumerate}[label=(M\arabic*)]
\item\label{item:1} $(M,\omega) = (T^*X,\dd{\lambda})$ where $X = \RM^2$
  or $X$ is a compact Riemannian surface and $\lambda$ is the
  Liouville one-form,
\item \label{item:2} $(M,\omega)$ is a quantizable (see Appendix
  \ref{subsec:def_BTO}) compact K\"ahler manifold of dimension four,
\item\label{item:3}
  $(M,\omega) = (\CM \times N, \omega_0 \oplus \omega_N)$ where
  $\omega_0$ is the standard symplectic form on $\CM$ and
  $(N,\omega_N)$ is a quantizable compact K\"ahler surface.
\end{enumerate}

These three situations occur in concrete examples coming from physical
problems. The coupled spin-oscillator system (see Example
  \ref{ex:spin-osc_class}), or Jaynes-Cummings model, is of great
relevance in quantum optics and quantum information
\cite{jaynes-cummings, shore-knight, raimond-brune-haroche01,
  babelon-spin, babelon-doucot-focus, gutierrez-agarwal21} and has
also been studied from the mathematical viewpoint
\cite{san-alvaro-spin,ADH-spin}. Its classical phase space is
$\RM^2 \times S^2$, which corresponds to case \ref{item:3}. The
coupled angular momenta system (see Section \ref{sect:coupled_spins})
is defined on $S^2 \times S^2$, hence belongs to case \ref{item:2}. It
was used in~\cite{sadovski-zhilinski} in order to propose a systematic
way to describe energy rearrangement between spectral bands in
molecules, see also~\cite{dhont20}.

On $T^* S^2$, the spherical
pendulum~\cite{cushman-duist-pendulum} is not a semitoric system in
the strict acceptance that we took in
Section~\ref{sec:semitoric-systems} (because the Hamiltonian
generating the circle action is not proper) but possesses one
focus-focus singularity. The same situation occurs for the ``champagne
bottle'' on $T^*\RM^2$~\cite{child}. In fact, it is quite possible
that all strict semitoric systems on a cotangent bundle must be of
toric type (\emph{i.e.} they don't possess any focus-focus
singularity); we already know from~\cite{karshon-ziltener18} that such
a cotangent bundle must be $T^*\RM^2$. In the cotangent case, allowing
for a non-proper map $J$ would be important for future works
(see~\cite{san-pelayo-ratiu-affine}), and since many of our
constructions here are local, we believe that they should be adaptable
to that more general setting.

\begin{rema}
The cases \ref{item:1}, \ref{item:2} and \ref{item:3} cover all the semitoric systems with at least one focus-focus singularity that we know of. These three cases do not contain the case of $T^*S^1 \times N$ where $N$ is a smooth compact surface, but on such a manifold
every semitoric system must be of toric type. Indeed, it follows from
\cite[Corollary 5.5]{san-polytope} and \cite[Theorem
3.1]{godinho-sousa-dias} that the presence of a focus-focus
singularity forces the manifold to be simply connected.

  More generally, we do not include the case of a system on a
  non-compact symplectic manifold which is neither a cotangent bundle
  nor $\RM^2 \times N$ with $N$ compact; it is unclear how to quantize
  such a phase space, although some progress has recently been made in
  this direction \cite{KorMaMa}. But we are not aware of any
  concrete example of semitoric system with at least one focus-focus
  point in this setting.
\end{rema}

To each of the three geometric situations, we shall consider a quantum
version and its semiclassical limit. We will use the generic
terminology ``semiclassical operator'' to encompass all cases, and
refer to Appendix~\ref{sec:semicl-oper} for details.
\begin{defi}
\label{defi:semi_op}
  A \emph{semiclassical operator} is either:
  \begin{enumerate}
  \item In case~\ref{item:1}, a (possibly unbounded)
    $\h$-pseudodifferential operator acting on
    $\mathcal{H}_\h:=L^2(X)$.
  \item In case~\ref{item:2}, a Berezin-Toeplitz operator acting on
    $\mathcal{H}_\h:=H^0(M,\mathscr{L}^k \otimes \mathscr{K})$, the space of
    holomorphic sections of high tensor powers of a suitable line
    bundle, possibly twisted with another line bundle; there, the
    semiclassical parameter is $\h = \frac{1}{k}$.
  \item In case~\ref{item:3}, a (possibly unbounded) Berezin-Toeplitz
    operator acting on
    \begin{align*}
      \mathcal{H}_\h & :=
                       H^0(\CM \times N, \mathscr{L}_0^{k} \boxtimes (\mathscr{L}^k \otimes \mathscr{K})) \cap L^2(\CM \times N, \mathscr{L}_0^{k} \boxtimes (\mathscr{L}^k \otimes \mathscr{K})) \\
                     & \simeq  \mathcal{B}_k(\CM) \otimes H^0(N,\mathscr{L}^k \otimes \mathscr{K}),\label{eq:2}      
    \end{align*}
    still with $\h = \frac{1}{k}$. Here $\mathcal{B}_k(\CM)$ is the
    Bargmann space with weight $\exp(-k|z|^2)$.
  \end{enumerate}
\end{defi}
In fact, the three cases can be seen as instances of general (not
necessarily compact) Berezin-Toeplitz quantization. It is well-known,
for instance, that Weyl pseudodifferential quantization on $\RM^{2d}$
is equivalent to Berezin-Toeplitz quantization on $\CM^d$. Although a
fully general theory has not been developed yet, it is also known
since~\cite{BG} that, in a microlocal sense, contact Berezin-Toeplitz
quantization is always equivalent to homogeneous pseudodifferential
quantization.

In all cases, a semiclassical operator is actually a family of
operators $\hat H_\h$ indexed by the semiclassical parameter $\h$,
acting on a Hilbert space $\mathcal{H}_\h$ that may depend on $\h$ as
well. Most importantly for us, a selfadjoint semiclassical operator
$\hat H_\h$ has a \emph{principal symbol} $H\in\Cinf(M;\RM)$, which
does not depend on $\h$; conversely, for any classical Hamiltonian
$H\in\Cinf(M;\RM)$ (with suitable control at infinity in non-compact
cases) there exists a semiclassical operator $\hat H_\h$ whose
principal symbol is $H$. Any other semiclassical operator
$\hat{H}'_\h$ with principal symbol $H$ is $\O(\h)$-close to
$\hat{H}_\h$ in a suitable topology. See
Appendix~\ref{sec:semicl-oper}.

\begin{ex} We give below three examples of semiclassical operators corresponding to each of the three cases of Definition \ref{defi:semi_op}.
\begin{enumerate}
\item The semiclassical Schr\"odinger operator $-\h^2 \Delta + V$ with potential $V \in \Cinf(\RM^2,\RM)$ is a $\h$-pseudodifferential operator with principal symbol $T^* \RM^2 \to \RM, (x,\xi) \mapsto \| \xi \|^2 + V(x)$.
\item Let $\CM_{\leq d}[u]$ be the space of polynomials of degree at most $d$ in one complex variable $u$. The operator $\frac{1}{k+2} \left( 2 z \frac{d}{dz} - k \mathrm{Id} \right) \otimes \mathrm{Id} + \mathrm{Id} \otimes \frac{1}{k+2} \left( 2 w \frac{d}{dw} - k \mathrm{Id} \right) $ acting on $\CM_{\leq k}[z] \otimes \CM_{\leq k}[w]$ is a Berezin-Toeplitz operator with principal symbol $S^2 \times S^2 \to \RM, (x_1,y_1,z_1,x_2,y_2,z_2) \mapsto z_1 +  z_2$, see for instance \cite[Example 5.2.4]{yohann-book}.
\item The operator $k^{-1} \left( w \frac{d}{dw} + \frac{1}{2} \right) \otimes \mathrm{Id} + \mathrm{Id} \otimes \frac{1}{k+2} \left( 2 z \frac{d}{dz} - k \mathrm{Id} \right) $ acting on $\mathcal{B}_k(\CM) \otimes \CM_{\leq k}[z]$ is a Berezin-Toeplitz operator with principal symbol $\RM^2 \times S^2 \to \RM, (u,v,x,y,z) \mapsto \frac{1}{2} (u^2 + v^2) + z$.
\end{enumerate}
\end{ex}

Given two semiclassical operators $\hat J_\h$ and ${\hat H_\h}$, their
\emph{commutator} $\frac{i}{\h}[\hat J_\h, \hat H_\h]$ is again a
semiclassical operator, whose principal symbol is the Poisson bracket
$\{J,H\}$. We say that $\hat J_\h$ and ${\hat H_\h}$ commute if their
commutator vanishes. In this case, one can show that the spectral
measures of the selfadjoint operators $\hat J_\h$ and ${\hat H_\h}$
commute in the usual sense \cite{charbonnel}.

\begin{defi}
  A \emph{quantum integrable system} $(\hat{J}_\h, \hat{H}_\h)$ is the
  data of two commuting semiclassical operators acting on
  $\mathcal{H}_\h$ whose principal symbols $J, H$ form a Liouville
  integrable system. If moreover $(J,H)$ is a semitoric integrable
  system, we say that $(\hat{J}_\h, \hat{H}_\h)$ is a \emph{semitoric
    quantum integrable system}, or a \emph{quantum semitoric system}.
\end{defi}

\begin{ex} \label{ex:spin-osc_quant} The quantum Jaynes-Cummings
    model, or the system described in \cite[Section
    4]{san-alvaro-spin}, is certainly a semiclassical quantization of
    the system described in Example \ref{ex:spin-osc_class} in the
    sense of Appendix~\ref{sec:semicl-oper}, although this precise
    fact has, to the best of our knowledge, never been
    proven. Therefore, we will adopt a slightly different point of
    view and directly describe the quantum Hamiltonians
    $(\hat{J}_\h, \hat{H}_\h)$ as Berezin-Toeplitz operators, instead
    of a quantum reduction of $\h$-pseudodifferential operators by a
    circle action, which was the approach
    of~\cite{san-alvaro-spin}. Actually, it is expected that the
    quantum reduction of an $\h$-pseudodifferential operator by a
    torus action is always a Berezin-Toeplitz operator, but as far as
    we know this fact has not been established yet.

Hence we work in the case~\ref{item:3}. The quantization of
    the sphere is now quite standard; however, we will need a precise
    setting that has been explained in
    \cite{alvaro-yohann,yohann-book}. The hyperplane bundle
    $\mathcal{O}(1)$ is a prequantum line bundle for the symplectic
    manifold $(\mathbb C \mathbb P^1, \omega_{\text{FS}})$, where
    $\omega_{\text{FS}}$ is the Fubini-Study form, and the
    tautological line bundle $\mathcal{O}(-1)$ is a half-form bundle,
    so the Hilbert spaces
    $H^0(\mathbb C \mathbb P^1, \mathcal{O}(k) \otimes
    \mathcal{O}(-1))$, $k \geq 1$, yield a quantization of this phase
    space with metaplectic correction. Let $\pi_N$ be the
    stereographic projection from the north pole of
    $S^2 \subset \RM^3$ to its equatorial plane; then one readily
    checks that
    $\pi_N^* \omega_{\text{FS}} = -\frac{1}{2} \omega_{S^2}$. Hence,
    since we want to quantize $(S^2, \omega_{S^2})$, we consider
    instead the Hilbert spaces
    $\mathcal{H}_k = H^0(\mathbb C \mathbb P^1, \mathcal{O}(2k)
    \otimes \mathcal{O}(-1)) = H^0(\mathbb C \mathbb P^1,
    \mathcal{O}(2k-1))$, $k \geq 1$ and replace the coordinates
    $(x,y,z)$ on $S^2$ with $(x,-y,z)$ (which has the effect of
    changing the sign of the symplectic form). Hence, thanks to the
    results of \cite[Section 4.3]{alvaro-yohann}, we obtain the
    following. First, note that we have an isometry
  \[
    \mathcal{H}_k \simeq \CM_{\leq 2k-1}[z], \qquad \langle P, Q
    \rangle_k = \int_{\CM} \frac{P(z) \overline{Q(z)}}{(1 +
      |z|^2)^{2k+1}} |dz \wedge d\bar{z}| \] between $\mathcal{H}_k$
  and the space of polynomials of one complex variable with degree at
  most $2k-1$. Then the polynomials
  \[
    e_{\ell}: z \mapsto \sqrt{\frac{2k \binom{2k-1}{\ell}}{2\pi}} \
    z^{2k-1-\ell}, \qquad 0 \leq \ell \leq 2k-1 \] form an orthonormal
  basis of $\mathcal{H}_k$ and the operators
  $\hat{X}_k, \hat{Y}_k, \hat{Z}_k: \mathcal{H}_k \to \mathcal{H}_k$
  acting as
  \begin{equation} \begin{cases} \hat{X}_k e_{\ell} = \frac{1}{2k} \left( \sqrt{\ell (2k-\ell)} e_{\ell-1} + \sqrt{(\ell + 1) (2k-1-\ell)} e_{\ell+1} \right),\\[2mm]
      \hat{Y}_k e_{\ell} = \frac{i}{2k} \left( \sqrt{\ell (2k-\ell)} e_{\ell-1} - \sqrt{(\ell + 1) (2k-1-\ell)} e_{\ell+1} \right),\\[2mm]
      \hat{Z}_k e_{\ell} = \left(\frac{2(k-\ell)-1}{2k}\right)
      e_{\ell} \end{cases} \label{eq:BTO_sphere} \end{equation} on
  this basis are Berezin-Toeplitz operators with respective principal
  symbols $x, y$ and $z$.

 In order to quantize the Hamiltonians of
  Example~\ref{ex:spin-osc_class}, as in Appendix
  \ref{sec:semicl-oper}, we identify $\RM^2$ with $\CM$ by setting
  $w = \frac{1}{\sqrt{2}}(u - iv)$. Therefore, we consider the
  semiclassical parameter $\h = k^{-1}$, the Hilbert spaces
  $\mathcal{B}_k(\CM) \otimes \mathcal{H}_k$ and the operators
  \[
    \begin{cases} \hat{J}_{\h} = k^{-1} \left( w \frac{d}{dw} + \frac{1}{2} \right) \otimes \mathrm{Id} + \mathrm{Id} \otimes \hat{Z}_k , \\[2mm]
      \hat{H}_{\h} = \frac{1}{2\sqrt{2}} \left( \left( w + k^{-1}
          \frac{d}{dw} \right) \otimes \hat{X}_k + i \left( w - k^{-1}
          \frac{d}{dw} \right) \otimes \hat{Y}_k
      \right) \end{cases} \] acting on these spaces. These are
  commuting semiclassical operators in the sense of case~\ref{item:3}
  of Definition~\ref{defi:semi_op} with respective principal symbols
  $J$ and $H$.
\end{ex}

\begin{defi}
  The \emph{joint spectrum} of a quantum integrable system
  $(\hat{J}_\h, \hat{H}_\h)$ is the support of the joint spectral
  measure (see for instance~\cite[Section 6.5]{birman-solomjak-book})
  of $\hat J_\h$ and ${\hat H_\h}$.
\end{defi}

We shall only consider situations where the joint spectrum is
discrete: joint eigenvalues are isolated, with finite
multiplicity. This is of course automatic in the compact
Berezin-Toeplitz case, since the Hilbert spaces $\mathcal{H}_\h$ are
finite dimensional. In the non compact case, this can be seen as the
quantum analogue of the properness condition on the momentum map
$F=(J,H) : M \to \RM^2$; indeed the joint spectrum is discrete if and
only if, for any compact subset of $\RM^2$, the corresponding joint
spectral projection is compact (\emph{i.e.} of finite rank). In the
pseudodifferential case, a convenient assumption that guarantees
discreteness of the spectrum is the ellipticity at infinity of the
operator $\hat{J}_\h^2+ \hat{H}_\h^2$, see \cite{charbonnel}. If this
holds, we say that the quantum integrable system is \emph{proper}, and in
what follows we will always work with proper quantum integrable
systems.

\begin{ex}\label{ex:spin-osc_jsp} The joint spectrum of the
    quantum Jaynes-Cummings system $(\hat{J}_{\h}, \hat{H}_{\h})$
    defined in Example \ref{ex:spin-osc_quant} directly follows from
    \cite{san-alvaro-spin}. Indeed, one can check that, if we use the
    correspondence (with the notation of \cite{san-alvaro-spin} on the
    left and our notation on the right)
  \[
    \h \leftrightarrow k^{-1}, \quad n \leftrightarrow 2k-1, \quad k
    \leftrightarrow 2k - 1 - \ell, \] we find the exact same operator
  matrix as in the aforementioned paper (the last correspondence is
  here simply because the basis of the quantum space associated with
  the sphere was ordered the other way around in
  \cite{san-alvaro-spin}). So we can use Lemma 4.5 and Proposition
  4.7 in \cite{san-alvaro-spin} to compute this joint
  spectrum, which is the set
  \[
    \bigcup_{j \in \NM} \{ (-1 + k^{-1}(j+1),E_0(j)) , \ldots,
    (-1 + k^{-1}(j+1), E_{d(j)}(j)) \}
  \]
  where $d(j) = \min(j,2k-1)$ and $ E_0(j), \ldots, E_{d(j)}(j)$ are the eigenvalues of the matrix
  \[ A_j = \frac{1}{(2k)^{\frac{3}{2}}} \begin{pmatrix} 0 & \beta_1(j)
      & 0 & \ldots & 0 \\ \beta_1(j) & 0 & \beta_2(j) & \ddots &
      \vdots \\ 0 & \beta_2(j) & 0 & \ddots & 0 \\ \vdots & \ddots &
      \ddots & \ddots & \beta_{d(j)}(j) \\ 0 & \ldots & 0 &
      \beta_{d(j)}(j) & 0 \end{pmatrix} \] with
  $\beta_{\ell}(j) = \sqrt{\ell(j+1-\ell)(2k-\ell)}$ for
  $1 \leq \ell \leq d(j)$. In practice, we obtain the joint spectrum
  by numerically diagonalizing the matrices $A_j$.  Note that the
  above correspondence implies that if one wants to compare our
  results with those of \cite{san-alvaro-spin}, one should consider
  only odd values of $n$ in the latter. Part of the joint spectrum is
  displayed in Figure \ref{fig:jsp_spinosc}.
  \begin{figure}[H]
    \begin{center}
      \includegraphics[trim=60 10 90
      60,clip,scale=0.4]{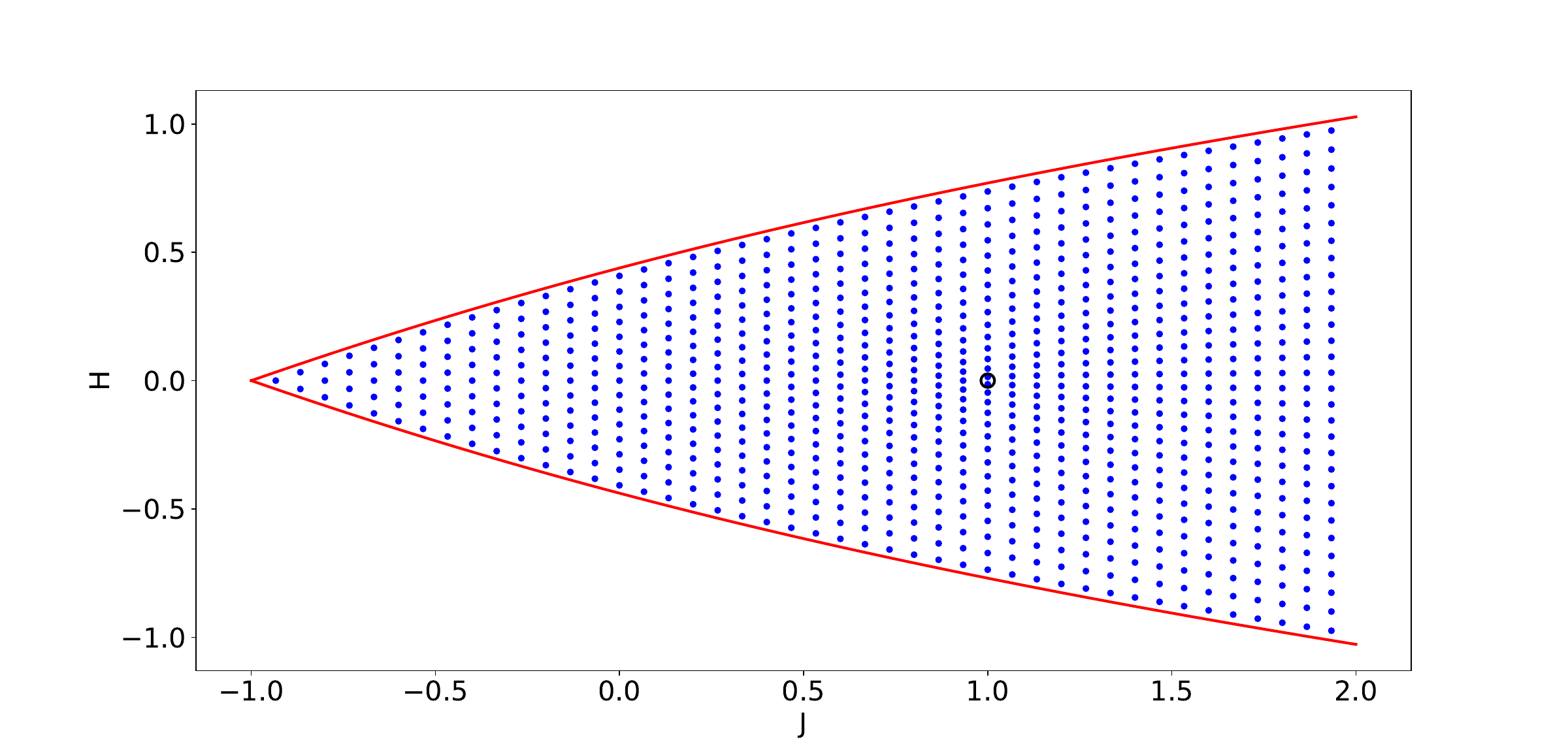}
    \end{center}
    \caption{\small The blue dots are the joint eigenvalues of the
      spin-oscillator system in the region $-1 \leq x \leq 2$ for
      $k = 15$. The red line corresponds to the boundary of the image
      of the momentum map, and the black circle indicates the position
      of the focus-focus value.}
    \label{fig:jsp_spinosc}
  \end{figure}
\end{ex}

We will need to compare families of spectra up to $\O(\h^2)$. By this
we mean the following (for an example of why this definition is
relevant, see Remark 3.9 in \cite{san-dauge-hall-rotation}).

\begin{defi}
  \label{defi:mod_h2}
  Let $A_\h, B_\h \subset \RM^2$ be two families of closed subsets
  indexed by $\h \in \mathcal{I}$, where $\mathcal{I} \subset \RM_+^*$
  is a set of positive real numbers for which zero is an accumulation
  point. We say that $A_\h = B_\h$ modulo $\O(\h^2)$ if for every
  compact set $K \subset \RM^2$, there exists $C > 0$ such that
  $d(A_\h \cap K, B_\h) \leq C \h^2$ and
  $d(B_\h \cap K, A_\h) \leq C \h^2$, where we recall that
  $d(A,B) = \max_{x \in A} d(x,B)$ if $A$ and $B$ are subsets of
  $\RM^2$ with $A$ compact.
\end{defi}

We are now in position to precisely state our main result.
\begin{theo}\label{theo:main}
  Let $(\Sigma_\h)_{\h\in \mathcal{I}}$ be a collection, indexed by
  $\h\in\mathcal{I}\subset\RM$, of point sets in $\RM^2$, that is
  assumed to be the joint spectrum of some unknown proper semitoric
  quantum integrable system $(\hat{J}_\h, \hat{H}_\h)$ with joint
  principal symbol $F$. Let $S\subset\RM^2$ be a vertical strip of
  bounded width. Then all symplectic invariants of the underlying classical semitoric system on $F^{-1}(S)$ can be explicitly recovered, in a constructive way, from the data of
  $(\Sigma_\h\cap S)_{\h\in \mathcal{I}}$ modulo $\O(\h^2)$.
  In particular, if two proper quantum semitoric systems have the same
  spectrum modulo $\O(\h^2)$, then their underlying classical systems
  are symplectically isomorphic.
\end{theo}
By assumption, the Hamiltonian $J$ is proper, and this implies that
the joint spectrum $\Sigma_\h$ may be unbounded only in the horizontal
direction. Thus, the restriction to the strip $S$ ensures that we are
looking at a compact region in $\RM^2$. Naturally, if
$(\hat{J}_\h, \hat{H}_\h)$ is known a priori to be associated with a
compact phase space, then the statement of the theorem holds without
the strip $S$.

The rest of the paper is devoted to the proof of
Theorem~\ref{theo:main}. By Theorem~\ref{theo:poly_twisting}, which
relies on Theorem~\ref{theo:recover-sigma_1}, we recover both the
twisting index and the polygon invariant. Moreover, we obtain the
position of each focus-focus critical value (see the second paragraph
of the proof of Theorem~\ref{theo:poly_twisting}). The height
invariant is then given by
Proposition~\ref{prop:height-invariant}. Finally, we recover the
Taylor series invariant by Theorem~\ref{theo:full_taylor}. Since we
have gathered the complete set of symplectic invariants of the
semitoric system, the triple $(M,\omega,F)$ is henceforth completely
determined up to isomorphism by the classification
result~\cite{san-alvaro-II}. This proves the theorem.

\begin{rema} As mentioned previously, Theorem~\ref{theo:main}
    holds for simple semitoric systems. The non simple case covers two
    situations: on the one hand, if we allow a focus-focus fiber to
    contain several critical points, we do not know whether the
    inverse spectral result holds; on the other hand, we conjecture
    that Theorem~\ref{theo:main} should extend to the situation where
    several simple focus-focus fibers belong to the same level set of
    $J$, using the classification of~\cite{palmer-pelayo-tang}. We
    expect that the techniques developed in our work apply with few
    modifications to this case.
\end{rema}

\section{Asymptotic lattices and half-lattices}
\label{sec:lattices}

The method we use to recover the polygonal invariant from the joint
spectrum of a proper quantum semitoric system is based on a detailed
analysis of the structure of this spectrum, not only near a regular value
of the underlying momentum map, but also near an elliptic critical value of
rank 1, and with a global point of view encompassing these two aspects. 

In this section we introduce the necessary tools to perform this program. First, we develop the theory of asymptotic half-lattices in order to generalize the notion of asymptotic lattices which was introduced in \cite{san-dauge-hall-rotation} to study the joint spectrum near a regular value. Second, we explain how and when one can extend families of asymptotic lattices and half-lattices to obtain a ``global asymptotic lattice''. Building on these results, we explain how to label such global asymptotic lattices (Theorem \ref{thm:construct_global_labelling}), which, for the joint spectrum of a quantum integrable system, corresponds to producing good global quantum numbers.

\subsection{Asymptotic lattices and labellings}
\label{subsect:asympt_latt}

Thanks to the Bohr-Sommerfeld quantization conditions (see Theorem
\ref{theo:BS_reg}), the joint spectrum in a neighborhood of a regular
value of the momentum map is an asymptotic lattice, using the
terminology of \cite{san-dauge-hall-rotation}. Roughly speaking, an
asymptotic lattice is just a semiclassical deformation of the straight
lattice $\h\ZM^d$ in a bounded domain. The precise definition,
restricted to the two-dimensional case, is as follows.
\begin{defi}[{\cite[Definition 3.6]{san-dauge-hall-rotation}}]
  \label{defi:asymp_latt} An \emph{asymptotic lattice} is a triple
  $(\mathcal{L}_\h,\mathcal{I},B)$ where $\mathcal{I} \subset \RM_+^*$
  is a set of positive real numbers for which zero is an accumulation
  point, $B \subset \RM^2$ is a simply connected bounded open set and
  $\h \in \mathcal{I} \mapsto \mathcal{L}_\h \subset B$ is a family of
  discrete sets, such that
  \begin{enumerate}
  \item\label{item:asla-separation} there exist $\h_0 > 0$,
    $\epsilon_0 > 0$ and $N_0 \geq 1$ such that for all
    $\h \in \mathcal{I} \cap \interval[open left]0{\hbar_0}$
    \[
      \h^{-N_0} \min_{\substack{(\lambda,\mu) \in \mathcal{L}_\h^2 \\
          \lambda \neq \mu}} \| \lambda - \mu \| \geq \epsilon_0,
    \]
  \item\label{item:asla-chart} there exist a bounded open set
    $U \subset \RM^2$ and a family of smooth maps
    $G_{\hbar}: U \to \RM^2$ such that
    \begin{itemize}
    \item there exist functions
      $G_0, G_1, G_2, \ldots \in C^{\infty}(U,\RM^2)$ such that $G_\h$
      has the asymptotic expansion
      \begin{equation} G_\h = G_0 + \h G_1 + \h^2 G_2 +
        \ldots \label{equ:asymp_chart}\end{equation} for the
      $C^{\infty}$ topology on $U$,
    \item $G_0$ is an orientation preserving diffeomorphism from $U$
      to a neighborhood of $\bar{B}$,
    \item $G_\h(\h \ZM^2 \cap U) = \mathcal{L}_\h + \O(\h^{\infty})$
      inside $B$, which means that there exists a sequence
      $(C_N)_{N \geq 0}$ of positive numbers such that
      \begin{itemize}
      \item for all $\h \in \mathcal{I}$, for all
        $\lambda \in \mathcal{L}_\h$, there exists $\zeta \in \ZM^2$
        such that $\h \zeta \in U$ and
        \begin{equation} \forall N \geq 0 \qquad \| \lambda - G_\h(\h
          \zeta) \| \leq C_N \h^N \label{equ:dist_hN}\end{equation}
      \item for every open set $U_0 \Subset G_0^{-1}(B)$ (here the
        notation $V \Subset W$ means that $\overline{V}$ is compact
        and contained in $W$), there exists $\h_1 > 0$ such that for
        all $\h \in \mathcal{I} \cap ]0,\h_1]$, for all
        $\zeta \in \ZM^2$ such that $\h \zeta \in U_0$, there exists
        $\lambda \in \mathcal{L}_\h$ such that Equation
        \eqref{equ:dist_hN} holds.
      \end{itemize}
    \end{itemize}
    The pair $(G_\h,U)$ is called an \emph{asymptotic chart} for
    $(\mathcal{L}_\h,\mathcal{I},B)$.
  \end{enumerate}
\end{defi}

\begin{figure}[h]
  \centering
  \includegraphics[width=\linewidth]{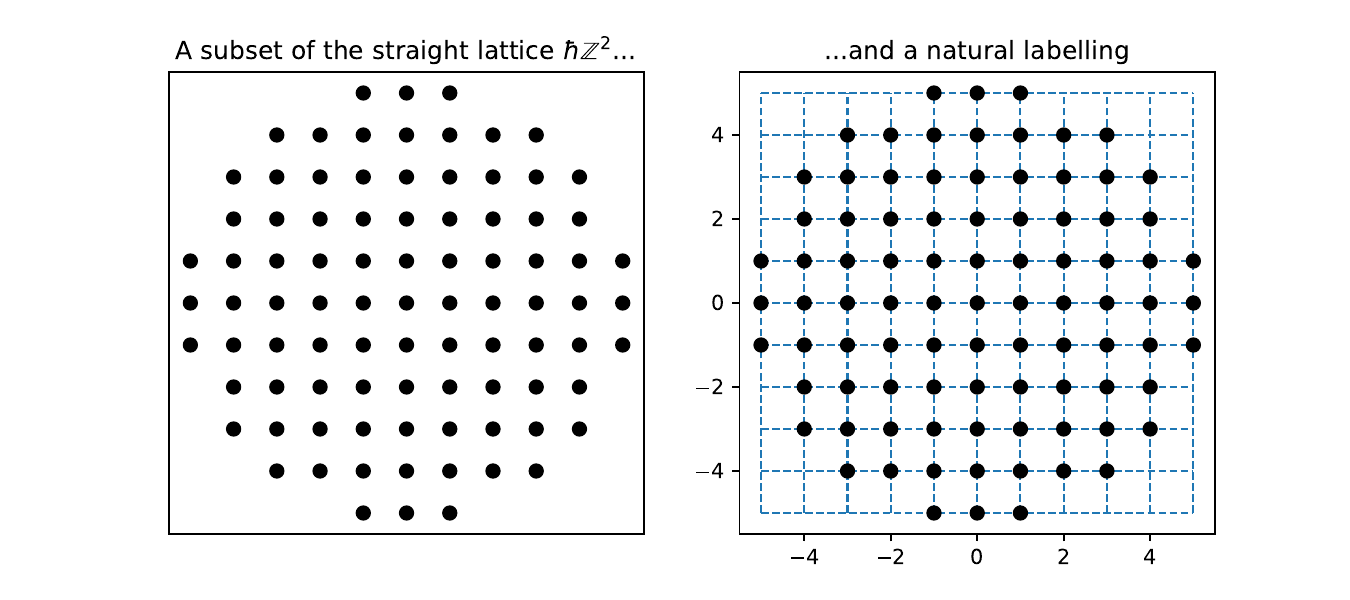}
  \caption{A trivial asymptotic lattice.}
  \label{fig:asympt_latt_trivial}
\end{figure}

\begin{figure}[h]
  \centering
  \includegraphics[width=\linewidth]{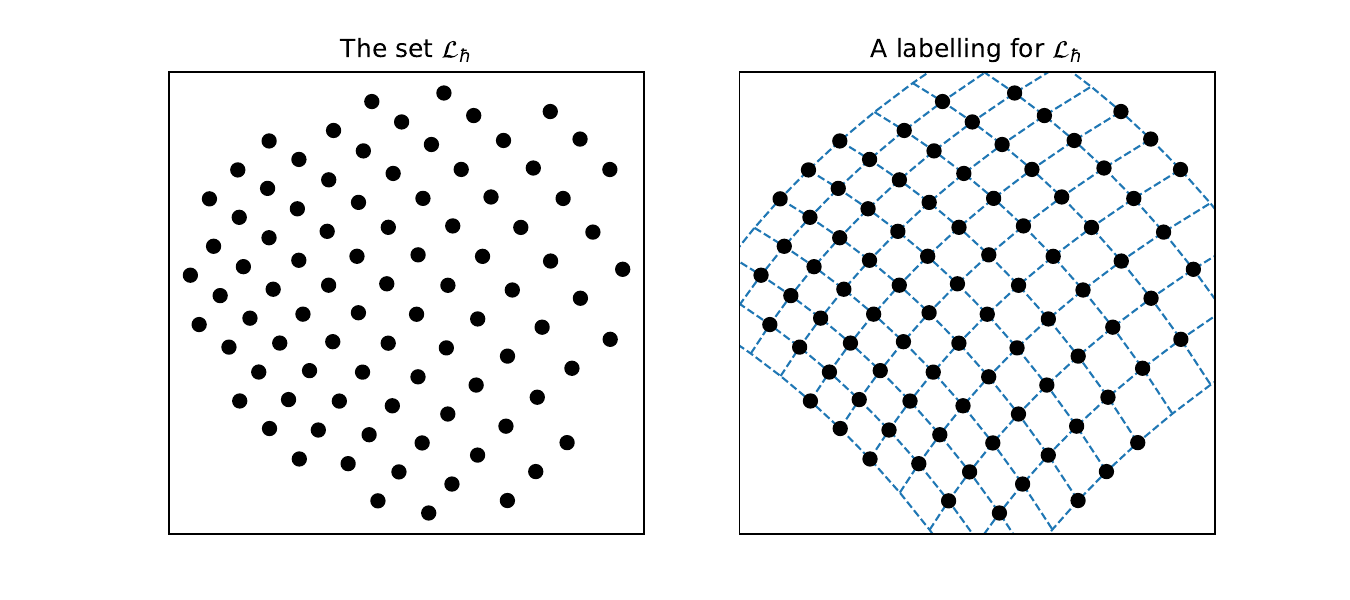}
  \caption{A general asymptotic lattice. The labelling is suggested by the dashed lines on the right-hand side figure.}
  \label{fig:asympt_latt_general}
\end{figure}

\begin{ex} Let $B$ be a bounded open subset of $\RM^2$ and let
    $U$ be a bounded open neighborood of $\bar{B}$ in $\RM^2$. The
    intersection $\h \ZM^2 \cap B$ of the rescaled square lattice with
    $B$ is an asymptotic lattice associated with the asymptotic chart
    $(\mathrm{Id},U)$, see Figure~\ref{fig:asympt_latt_trivial}. Given
    any $G_\h$ as in Equation \eqref{equ:asymp_chart},
    $G_\h(\h \ZM^2)$ is an asymptotic lattice in any compact set
    contained in $G_0(U)$, see Figure~\ref{fig:asympt_latt_general}.
\end{ex}

\begin{theo}
  \label{theo:BS_reg}
  Let $(A_\h, B_\h)$, $\h \in \mathcal{I}$, be a proper quantum
  integrable system with joint principal symbol $F = (a_0, b_0)$, and
  let $\Sigma_{\h}$ be its joint spectrum. Let $c_0 \in \RM^2$ be a
  regular value of $F$ such that $F^{-1}(c_0)$ is connected. Then
  there exists an open ball $B \subset \RM^2$ containing $c_0$ such
  that $(\Sigma_\h,\mathcal{I},B)$ is an asymptotic lattice, and
  admits an asymptotic chart of the form \eqref{equ:asymp_chart} with
  ${\rm d}G_0 = {\rm d}\tilde{G}_0$ where $\tilde{G}_0^{-1}$ is an
  action diffeomorphism.
\end{theo}

\begin{demo}
  This is well-known in the case where $A_\h$ and $B_\h$ are
  $\h$-pseudodifferential operators, see \cite[Theorem 3.2, Theorem
  3.7]{san-dauge-hall-rotation} and the references therein. Assume
  that $\h = k^{-1}$ for some $k\in\NM^*$, and (with the natural abuse
  of notation) that $A_\h = A_k$, $B_\h = B_k$ are Berezin-Toeplitz
  operators on a compact manifold equipped with a prequantum line
  bundle $(\mathscr{L},\nabla)$. Then from \cite[Theorem
  3.1]{laurent-BS} (see also \cite[Section 3.2]{laurent-half}) we know
  that the joint spectrum near a regular value $c_0$ of $F$ coincides
  modulo $\O(k^{-\infty})$ with the set of solutions $\lambda$ to the
  equation
  \[   g(\lambda,k) \in k^{-1} \ZM^2 \]
  where $g(\cdot,k)$ has an asymptotic expansion of the form
  $g(\cdot,k) = g_0 + k^{-1} g_1 + \ldots$ and $g_0 = (g_0^{(1)}, g_0^{(2)})$
  is computed as follows. For $c$ close to $c_0$, let $\Lambda_c$ be
  the Lagrangian torus $F^{-1}(c)$, and choose two loops
  $\gamma_1(c), \gamma_2(c)$, depending continuously on $c$, whose
  classes form a basis of $H_1(\Lambda_c,\ZM)$. Then for $i = 1, 2$,
  $2 \pi g_0^{(i)}(c) = \text{hol}(\gamma_i(c),\mathscr{L},\nabla)$ is
  the holonomy of $\gamma_i(c)$ in $(\mathscr{L},\nabla)$.
	
  In fact, the proof of this result can easily be adapted for
  Berezin-Toeplitz operators on a manifold of the form $\CM \times M$
  with $M$ compact, since the properness of $F$ implies that the
  fibers near $F^{-1}(c)$ are compact, the microlocal normal form used
  in \cite{laurent-BS} can still be achieved in this case, and the
  properness of $(A_k,B_k)$ implies that the corresponding joint
  eigenfunctions are localized near $F^{-1}(c)$. Consequently, the
  rest of the proof below applies to both cases \ref{item:2} and
  \ref{item:3}.

  It remains to show that $g_0$ has the required property. We endow
  $\mathbb{T}^2 \times \RM^2 = (\RM \slash 2\pi\ZM)^2 \times \RM^2$
  with coordinates $(\theta_1, \theta_2, I_1, I_2)$ and symplectic
  form $\omega_0 = dI_1 \wedge d\theta_1 + dI_2 \wedge d\theta_2$. The
  action-angle theorem yields a symplectomorphism $\phi$ from a
  neighborhood of $\Lambda_{c_0}$ in $M$ to a neighborhood of the zero
  section in $\mathbb{T}^2 \times \RM^2$ and a local diffeomorphism
  $G_0: \RM^2 \to \RM^2$ such that $F \circ \phi^{-1} =
  G_0(I_1,I_2)$. In what follows, we will write $\psi = \phi^{-1}$ and
  $H_0 = (H_0^{(1)},H_0^{(2)}) = G_0^{-1}$. We can choose
  $\gamma_1, \gamma_2$ satisfying the above condition as follows. Let
  $\tilde \gamma_1(c), \tilde \gamma_2(c)$ be the loops inside
  $\mathbb{T}^2 \times \RM^2$ defined as
  \[
    \tilde\gamma_1(c) = \left\{ (\theta_1, 0, G_0^{-1}(c)) \ | \ 0
      \leq \theta_1 \leq 2\pi \right\}, \quad \tilde\gamma_2(c) =
    \left\{ (0,\theta_2, G_0^{-1}(c)) \ | \ 0 \leq \theta_1 \leq 2\pi
    \right\}.
  \]
  Then we set $\gamma_1 = \phi^* \tilde \gamma_1$ and
  $\gamma_2 = \phi^* \gamma_2$. Then for $i = 1, 2$,
  \[
    g_0^{(i)}(c) = \text{hol}(\gamma_i(c),\mathscr{L},\nabla) =
    \text{hol}(\tilde\gamma_i(c),\psi^*\mathscr{L},\psi^*\nabla).
  \]
  But the curvature of $\psi^* \nabla$ is
  $ \text{curv}(\psi^* \nabla) = -i\psi^* \text{curv}(\nabla) =
  -i\psi^* \omega = -i\omega_0$, so
  $\psi^* \mathscr{L} = \mathscr{L}_0 \otimes \mathscr{P}$ where
  $\mathscr{L}_0 = \mathbb{T}^2 \times \RM^2 \times \CM$ with
  connection $\nabla_0 = d - i \alpha_0$ with
  $\alpha_0 = I_1 d\theta_1 + I_2 d\theta_2$ and
  $(\mathscr{P},\nabla_{\mathscr{P}})$ is a flat line bundle over
  $\mathbb{T}^2 \times \RM^2$. Consequently
  \[
    \text{hol}(\tilde\gamma_i(c),\psi^*\mathscr{L},\psi^*\nabla) =
    \text{hol}(\tilde\gamma_i(c),\mathscr{L}_0,\nabla_0) +
    \text{hol}(\tilde\gamma_i(c),\mathscr{P},\nabla_{\mathscr{P}}).
  \]
  On the one hand, by the Ambrose-Singer theorem, the holonomy group
  of $(\mathscr{P},\nabla_{\mathscr{P}})$ is a discrete subgroup of
  $\RM$, so
  $\text{hol}(\tilde\gamma_i(c),\mathscr{P},\nabla_{\mathscr{P}}) :=
  C_i$ does not depend on $c$. On the other hand,
  \[
    \text{hol}(\tilde\gamma_i(c),\mathscr{L}_0,\nabla_0) =
    \frac{1}{2\pi} \int_{\tilde\gamma_i(c)} \alpha_0 = H_0^{(i)}(c).
  \]
  Hence we finally obtain that $g_0 = H_0 + (C_1,C_2)$, so
  $\dd g_0 = \dd H_0$. This implies that $g_0$ is invertible and so we
  can construct an asymptotic chart for the joint spectrum near $c_0$
  by inverting $g_k$, and the second part of the statement is now
  immediate.
\end{demo}

In \cite{san-dauge-hall-rotation}, the authors studied the question of
labelling the elements of an asymptotic lattice in a consistent way.

\begin{defi}
  \label{defi:good_lab_lattice}
  Let $(\mathcal{L}_\h,\mathcal{I},B)$ be an asymptotic lattice with
  asymptotic chart $(G_\h,U)$. A \emph{good labelling} of
  $(\mathcal{L}_\h,\mathcal{I},B)$ associated with $G_\h$ is a family
  of maps $\ell_\h: \mathcal{L}_\h \to \ZM^2$, $\h \in \mathcal{I}$,
  such that for every $\lambda \in \mathcal{L}_\h$,
  $\h \ell_\h(\lambda) \in U$ and
  \[
    \forall N \geq 0 \qquad \| G_\h(\h \ell_\h(\lambda)) - \lambda \|
    \leq C_N \h^N
  \]
  where $(C_N)_{N \geq 0}$ is as in Definition \ref{defi:asymp_latt}.
\end{defi}
Examples of asymptotic lattices with labellings are given in
Figures~\ref{fig:asympt_latt_trivial}
and~\ref{fig:asympt_latt_general}.

\begin{rema}
  \label{rema:def_label}
  Having a good labelling amounts to presenting the set
  $\mathcal{L}_\h$ ``in a natural way'' as the set of
  $\lambda_{m,n}(\h)$ for $(m,n)$ in some finite subset of $\ZM^2$
  which depends on $\h$. The correspondence is given by
  $\ell_\h(\lambda_{m,n}(\h)) = (m,n)$.
\end{rema}

It was shown in \cite[Lemma 3.11]{san-dauge-hall-rotation} that given
an asymptotic chart $(G_\h,U)$ for the asymptotic lattice
$(\mathcal{L}_\h,\mathcal{I},B)$, there exists a (unique for $\h$
small enough) associated good labelling $\ell_\h$. Moreover, for fixed
$\h$, the map $\ell_\h$ is injective.

It is important to notice that a given asymptotic lattice does not
possess a unique asymptotic chart (so the same holds for good
labellings). Indeed, as observed in \cite[Lemma
3.19]{san-dauge-hall-rotation}, if $(G_\h,U)$ is an asymptotic chart
for the asymptotic lattice $(\mathcal{L}_\h,\mathcal{I},B)$ and if
$A \in \mathrm{SL}(2,\ZM)$, then $(G_\h \circ A, A^{-1}U)$ is another
asymptotic chart for this asymptotic lattice. If $\ell_\h$ is the good
labelling associated with $(G_\h,U)$, then the good labelling
associated with $(G_\h \circ A, A^{-1}U)$ is $A^{-1} \circ \ell_\h$.

In fact, it was proved in \cite[Proposition
3.22]{san-dauge-hall-rotation} that if
$(\mathcal{L}_{\h}, \mathcal{I}, B)$ satisfies a continuity property
with respect to $\h$ (see \cite[Definition
3.21]{san-dauge-hall-rotation}) and $\ell_{\h}, \tilde \ell_{\h}$ are
two good labellings for $\mathcal{L}_{\h}$, then there exists a unique
$\tau \in \text{GA}^+(2,\ZM)$ and $\h_0 > 0$ such that for every
$\h \in (0,\h_0] \cap \mathcal{I}$,
$\tilde \ell_{\h} = \tau \circ \ell_{\h}$. Here
$\text{GA}^+(2,\ZM) = \text{SL}(2,\ZM) \ltimes \ZM^2$ is the group of
orientation-preserving integral affine transformations. Unfortunately,
the joint spectrum of a quantum semitoric system formed by
Berezin-Toeplitz operators does not satisfy this continuity property,
so we cannot apply the aforementioned proposition as is. However, we
can use a slightly less restrictive definition of labelling.

\begin{defi}[{\cite[Definition 3.16]{san-dauge-hall-rotation}}]
  \label{defi:linear_labelling}
  Given an asymptotic lattice $(\mathcal{L}_{\h}, \mathcal{I}, B)$, a
  \emph{linear labelling} is a family of maps
  $\bar{\ell}_\h: \mathcal{L}_\h \to \ZM^2$, $\h \in \mathcal{I}$ of
  the form $\bar{\ell}_\h = \ell_\h + \kappa_\h$ where $\ell_\h$ is a
  good labelling and $(\kappa_\h)_{\h \in \mathcal{I}}$ is a family of
  vectors in $\ZM^2$.
\end{defi}

It was shown in \cite[Proposition 3.20]{san-dauge-hall-rotation} that
if $\bar{\ell}_\h^{(1)}$ and $\bar{\ell}_\h^{(2)}$ are two linear
labellings for a given asymptotic lattice
$(\mathcal{L}_{\h}, \mathcal{I}, B)$, then for any open set
$\tilde{B} \Subset B$, there exists a unique matrix
$A \in \mathrm{SL}(2,\ZM)$, $\h_0 > 0$ and a family
$(\kappa_\h)_{\h \in \mathcal{I} \cap [0,\h_0]}$ of vectors in $\ZM^2$
such that
\[
  \forall \h \in \mathcal{I} \cap [0,\h_0] \qquad \bar{\ell}_\h^{(2)}
  = A \circ \bar{\ell}_\h^{(1)} + \kappa_\h \quad \text{ on }
  \mathcal{L}_\h \cap \tilde{B}.
\]
This result does not require the continuity property mentioned above;
therefore, it is still valid in the context of Berezin-Toeplitz
operators.

\begin{rema}
  \label{rema:change_A}
  For the asymptotic lattice given by the joint spectrum of a quantum
  integrable system near a regular value of the joint principal symbol
  (Theorem~\ref{theo:BS_reg}), the matrix $A$ above corresponds to a
  change of action variables, as in~\eqref{equ:change-action}.
\end{rema}

Let $(\hat{J}_\h, \hat{H}_\h)_{\h \in (0,\h_0]}$ be a semitoric proper
quantum integrable system with joint principal symbol $F$, and let
$c \in \RM^2$ be a regular value of $F$. Let $\Sigma_\h$ be the joint
spectrum of $(\hat{J}_\h, \hat{H}_\h)$, and let $B$ be a bounded,
simply connected open subset of regular values of $F$ around $c$ such
that $(\Sigma_\h,(0,\h_0],B)$ is an asymptotic lattice. By \cite[Lemma
3.34]{san-dauge-hall-rotation}, this lattice admits an asymptotic
chart $G_\h \sim G_0 + \h G_1 + \ldots$ such that $G_0: U \to \RM^2$
is of the form $G_0 = (G_0^{(1)},G_0^{(2)})$ where
\begin{equation}
  \forall (\xi_1,\xi_2) \in U \qquad \dd G_0^{(1)}(\xi_1,\xi_2) = \dd \xi_1.
  \label{equ:semitoric_G0}
\end{equation}
Such an asymptotic chart is called a \emph{semitoric asymptotic
  chart}. By \cite[Proposition 3.33]{san-dauge-hall-rotation}, there
exists an open ball $\tilde{B} \subset B$ containing $c$ such that
$(\Sigma_\h,(0,\h_0],\tilde{B})$ admits a \emph{semitoric good
  labelling}, that is, a good labelling
$\ell_\h: \lambda \mapsto (j,\ell)$ such that
\begin{displaymath}
  J_{j,\ell}(\h) = \alpha_0 + \h (j + \alpha_1 + \O(\lambda - c)) + \O(\h^2)
  \label{equ:semitoric_labelling}
\end{displaymath}
uniformly for
$\lambda = (J_{j,\ell}(\h),E_{j,\ell}(\hbar)) \in \Sigma_\h \cap
\tilde{B}$, with $\alpha_0, \alpha_1 \in \RM$. As usual, a
\emph{semitoric linear labelling} will be a labelling that differs
from a semitoric good labelling by the translation by a vector
$\kappa_\h\in \ZM^2$. The proof of these results only uses general
properties of asymptotic lattices and asymptotic charts, so they are
also valid for Berezin-Toeplitz operators.

Given an asymptotic lattice $(\mathcal{L}_{\h}, \mathcal{I}, B)$ and a
decreasing sequence $(\h_n)_{n \geq 1}$ of elements of $\mathcal{I}$
converging to 0, the algorithm described in \cite[Section
3.5]{san-dauge-hall-rotation} (and more specifically \cite[Theorem
3.48]{san-dauge-hall-rotation}) produces a linear labelling of the
asymptotic lattice $(\mathcal{L}_\h,\{\h_n, n \geq 1\},B)$. Let us
describe informally how this works, referring
to~\cite{san-dauge-hall-rotation} for details. The result is actually
the combination of two algorithms, which we call here ``Algorithm 1''
and ``Algorithm 2''.

Algorithm 1 from \cite[Section 3.5.1]{san-dauge-hall-rotation}
works for any fixed value of $\h$. It consists first in selecting an
affine basis of the asymptotic lattice, which is a triple
$(\lambda_{(0,0)}, \lambda_{(1,0)}, \lambda_{(0,1)})$ of points of
$\mathcal{L}_\h$ corresponding, through any (unknown) asymptotic
chart, to an affine basis of $\ZM^2$. Then, it uses a ``discrete
parallel transport'' along the directions
$v_1:=\lambda_{(1,0)} - \lambda_{(0,0)}$ and
$v_2:=\lambda_{(0,1)} - \lambda_{(0,0)}$, to label all points, in a
possibly smaller open set $B'\subset B$, as $\lambda_{(n,m)}$. This
parallel transport by definition has to coincide with the usual
addition on $\ZM^2$ on the chart side, provided we use small enough
charts with small enough values of $\h$.

Algorithm 2 from \cite[Section 3.5.3]{san-dauge-hall-rotation}
works with a given sequence $(\h_n)_{n\geq 1}$, converging to zero. It
consists in a post-correction of Algorithm 1 in order to make all
choices ``continuous with respect to $\h$''. In general, Algorithm 1
will produce discontinuous labellings, and only through Algorithm 2
can one ensure that the result will be a correct linear labelling;
see~\cite[Theorem 3.48]{san-dauge-hall-rotation}.

In the semitoric case, the specialization of Algorithm 1
  indicated in~\cite[Section 3.5.2]{san-dauge-hall-rotation} ensures
  that the produced linear labelling is semitoric. 

\subsection{Asymptotic half-lattices}
\label{sec:asympt-half-latt}

Presenting the joint spectrum $\Sigma_\h$ of a proper quantum
integrable system as an asymptotic lattice, as above, will be
instrumental in recovering symplectic invariants defined near a
regular value of the momentum map.  However, in order to recover the
polygonal invariant (Section \ref{subsect:polygon}), we will also need
to work in a neighborhood of a critical value of elliptic-transverse
type. In this region, the joint spectrum is not an asymptotic lattice
anymore, but rather an asymptotic half-lattice, which, roughly
speaking, is a deformation of $\h(\ZM \times \NM)$ in a bounded
domain. This motivates the following definition, a simple adaptation
of Definition \ref{defi:asymp_latt}.

\begin{defi}
  \label{defi:asymp_half_latt}
  An \emph{asymptotic half-lattice} is the data of a triple
  $(\mathcal{L}_\h,\mathcal{I},B)$ where $\mathcal{I} \subset \RM_+^*$
  is a set of positive real numbers for which zero is an accumulation
  point, $B \subset \RM^2$ is a simply connected bounded open set and
  $\h \in \mathcal{I} \mapsto \mathcal{L}_\h \subset B$ is a family of
  discrete sets, such that
  \begin{enumerate}
  \item there exist $\h_0 > 0$, $\epsilon_0 > 0$ and $N_0 \geq 1$ such
    that for all $\h \in \mathcal{I} \cap ]0,\hbar_0]$
    \[
      \h^{-N_0} \min_{\substack{(\lambda,\mu) \in \mathcal{L}_\h^2 \\
          \lambda \neq \mu}} \| \lambda - \mu \| \geq \epsilon_0,
    \]
  \item there exist a bounded open set $U \subset \RM^2$ and a family
    of smooth maps $G_{\hbar}: U \to \RM^2$, such that
    \begin{itemize}
    \item there exist functions
      $G_0, G_1, G_2, \ldots \in C^{\infty}(U,\RM^2)$ such that $G_\h$
      has the asymptotic expansion
      \begin{equation} G_\h = G_0 + \h G_1 + \h^2 G_2 +
        \ldots \label{equ:asymp_chart_half}\end{equation} for the
      $C^{\infty}$ topology on $U$,
    \item $G_0$ is an orientation preserving diffeomorphism from $U$
      to a neighborhood of $\bar{B}$,
    \item $G_0^{-1}(B) \subset U$ is a convex set containing a point
      of the form $(x,0)$ for some $x \in \RM$,
    \item
      $G_\h(\h (\ZM \times \NM) \cap U) = \mathcal{L}_\h +
      \O(\h^{\infty})$ inside $B$, which means that there exists a
      sequence $(C_N)_{N \geq 0}$ of positive numbers such that
      \begin{itemize}
      \item for all $\h \in \mathcal{I}$, for all
        $\lambda \in \mathcal{L}_\h$, there exists
        $\ell \in \ZM \times \NM$ such that $\h \ell \in U$ and
        \begin{equation} \forall N \geq 0 \qquad \| \lambda - G_\h(\h
          \ell) \| \leq C_N
          \h^N \label{equ:dist_hN_half}\end{equation}
      \item for every open set $U_0 \Subset G_0^{-1}(B)$, there exists
        $\h_1 > 0$ such that for all
        $\h \in \mathcal{I} \cap ]0,\h_1]$, for all
        $\ell \in \ZM \times \NM$ such that $\h \ell \in U_0 $, there
        exists $\lambda \in \mathcal{L}_\h$ such that Equation
        \eqref{equ:dist_hN_half} holds;
      \end{itemize}
    \end{itemize}
    as before, the pair $(G_\h,U)$ is called an \emph{asymptotic
      chart} for $(\mathcal{L}_\h,\mathcal{I},B)$.
  \end{enumerate}
\end{defi}

\begin{figure}[H]
  \begin{center}
    \includegraphics[width=0.7\linewidth]{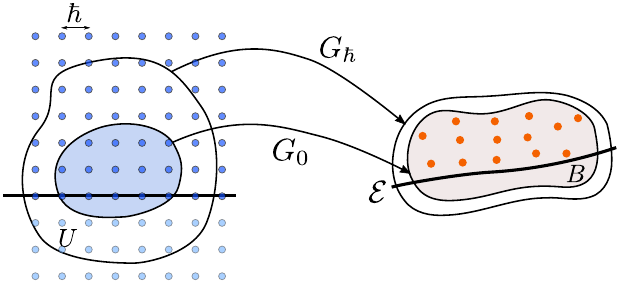}
  \end{center}
  \caption{\small An example of asymptotic half-lattice.}
  \label{fig:half_lattice}
\end{figure}

For $\h$ small enough, $G_\h$ is a diffeomorphism onto its image;
hence the image by $G_\h$ of the line segment
$\{y=0\}\cap G_0^{-1}(B)$ is a smooth curve that separates $B$ into
two connected components.  The asymptotic half-lattice is, modulo an
error of size $\O(\h^\infty)$, contained in one of these
components. In fact this curve converges when $\h \to 0$ to
$\mathcal{E} = G_0(\{y=0\} \cap G_0^{-1}(B))$.
\begin{defi}
  \label{defi:boundary_half}
  We call $\mathcal{E}$ the \emph{boundary} of the asymptotic
  half-lattice $(\mathcal{L}_\h,\mathcal{I},B)$.
\end{defi}
This boundary is defined intrinsically since it coincides with the
topological boundary in $B$ of the set of accumulation points of
$(\mathcal{L}_\h)_{\h \in \mathcal{I}}$. Since $U$ is convex, the
boundary $\mathcal{E}$ is connected. See Figure
\ref{fig:half_lattice}.

Let $(\hat{J}_\h, \hat{H}_\h)$ be a proper quantum integrable system
with joint principal symbol $F = (J,H)$. Let $c = (c_1,c_2)$ be a
$J$-transversally elliptic critical value of $F$: this is a critical
value of elliptic-transverse type of $F$ such that $c_1$ is a regular
value of $J$ and $c_2$ is a non-degenerate critical value of $H$
restricted to the level set $J^{-1}(c_1)$. Assume that $F^{-1}(c)$ is
connected.

By Theorem \ref{theo:BS_ell_trans}, the joint spectrum of
$(\hat{J}_\h, \hat{H}_\h)$ near $c$ is an asymptotic half-lattice,
whose boundary is the boundary of $F(M)$, with asymptotic chart
$G_\h = G_0 + \h G_1 + \ldots$ where $G_0$ is such that
\[
  (F \circ \phi^{-1})(\theta_1,\xi_1,x_2,\xi_2) = G_0(\xi_1,
  q(x_2,\xi_2))
\]
where $(\theta_1,\xi_1,x_2,\xi_2)$ are coordinates on
$T^*S^1 \times T^*\RM$ endowed with the symplectic form
$\omega_0 = d\xi_1 \wedge d\theta_1 + d\xi_2 \wedge dx_2$, $\phi$ is a
symplectomorphism from a neighborhood of $F^{-1}(c)$ in $M$ to a
neighborhood of the zero section times $T^*\RM$ in
$T^* S^1 \times T^*\RM$, and
$q(x_2, \xi_2) = \frac{1}{2} (x_2^2 + \xi_2^2)$. While this statement,
which was stated without proof in \cite[Theorem
3.38]{san-dauge-hall-rotation}, is sometimes considered ``well known''
(at least for $\h$-pseudodifferential operators), we couldn't find a
proof in the literature; hence we devote Section \ref{sect:BScrit} to
filling this gap.

In view of the inverse problem, we will need to label these asymptotic
half-lattices. Hence we have to show that they admit a labelling, and
to give an algorithm to obtain such a labelling.
 
\begin{defi}
  \label{defi:good_lab_half}
  Let $(\mathcal{L}_\h,\mathcal{I},B)$ be an asymptotic half-lattice
  with asymptotic chart $(G_\h,U)$. A \emph{good labelling} of
  $(\mathcal{L}_\h,\mathcal{I},B)$ is a family of maps
  $\ell_\h: \mathcal{L}_\h \to \ZM^2$, $\h \in \mathcal{I}$, such that
  for every $\lambda \in \mathcal{L}_\h$, $\h \ell_\h(\lambda) \in U$
  and
  \[
    \forall N \geq 0 \qquad \| G_\h (\h \ell_\h(\lambda)) - \lambda \|
    \leq C_N \h^N
  \]
  where $(C_N)_{N \geq 0}$ is as in Definition
  \ref{defi:asymp_half_latt}.
\end{defi}

This definition is similar to Definition \ref{defi:good_lab_lattice},
but there is an important difference. Because the labels along the
boundary are of the form $(m,0)$, $m \in \ZM$, there can only be a
drift (see \cite[Definition 3.25]{san-dauge-hall-rotation}) in the
horizontal direction. The proof of the following result is similar to
the proof of Lemma 3.11 in \cite{san-dauge-hall-rotation}.

\begin{prop}
  \label{prop:good_labelling_half}
  Let $(\mathcal{L}_\h,\mathcal{I},B)$ be an asymptotic half-lattice
  with asymptotic chart $(G_\h,U)$. There exists a good labelling of
  $(\mathcal{L}_\h,\mathcal{I},B)$ associated with $G_\h$.
\end{prop}

\begin{figure}[H]
  \begin{center}
    \includegraphics[width=0.4\linewidth]{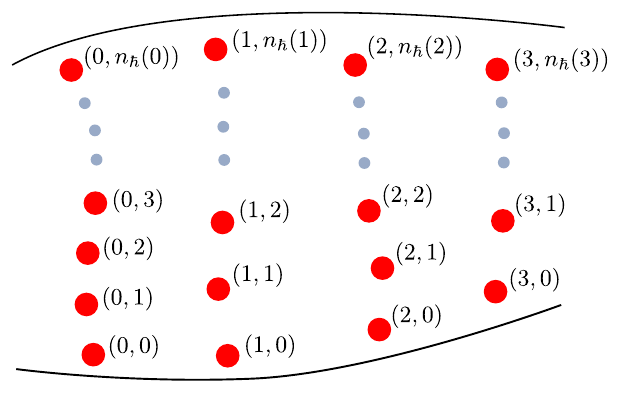}
  \end{center}
  \caption{\small An example of what would be a global labelling of a
    ``global asymptotic lattice''. The labels near the lower boundary
    correspond to a good labelling, which forces the labels near the
    upper boundary to be of the form $(m,n_\h(m))$ with $n_\h(m)$ of
    order $\O(1/\h)$, so in particular cannot constitute a good
    labelling.}
  \label{fig:problem_global_labelling}
\end{figure}

This notion of good labelling is a relevant local notion, but is not
sufficient when dealing with global situations. More precisely, it is
attached to each component of the boundary $\mathcal{E}$, and cannot
be globalised if this boundary is disconnected, see Figure
\ref{fig:problem_global_labelling}.

A good labelling is a special case of linear labelling, the definition
of which is similar to the one for asymptotic lattices. However there
is a crucial distinction: we need to relax the condition that the
labels along the boundary are of the form $(m,0)$, $m \in \ZM$. This
will be useful when constructing a ``global labelling'' on the union
of an asymptotic lattice and an asymptotic half-lattice, see Lemma
\ref{lemm:ext_label_half}.

\begin{defi}
  \label{defi:linear_lab_half}
  A \emph{linear labelling} of an asymptotic half-lattice
  $(\mathcal{L}_{\h}, \mathcal{I}, B)$ is a family of maps
  $\bar{\ell}_\h: \mathcal{L}_\h \to \ZM^2$, $h \in \mathcal{I}$ of
  the form $\bar{\ell}_\h = A \circ \ell_\h + \kappa_\h$ where
  $\ell_\h$ is a good labelling, $A \in \mathrm{SL}(2,\ZM)$ and
  $(\kappa_\h)_{\h \in \mathcal{I}}$ is a family of vectors in
  $\ZM^2$.
\end{defi}

The following analogue of \cite[Proposition
3.20]{san-dauge-hall-rotation} holds for asymptotic half-lattices.

\begin{lemm}
  \label{lemm:transition_labellings_half}
  Let $\bar{\ell}_\h^{(1)}$ and $\bar{\ell}_\h^{(2)}$ be two linear
  labellings for a given asymptotic half-lattice
  $(\mathcal{L}_{\h}, \mathcal{I}, B)$, then for any open set
  $\tilde{B} \Subset B$, there exists a unique matrix
  $A \in \mathrm{SL}(2,\ZM)$, $\h_0 > 0$ and a family
  $(\kappa_\h)_{\h \in \mathcal{I} \cap [0,\h_0]}$ of vectors in
  $\ZM^2$ such that
  \[
    \forall \h \in \mathcal{I} \cap [0,\h_0] \qquad
    \bar{\ell}_\h^{(2)} = A \circ \bar{\ell}_\h^{(1)} + \kappa_\h
    \quad \text{ on } \mathcal{L}_\h \cap \tilde{B}.
  \]
\end{lemm}

\begin{demo}
  The proof of the analogous result for asymptotic lattices can be
  adapted by following the same strategy as in the proof of
  \cite[Proposition 3.20]{san-dauge-hall-rotation}. We construct two
  affine bases which are adapted to the boundary of the half-lattice
  by first choosing $\lambda_0 \in \mathcal{L}_\h$ so that
  $\lambda_0 = G_\h^{(1)}(n_1,0) = G_\h^{(2)}(n_2,0)$ for some
  $n_1, n_2 \in \ZM$, and then by considering the images of the
  canonical basis of $\RM^2$ by the two labellings. Then as in the
  aforementioned proof, the action of $\{-1,0,1\}^2$ is transitive,
  and the analogue of \cite[Lemma 3.18]{san-dauge-hall-rotation},
  which still holds in this context, allows us to conclude.
\end{demo}

Similarly to the case of usual asymptotic lattices, we define
\emph{semitoric} asymptotic half-lattices by
enforcing~\eqref{equ:semitoric_G0}. A consequence of this restriction
is that we now have to distinguish between ``upper'' half-lattices and
``lower'' half-lattices: the current
Definition~\ref{defi:asymp_half_latt} only deals with ``upper''
half-lattices, while ``lower'' half-lattices need either replacing
$\NM$ in that definition by $\ZM_-$, or requiring $G_0$ to be
orientation reversing (because switching to the ``upper'' case amounts
to composing by $(x,y)\mapsto(x,-y)$). Since these modifications are
rather obvious, for the sake of simplicity we shall discuss only the
``upper'' case.

Let $(\hat{J}_\h, \hat{H}_\h)$ be a proper semitoric quantum
integrable system, and let $c=(c_1,c_2)$ be a $J$-transversally
elliptic critical value of the underlying integrable system $(J,H)$.

Let $(\Sigma_\h, \mathcal{I}, B)$ be the semitoric asymptotic
half-lattice formed by the joint spectrum of
$(\hat{J}_\h, \hat{H}_\h)$ where $B \subset \RM^2$ is a neighborhood
of $c$. We propose here an algorithm to construct a linear semitoric
labelling of this joint spectrum (it would also be interesting to have
an algorithm for general asymptotic half-lattices, but in this work we
only need the semitoric case). Our algorithm proceeds as follows.

\begin{algo}
  \label{algo:algo_labelling_half}
  First, choose an open subset $B_0 \Subset B$ containing $c$.  Then,
  for any given $\h$, follow the steps below.
  \begin{enumerate}
  \item \label{step:mu} Choose $\mu$, an element of $\Sigma_\h$ with
    minimal Euclidean distance to $c$. This element is not necessarily
    unique.

  \item \label{step:00} Consider the vertical strip $\mathcal{S}_0$ of
    width $\h^{\frac{3}{2}}$ around $\mu$. Let
    $\lambda_{(0,0)} \in \Sigma_\h$ be an element with lowest ordinate
    in that strip. Such an element always exists and, once $\mu$ is
    chosen (which we assume at this step), $\lambda_{(0,0)}$ is unique
    if $\h$ is small enough.
  
  \item \label{step:01} Let
    $\lambda_{(0,1)} \in \Sigma_\h\cap \mathcal{S}_0$ be the (unique
    if $\h$ is small enough) nearest point to $\lambda_{(0,0)}$
    located above $\lambda_{(0,0)}$.

  \item \label{step:10} Consider now the translated strip
    $\mathcal{S}_1:= \mathcal{S}_0+(\h,0)$, and choose an element
    $\lambda_{(1,0)} \in \Sigma_\h\cap \mathcal{S}_1$ with lowest
    ordinate.
  
  \item Given the triple
    $(\lambda_{(0,0)}, \lambda_{(1,0)}, \lambda_{(0,1)})$ (which, for
    $\h$ small enough, will be an affine basis of the asymptotic
    lattice), we complete the labelling $\lambda_{n,m}$ as in the
    usual algorithm, but restricting to $m\geq 0$ (thus, we skip steps
    10, 11, and 12 of that algorithm).
  \end{enumerate}

\end{algo}

\begin{figure}[H]
  \begin{center}
    \includegraphics[width=0.7\linewidth]{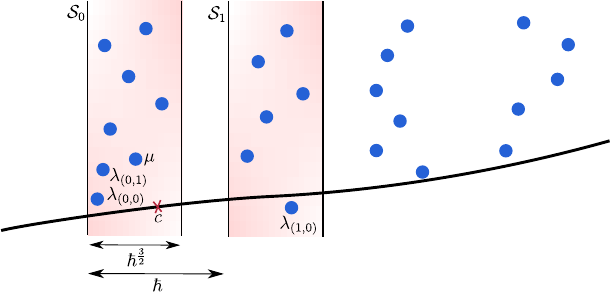}
  \end{center}
  \caption{\small Illustration of the first few steps of Algorithm \ref{algo:algo_labelling_half}. The elements of $\Sigma_\h$ are represented by blue dots.}
  \label{fig:semitoric_algorithm}
\end{figure}

Notice that, contrary to the way the general algorithm
from~\cite{san-dauge-hall-rotation} works, in the semitoric case it
makes more sense to label ``vertically'', that is first obtain all the
labels $(0,m)$, then all the labels $(1,m)$, and so on. An interesting
feature of this algorithm, compared to the algorithm for asymptotic
lattices given in~\cite{san-dauge-hall-rotation}, is that it does not
necessitate a second, correcting, algorithm; thanks to the presence of
the boundary, all steps (but the first one) have unique solutions for
$\h$ small enough.

\begin{prop}
  \label{prop:algo_half_ok}
  The algorithm above produces a linear labelling of
  $(\Sigma_\h, \mathcal{I}, B)$ associated with a semitoric asymptotic
  chart $(G_\h, U)$, that is an asymptotic chart such that the first
  component $G_0^{(1)}$ of $G_0 = (G_0^{(1)},G_0^{(2)})$ satisfies
  ${\rm d} G_0^{(1)} = {\rm d} \xi_1$.
\end{prop}

\begin{rema}
  This linear labelling, call it $\ell_\h$, has the nice property that
  the eigenvalues which are the closest to the line of critical values
  (which is the boundary of the asymptotic half-lattice
  $(\Sigma_\h, \mathcal{I}, B)$) are labelled as $(n,0)$ with
  $n \in \ZM$. In other words, the only matrix $A$ such that
  $A \circ \ell_\h + \kappa_\h$ is good for some $\kappa_\h \in \ZM^2$
  (see Definition \ref{defi:linear_lab_half}) is the identity.
\end{rema}

\begin{demo}
  From~\cite[Proposition 3.37]{san-dauge-hall-rotation} we know that
  the joint eigenvalues $\lambda$ in a small neighborhood of $c$ are
  contained in a union of vertical strips $V_j$ given by the equation
  \[
    x = \alpha_0 + \h (j + \alpha_1 + \O(\lambda-c)) + \O(\h^2), \quad
    j \in\ZM,
  \]
  where $\alpha_0,\alpha_1\in\RM$ are fixed. Let $V_{j_0}$ be the
  strip containing the point $\mu$ of Step~\ref{step:mu} (of course,
  $j_0$ depends on $\h$). By~\cite[Proposition
  3.37]{san-dauge-hall-rotation} the eigenvalues in each strip have,
  for $\h$ small enough, pairwise distinct ordinates, and we may
  choose the unique lowest one $\lambda_{(0,0)}$ (Step~\ref{step:00},
  with $\mathcal{S}_0\subset V_{j_0}$), and the next lowest one
  $\lambda_{(0,1)}$ (Step~\ref{step:01}).  Since
  $\mathcal{S}_0+(\h,0)\subset V_{j_0+1}$, Step~\ref{step:10}
  similarly defines a unique element $\lambda_{(1,0)}$.
  
  In order to show that the algorithm constructs a linear labelling,
  we use some details of the proof of~\cite[Proposition
  3.37]{san-dauge-hall-rotation}. In particular, there exists an
  asymptotic chart $G_\h$ for the asymptotic half-lattice $\Sigma_\h$
  such that
  \[
    G_0(\xi_1,\xi_2) = (\xi_1+\alpha_0, f(\xi_1,\xi_2)), \quad
    \partial_{\xi_2} f >0.
  \]
  Let $(\ell_1,\ell_2)$ be the good labelling associated with
  $G_\h$. The image by $G_\h$ of $\{\h \ell_1\}\times \h\NM$
  (restricted to its domain of definition, of course) is contained in
  one of the strips $V_j$, hence, up to a constant
  $\kappa_1(\h)\in\ZM$, we must have, for each joint eigenvalue
  $\lambda\in V_j$, $\ell_1(\lambda) = j + \kappa_1(\h)$. Since
  $ \partial_y f >0$, the joint eigenvalue with label
  $(\ell_1,\ell_2=0)$ is the lowest of its strip $V_j$ and hence must
  coincide with $\lambda_{(0,0)}$ when $j=j_0$, and with
  $\lambda_{(1,0)}$ when $j=j_0+1$. Similarly, the labels of
  $\lambda_{(0,1)}$ must be $\ell_1=j_0+\kappa_1(\h)$,
  $\ell_2=1$. This shows that the triple
  $(\lambda_{(0,0)}, \lambda_{(1,0)}, \lambda_{(0,1)})$ is an affine
  basis of $\Sigma_\h$, and hence, by parallel
  transport~\cite[Proposition 3.17]{san-dauge-hall-rotation}, the
  labelling $\lambda_{(n,m)}\mapsto(n,m)$ of the algorithm must
  coincide with the linear labelling
  \[
    \lambda \mapsto (\ell_1(\lambda) + \kappa_1(\h), \ell_2(\lambda)).
  \]
\end{demo}

\begin{rema}
  \label{rema:position-c}
  One could argue that one does not know \emph{a priori} how to choose
  a singular value $c$. But $c$ was used to simplify the presentation,
  and actually its knowledge is not necessary, since the position of a
  transversally-elliptic value can be obtained up to $\O(\h)$ by
  considering any point in the half-lattice and by finding a point
  with minimal ordinate in a strip of width $\h^{2/3}$ around this
  point.
\end{rema}

\begin{rema}
  There are two other situations regarding elliptic
    singularities that can occur to an integrable system on a
  four-dimensional manifold.
  \begin{itemize}
  \item The image of the momentum map $F = (J,H)$ could display
    so-called vertical walls, which correspond to images of
    $H$-transversally elliptic critical values of $F$. In the case of
    a semitoric system $(J,H)$, such a vertical wall can only appear
    at a global minimum or maximum of $J$. It turns out that, although
    we will have to deal with these vertical walls later on, we will
    avoid describing the structure of the joint spectrum of
    $(\hat{J}_\h, \hat{H}_\h)$ near any of their points. Nevertheless,
    this joint spectrum simply forms a ``vertical half-lattice''.
  \item The image of $F$ may also display ``corners'' where two lines
    of transversally elliptic critical values intersect, corresponding
    to images of singularities of $F$ of elliptic-elliptic
    type. Again, we will explain below (see Section
    \ref{subsect:polygon}) why we do not need to understand the
    structure of the joint spectrum near such a point. This joint
    spectrum is neither an asymptotic lattice nor an asymptotic
    half-lattice, but rather an ``asymptotic quarter-lattice''
    modelled on $\NM \times \NM$. In the setting of homogeneous
    pseudodifferential operators, this was the situation studied
    in~\cite{colinII}.
  \end{itemize}
\end{rema}

\subsection{Extension of an asymptotic lattice}

An important property of asymptotic lattices, which will be key in
reconstructing the polygon invariant from the joint spectrum of a
quantum semitoric system, is that they behave like flat sheaves.

\begin{lemm}[restriction of asymptotic lattices]
  \label{lemm:asla-restriction}
  If $(\mathcal{L}_\h,\mathcal{I},B)$ is an asymptotic lattice, and
  $\tilde B \subset B$ is a simply connected open subset of $B$, then
  $(\mathcal{L}_\h\cap \tilde B,\mathcal{I},\tilde B)$ is also an
  asymptotic lattice. Moreover, if $\ell_\h$ is a good (respectively
  linear) labelling for $\mathcal{L}_\h$, then the restriction of
  ${\ell}_\h$ to $\tilde{\mathcal{L}}_\h$ is a good (respectively
  linear) labelling for $\tilde{\mathcal{L}}_\h$.
\end{lemm}
\begin{demo}
  We check the various items of Definition~\ref{defi:asymp_latt}.
  Item~\ref{item:asla-separation} is automatically inherited if we
  replace $\mathcal{L}_\h$ by
  $\tilde{\mathcal{L}}_\h := \mathcal{L}_\h\cap \tilde B$. Concerning
  item~\ref{item:asla-chart}, we claim that the same chart $G_\h$
  (\emph{i.e} with domain $\tilde U:= U$) is valid: it suffices to
  check the last property stated below~\eqref{equ:dist_hN}. If
  $\tilde U_0\subset G_0^{-1}(\tilde B)$ is given, since
  $\tilde U_0 \subset U_0$, by assumption we find a corresponding
  $\lambda\in\mathcal{L}_\h$. We also know that
  $\lambda \in G_\h(\tilde U_0) + \O(\h^\infty) \subset G_0(\tilde
  U_0) + \O(\h)\subset \check B+\O(\h)$, for some
  $\check B\Subset\tilde B$. Hence if $\h_1$ is small enough, for all
  $\h\leq \h_1$, $\lambda \in \tilde B$.
\end{demo}

We now prove the unique extension property (which is related to the
parallel transport of~\cite{san-dauge-hall-rotation}).
\begin{lemm}
  \label{lemm:asla-extension}
  Let $(\mathcal{L}_\h,\mathcal{I},B)$ be an asymptotic lattice. Let
  $\tilde B \subset B$ such that $\tilde{B}$ is simply
  connected. Given any linear labelling $\tilde{\ell}_\h$ for the
  asymptotic lattice
  $(\tilde{\mathcal{L}}_\h = \mathcal{L}_\h \cap
  \tilde{B},\mathcal{I},\tilde B)$, there exists a linear labelling
  $\ell_\h$ for $(\mathcal{L}_\h,\mathcal{I},B)$ which agrees with
  $\tilde{\ell}_\h$ on $\mathcal{L}_\h \cap \check{B}$ for every
  $\check{B} \Subset \tilde{B}$. Moreover, for any $\hat{B} \Subset B$
  containing $\tilde{B}$, the restriction of $\ell_\h$ to
  $\mathcal{L}_\h \cap \hat{B}$ is unique for $\h$ small
  enough. Furthermore, if $\tilde{\ell}_\h$ is a good labelling, then
  $\ell_\h$ is a good labelling as well; in that case, if $G_\h$ is an
  asymptotic chart associated with ${\ell}_\h$, and $\tilde G_\h$ is
  an asymptotic chart associated with $\tilde{\ell}_\h$, then
  $G_0^{-1} = \tilde G_0^{-1}$ on $\tilde B$.
\end{lemm}

\begin{demo}
  If $\tilde{B} = B$ or $\tilde{B} = \emptyset$, the statement is
  trivial, so from now on we assume that
  $\emptyset \subsetneq \tilde{B} \subsetneq B$. We start with the
  uniqueness statement. Let $\check{B} \Subset \tilde{B}$ and let
  $\ell_\h^{(1)}$, $\ell_\h^{(2)}$ be two linear labellings agreeing
  with $\tilde{\ell}_\h$ on $\mathcal{L}_\h \cap \check{B}$. Let
  $\hat{B} \Subset B$ containing $\tilde{B}$; then by
  \cite[Proposition 3.20]{san-dauge-hall-rotation}, there exists a
  unique matrix $A \in \mathrm{SL}(2,\ZM)$, $\h_0 > 0$ and a unique
  family $(\kappa_\h)_{\h \in \mathcal{I} \cap [0,\h_0]}$ of vectors
  in $\ZM^2$ such that
  \[
    \forall \h \in \mathcal{I} \cap [0,\h_0] \qquad \ell_\h^{(1)} = A
    \circ \ell_\h^{(2)} + \kappa_\h \quad \text{ on } \mathcal{L}_\h
    \cap \hat{B}.
  \]
  Since $\ell_\h^{(1)}$ and $\ell_\h^{(2)}$ agree on
  $\tilde{\mathcal{L}}_\h$, necessarily $A = \mathrm{Id}$ and
  $\kappa_\h = 0$ (as long as $\tilde{\mathcal{L}}_\h$ contains three
  elements whose images by $\ell_\h^{(1)}$ form an affine basis of
  $\ZM^2$, which is true for $\h$ small enough) and
  $\ell_\h^{(1)} = \ell_\h^{(2)}$ on $\mathcal{L}_\h \cap \hat{B}$.
 
  For the existence part, note that by \cite[Lemma
  3.11]{san-dauge-hall-rotation}, there exists a linear labelling
  $\check{\ell}_\h$ for $(\mathcal{L}_\h,\mathcal{I},B)$. Then the
  restriction of $\check{\ell}_\h$ to $\tilde{\mathcal{L}}_\h$ is a
  linear labelling for
  $(\tilde{\mathcal{L}}_\h,\mathcal{I},\tilde B)$. Hence by
  \cite[Proposition 3.20]{san-dauge-hall-rotation} again, for any
  $\check{B} \Subset \tilde B$, there exists a unique matrix
  $C \in \mathrm{SL}(2,\ZM)$, $\h_1 > 0$ and a unique family
  $(\nu_\h)_{\h \in \mathcal{I} \cap [0,\h_1]}$ of vectors in $\ZM^2$
  such that
  \[
    \forall \h \in \mathcal{I} \cap [0,\h_1] \qquad \check{\ell}_\h =
    C \circ \tilde{\ell}_\h + \nu_\h \quad \text{ on }
    \tilde{\mathcal{L}}_\h \cap \check{B}.
  \]
  Note that by the uniqueness statement, the matrix $C$ does not
  depend on $\check{B}$ as long as $\check{B} \neq \emptyset$.

  Assume that $\tilde{\ell}_\h$ is a good labelling, and let
  $\tilde{G}_\h$ be the associated asymptotic chart. Since $\ell_\h$
  is a linear labelling, there exists a family $(\kappa_\h)$ of
  vectors in $\ZM^2$ such that $\ell_\h + \kappa_\h$ is a good
  labelling. Let $\hat{G}_\h$ be the asymptotic chart associated with
  this good labelling. It follows from the proof of \cite[Proposition
  3.20]{san-dauge-hall-rotation} that
  $\dd {\hat{G}_\h}^{-1} = \dd {\tilde{G}_\h}^{-1} + \O(\h^{\infty})$
  on $\check{B}$. Hence there exists a family
  $(\nu_\h)_{\h \in \mathcal{I}}$ of elements of $\RM^2$ with an
  asymptotic expansion in non negative powers of $\h$ such that
  $\hat{G}_\h^{-1} = {\tilde{G}_\h}^{-1} + \nu_\h + \O(\h^{\infty})$
  on $\check{B}$. Since $\ell_\h = \tilde{\ell}_\h$ on
  $\mathcal{L}_\h \cap \check{B}$, using Equation \eqref{equ:dist_hN}
  then yields $\h \kappa_\h = - \nu_\h + \O(\h^{\infty})$. Now, let
  $G_\h: \xi \mapsto \hat{G}_\h(\xi + \nu_\h)$; then $G_\h$ is an
  asymptotic chart for $\mathcal{L}_\h$ and the corresponding good
  labelling coincides with $\tilde{\ell}_\h$ on
  $\mathcal{L}_\h \cap \check{B}$.
\end{demo}

\begin{lemm}
  \label{lemm:asla-glueing}
  Let $(\mathcal{L}^{(1)}_\h\!,\mathcal{I},B_1)$ and
  $(\mathcal{L}^{(2)}_\h\!,\mathcal{I},B_2)$ be two asymptotic
  lattices, sharing the same parameter set $\mathcal{I}$. Assume that
  $B_1\cap B_2$ is simply connected and non empty, and that
  \[
    \forall \h\in \mathcal{I}, \quad \mathcal{L}^{(1)}_\h \cap B_2 =
    \mathcal{L}^{(2)}_\h\cap B_1.
  \]
  Then for any simply connected open set $B\Subset B_1\cup B_2$,
  $\left((\mathcal{L}^{(1)}_\h \cup \mathcal{L}^{(2)}_\h)\cap B,
    \mathcal{I},B\right)$ is an asymptotic lattice.
\end{lemm}
	
\begin{demo} 
  First note that, since $B_1, B_2$ and $B_1\cap B_2$ are open and
  connected, they are path connected, and the Seifert-van Kampen
  theorem implies that $B_1\cup B_2$ is also simply connected.  Since
  $\mathcal{L}^{(1)}_\h \cap B_2 = \mathcal{L}^{(1)}_\h \cap B_1\cap
  B_2$, we may apply Lemma~\ref{lemm:asla-restriction} to conclude
  that $(\mathcal{L}^{(1)}_\h \cap B_2, \mathcal{I}, B_1\cap B_2)$ is
  an asymptotic lattice. Let $\tilde{\ell}_\h$ be a good labelling for
  it. Let $\tilde{B}_1 \Subset B_1$ and $\tilde{B}_2 \Subset B_2$ such
  that
  $\bar{B} \subset \tilde{B} = \tilde{B}_1 \cup \tilde{B}_2 \Subset
  B_1 \cup B_2$, and let $W = \tilde{B}_1 \cap \tilde{B}_2$. By
  Lemma~\ref{lemm:asla-extension}, we construct a good labelling
  $\ell^{(1)}_\h$ for $\mathcal{L}_\h^{(1)}$ which coincides with
  $\tilde{\ell}_\h$ on $\mathcal{L}_\h^{(1)} \cap W$. Similarly, we
  construct a good labelling $\ell^{(2)}_\h$ for
  $\mathcal{L}_\h^{(2)}$ which coincides with $\tilde{\ell}_\h$ on
  $\mathcal{L}_\h^{(2)} \cap W$. We may now define the map
  $\ell_\h: (\mathcal{L}^{(1)}_\h \cup \mathcal{L}^{(2)}_\h) \cap
  \tilde{B} \to \ZM$, for all $\h\in\mathcal{I}$, by
  \[
    \ell_\h =
    \begin{cases}
      \ell^{(1)}_\h & \text{ on } \mathcal{L}^{(1)}_\h \cap \tilde{B}_1,\\[2mm]
      \ell^{(2)}_\h & \text{ on } \mathcal{L}^{(2)}_\h \cap
      \tilde{B}_2.
    \end{cases}
  \]
  The labellings $\ell^{(j)}_\h$, $j=1,2$ are associated with
  asymptotic charts $G_\h^{(j)}$, defined on open sets $U_j$. By
  uniqueness of the asymptotic chart associated with a good labelling,
  $G_\h^{(1)}= G_\h^{(2)} + \O(\h^\infty)$ on
  $V = (G_0^{(1)})^{-1}(W)$ (recall that
  $(G_0^{(1)})^{-1}= (G_0^{(2)})^{-1}$ on
  $\tilde{B}_1 \cap \tilde{B}_2$). Hence, $G_\h^{(1)}$ and
  $G_\h^{(2)}$ share the same asymptotic expansion on $V$. We define
  the family $(G_\h)_{\h\in\mathcal{I}}$ on $U:=V_1\cup V_2$, where
  $V_j:=(G_0^{(j)})^{-1}(\tilde{B}_j)$ by gluing the asymptotic
  expansions of $G_\h^{(1)}$ and $G_\h^{(2)}$ and applying a Borel
  summation. It remains to prove that the principal term $G_0$ is a
  diffeomorphism into its image $G_0(U) = \tilde{B}$. Since it is a
  local diffeomorphism, we just need to show injectivity. Let $\xi_1$,
  $\xi_2 \in U$ be such that $G_0(\xi_1) = G_0(\xi_2)$. Notice that
  $V_j \subset U_j$, and we know that $G_0$ is the restriction of
  $G_0^{(j)}$ on that subset, and hence is injective there. Hence we
  may assume that $\xi_j \in V_j$, for $j=1,2$. Hence
  $G_0(\xi_j) \in \tilde{B}_j$; therefore, for $j=1,2$,
  $G_0(\xi_j)\in \tilde{B}_1 \cap \tilde{B}_2$. Hence $\xi_j\in V$,
  which is contained in, for instance, $V_1$, and we can conclude by
  the injectivity of $G_0$ there, that $\xi_1 = \xi_2$.
 
\end{demo}

\subsection{Extension of an asymptotic half-lattice}

\begin{figure}[H]
  \begin{center}
    \includegraphics{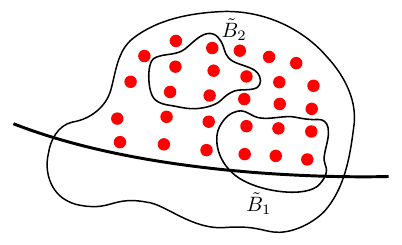}
  \end{center}
  \caption{\small The restriction of this asymptotic half-lattice to
    $\tilde B_1$ will be an asymptotic half-lattice, whereas its
    restriction to $\tilde B_2$ will be an asymptotic lattice.}
  \label{fig:half_lattice_restriction}
\end{figure}

We need similar statements for asymptotic half-lattices; but
additional difficulties appear. For instance, in the following
results, which are the analogues of Lemma \ref{lemm:asla-restriction},
we must take into account the fact that the restriction of an
asymptotic half-lattice can be either an asymptotic lattice or an
asymptotic half-lattice, see Figure
\ref{fig:half_lattice_restriction}.

\begin{lemm}
  \label{lemm:restriction_half_lattice_far}
  Let $(\mathcal{L}_\h,\mathcal{I},B)$ be an asymptotic half-lattice,
  and let $(\ell_\h)_{\h \in \mathcal{I}}$ be a good (respectively
  linear) labelling for $(\mathcal{L}_\h,\mathcal{I},B)$. Let
  $\tilde{B} \subset \mathrm{int}(\underline{\mathcal{L}_\h})$ be any
  simply connected open set, where $\underline{\mathcal{L}_\h}$ is the
  set of accumulation points of
  $\bigcup_{\h \in \mathcal{I}} \mathcal{L}_\h$ in $B$. Then
  $(\mathcal{L}_\h \cap \tilde{B},\mathcal{I},\tilde{B})$ is an
  asymptotic lattice, and the restriction of
  $(\ell_\h)_{\h \in \mathcal{I}}$ to $\mathcal{L}_\h \cap \tilde{B}$
  is a good (respectively linear) labelling for
  $(\mathcal{L}_\h \cap \tilde{B},\mathcal{I},\tilde{B})$.
\end{lemm}

In the case of the restriction to a subset intersecting the boundary
of an asymptotic half-lattice, we need to be a little bit more
careful.

\begin{defi}
  Let $(\mathcal{L}_\h,\mathcal{I},B)$ be an asymptotic
  half-lattice. A set $\tilde{B} \subset B$ is called an
  \emph{admissible domain} if there exists an asymptotic chart $G_\h$
  such that $\tilde{B}$ is the image by $G_0$ of a convex set
  $K \subset G_0^{-1}(B)$ containing a point of the form $(x,0)$,
  $x \in \RM$ (hence this is true for any asymptotic chart).
\end{defi}

Note that by definition, $B$ itself is an admissible domain. The proof
of the following lemma is similar to the proof of Lemma
\ref{lemm:asla-restriction}.

\begin{lemm}
  \label{lemm:restriction_half_lattice}
  Let $(\mathcal{L}_\h,\mathcal{I},B)$ be an asymptotic half-lattice,
  and let $(\ell_\h)_{\h \in \mathcal{I}}$ be a good (respectively
  linear) labelling for $(\mathcal{L}_\h,\mathcal{I},B)$. Let
  $\tilde{B} \subset B$ be an admissible domain. Then
  $(\mathcal{L}_\h \cap \tilde{B},\mathcal{I},\tilde{B})$ is an
  asymptotic half-lattice, and the restriction of
  $(\ell_\h)_{\h \in \mathcal{I}}$ to $\mathcal{L}_\h \cap \tilde{B}$
  is a good (respectively linear) labelling for
  $(\mathcal{L}_\h \cap \tilde{B},\mathcal{I},\tilde{B})$.
\end{lemm}

\begin{lemm}
  \label{lemm:half-asla-extension}
  Let $(\mathcal{L}_\h,\mathcal{I},B)$ be an asymptotic
  half-lattice. Let $\tilde{B} \subset B$ be an admissible domain and
  let
  $(\tilde{\mathcal{L}}_\h = \mathcal{L}_\h \cap
  \tilde{B},\mathcal{I},\tilde B)$ be the corresponding asymptotic
  half-lattice. Given any linear labelling $\tilde{\ell}_\h$ for
  $(\tilde{\mathcal{L}}_\h,\mathcal{I},\tilde B)$, there exists a
  linear labelling $\ell_\h$ for $\mathcal{L}_\h$ which agrees with
  $\tilde{\ell}_\h$ on $\tilde{\mathcal{L}}_\h \cap \check{B}$ for
  every $\check{B} \Subset \tilde{B}$. Moreover, for any
  $\hat{B} \Subset B$ containing $\tilde{B}$, the restriction of
  $\ell_\h$ to $\mathcal{L}_\h \cap \hat{B}$ is unique for $\h$ small
  enough. Furthermore, if $\tilde{\ell}_\h$ is a good labelling, then
  $\ell_\h$ is a good labelling as well; in that case, if $G_\h$ is an
  asymptotic chart associated with ${\ell}_\h$, and $\tilde G_\h$ is
  an asymptotic chart associated with $\tilde{\ell}_\h$, then
  $G_0^{-1} = \tilde G_0^{-1}$ on
  $\tilde B \cap \underline{\mathcal{L}_\h}$.
\end{lemm}

\begin{demo}
  The proof is essentially the same as the proof of Lemma
  \ref{lemm:asla-extension}. When dealing with a half-lattice, one has
  to use Lemma \ref{lemm:transition_labellings_half} instead of
  \cite[Proposition 3.20]{san-dauge-hall-rotation}. However, one has
  to be careful because in that case, given two good labellings which
  coincide on $\mathcal{L}_\h \cap \check{B}$ and associated with
  asymptotic charts $G_\h$ and $\tilde{G}_\h$, the equality
  $\dd G_\h^{-1} = \dd {\tilde{G}_\h}^{-1} + \O(\h^{\infty})$ only
  holds on $\check{B}^+ = \underline{\mathcal{L}_\h} \cap
  \check{B}$. This implies that on $\check{B}$,
  $G_\h^{-1} = {\tilde{G}_\h}^{-1} + \nu_\h + \O(\h^{\infty})$ modulo
  a term which vanishes on $\check{B}^+$. But since we use this
  equality on $\mathcal{L}_\h$, which is at distance at most $\O(\h)$
  of $\check{B}^+$, the proof still works since the additional term
  only adds a $\O(\h^{\infty})$. Furthermore, there is also a slight
  difference with the aforementioned proof coming from the fact that
  if $\ell_\h$ is a linear labelling, then there exists
  $A \in \text{SL}(2,\ZM)$ and $\kappa_\h \in \ZM^2$ such that
  $A \circ \ell_\h + \kappa_\h$ is good. But the fact that the above
  equality only holds on $\check{B}^+$ is enough to prove that $A$ is
  the identity.
\end{demo}

\begin{ex} To illustrate the difference between the above proof and the proof of Lemma
  \ref{lemm:asla-extension}, consider for instance the asymptotic lattice $\mathcal{L}_\h = \h \ZM^2 \cap ([-1,1] \times [-C,1])$ for some $C > 0$. Let $r: \RM^2 \to \RM$ be the smooth function such that $r(x,y) = 0$ if $y \geq 0$ and $r(x,y) = e^{-\frac{1}{y^2}}$ otherwise. Then $\mathrm{Id}$ and $\mathrm{Id} + r$ are two asymptotic charts for $\mathcal{L}_\h$, but we see that $r(\mathcal{L}_\h) = \O(\h^{\infty})$.
\end{ex}

\begin{lemm}
  \label{lemm:half-asla-glueing}
  Let $(\mathcal{L}^{(1)}_\h\!,\mathcal{I},B_1)$ and
  $(\mathcal{L}^{(2)}_\h\!,\mathcal{I},B_2)$ be two asymptotic
  half-lattices, sharing the same parameter set $\mathcal{I}$, with
  respective boundaries $\mathcal{E}_1$ and $\mathcal{E}_2$ (see
  Definition \ref{defi:boundary_half}) and asymptotic charts
  $G_\h^{(1)}$ and $G_\h^{(2)}$. Let
  $\mathcal{L}_\h = \mathcal{L}^{(1)}_\h \cup
  \mathcal{L}^{(2)}_\h$. Assume that $B_1\cap B_2$ is simply connected
  and non empty, that $\mathcal{E}_1 \cap B_2$ is connected and that
  \[
    \forall \h\in \mathcal{I}, \quad \mathcal{L}^{(1)}_\h \cap B_2 =
    \mathcal{L}^{(2)}_\h\cap B_1.
  \]
  Then $\mathcal{E}_1 \cap B_2 = \mathcal{E}_2 \cap B_1$ and for any
  admissible domain $\tilde{B} \Subset B_1 \cup B_2$,
  $\left(\mathcal{L}_\h \cap \tilde{B},\mathcal{I},\tilde{B}\right)$
  is an asymptotic half-lattice. Moreover, for $i=1,2$, let
  $\tilde{B}_i \Subset B_i$ be an admissible domain. Then there exists
  a family of maps
  $(\ell_\h: \mathcal{L}_\h \cap (\tilde{B}_1 \cup \tilde{B}_2) \to
  \ZM^2)_{\h \in \mathcal{I}}$ such that
  ${\ell_\h}_{|\mathcal{L}^{(1)}_\h \cap \tilde{B}_1}$ and
  ${\ell_\h}_{|\mathcal{L}^{(2)}_\h \cap \tilde{B}_2}$ are linear
  labellings for
  $(\mathcal{L}^{(1)}_\h\! \cap \tilde{B}_1,\mathcal{I}, \tilde{B}_1)$
  and
  $(\mathcal{L}^{(2)}_\h\! \cap \tilde{B}_2,\mathcal{I},\tilde{B}_2)$
  respectively. Furthermore, $\ell_\h$ is uniquely defined from its
  restriction to $\mathcal{L}^{(1)}_\h\! \cap \tilde{B}_1$ or
  $\mathcal{L}^{(2)}_\h\! \cap \tilde{B}_2$.
\end{lemm}

\begin{demo}
  The proof is similar to the proof of Lemma
  \ref{lemm:asla-glueing}. The main differences are the following:
  \begin{itemize}
  \item instead of Lemmas \ref{lemm:asla-restriction} and Lemma
    \ref{lemm:asla-extension}, we use Lemmas
    \ref{lemm:restriction_half_lattice} and
    \ref{lemm:half-asla-extension},
  \item as in the proof of Lemma \ref{lemm:half-asla-extension}, the
    two charts $G_\h^{(1)}$ and $G_\h^{(2)}$ will coincide only up to
    $\O(\h^{\infty})$ and a term which vanishes on
    $\tilde{B}_1^+ \cup \tilde{B}_2^+$. But then we can still define a
    common chart $G_\h$ which coincides with each one of them where it
    should and which is a diffeomorphism on a neighborhood of
    $\overline{\tilde{B}_1 \cup \tilde{B}_2}$.
  \end{itemize}
  For the last assertion, let
  $\tilde{B}_i \Subset \hat{B}_i \Subset B_i$ for $i=1,2$. By the
  first part, there exists a family of maps
  $({\hat{\ell}}_\h: \mathcal{L}_\h \cap (\hat{B}_1 \cup \hat{B}_2)
  \to \ZM^2)_{\h \in \mathcal{I}}$ such that the restrictions of
  $\hat{\ell}_\h$ to $\mathcal{L}^{(1)}_\h \cap \hat{B}_1$ and
  $\mathcal{L}^{(2)}_\h \cap \hat{B}_2$ are linear labellings for
  $(\mathcal{L}^{(1)}_\h\! \cap \hat{B}_1,\mathcal{I}, \hat{B}_1)$ and
  $(\mathcal{L}^{(2)}_\h\! \cap \hat{B}_2,\mathcal{I},\hat{B}_2)$
  respectively. By Lemma \ref{lemm:transition_labellings_half}, there
  exists a unique matrix $A \in \mathrm{SL}(2,\ZM)$, $\h_0 > 0$ and a
  family $(\kappa_\h)_{\h \in \mathcal{I} \cap [0,\h_0]}$ of vectors
  in $\ZM^2$ such that
  \[
    \forall \h \in \mathcal{I} \cap [0,\h_0] \qquad \hat{\ell}_\h = A
    \circ \ell_\h + \kappa_\h \quad \text{ on } \mathcal{L}_\h^{(1)}
    \cap \tilde{B_1}.
  \]
  So $\check{\ell}_\h = A^{-1} (\hat{\ell}_\h - \kappa_\h)$ is a
  linear labelling on $\mathcal{L}_\h \cap (\hat{B}_1 \cup \hat{B}_2)$
  which coincides with $\ell_\h$ on
  $\mathcal{L}_\h^{(1)} \cap \tilde{B_1}$. Let
  $(\ell'_\h: \mathcal{L}_\h \cap (\tilde{B}_1 \cup \tilde{B}_2) \to
  \ZM^2)_{\h \in \mathcal{I}}$ be another family of maps such that
  ${\ell'_\h}_{|\mathcal{L}^{(1)}_\h \cap \tilde{B}_1}$ and
  ${\ell'_\h}_{|\mathcal{L}^{(2)}_\h \cap \tilde{B}_2}$ are linear
  labellings for
  $(\mathcal{L}^{(1)}_\h\! \cap \tilde{B}_1,\mathcal{I}, \tilde{B}_1)$
  and
  $(\mathcal{L}^{(2)}_\h\! \cap \tilde{B}_2,\mathcal{I},\tilde{B}_2)$
  respectively, and such that the restrictions of $\ell_\h$ and
  $\ell'_\h$ to $\mathcal{L}^{(1)}_\h \cap \tilde{B}_1$ coincide. Let
  $\hat{\ell}'_\h$ be its extension to
  $\mathcal{L}_\h \cap (\hat{B}_1 \cup \hat{B}_2)$ as before. By
  applying Lemma \ref{lemm:transition_labellings_half} again, there
  exists a unique matrix $A \in \mathrm{SL}(2,\ZM)$, $\h_1 > 0$ and a
  family $(\kappa_\h)_{\h \in \mathcal{I} \cap [0,\h_1]}$ of vectors
  in $\ZM^2$ such that
  \[
    \forall \h \in \mathcal{I} \cap [0,\h_1] \qquad \hat{\ell}'_\h = A
    \circ \hat{\ell}_\h + \kappa_\h \quad \text{ on }
    \mathcal{L}_\h^{(2)} \cap \tilde{B_2}.
  \]
  But since $\hat{\ell}'_\h$ and $\hat{\ell}_\h$ coincide on
  $\mathcal{L}_\h^{(2)} \cap \tilde{B_1} \cap \tilde{B_2}$ by
  construction, we obtain that $\kappa_\h = 0$ and $A = I$. So
  $\hat{\ell}'_\h = \hat{\ell}_\h$ on
  $\mathcal{L}_\h \cap (\tilde{B}_1 \cup \tilde{B}_2)$.
\end{demo}

In fact, we can also consider the union of one asymptotic lattice and
one asymptotic half-lattice.

\begin{lemm}
  \label{lemm:ext_label_half}
  Let $(\mathcal{L}^{(1)}_\h\!,\mathcal{I},B_1)$ be an asymptotic
  half-lattice and $(\mathcal{L}^{(2)}_\h\!,\mathcal{I},B_2)$ be an
  asymptotic lattice, sharing the same parameter set $\mathcal{I}$,
  with respective asymptotic charts $G_\h^{(1)}$ and $G_\h^{(2)}$. Let
  $\mathcal{L}_\h = \mathcal{L}^{(1)}_\h \cup
  \mathcal{L}^{(2)}_\h$. Assume that $B_1\cap B_2$ is simply connected
  and non empty, and that
  \[
    \forall \h\in \mathcal{I}, \quad \mathcal{L}^{(1)}_\h \cap B_2 =
    \mathcal{L}^{(2)}_\h\cap B_1.
  \]
  Let $\tilde{B}_1 \Subset B_1$ be an admissible domain and let
  $\tilde{B}_2 \Subset B_2$ be simply connected. Then there exists a
  family of maps
  $(\ell_\h: \mathcal{L}_\h \cap (\tilde{B}_1 \cup \tilde{B}_2) \to
  \ZM^2)_{\h \in \mathcal{I}}$ such that
  ${\ell_\h}_{|\mathcal{L}^{(1)}_\h \cap \tilde{B}_1}$ and
  ${\ell_\h}_{|\mathcal{L}^{(2)}_\h \cap \tilde{B}_2}$ are linear
  labellings for
  $(\mathcal{L}^{(1)}_\h\! \cap \tilde{B}_1,\mathcal{I}, \tilde{B}_1)$
  and
  $(\mathcal{L}^{(2)}_\h\! \cap \tilde{B}_2,\mathcal{I},\tilde{B}_2)$
  respectively. Moreover, $\ell_\h$ is uniquely defined from its
  restriction to $\mathcal{L}^{(1)}_\h\! \cap \tilde{B}_1$ or
  $\mathcal{L}^{(2)}_\h\! \cap \tilde{B}_2$. Furthermore, if
  $\tilde{B} \Subset B_1 \cup B_2$ is admissible, then
  $\left(\mathcal{L}_\h \cap \tilde{B},\mathcal{I},\tilde{B}\right)$
  is an asymptotic half-lattice.
\end{lemm}

\begin{demo}
  The proof is similar to the proof of Lemma
  \ref{lemm:half-asla-glueing}, taking into account the definition of
  linear labelling for an asymptotic half-lattice (Definition
  \ref{defi:linear_lab_half}), which allows the composition with an
  arbitrary element of $\text{SL}(2,\ZM)$.
\end{demo}

\subsection{Global labellings}
\label{sec:global}

Using the previous results, we can construct a ``global labelling''
for the union of several asymptotic lattices and asymptotic
half-lattices, under suitable assumptions.

When working with a union of asymptotic lattices (respectively a union
of half-lattices sharing a connected boundary), we can directly apply
Lemmas \ref{lemm:asla-glueing} and \ref{lemm:half-asla-glueing} to
obtain the following.

\begin{coro}
  Let $B_1, \ldots, B_p \subset \RM^2$ be simply connected open sets
  such that for every $ i \in \{1, \ldots, p\}$,
  $(\mathcal{L}^{(i)}_\h\!,\mathcal{I},B_i)$ is an asymptotic
  lattice. Assume that $B = \bigcup_{i=1}^q B_i$ is simply connected
  and that for every $i,j$ such that $B_i \cap B_j \neq \emptyset$:
  \begin{itemize}
  \item $B_i \cap B_j$ is simply connected,
  \item $\forall \h \in \mathcal{I}$,
    $\mathcal{L}^{(i)}_\h \cap B_j = \mathcal{L}^{(j)}_\h\cap B_i$.
  \end{itemize}
  Let $\mathcal{L}_\h := \bigcup_{i=1}^p \mathcal{L}^{(i)}_\h$. Then
  for every simply connected set $\tilde{B} \Subset B$,
  $(\mathcal{L}_\h \cap \tilde{B}, \mathcal{I}, \tilde{B})$ is an
  asymptotic lattice.
\end{coro}

\begin{coro}
  \label{coro:construct_global_half_labelling}
  Let $B_1, \ldots, B_p \subset \RM^2$ be simply connected open sets
  such that for every $ i \in \{1, \ldots, p\}$,
  $(\mathcal{L}^{(i)}_\h\!,\mathcal{I},B_i)$ is an asymptotic
  half-lattice. Assume that $\bigcup_{i=1}^q B_i$ is simply connected
  and that for every $i,j$ such that $B_i \cap B_j \neq \emptyset$:
  \begin{itemize}
  \item $B_i \cap B_j$ is simply connected,
  \item $\forall \h \in \mathcal{I}$,
    $\mathcal{L}^{(i)}_\h \cap B_j = \mathcal{L}^{(j)}_\h\cap B_i$,
  \item $B_i \cap B_j$ is admissible and $\mathcal{E}_i \cap B_j$ is
    connected (recall that $\mathcal{E}_i$ stands for the boundary of
    $(\mathcal{L}^{(i)}_\h\!,\mathcal{I},B_i)$).
  \end{itemize}
  Let $\mathcal{L}_\h := \bigcup_{i=1}^q \mathcal{L}^{(i)}_\h$. Then
  for every admissible $\tilde{B} \Subset B$,
  $(\mathcal{L}_\h \cap \tilde{B}, \mathcal{I}, \tilde{B})$ is an
  asymptotic half-lattice.
\end{coro}

In particular, when working with the joint spectrum $\Sigma_\h$ of a
proper quantum integrable system, the first of these results implies
that we can obtain good quantum numbers for the intersection of
$\Sigma_\h$ with any simply connected subset of the set of regular
values of the momentum map. Similarly, the second result implies that
we can label this joint spectrum along a line of
transversally-elliptic values, as long as we do not encounter any
elliptic-elliptic value.

\begin{rema}
  It follows from Lemma \ref{lemm:half-asla-glueing} that the union of
  the boundaries of the asymptotic half-lattices in Corollary
  \ref{coro:construct_global_half_labelling} is a one-dimensional
  manifold $\mathcal{E}$. This corollary implies the interesting
  topological fact that no component of $\mathcal{E}$ is
  closed. Indeed, $\Phi(\mathcal{E})$ is an affine line with a natural
  orientation coming from the fact that the points of the union of the
  half-lattices always stand on the same side of this line. This is
  the quantum version of a result that also holds classically, see for
  instance \cite[Theorem 3.4]{san-polytope}.
\end{rema}

Now, if we want to obtain a global labelling for a union of both
asymptotic lattices and half-lattices, we need to work a little bit
more. This will be crucial in the next section since we will want to
produce good quantum numbers near both regular and transversally
elliptic values.

\begin{theo}
  \label{thm:construct_global_labelling}
  Let $B_1, \ldots, B_p, B_{p+1}, \ldots, B_q \subset \RM^2$ be simply
  connected open sets such that for every $ i \in \{1, \ldots, p\}$,
  $(\mathcal{L}^{(i)}_\h\!,\mathcal{I},B_i)$ is an asymptotic lattice
  and for every $i \in \{ p+1, \ldots, q\}$,
  $(\mathcal{L}^{(i)}_\h\!,\mathcal{I},B_i)$ is an asymptotic
  half-lattice. Assume that $\bigcup_{i=1}^q B_i$ is simply connected
  and that for every $i,j$ such that $B_i \cap B_j \neq \emptyset$:
  \begin{itemize}
  \item $B_i \cap B_j$ is simply connected,
  \item $\forall \h \in \mathcal{I}$,
    $\mathcal{L}^{(i)}_\h \cap B_j = \mathcal{L}^{(j)}_\h\cap B_i$,
  \item if $i,j \in \{ p+1, \ldots, q\}$, $B_i \cap B_j$ is admissible
    and $\mathcal{E}_i \cap B_j$ is connected.
  \end{itemize}
  Let $(\tilde{B}_i)_{1 \leq i \leq q}$ be a family of open sets
  satisfying the same assumptions as the $B_i$ and such that for every
  $i$, $\tilde{B}_i \Subset B_i$ and for every
  $i \in \{p+1, \ldots, q\}$, $\tilde{B}_i$ is admissible. Let
  $\mathcal{L}_\h := \bigcup_{i=1}^q \mathcal{L}^{(i)}_\h$. Then there
  exists a family of maps
  $\ell_\h: \mathcal{L}_\h \cap \bigcup_{i=1}^q \tilde{B}_i \to \ZM^2$
  such that for every $i \in \{1, \ldots, q\}$,
  ${\ell_\h}_{|\mathcal{L}^{(i)}_\h \cap \tilde{B}_i}$ is a linear
  labelling. Moreover, $\ell_\h$ is uniquely defined by its
  restriction to any of the $\mathcal{L}^{(i)}_\h \cap
  \tilde{B}_i$. Furthermore, there exist a smooth function $\Phi$ from
  a neighborhood of $\overline{\bigcup_{i=1}^q B_i}$ to $\RM^2$, a
  family $(\nu_\h)$ of vectors in $\ZM^2$ and a constant $K > 0$ such
  that for every $\h$ small enough and for every
  $\lambda \in \mathcal{L}_\h \cap \bigcup_{i=1}^q \tilde{B}_i$,
  \[
    \| \Phi(\lambda) - \h \ell_\h(\lambda) - \nu_\h \| \leq K \h.
  \]
\end{theo}

\begin{demo}
  Let $\ell_\h^{(1)}, \ldots, \ell_\h^{(q)}$ be linear labellings for
  $(\mathcal{L}^{(1)}_\h\!,\mathcal{I},B_1), \ldots,
  (\mathcal{L}^{(q)}_\h\!,\mathcal{I},B_q)$. For every pair $(i,j)$
  such that $\tilde{B}_i \cap \tilde{B}_j \neq \emptyset$,
  $\mathcal{L}_\h \cap \tilde{B}_i \cap \tilde{B}_j$ is an asymptotic
  lattice or half-lattice thanks to Lemmas
  \ref{lemm:asla-restriction}, \ref{lemm:restriction_half_lattice_far}
  and \ref{lemm:restriction_half_lattice}, so by \cite[Proposition
  3.20]{san-dauge-hall-rotation} and Lemma
  \ref{lemm:transition_labellings_half}, there exists a unique matrix
  $A_{ji} \in \mathrm{SL}(2,\ZM)$, $\h_{ji} > 0$ and a family
  $(\kappa_\h^{(ji)})_{\h \in \mathcal{I} \cap [0,\h_{ji}]}$ of
  vectors in $\ZM^2$ such that
  \[
    \forall \h \in \mathcal{I} \cap [0,\h_0] \qquad \ell_\h^{(j)} =
    A_{ji} \circ \ell_\h^{(i)} + \kappa_\h^{(ji)} \quad \text{ on }
    \mathcal{L}_\h^{(i)} \cap \tilde{B}_j.
  \]
  Let $\h_0 = \min_{i,j} \h_{ji} > 0$. Then we obtain a family of
  affine maps $\phi_{ji} = A_{ji} \cdot + \kappa_\h^{(ji)}$ defined
  for $\h \in \mathcal{I} \cap (0,\h_0]$, which is in fact a cocycle:
  whenever
  $\tilde{B}_i \cap \tilde{B}_j \cap \tilde{B}_k \neq \emptyset$,
  $\phi_{ij} \circ \phi_{jk} \circ \phi_{ki} = I$. Let
  $c_0 \in \bigcup_{i=1}^q \tilde{B}_i$. For every
  $c \in \bigcup_{i=1}^q \tilde{B}_i$, we construct the map $\ell_\h$
  in a neighborhood of $c$ as follows. Let $\gamma$ be a continuous
  path from $c_0$ to $c$. Let
  $\tilde{B}_{i_0}, \ldots, \tilde{B}_{i_n}$ be a chain from $c_0$ to
  $c$ covering $\gamma$, that is
  $\tilde{B}_{i_p} \cap \tilde{B}_{i_{p+1}} \neq \emptyset$ for every
  $p$, $c_0 \in \tilde{B}_{i_0}$ and $c \in \tilde{B}_{i_n}$. Then we
  construct $\ell_\h$ near $c$ by setting
  $\ell_\h = \phi_{i_0 i_1} \circ \ldots \circ \phi_{i_{n-1} i_n}
  \circ \ell_\h^{(n)}$. Then by a standard argument (for instance by
  induction on $n$) using the cocycle condition, $\ell_\h$ does not
  depend on the choice of such a chain. The same argument implies that
  $\ell_\h$ does not depend on the choice of $\gamma$ up to homotopy
  with fixed endpoints. Since $\bigcup_{i=1}^q \tilde{B}_i $ is simply
  connected, $\ell_\h$ is well-defined.

  Now, let $j \in \{ 1, \ldots, q \}$; then
  ${\ell_\h}_{|\tilde{B}_j \cap \mathcal{L}^{(j)}_\h}$ is a linear
  labelling, so there exist a matrix $A_j \in \text{SL}(2,\ZM)$ and a
  vector $\kappa_\h^{(j)}$ such that
  $\tilde{\ell}_\h^{(j)} := A_j \circ \ell_\h + \kappa_\h^{(j)}$ is a
  good labelling for $\tilde{B}_j \cap \mathcal{L}^{(j)}_\h$ (recall
  Definition \ref{defi:linear_lab_half} for half-lattices). Let
  $G_\h^{(j)}$ be a corresponding asymptotic chart, and let
  $U_j = (G_0^{(j)})^{-1}(\tilde{B}_j) \subset
  (G_0^{(j)})^{-1}(B_j)$. Let $\h_0 > 0$ and for every
  $j \in \{1, \ldots, q\}$, let $C_j, M_j, L_j$ be the constants such
  that $\forall \h \in \mathcal{I} \cap [0,h_0]$
  \[
    \begin{cases} \forall \lambda_\h \in \mathcal{L}_\h^{(j)} \cap \tilde{B}_j \qquad \snorm{ G_\h^{(j)}(\h \tilde{\ell}_\h^{(j)}(\lambda_\h)) - \lambda_\h} \leq C_j \h, \\
      \sup_{U_j} \snorm{G_\h^{(j)} - G_0^{(j)}} \leq M_j \h,
    \end{cases}
  \]
  and let $D_j = \sup_{U_j} \| \dd (G_0^{(j)})^{-1} \|$. The existence
  of $C_j$ comes from the definitions of asymptotic lattices and
  half-lattices, and the existence of $M_j$ comes from item $2.$ in
  \cite[Lemma 3.8]{san-dauge-hall-rotation} for the case of asymptotic
  lattices; this still holds for asymptotic half-lattices, since this
  is a direct consequence of the asymptotic expansion
  \eqref{equ:asymp_chart} which also holds for asymptotic
  half-lattices, see Equation \eqref{equ:asymp_chart_half}. Finally,
  let $C = \max_{j \in \{1, \ldots, q\}} C_j$,
  $M = \max_{j \in \{1, \ldots, q\}} M_j$,
  $D = \max_{j \in \{1, \ldots, q\}} D_j$ and
  $\alpha = \max_j \|A_j^{-1}\|$. Then for
  $\lambda_\h \in \mathcal{L}_\h^{(j)} \cap \tilde{B}_j $
  \[
    \begin{split} \left\| G_0^{(j)}(\h
        \tilde{\ell}_\h^{(j)}(\lambda_\h)) - \lambda_\h \right\| &
      \leq \left\| G_0^{(j)}(\h \tilde{\ell}_\h^{(j)}(\lambda_\h)) -
        G_\h^{(j)}(\h \tilde{\ell}_\h^{(j)}(\lambda_\h)) \right\| +
      \left\| G_\h^{(j)}(\h \tilde{\ell}_\h^{(j)}(\lambda_\h)) -
        \lambda_\h \right\| \\ & \leq (M + C) \h.
    \end{split}
  \]
  This implies that
  \[
    \left\| \h \tilde{\ell}_\h^{(j)}(\lambda_\h) -
      (G_0^{(j)})^{-1}(\lambda_\h) \right\| = \left\|
      (G_0^{(j)})^{-1}( G_0^{(j)}(\h
      \tilde{\ell}_\h^{(j)}(\lambda_\h)) -
      (G_0^{(j)})^{-1}(\lambda_\h) \right\| \leq D (M + C) \h.
  \]
  So if we set $\Phi_j := (G_0^{(j)} \circ A_j)^{-1}$, we finally
  obtain that
  \begin{equation} \left\| \h \ell_\h(\lambda_\h) + \h A_j^{-1}
      \kappa_\h^{(j)} - \Phi_j(\lambda_\h) \right\| \leq \alpha D (M +
    C) \h. \label{eq:Phi_j}
  \end{equation}
  
  Now, assume that $\tilde{B}_i \cap \tilde{B}_j \neq \emptyset$. Then
  \eqref{eq:Phi_j} implies that for every
  $\lambda_\h \in \mathcal{L}_\h^{(j)} \cap \tilde{B}_i =
  \mathcal{L}_\h^{(i)} \cap \tilde{B}_j$,
  \begin{equation} \left\| \h A_j^{-1} \kappa_\h^{(j)} - \h A_i^{-1}
      \kappa_\h^{(i)} + \Phi_i(\lambda_\h) - \Phi_j(\lambda_\h)
    \right\| \leq 2\alpha D (M + C) \h. \label{eq:Phi_i_Phi_j}
  \end{equation}
  Let $c \in \tilde{B}_i \cap \tilde{B}_j$. From \cite[Lemma
  3.15]{san-dauge-hall-rotation} and its proof (which works similarly
  for half-lattices), there exists $\h_1 \leq \h_0$ and a family
  $(\lambda_\h)_{\h \in \mathcal{I} \cap [0,\h_1]}$ such that
  $ \forall \h \in \mathcal{I} \cap [0,\h_1]$,
  $\lambda_\h \in \tilde{B}_i \cap \tilde{B}_j$ and
  $\lambda_\h \underset{\h \to 0}{\longrightarrow} c$. Then the above
  equation implies that
  $\h A_j^{-1} \kappa_\h^{(j)} - \h A_i^{-1} \kappa_\h^{(i)}$ has a
  limit when $\h \to 0$, and that $\Phi_i(c) - \Phi_j(c)$ does not
  depend on $c$. Hence
  ${\dd \Phi_i}_{|\tilde{B}_i \cap \tilde{B}_j} = {\dd
    \Phi_j}_{|\tilde{B}_i \cap \tilde{B}_j}$, so by connectedness
  there exists a global $\Phi$ such that
  ${\dd \Phi}_{|\tilde{B}_j} = {\dd \Phi_j}_{|\tilde{B}_j}$ for every
  $j$.

  So let $v_j \in \RM^2$ be the constant such that
  $\Phi(c) = \Phi_j(c) + v_j$ for every $c \in \tilde{B}_j$, and set
  $\nu_\h^{(j)} := \h A_j^{-1} \kappa_\h^{(j)} + v_j$. Equation
  \eqref{eq:Phi_i_Phi_j} yields
  \[
    \left\| \nu_\h^{(j)} - \nu_\h^{(i)} \right\| \leq \alpha D (M + C)
    \h
  \]
  whenever $\tilde{B}_i \cap \tilde{B}_j \neq \emptyset$. So by
  considering chains of $\tilde{B}_m$, we obtain that for any $i,j$,
  $\| \nu_\h^{(j)} - \nu_\h^{(i)} \| \leq q \alpha D (M + C) \h
  $. Thus, if $\nu_\h := \frac{1}{q} \sum_{i=1}^q \nu_\h^{(i)}$, we
  have $\| \nu_\h - \nu_\h^{(j)} \| \leq q \alpha D (M + C) \h $ for
  every $j$, and Equation \eqref{eq:Phi_j} implies that for every
  $\lambda_\h \in \mathcal{L}_\h \cap \bigcup_{i=1}^q \tilde{B}_i$,
  \[
    \left\| \h \ell_\h(\lambda_\h) + \nu_\h - \Phi(\lambda_\h)
    \right\| \leq (q + 1) \alpha D (M + C) \h.
  \]
\end{demo}

\begin{rema}
  Depending on the topology of the union in the previous proposition,
  one may be able to find a family of vectors $\kappa_\h \in \ZM^2$
  such that $\ell_\h + \kappa_\h$ is a ``good global labelling'' in
  the sense that its restriction to every $\tilde{B}_i$ is a good
  labelling. This is for instance the case when the boundary attached
  to the half-lattices is connected. However, as explained in the
  discussion preceding Figure \ref{fig:problem_global_labelling},
  there is no chance to obtain the same result in all generality.
\end{rema}

\begin{rema}
  \label{rema:explicit_global}
  In view of the above proposition and its proof, it suffices to be
  able to produce any linear labelling on each of the asymptotic
  lattices and half-lattices to be able to construct the global
  labelling $\ell_\h$. Indeed once such labellings are given, the
  affine maps relating them and coming from \cite[Proposition
  3.20]{san-dauge-hall-rotation} and Lemma
  \ref{lemm:transition_labellings_half} can be explicitly
  recovered. In the case where these lattices are of the form
  $\mathcal{L}^{(i)}_\h = \Sigma_\h \cap B_i$ where $\Sigma_\h$ is the
  joint spectrum of a proper quantum semitoric system and $B_i$ is a
  small neighborhood of a regular or $J$-transversally elliptic value,
  the algorithms described at the end of Sections
  \ref{subsect:asympt_latt} and \ref{sec:asympt-half-latt} allow to
  obtain such linear labellings $\ell_\h^{(i)}$, hence $\ell_\h$ can
  be fully constructed algorithmically.

  Note that, on the contrary, this proof does not give any procedure
  to recover the family of vectors $\nu_\h$. But as we will see in the
  next section, knowing the value of $\nu_\h$ is not necessary for our
  inverse spectral result for semitoric systems.
\end{rema}

\section{Recovering the twisting index from the joint spectrum}
\label{sec:twisting}

In this section we are given a proper quantum semitoric system
$(\hat{J}_\h, \hat{H}_\h)$, and we explain how to use the results of the previous section to recover the twisting
index invariant from its joint spectrum. Since the twisting index is
the data of the polygon invariant decorated with the corresponding
twisting numbers for each focus-focus value, we proceed in two steps:
the first step is to recover the invariants $\sigma_1^{\textup{p}}(0)$ associated with the focus-focus values
(see Section~\ref{sec:twist-numb-twist}), and the second step is to
recover the polygonal invariant and all the attached twisting numbers.

\subsection{Recovering $\sigma_1(0)$}

Recall that the quantity $\sigma_1(0)$, defined in
Section~\ref{sec:twist-numb-twist}, depends only on the choice of an
action variable $L$ independent of $J$, see the discussion after
Remark~\ref{rema:gamma}.

\begin{theo}
  \label{theo:recover-sigma_1}
  From the $\h$-family of joint spectra $\Sigma_\h$ of a proper
  quantum semitoric system in a neighborhood of a focus-focus critical
  value $c_0=0$, one can recover, in a constructive way, the
  symplectic invariant $[S_{1,0}]$ (see Section
  \ref{sec:tayl-seri-invar}). More precisely, given a semitoric
  labelling on a small ball near $0$ containing only regular values of
  $F$, associated with an action variable $L$, one can recover, in a
  constructive way, the quantity $\sigma_1(0)$ associated with $L$.
\end{theo}

This theorem is the key result used in the proof of Theorem \ref{theo:poly_twisting} to recover the twisting numbers.

Because a uniform description of the joint spectrum in a neighborhood
of the focus-focus value exists only for pseudodifferential
operators~\cite{san-focus}, we shall here prove this theorem by a
simpler, albeit less efficient (from a numerical viewpoint), approach,
which consists in considering regular values $c$ close to $0$, before
letting them tend to the origin.

Thus, let $c$ be a regular value of $F$ and let $B \subset \RM^2$ be
an open ball containing $c$, small enough to contain only regular values of $F$. Here we assume that $c$ is sufficiently close to $0$ so that $B$ is contained in the simply connected set $U$ contained in a punctured neighborhood of $0$ defined in Section \ref{sec:twist-numb-twist}.
Let $\lambda\mapsto (j,\ell)\in\ZM^2$ be a linear labelling of
  $\mathcal{L}_{\h} \cap B$ (see Definition
  \ref{defi:linear_labelling}), associated with a semitoric
asymptotic chart $G_{\hbar}: \tilde U \to \RM^2$, where
$\tilde U \subset \RM^2$ is some bounded open set; such a labelling is
given in \cite[Section 3.5.2]{san-dauge-hall-rotation} and is
constructed from $\mathcal{L}_{\h}$ thanks to an explicit algorithm,
see the discussion at the end of Section \ref{subsect:asympt_latt}.
By definition, $G_{\h}$ has an asymptotic expansion of the form
\[
  G_{\h} = G_0 + \h G_1 + \h^2 G_2 + \ldots
\]
in the $C^{\infty}$ topology, where $G_0^{-1}$ is an action
diffeomorphism (see Section \ref{sec:symp_prel}); there exists a
choice of action variables $(J,L)$ defined near $F^{-1}(c)$ such that
\begin{equation} F = (J,H) = G_0(J,L). \label{equ:F_G0} \end{equation}

As in Remark~\ref{rema:def_label}, we will use the notation
  $\lambda_{j,\ell}$ for the joint eigenvalue labelled by $(j,\ell)$,
  and we let
\begin{equation}
  \lambda_{j,\ell}(\h)= (J_{j,\ell}(\h), E_{j,\ell}(\h) )\,.
  \label{equ:semitoric-labelling}
\end{equation}

\begin{lemm}\label{lemm:quantum-a}
  Let $j,\ell$ be $\h$-dependent integers such that the joint
  eigenvalues $\lambda_{j,\ell}$, $\lambda_{j+1,\ell}$ and
  $\lambda_{j,\ell+1}$ are well-defined in an $\O(\h)$-neighborhood of
  $c$. We have:
  \begin{enumerate}
  \item
    $\displaystyle \frac{E_{j,\ell}(\h) - E_{j+1,\ell}(\h)}{\h} =
    \frac{a_1(c)}{a_2(c)} + \O_c(\h)$,
  \item
    $\displaystyle \frac{\h}{E_{j,\ell+1}(\h) - E_{j,\ell}(\h)} = a_2(c) +
    \O_c(\h)$,
  \end{enumerate}
  where $a_1, a_2$ are the functions that appear in the decomposition
  \eqref{equ:a} of $L$, and $E_{j,\ell}$ is the second component of
  $\lambda_{j,\ell}$, see~\eqref{equ:semitoric-labelling}.
\end{lemm}

\begin{demo}
  From~\eqref{equ:F_G0}, we have that $(J,L) = G_0^{-1}(J,H)$. It
  follows (with a slight abuse of notation) that
  \[
    \begin{pmatrix} \ham{J} \\ \ham{L} \end{pmatrix} = \dd
    G_0^{-1}(J,H) \begin{pmatrix} \ham{J} \\ \ham{H} \end{pmatrix};
  \]
  comparing this with Equation \eqref{equ:a} yields, with
  $\xi_0 = G_0^{-1}(c)$,
  \[
    \dd G_0^{-1}(c) = \begin{pmatrix} 1 & 0 \\ a_1(c) &
      a_2(c) \end{pmatrix}, \quad \dd G_0(\xi_0) = \begin{pmatrix} 1 &
      0 \\ -\frac{a_1(c)}{a_2(c)} & \frac{1}{a_2(c)} \end{pmatrix}.
  \]

Let $\lambda'_{j', \ell'}$ be the good labelling associated with $G_\h$, so that there exist constants $\kappa_1(\h), \kappa_2(\h)$ in $\ZM$ such that
  \[
\lambda_{j,\ell} = \lambda'_{j-\kappa_1(h), \ell-\kappa_2(\h)}\,,
\]
and hence
  \[
    \lambda_{j,\ell} = G_{\h}(\h j',\h \ell') + \O(\h^{\infty}) = G_0(\h
    j', \h \ell') + \h G_1(\h j', \h \ell') + \O(\h^2),
  \]
  where we use $j' = j-\kappa_1(h), \ell' = \ell-\kappa_2(\h)$, and
   the $\O(\h^2)$ is uniform assuming $(\h j', \h \ell')$ stays in
  a compact set. On the other hand, we have by assumption
  $\lambda_{j,\ell}-c=\O(\h)$, hence by invertibility of
  $G_\h$~\cite[Lemma 3.8]{san-dauge-hall-rotation}, we obtain
  $(\h j', \h \ell')=\O(\h)$ as well.  Therefore, Taylor's formula gives
  \[
    \begin{cases} \lambda_{j,\ell} - \lambda_{j+1,\ell} =
      - \h\dd G_0 (\h j', \h\ell')\cdot \vec z_1 + \O(\h^2)\\
      \lambda_{j,\ell+1} - \lambda_{j,\ell} = \h\dd G_0 (\h j',
      \h\ell')\cdot \vec z_2 + \O(\h^2)
    \end{cases}
  \]
  where $\vec z_1 := (\begin{smallmatrix} 1 \\ 0
  \end{smallmatrix}) $,
  $\textstyle\vec z_2:= (\begin{smallmatrix} 0\\1
  \end{smallmatrix}) $.  Since $\lambda_{j,\ell}= c + \O(\h)$, we have
  $(\h j, \h \ell) = \xi_0 + \O(\h)$ where $\xi_0 =
  G_0^{-1}(c)$. Hence
  \[
    \lambda_{j,\ell} - \lambda_{j+1,\ell} = - \h \ \dd G_0(\xi_0)
    \cdot \vec{z}_1 + \O(\h^2),
  \]
  so finally, taking the second component,
  \[
    \frac{E_{j,\ell} - E_{j+1,\ell}}{\h} = \frac{a_1(c)}{a_2(c)} +
    \O(\h).
  \]
  Similarly,
  \[
    \frac{E_{j,\ell+1} - E_{j,\ell}}{\h} = \dd G_0(\xi_0) \cdot
    \vec{z}_2 + \O(\h^2) = \frac{1}{a_2(c)} + \O(\h).
  \]
\end{demo}

In the proof above and in most of the remaining of the text, we
  write $E_{j,\ell}$ instead of $E_{j,\ell}(\h)$ for notational
  simplicity.

\begin{rema}
  By multiplying the two lines of Lemma~\ref{lemm:quantum-a}, we
  obtain
  \[
    \frac{E_{j,\ell} - E_{j+1,\ell}}{E_{j,\ell+1} - E_{j,\ell}} =
    a_1(c) + \O_c(\h)
  \]
  and $a_1$ is precisely the rotation number of $H$ with respect to
  $(J,L)$, which recovers a result of~\cite{san-dauge-hall-rotation}.
\end{rema}

\begin{lemm}
  \label{lemm:tangent}
  In order to compute $\sigma_1(0)$ (see Lemma
  \ref{lemm:def_sigma1_zero}), the radial curve $\gamma_r$
  (see~Definition~\ref{defi:radial-curve}) can be replaced by any
  curve $\gamma$ that is tangent to $\gamma_r$ at the origin. In other words, $\sigma_1(0)$ is the limit of $\tau_1(c)$ when $c$ tends to the origin along $\gamma$.
\end{lemm}

\begin{demo}
  Let $\gamma$ be any curve that is tangent to $\gamma_r$ at the
  origin; then $\gamma$ is also the graph of a smooth function
  $\psi$. Keeping the notation introduced in the proof of Lemma
  \ref{lemm:def_sigma1_zero}, we have that
  \[
    x \mapsto \sigma_1(x, \psi(x)) = \tau_1(x, \psi(x)) +
    \frac{1}{2\pi} \Im(\log(x + i f_r(x,\psi(x))) \] is smooth at $x = 0$, and
  that
  \[
    \sigma_1(x, \psi(x)) - \sigma_1(x,\varphi(x)) \underset{x \to
      0}{\longrightarrow} 0
  \]
  since $\psi(0) = 0 = \varphi(0)$. By Lemma \ref{lemm:def_sigma1_zero}, we know that $\tau_1(x, \varphi(x))$ goes to $\sigma_1(0)$ when $x$ goes to zero. So in view of the above equations, to prove that $\tau_1(x, \psi(x))$ goes to $\sigma_1(0)$ when $x$ goes to zero, it suffices to show that
  \[
    \Im(\log(x + i f_r(x,\psi(x))) = \arctan\left( \frac{f_r(x,\psi(x))}{x}
    \right) \underset{x \to 0^+}{\longrightarrow} 0.
  \]
  Let $\delta > 0$ be small enough and let $B$ be a closed ball
  containing $[0,\delta] \times \varphi([0,\delta])$ and
  $[0,\delta] \times \psi([0,\delta])$. Then for $x \in [0,\delta]$,
  we have that
  \[
    | f_r(x,\psi(x)) | = | f_r(x,\psi(x)) - f_r(x,\varphi(x)) | \leq
    \left( \sup_B \| \dd f_r \| \right) | \psi(x) - \varphi(x) |.
  \]
  But $\psi(0) = \varphi(0)$ and $\psi'(0) = \varphi'(0)$ so
  $\psi(x) - \varphi(x) = \O(x^2)$. Thus,
  \[
    \frac{f_r(x,\psi(x))}{x} \underset{x \to 0^+}{\longrightarrow} 0
  \]
  and the same holds for $\Im(\log(x + i f_r(x,\psi(x)))$.
\end{demo}

Therefore, the first thing that we want to do is to recover the slope
$s(0)$ of $\gamma_r$ from the joint spectrum. In fact, we can do
better and recover both linear terms in the Taylor series expansion of
Eliasson's function $f_r$ (and we will see in Section
\ref{sec:taylor_series} how to recover the higher order terms in this
expansion).

\begin{lemm}
  \label{lemm:dxfr_dyfr}
  One can recover $\partial_x f_r(0)$ and $\partial_y f_r(0)$ from the
  knowledge of $a_1$ and $a_2$ on $\Gamma$. More precisely, for any
  fixed $\mu > 1$, we have the explicit asymptotics
  \[
    \begin{cases} \partial_x f_r(0) = \frac{2\pi (a_1(x,0) - a_1(\mu x,0))}{\ln \mu} + \O(x \ln x), \\[2mm]
      \partial_y f_r(0) = \frac{2\pi (a_2(x,0) - a_2(\mu x,0))}{\ln
        \mu} + \O(x \ln x)
    \end{cases}
  \]
  when $x \to 0^+$.
\end{lemm}

\begin{demo}
  We start with $\partial_y f_r(0)$. We know from Proposition
  \ref{prop:sigma_smooth} that
  \[
    \tau_2(x,0) + \frac{1}{2\pi} \Re(\log(x + i f_r(x,0))) =
    \sigma_2(0) + \O(x)
  \]
  when $x \to 0^+$. But
  \[
    \frac{1}{2\pi} \Re(\log(x + i f_r(x,0))) = \frac{1}{4\pi} \ln(x^2
    + f_r(x,0)^2) = \frac{1}{2\pi} \ln x + \frac{1}{4\pi} \ln\left(1 +
      \frac{f_r(x,0)^2}{x^2}\right).
  \]
  This already suffices to obtain, using Equation \eqref{equ:a1_a2},
  that
  \[
    a_2(x,0) = -\frac{1}{2\pi} \partial_y f_r(0) \ln x + \O(1)
  \]
  which implies that
  \[
    -\frac{2\pi a_2(x,0)}{\ln x} = \partial_y f_r(0) + \O\left(
      \frac{1}{\ln x} \right) \underset{x \to 0^+}{\longrightarrow}
    \partial_y f_r(0).
  \]
  However, this convergence is slow because of the remainder in
  $\frac{1}{\ln x}$, and we can improve its speed by going further
  into the expansion of $a_2$ and using the same trick as in
  \cite[Section 5.3]{san-alvaro-spin}. More precisely, we can write
  \[
    \tau_2(x,0) + \frac{1}{2\pi} \Re(\log(x + i f_r(x,0))) = \tilde
    \tau_2(x,0) + \frac{1}{2\pi} \ln x + \frac{1}{4\pi} \ln(1 +
    \partial_x f_r(0)^2) + \O(x)
  \]
  since $f_r(0) = 0$; this implies, using again Equation
  \eqref{equ:a1_a2}, that
  \[
     a_2(x,0) = \left( -\frac{1}{2\pi} \ln x + \sigma_2(0) -
      \frac{1}{4\pi} \ln(1 + \partial_x f_r(0)^2) \right) \partial_y
    f_r(0) + \O(x \ln x). \] After writing this equation for another
  $\tilde x$ and subtracting both equations, we obtain that
  \[
    a_2(x,0) - a_2(\tilde x,0) = \frac{\partial_y f_r(0)}{2\pi}
    \ln\left( \frac{\tilde x}{x} \right) + \O(x \ln x) + \O(\tilde x
    \ln \tilde x).
  \]
  In particular, if we choose $\tilde x = \mu x$ for some fixed
  $\mu > 1$, this yields
  \[
    \partial_y f_r(0) = \frac{2\pi (a_2(x,0) - a_2(\mu x,0))}{\ln \mu}
    + \O(x \ln x).
  \]

  The case of $\partial_x f_r(0)$ is similar, so we only give a few
  details. We use once again Proposition \ref{prop:sigma_smooth} to
  write
  \[
    \tau_1(x,0) + \frac{1}{2\pi} \Im(\log(x + i f_r(x,0))) =
    \sigma_1(0) + \O(x), \] and we expand
  \[
    \Im(\log(x + i f_r(x,0))) = \arctan\left( \frac{f_r(x,0)}{x}
    \right) = \arctan\left( \partial_x f_r(0) \right) + \O(x).
  \]
  Using this, Equation \eqref{equ:a1_a2} and the Taylor expansion of
  $a_2$ given above, we obtain that
  \begin{multline*} a_1(x,0) = \sigma_1(0) - \frac{1}{2\pi}
    \arctan\left( \partial_x f_r(0) \right) + \left(\sigma_2(0) -
      \frac{\ln x}{2\pi} - \frac{1}{4\pi} \ln(1 + \partial_x f_r(0)^2)
    \right) \partial_x f_r(0) \\ + \O(x \ln x) \end{multline*} and by
  the same reasoning as above, we get
  \[
    \partial_x f_r(0) = \frac{2\pi (a_1(x,0) - a_1(\mu x,0))}{\ln \mu}
    + \O(x \ln x) \] for any given $\mu > 1$.
\end{demo}

Using Lemma \ref{lemm:quantum-a} to obtain $a_1$ and $a_2$ from the
spectrum, Lemma~\ref{lemm:dxfr_dyfr} yields the following formulas.
\begin{coro} Consider a labelling covering both $(x,0)$ and
    $(\mu x, 0)$, and let $E_{j,\ell}$ be the second component of
    $\lambda_{j,\ell}$, see~\eqref{equ:semitoric-labelling}. Let
    $j_1,\ell_1$ (respectively $j_2,\ell_2$) be $\h$-dependent
    integers such that the joint eigenvalues $\lambda_{j_1,\ell_1}$,
    $\lambda_{j_1+1,\ell_1}$ and $\lambda_{j_1,\ell_1+1}$
    (respectively $\lambda_{j_2,\ell_2}$, $\lambda_{j_2+1,\ell_2}$ and
    $\lambda_{j_2,\ell_2+1}$) are well-defined in an
    $\O(\h)$-neighborhood of $(x,0)$ (respectively $(\mu x,
    0)$). Then,
    \begin{equation} \partial_x f_r(0) = \lim_{x \to 0^+} \lim_{\h \to
        0} \frac{2\pi}{\ln \mu} \left( \frac{E_{j_1,\ell_1} -
          E_{j_1+1,\ell_1}}{E_{j_1,\ell_1+1} - E_{j_1,\ell_1}} -
        \frac{E_{j_2,\ell_2} - E_{j_2+1,\ell_2}}{E_{j_2,\ell_2+1} -
          E_{j_2,\ell_2}} \right) \label{eq:dxfr_lim}\end{equation}
    and
    \begin{equation} \partial_y f_r(0) = \lim_{x \to 0^+} \lim_{\h \to
        0} \frac{2\pi \h}{\ln \mu} \left( \frac{1}{E_{j_1,\ell_1+1} -
          E_{j_1,\ell_1}} - \frac{1}{E_{j_2,\ell_2+1} -
          E_{j_2,\ell_2}} \right) \,. \label{eq:dyfr_lim}\end{equation}
  \end{coro}

\begin{rema}\label{rema:double_limit} We see in the above
    corollary that some of the quantities that we recover from the
    joint spectrum should be obtained by first taking the limit
    $\h \to 0$ for a quantity defined at a regular value $c$ and then
    the limit where $c$ tends to the focus-focus critical value. Hence
    for numerical purposes, it is important to fix some $c$ close to
    the critical value and let $\h$ vary for this given $c$. If
    instead one fixes a small value of $\h$ and lets $c$ vary, one
    could get less convincing results since it may happen that $\h$
    should be taken smaller and smaller as $c$ becomes closer to the
    singular value, see Example~\ref{ex:double_limit_spin-osc}.
\end{rema}

\begin{ex}\label{ex:dfr_spin-osc} For the spin-oscillator system
    (see Example \ref{ex:spin-osc_quant}), we recover
    $\partial_x f_r(0)$ in Figure \ref{fig:dxfr_spinosc} and
    $\partial_y f_r(0)$ in Figure \ref{fig:dyfr_spinosc} using
    Formulas \eqref{eq:dxfr_lim} and \eqref{eq:dyfr_lim} (and hence we
    obtain $s(0)$). In these formulas, the quantities $E_{j,\ell}$
    depend on the semiclassical parameter $\h = k^{-1}$. Recall from
    Equation \eqref{eq:fr_elia_spinosc} that $\partial_x f_r(0) = 0$
    and $\partial_y f_r(0) = 2$, which implies $s(0)=0$.
    \begin{figure}[H]
  \begin{center}
    \includegraphics[trim=80 10 80
    60,clip,width=0.75\textwidth]{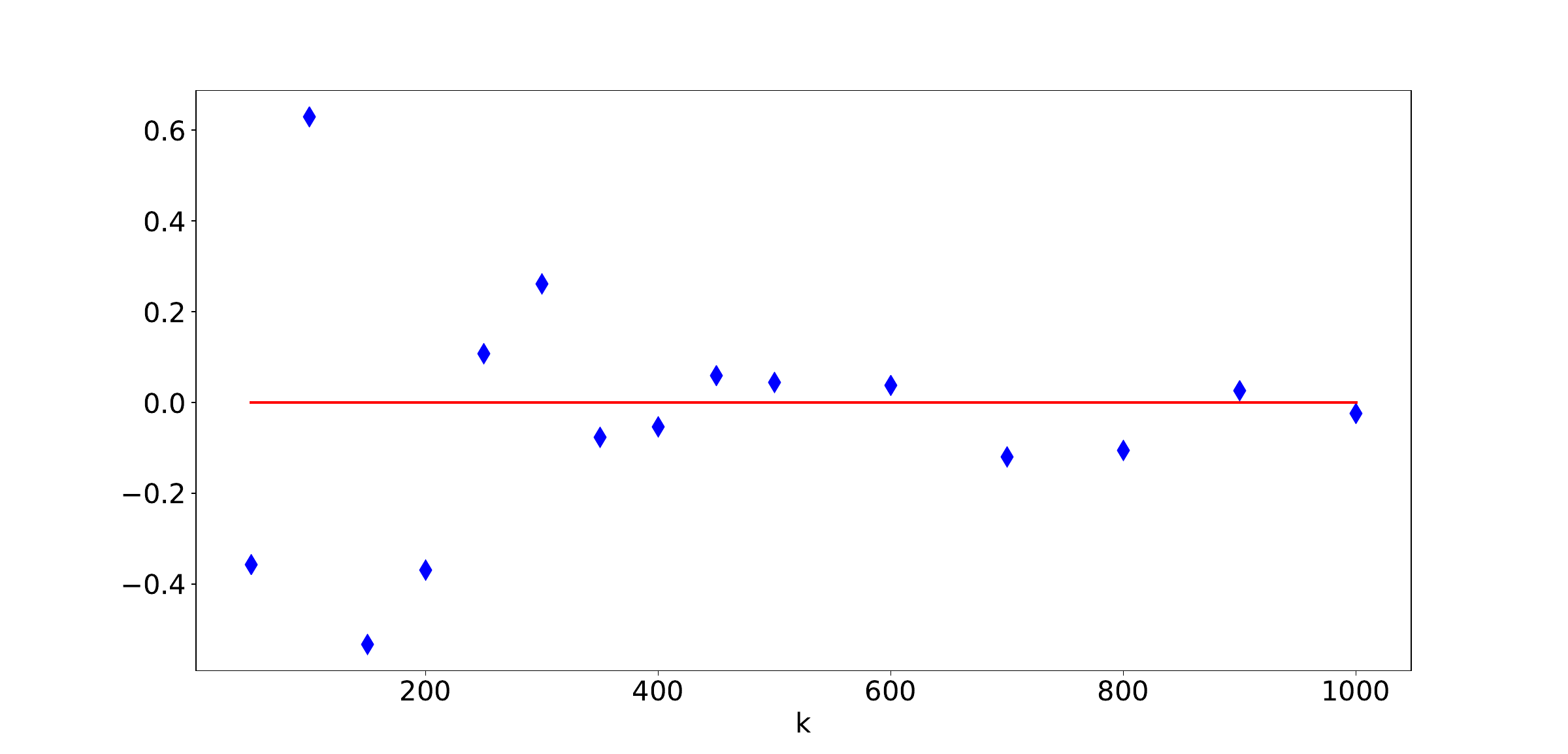}
  \end{center}
  \caption{\small Determination of $\partial_x f_r(0)$ for the
    spin-oscillator using Formula \eqref{eq:dxfr_lim} with $x = 0.01$,
    $\mu = 2$, $(j_1,\ell_1) = (0,0)$, and $ (j_2,\ell_2)$
    corresponding to the closest eigenvalue to $(\mu x, 0)$, for
    different values of $k$. The red line corresponds to the
    theoretical result $\partial_x f_r(0) = 0$.}
  \label{fig:dxfr_spinosc}
\end{figure}
\begin{figure}[H]
  \begin{center}
    \includegraphics[trim=80 10 80
    60,clip,width=0.75\textwidth]{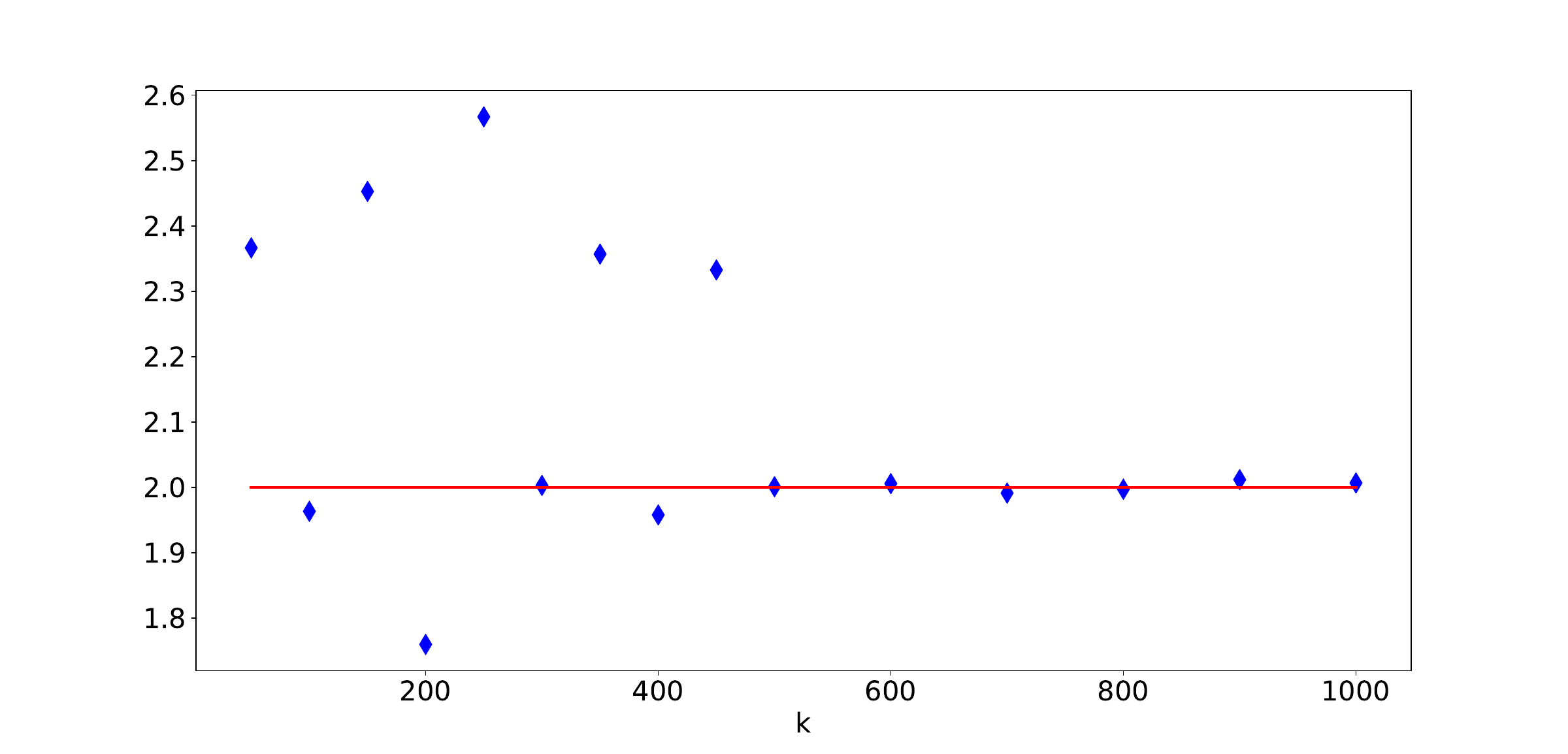}
  \end{center}
  \caption{\small Determination of $\partial_y f_r(0)$ for the
    spin-oscillator using Formula \eqref{eq:dyfr_lim} with $x = 0.01$,
    $\mu = 2$, $(j_1,\ell_1) = (0,0)$, and $(j_2,\ell_2)$
    corresponding to the closest eigenvalue to $(\mu x, 0)$, for
    different values of $k$. The red line corresponds to the
    theoretical result $\partial_y f_r(0) = 2$.}
  \label{fig:dyfr_spinosc}
\end{figure}
\end{ex}

\begin{demo}[of Theorem~\ref{theo:recover-sigma_1}]
  \paragraph{Step 1.} There are no well-defined action-angle
  coordinates at the origin, but the idea is to choose action
  variables near regular values $c$ on the radial curve $\gamma_r$
  (Definition~\ref{defi:radial-curve}) in a continuous way. For any
  $c\neq0$ in a sectorial neighborhood of $\gamma_r$, we choose a
  semitoric linear labelling $\lambda_{j, \ell}(\h)$, in such a way that the
  corresponding action variable $L$ does not depend on $c$ (this is
  always possible; from a practical viewpoint, a discontinuity in this
  action would be reflected by the composition of the labelling with a
  matrix of the form $\begin{pmatrix} 1 & 0 \\ n & 1 \end{pmatrix}$
  for a fixed integer $n$, and thus would be easily detectable), apply
  Lemma~\ref{lemm:quantum-a}, and let $\h\to0$; thus, we are able to
  recover from the joint spectrum the functions $a_1$ and $a_2$ on a
  punctured neighborhood of the origin. These functions are
  single-valued if we stick to a simply connected open subset of the
  punctured neighborhood (here, the right-half plane $\Gamma$ as in
  Section \ref{sec:twist-numb-twist}).

  \paragraph{Step 2.} 
  Thanks to Lemma \ref{lemm:dxfr_dyfr} (see also Formulas
  \eqref{eq:dxfr_lim} and \eqref{eq:dyfr_lim}), we can then recover
  the slope $s(0) = -\frac{\partial_x f_r(0)}{\partial_y f_r(0)}$ from
  the joint spectrum.

  \paragraph{Step 3.}
  We will approximate $\gamma_r$ by the line through the origin with
  slope $s(0)$. Thus, we define $\tilde\sigma(x) := \tau_1(x,s(0)x)$
  for $x > 0$.  Using~\eqref{equ:alpha} again, we have, when
  $c=(x,s(0)x)$,
  \[
    \tilde\sigma(x) = a_1(c) + s(c) a_2(c) = a_1(c) + s(0)a_2(c) +
    \O(x \ln x) \] because $a_2(c) \sim C \ln x$ for some constant
  $C \neq 0$ by Equation \eqref{equ:a1_a2} and Proposition
  \ref{prop:sigma_smooth}.  Since
  $\sigma_1(0) = \lim_{x\to 0^+}\tilde\sigma(x)$ (see
  Lemma~\ref{lemm:tangent}), we get
  \[
    \sigma_1(0) = \lim_{x\to 0^+} a_1(c) + s(0)a_2(c), \quad
    c=(x,s(0)x).
  \]
  In view of Steps 1 and 2, this shows that $\sigma_1(0)$ can be
  recovered from the joint spectrum.
  
\end{demo}

\begin{rema} In practice, applying Lemma \ref{lemm:quantum-a}, we have
  for $c = (x, s(0) x)$
  \begin{equation} \tilde\sigma_1(x) := a_1(c) + s(0) a_2(c) =
    \frac{E_{j,\ell} - E_{j+1,\ell}}{E_{j,\ell+1} - E_{j,\ell}} +
    \frac{\h s(0)}{E_{j,\ell+1} - E_{j,\ell}} + \O_c(\h)
    \label{equ:sigma} \end{equation}
  once $s(0)$ is known, and $\sigma_1(0) = \lim_{x \to 0^+} \tilde\sigma_1(x)$.
\end{rema}

\begin{ex}\label{ex:S10_spin-osc} We consider once again the
    spin-oscillator system of Example \ref{ex:spin-osc_quant}. We
    obtain a good labelling and recover the associated $\sigma_1(0)$
    by using Formula \eqref{equ:sigma} for a fixed regular value $c$
    close to the focus-focus value and varying $k$ in Figure
    \ref{fig:sigma_spinosc_kvar}, as follows: since $s(0)=0$ (see
    Example~\ref{ex:dfr_spin-osc}), we have
    \begin{equation}
    \sigma_1(0) = \lim_{x \to 0^+}\lim_{k\to+\infty} \frac{E_{0,0} -
      E_{1,0}}{E_{0,1} - E_{0,0}} \,,\label{equ:sigma_spin-osc}
  \end{equation}
    where $E_{0,0}, E_{1,0}$ and $E_{0,1}$ are obtained from the joint
    eigenvalues close to $c=(x,0)$
    (see~\eqref{equ:semitoric-labelling}) using the labelling given by
    the algorithm of~\cite[Section 3.5.2]{san-dauge-hall-rotation},
    see Section~\ref{subsect:asympt_latt}; in this case, numerics
    suggest $\sigma_1(0)=0$, which is an unstable case, see
    Remark~\ref{rema:taylor2}. We investigate the corresponding error
    term in Figure \ref{fig:sigma_spinosc_kvar_err}; recall that the
    integer part of $\sigma_1(0)$ gives the twisting number $p$
    associated with the action variable selected by the labelling,
    while its fractional part gives the coefficient $[S_{1,0}]$ of the
    Taylor series invariant.  From the numerics only, we cannot tell
    whether $(p=0, \sigma_1(0) \geq 0)$ or
    $(p=-1, \sigma_1(0) < 0)$.  On the other hand, the
    invariant $[S_{1,0}]$, being defined modulo $\ZM$, is actually
    stable and close to zero, as predicted theoretically from Equation
    \eqref{eq:taylor_series_spinosc}. The twisting index is equally
    stable, since the two cases mentioned above are related by an
    action of $\mathcal{T}$, see Section~\ref{sec:semitoric-polygons}.
    \begin{figure}[H]
  \begin{center}
    \includegraphics[trim=80 10 80
    60,clip,width=0.75\textwidth]{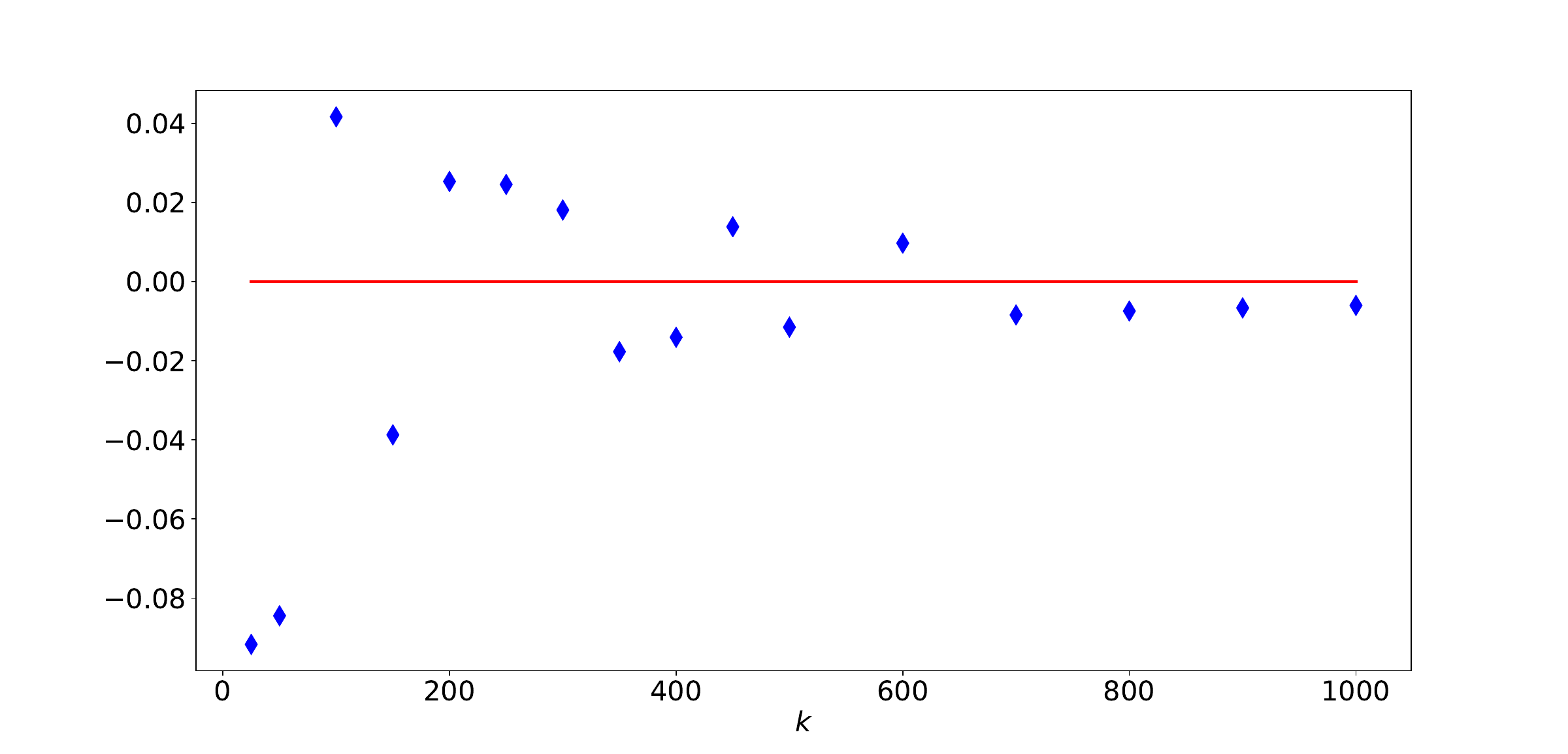}
  \end{center}
  \caption{\small Determination of $\sigma_1(0)$ for the
    spin-oscillator system, see Example \ref{ex:S10_spin-osc}. The
    blue diamonds correspond to Formula \eqref{equ:sigma} evaluated at
    $(j,\ell) = (0,0)$ with $x = 0.01$, for different values of
    $k$. The red line corresponds to the theoretical value
    $\sigma_1^{\textup{p}}(0) = 0$. Since the value of the invariant
    is integer, this is an example where the twisting number is
    unstable.}
  \label{fig:sigma_spinosc_kvar}
\end{figure}
\begin{figure}[H]
  \begin{center}
    \includegraphics[trim=80 10 80
    60,clip,width=0.75\textwidth]{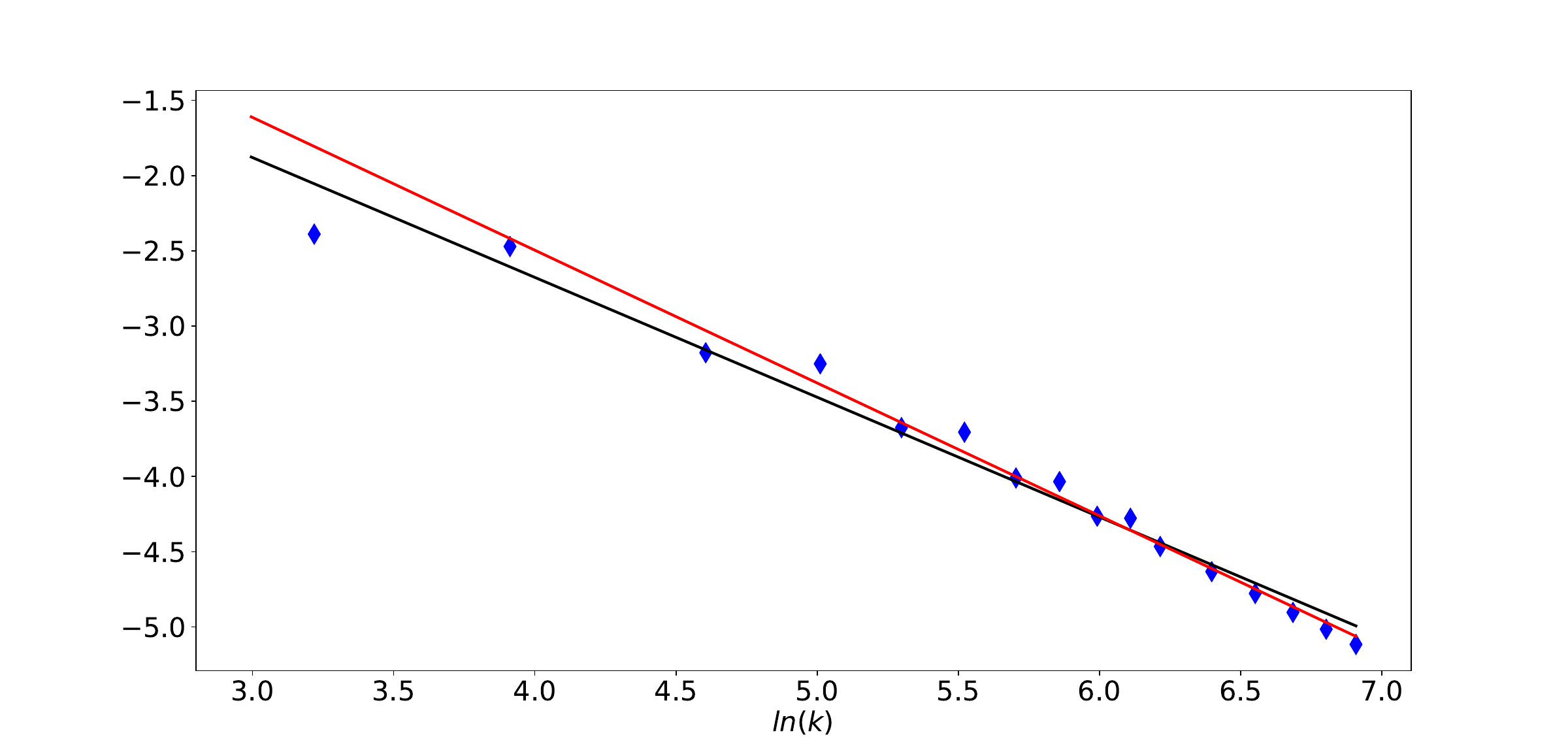}
  \end{center}
  \caption{\small Error in the determination of
    $\sigma_1(0)$ for the spin-oscillator system. The
    blue diamonds correspond to the logarithm of the error between
    Formula \eqref{equ:sigma_spin-osc} evaluated at $(j,\ell) = (0,0)$
    and $\sigma_1(0)=0$ with $x = 0.01$, for different
    values of $\ln k$. In black, the line of linear regression; in
    red, the line of linear regression computed after discarding the
    first two points. }
  \label{fig:sigma_spinosc_kvar_err}
\end{figure}
\end{ex}

\begin{figure}[H]
  \begin{subfigure}{.49\textwidth}
    \centering \includegraphics[trim=80 10 80
    60,clip,width=\textwidth]{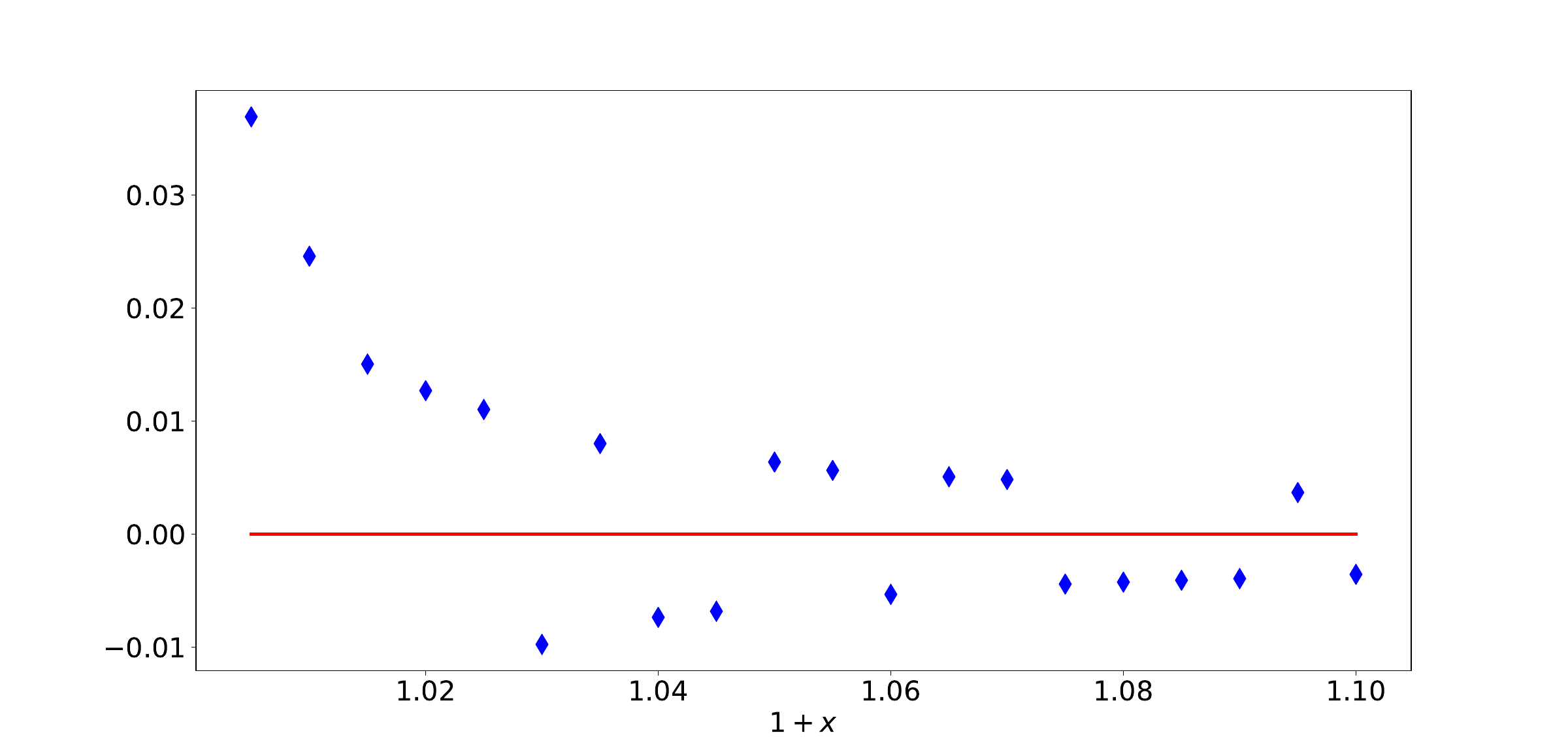}
    \caption{\small $k=250$.}
  \end{subfigure}
  \begin{subfigure}{.49\textwidth}
    \centering \includegraphics[trim=80 10 80
    60,clip,width=\textwidth]{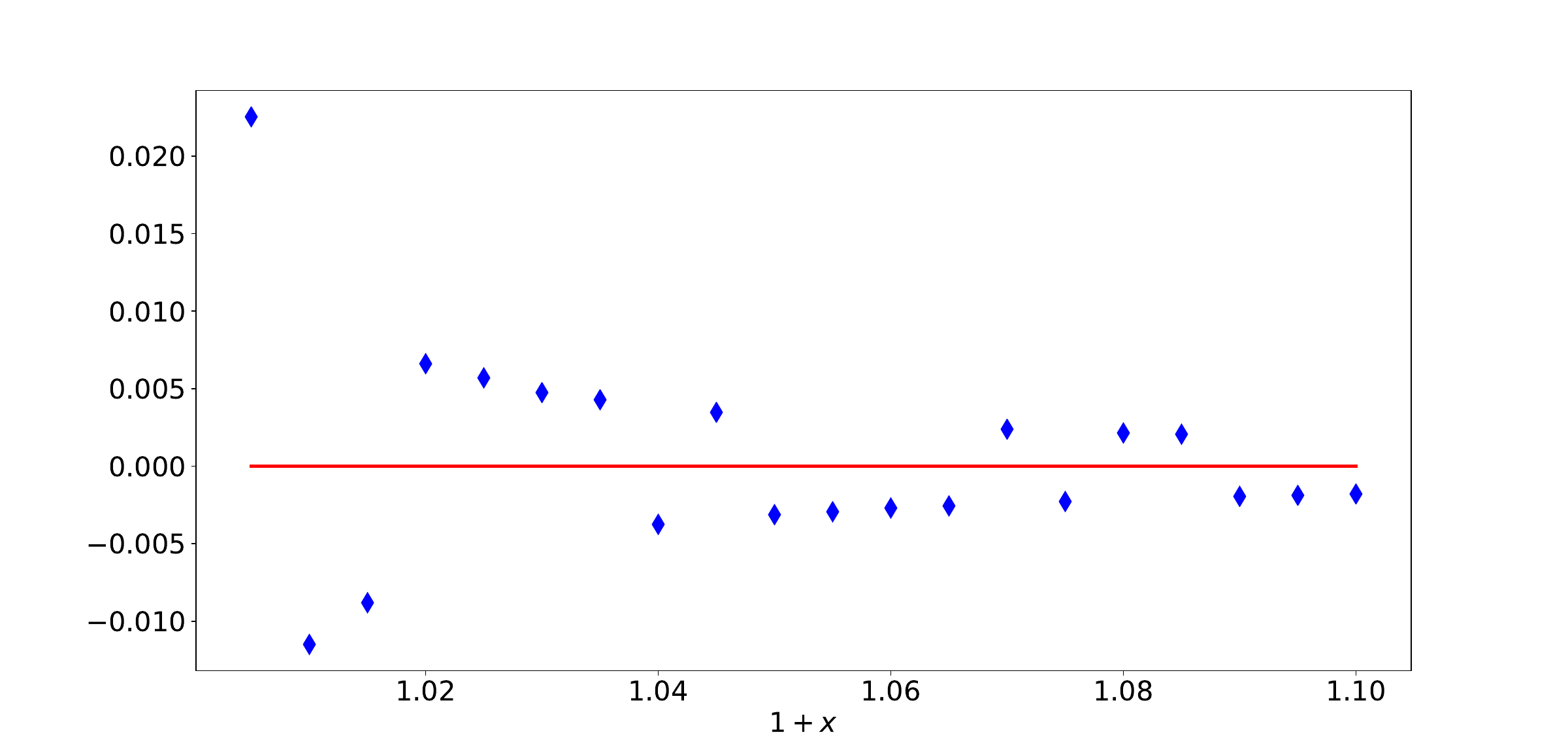}
    \caption{\small $k=500$.}
  \end{subfigure}
  \caption{\small Determination of $\sigma_1(0)$ for the
    spin-oscillator system. The blue diamonds correspond to Formula
    \eqref{equ:sigma} evaluated at $(j,\ell) = (0,0)$ for a given $k$
    and different values of $x$. The red line corresponds to the
    theoretical value $\sigma_1^{\text{p}}(0) = 0$. The focus-focus
    critical value corresponds to $x=0$.}
  \label{fig:sigma_spin_osc_fixed_k}
\end{figure}

\begin{ex}\label{ex:double_limit_spin-osc}
    Formula~\eqref{equ:sigma_spin-osc} for the numerical computation
    of $\sigma_1(0)$ for the spin-oscillator is an instance of the
    numerical issue concerning double limits mentioned in
    Remark~\ref{rema:double_limit}. We illustrate this point in Figure
    \ref{fig:sigma_spin_osc_fixed_k}.
\end{ex}

As a consequence of Theorem~\ref{theo:recover-sigma_1}, we obtain the following interesting fact.
\begin{prop}
  \label{prop:priv_label}
  Given the $\h$-family of joint spectra $\Sigma_\h$ of a proper
  quantum semitoric system in a neighborhood of a focus-focus critical
  value $c_0=0$, one can recover a semitoric linear labelling
  associated with the privileged momentum map $(J,L_{\textup{priv}})$,
  in a constructive way.
\end{prop}

\begin{demo}
  Starting with an arbitrary semitoric linear labelling
  $(\lambda_{j,\ell})$, associated with some action variables $(J,L)$,
  we apply Theorem~\ref{theo:recover-sigma_1} to recover
  $\sigma_1(0)$. Now let $p=\floor{\sigma_1(0)}$.  The privileged
  action is $L_\textup{priv} = L - p J$; the new set of action variables $(J,L_\textup{priv})$ is obtained from $(J,L)$ by acting by the matrix
$A=
\begin{pmatrix}
  1&0\\-p&1
\end{pmatrix}
$, and
  hence (see Remark~\ref{rema:change_A}) the privileged
  labelling $\lambda_{j,\ell}^\textup{p}$ is given by
  $\lambda_{j,\ell}(\h) = \lambda^\textup{p}_{j, \ell - p j}(\h)$,
  \emph{i.e.}
  \[
    \lambda^\textup{p}_{j,\ell}(\h) = \lambda_{j, \ell + p j}(\h).
  \]
\end{demo}

\begin{ex} To illustrate Proposition \ref{prop:priv_label}, we
    consider the example of the spin-oscillator (Example
    \ref{ex:spin-osc_class}) and replace $H$ by $H + J$; this gives us
    a new semitoric system $(J,H+J)$ with one focus-focus singularity
    with singular value $(1,1)$. The labelling $\lambda_{j, \ell}(\h)$
    provided by the algorithm described at the end of Section
    \ref{subsect:asympt_latt} yields $p=3$, thus the privileged
    labelling satisfies
    $\lambda^\textup{p}_{j,\ell}(\h) = \lambda_{j, \ell + 3
      j}(\h)$. This is depicted in Figure \ref{fig:label_priv}. Note
    that $s(0) = 1$ in this example, since Eliasson's functions
    $\tilde{f}_r$ for this system and $f_r$ for the spin-oscillator
    system are related by $\tilde{f}_r(x,y) = f_r(x,y-x)$.
\begin{figure}[H]
  \begin{subfigure}{.48\textwidth}
    \centering \includegraphics[scale=0.38]{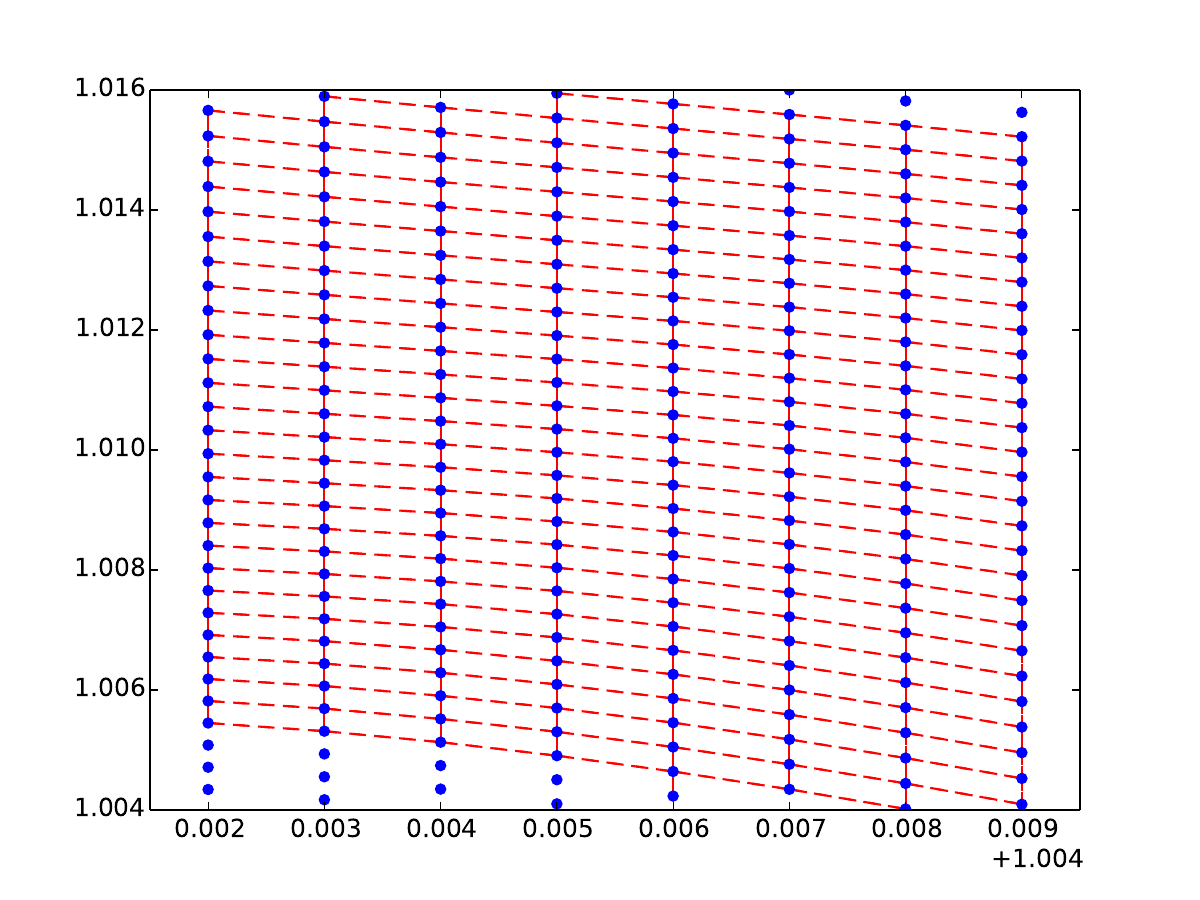}
    \caption{\small The labelling $\lambda_{j, \ell}(\h)$.}
  \end{subfigure}
  \begin{subfigure}{.48\textwidth}
    \centering \includegraphics[scale=0.38]{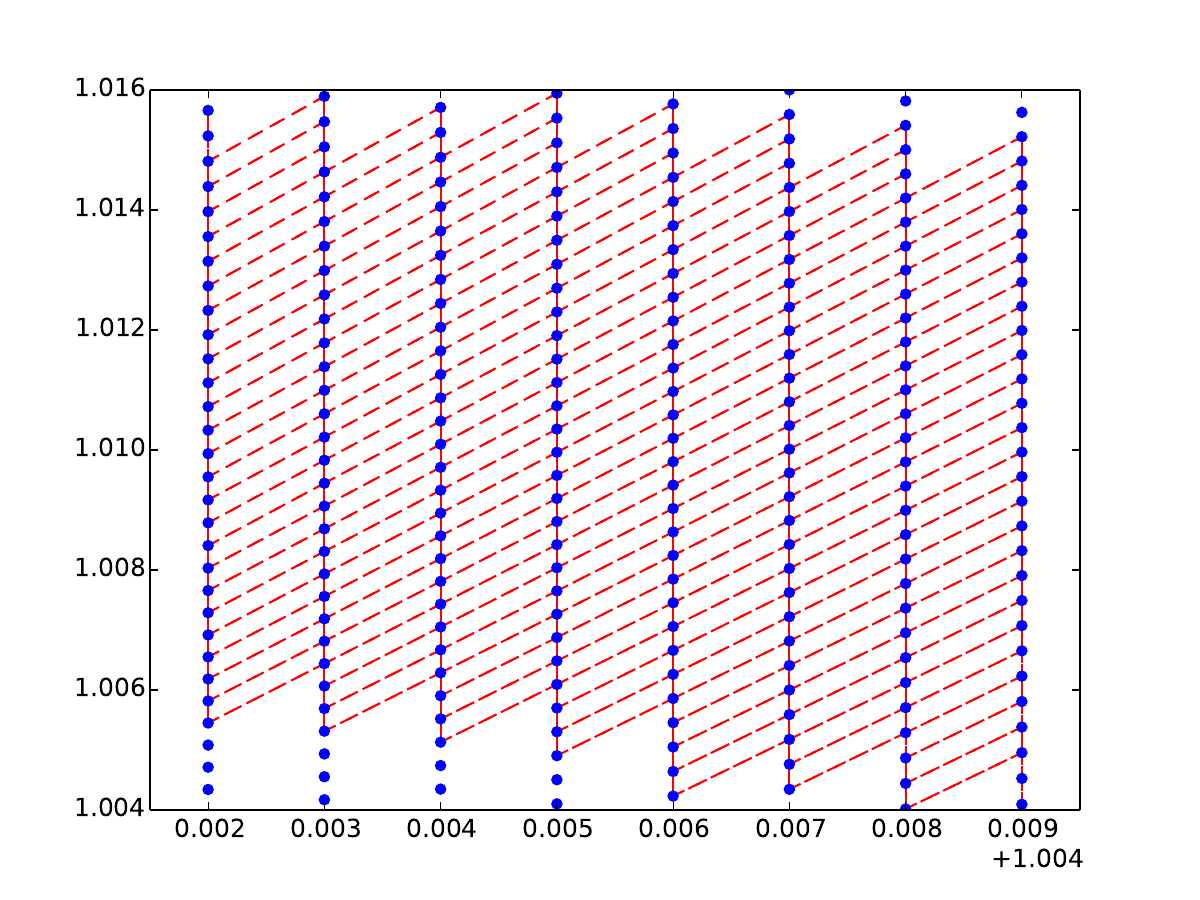}
    \caption{\small The labelling $\lambda^\textup{p}_{j,\ell}(\h)$.}
  \end{subfigure}
  \caption{\small Two labellings of the joint spectrum near
    $c = (1.01,1.01)$ for the system $(J,H+J)$ where $(J,H)$ is the
    spin-oscillator system, for $k=1000$. Here $s(0) = 1$ and the
    approximate value of $\sigma_1(0)$ given by
    Formula~\eqref{equ:sigma} evaluated at $x = 0.01$ with
    $(j,\ell) = (0,0)$ is
    $\frac{E_{0,0} - E_{1,0}}{E_{0,1} - E_{0,0}} + \frac{1}{k(E_{0,1}
      - E_{0,0})} \approx 3.046$. The labellings are suggested by the
    red dashed lines.}
  \label{fig:label_priv}
\end{figure}
\end{ex}

\subsection{Recovering the twisting index invariant}
\label{subsect:polygon}

Let $\Sigma_\h$ be the joint spectrum of a proper semitoric quantum
integrable system $(\hat{J}_\h, \hat{H}_\h)$ with joint principal
symbol $F = (J,H)$. Let $(x_1, y_1), \ldots, (x_{m_f}, y_{m_f})$ be
the focus-focus values of $F$ and let $V_1, \ldots, V_{m_f}$ be the
vertical half-lines defined as
\begin{equation}
  \forall j \in \{1, \ldots, m_f \}, \qquad V_j = \{ (x_j, y) \ | \ y
  \geq y_j \}.
  \label{equ:Vj}
\end{equation}
Moreover, let $E$ be the set of elliptic-elliptic
values of $F$ and let $W$ be the set of vertical walls of $F$; note
that $E$ and $W$ may be empty. Actually we can also have $m_f = 0$, in
which case there is no $L_j$; we include this case in what follows,
even though we slightly abuse notation for the sake of simplicity.

\begin{figure}[H]
  \begin{subfigure}{.49\textwidth}
    \centering
    \includegraphics[width=0.9\linewidth]{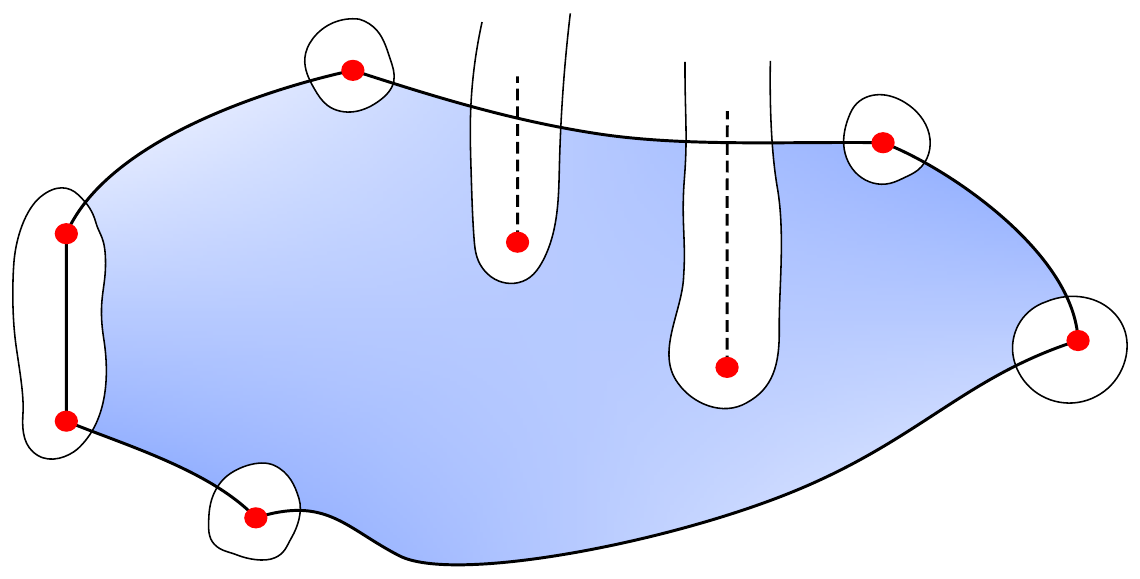}
    \caption{\small A good choice of $\mathcal{U}$. }
  \end{subfigure}
  \begin{subfigure}{.49\textwidth}
    \centering
    \includegraphics[width=0.9\linewidth]{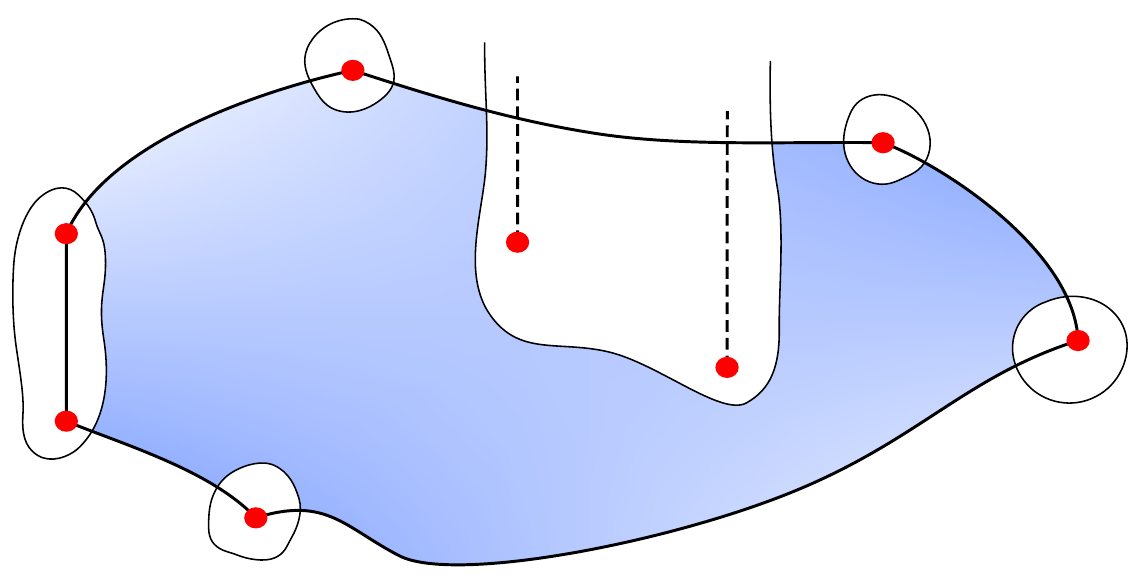}
    \caption{\small A bad choice of $\mathcal{U}$, where we will miss the
      edge of the polygon with vertices coming from the two
      focus-focus values.}
  \end{subfigure}
  \caption{\small Two examples of choice of $\mathcal{U}$. In these examples
    $F(M)$ is compact and we take
    $S = \{ (x,y) \in \RM^2 \ | \ u_1 \leq x \leq u_2 \}$ with
    $J(M) \subset [u_1,u_2]$. The rank zero (elliptic-elliptic and focus-focus) critical points of $F$ are
    indicated by red dots, and $F(M) \cap \mathcal{U}^c$ is the blue
    filled region.}
  \label{fig:choice_compact_labelling}
\end{figure}

We will now explain how to recover a representative of the polygon
invariant from the joint spectrum. In fact, since this polygon may not
be bounded, what we really recover is its intersection with any
vertical strip. So we consider a pair $(S,\mathcal{U})$ such that
$S \subset \RM^2$ is a vertical strip
$S = \{(x,y) \ | \ u_1 \leq x \leq u_2 \}$ where
$u_1, u_2 \notin \{x_1, \ldots, x_{m_f} \}$, $\mathcal{U}$ and
$\mathcal{V}$ are open neighborhoods of
$V_1 \cup \ldots \cup V_{m_f} \cup E \cup W$ with
$\mathcal{U} \Subset \mathcal{V}$,  and
$\mathcal{K}(S,\mathcal{U}) := F(M) \cap S \cap \mathcal{U}^c$ is
simply connected (note that $\mathcal{K}(S,\mathcal{U})$ is compact
since $J$ is proper), and with $\mathcal{U}$ small enough to avoid
problems with consecutive critical values of $F$ (see Figure
\ref{fig:choice_compact_labelling}). For instance one can construct
$\mathcal{U}$ as the union of $\varepsilon$-neighborhoods of every
$V_i$, of every element of $E$ and of $W$ for some $\varepsilon > 0$
small enough.

Let $c \in \mathcal{K}(S,\mathcal{U})$. By construction, $c$ is either
a regular value of $F$, in which case Theorem \ref{theo:BS_reg} states
that there exists an open ball $B_c \subset \mathcal{V}$ containing
$c$ such that $(\Sigma_\h \cap B_c,\mathcal{I},B_c)$ is an asymptotic
lattice, or a $J$-transversally elliptic value of $F$, in which case
by Theorem \ref{theo:BS_ell_trans}, there exists an open ball
$B_c \subset \mathcal{V}$ containing $c$ such that
$(\Sigma_\h \cap B_c,\mathcal{I},B_c)$ is an asymptotic
half-lattice. Since $\mathcal{K}(S,\mathcal{U})$ is compact, we can
extract from the open cover
$\bigcup_{c \in \mathcal{K}(S,\mathcal{U})} B_c \supset
\mathcal{K}(S,\mathcal{U})$ a finite open cover
$\bigcup_{\ell=1}^q B_j \supset \mathcal{K}(S,\mathcal{U})$ so that
there exists $p \in \{0, \ldots, q\}$ such that for every
$j \in \{1, \ldots, p\}$, $(\Sigma_\h \cap B_j, \mathcal{I}, B_j)$ is
an asymptotic lattice and for every $j \in \{p+1, \ldots, q\}$,
$(\Sigma_\h \cap B_j, \mathcal{I}, B_j)$ is an asymptotic
half-lattice. Let $(\tilde{B}_j)_{1 \leq j \leq q}$ be a refinement of
$(B_j)_{1 \leq j \leq q}$ satisfying the assumptions of Theorem
\ref{thm:construct_global_labelling} with respect to
$\mathcal{L}_\h^{(i)} = \Sigma_\h \cap B_i$. Using this theorem,
we construct two maps
$\ell_\h: \Sigma_\h \cap \mathcal{K}(S,\mathcal{U}) \to \ZM^2$ and
$\Phi: \mathcal{K}(S,\mathcal{U}) \to \RM^2$ and a vector
$\nu_\h \in \ZM^2$ such that for every $j \in \{1, \ldots, q \}$,
${\ell_\h}_{|\Sigma_\h \cap \tilde{B}_j}$ is a linear labelling and
$\| \Phi(\lambda) - \h \ell_\h(\lambda) - \nu_\h \| \leq K \h$.

\begin{lemm}
  \label{lm:phi_semitoric}
  The maps $\ell_\h$ and $\Phi$ can be chosen to be semitoric,
  i.e. such that for every $j \in \{1, \ldots, q\}$,
  ${\rm d}(\Phi_{|\tilde{B}_j})^{-1}(\xi_1,\xi_2) = {\rm d}\xi_1$.
\end{lemm}

\begin{demo}
  Let $j \in \{1, \ldots, q\}$. We can choose a semitoric labelling
  $\ell_\h^{(j)}$ for $\Sigma_\h \cap B_j$. This comes from
  \cite[Lemma 3.34]{san-dauge-hall-rotation} if $1 \leq j \leq p$ and
  from the fact that $G_0^{(j)}$ in Theorem \ref{theo:BS_ell_trans}
  can be chosen semitoric, see Lemma \ref{lemm:dufour-molino}, if
  $p+1 \leq j \leq q$. By Theorem
  \ref{thm:construct_global_labelling}, the choice of this labelling
  $\ell_\h^{(j)}$ determines $\ell_\h$ such that $\ell_\h^{(j)}$
  becomes the restriction of $\ell_\h$ to
  $\Sigma_\h \cap \tilde{B}_j$. And since $G_0^{(j)}$ is semitoric,
  for all $i \in \{1,\dots,q\}$ the map $G_0^{(i)}$ associated with
  the asymptotic chart of the restriction of $\ell_\h$ to
  $\Sigma_\h \cap \tilde{B}_i$ must be semitoric as well.
\end{demo}

In what follows, we will always assume that $\ell_\h$ and $\Phi$ are
semitoric.

\begin{defi}
  We call $\ell_\h$ a \emph{quantum cartographic map} associated with
  $S$ and $\mathcal{U}$.
\end{defi}

By construction of the cartographic homeomorphism (see \cite[Section
5.2.2]{san-daniele}), we have the following.

\begin{lemm}
  \label{lm:carto}
  The map $\Phi$ uniquely extends to $F(M)$ to a cartographic
  homeomorphism for $(M,\omega,F)$, that we still call $\Phi$. This
  cartographic homeomorphism corresponds to
  $\vec{\epsilon} = (1, \ldots, 1)$, using notation from Section
  \ref{sec:semitoric-polygons}.
\end{lemm}

Let $d$ be the Euclidean distance on $\RM^2$. Recall that the
Hausdorff distance between two compact sets $K_1, K_2 \subset \RM^2$
is defined as
\[
  d_H(K_1,K_2) = \inf\{ \varepsilon \geq 0 \ | \ K_1 \subset
  K_2^{\varepsilon} \text{ and } K_2 \subset K_1^{\varepsilon} \}
\]
where
$K_j^{\varepsilon} = \bigcup_{x \in K_j} \{ y \in \RM^2 \ | \ d(x,y)
\leq \varepsilon \}$.

\begin{prop}
  \label{prop:poly_num}
  Let $\ell_\h$ be a quantum cartographic map associated with $S$ and
  $\mathcal{U}$, let $\Phi$ be the corresponding cartographic
  homeomorphism, let $\nu_\h$ be the corresponding vector (see
  Theorem \ref{thm:construct_global_labelling}) and let
  $\Delta_\h(\mathcal{K}(S,\mathcal{U})) := \h
  \ell_\h(\mathcal{K}(S,\mathcal{U}))$. Then the set
  $\nu_\h + \Delta_\h(\mathcal{K}(S,\mathcal{U}))$ converges to
  $\Phi(\mathcal{K}(S,\mathcal{U}))$ when $\h$ goes to $0$ in the
  sense of the Hausdorff distance. More precisely, there exists
  $\h_1 > 0$ and a constant $C > 0$ such that
  \[
    \forall \h \in \mathcal{I} \cap [0,\h_1] \qquad d_H(\nu_\h +
    \Delta_\h(\mathcal{K}(S,\mathcal{U})),
    \Phi(\mathcal{K}(S,\mathcal{U}))) \leq C \h.
  \]
\end{prop}

\begin{demo}
  Let $M$ be the maximum of $\| \dd \Phi \|$ on
  $\overline{\bigcup_{i=1}^q B_i}$. Let
  $\xi \in \Phi(\mathcal{K}(S,\mathcal{U}))$; there exists
  $c \in \mathcal{K}(S,\mathcal{U})$ such that $\xi = \Phi(c)$. From
  \cite[Lemma 3.15]{san-dauge-hall-rotation} and its proof (which
  works similarly for half-lattices), there exist a constant $\alpha > 0$,
  $\h_1 \in \mathcal{I}$ and a family
  $(\lambda_\h)_{\h \in \mathcal{I} \cap [0,\h_1]}$ such that
  $ \forall \h \in \mathcal{I} \cap [0,\h_1]$,
  $\lambda_\h \in \mathcal{K}(S,U)$ and
  $\| \lambda_\h - c \| \leq \alpha \h $. Let
  $\xi_\h = \h \ell_\h(\lambda_\h)$; then
  $\xi_\h \in \Delta_\h(\mathcal{K}(S,\mathcal{U}))$ and
  \[
    \| \nu_\h + \xi_\h - \xi \| \leq \| \nu_\h + \xi_h -
    \Phi(\lambda_\h) \| + \| \Phi(\lambda_\h) - \Phi(c) \| \leq (K + \alpha
    M) \h \] where $K$ is as in Theorem
  \ref{thm:construct_global_labelling}.

  Conversely, let $(\xi_\h)_{\h \in \mathcal{I}}$ be a family of
  elements of $\Delta_\h(\mathcal{K}(S,\mathcal{U}))$. Then there
  exists a family $(\lambda_\h)_{\h \in \mathcal{I}}$ of elements of
  $\Sigma_\h \cap \mathcal{K}(S,\mathcal{U})$ such that for every
  $\h \in \mathcal{I}$, $\xi_\h = \h \ell_\h(\lambda_\h)$. Then
  $\Phi(\lambda_\h) \in \Phi(\mathcal{K}(S,\mathcal{U}))$ and
  \[
    \| \nu_\h + \xi_\h - \Phi(\lambda_\h) \| \leq K \h \leq (K + \alpha M)
    \h.
  \]

  Hence
  $d_H(\nu_\h + \Delta_\h(\mathcal{K}(S,\mathcal{U})),
  \Phi(\mathcal{K}(S,\mathcal{U}))) \leq C \h$ with
  $C = K + \alpha M$.
\end{demo}

As a corollary, from the joint spectrum of a proper quantum semitoric
system, we can recover the twisting index invariant of the underlying
classical semitoric system.

\begin{figure}[H]
  \begin{center}
    \includegraphics[trim=80 10 80
    60,clip,width=0.75\textwidth]{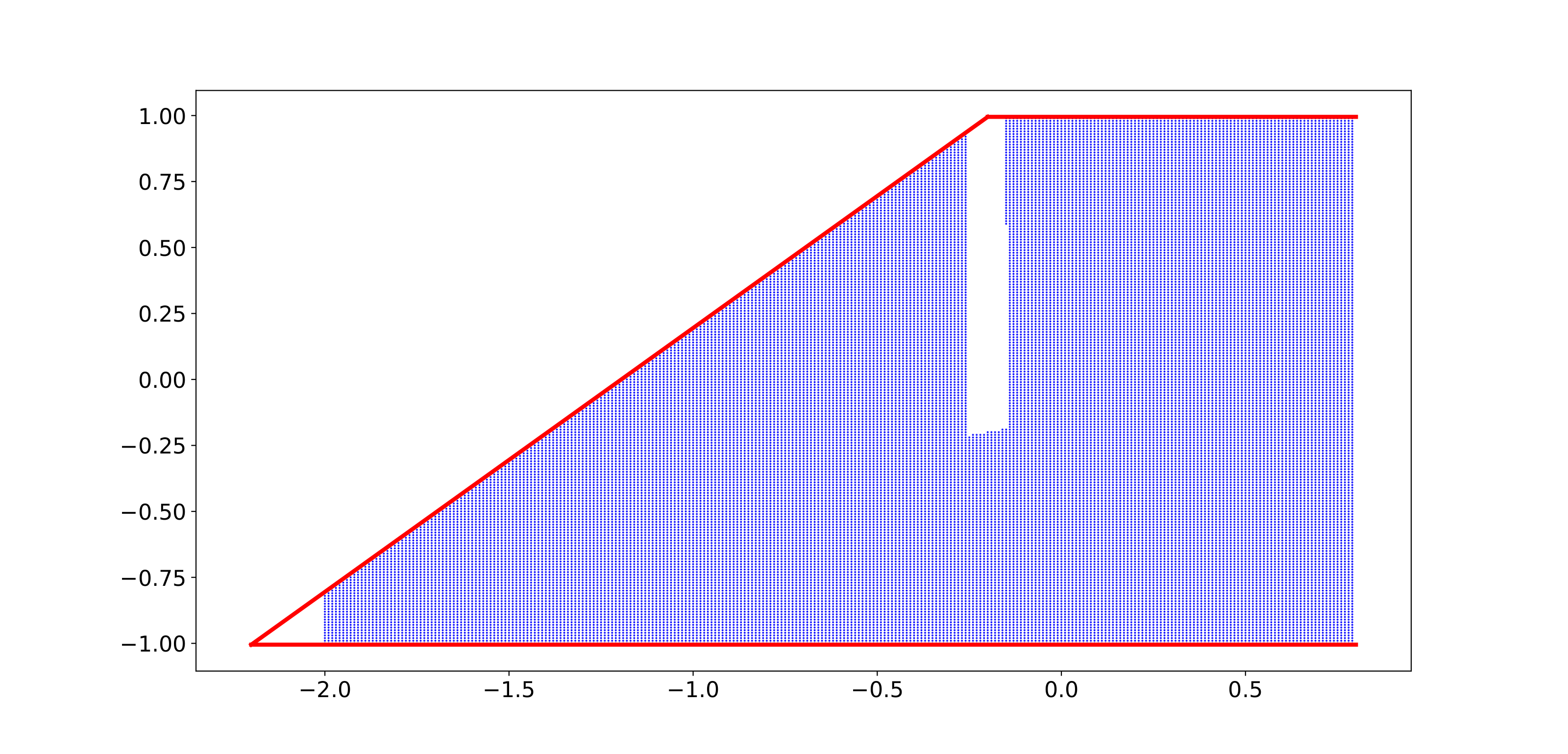}
  \end{center}
  \caption{\small Determination of the privileged polygon for the
    spin-oscillator, as described in Example~\ref{ex:poly_spin-osc};
    the blue dots represent the set
    $\Delta_\h(\mathcal{K}(S,\mathcal{U}))$ where
    $S = \{ (x,y) \in \RM^2 \ | \ -0.8 \leq x \leq 2 \}$ for $k=100$,
    while the solid red lines represent a translation of the
    privileged semitoric polygon shown in Figure
    \ref{fig:polygon_spinosc}. Note that this translation is
    unavoidable because of the vector $\nu_\h$ in Proposition
    \ref{prop:poly_num}.}
  \label{fig:polygon_spinosc_num}
\end{figure}

\begin{ex} \label{ex:poly_spin-osc} In Figure
    \ref{fig:polygon_spinosc_num}, we recover the privileged semitoric
    polygon for the spin-oscillator system of Example
    \ref{ex:spin-osc_class}; this amounts, on the one hand, to
    extending the labelling and plotting the corresponding set
    $\Delta_\h(\mathcal{K}(S,\mathcal{U}))$, according to Proposition
    \ref{prop:poly_num} (see also Remark \ref{rema:explicit_global}),
    and on the other hand to computing an approximation of
    $\sigma_1(0)$. We can then apply Proposition
    \ref{prop:priv_label}. In our case, the approximate value for
    $\sigma_1(0)$ given by~\eqref{equ:sigma_spin-osc} belongs to
    $[0,\frac12]$, which suggests that the twisting number is $p=0$
    and that the privileged polygon is as displayed in Figure
    \ref{fig:polygon_spinosc}. Of course, the limit $x\to 0^+$ is not
    guaranteed to preserve the non-negativity of $\sigma_1(0)$, and it
    would be more robust to have a direct semiclassical formula at the
    critical value $x=0$. Such a formula has not yet been established
    for Berezin-Toeplitz operators.
  \end{ex}
    
\begin{theo}
  \label{theo:poly_twisting}
  Let $F = (J,H)$ be a semitoric system and let $S \subset \RM^2$ be a
  vertical strip of the form
  $S = \{(x,y) \in \RM^2 \ | \ u_1 \leq x \leq u_2 \}$ where
  $u_1, u_2 \notin \{x_1, \ldots, x_{m_f} \}$. Given the $\h$-family
  of joint spectra $\Sigma_\h$ modulo $\O(\h^2)$ (see Definition
  \ref{defi:mod_h2}) of a proper quantum semitoric system
  $(\hat{J}_\h, \hat{H}_\h)$ quantizing $(J,H)$, one can recover, in a
  constructive way, the set $\Phi(S \cap F(M))$ and the twisting
  numbers associated with the polygon $\Phi(F(M))$ of the focus-focus
  values contained in $S$, where $\Phi$ is some cartographic
  homeomorphism corresponding to upwards cuts only,
  i.e. $\vec{\epsilon} = (1, \ldots, 1)$. In particular, the knowledge
  of $\Sigma_\h$ modulo $\O(\h^2)$ allows to recover the complete
  twisting index invariant of $F$. Moreover if $M$ is compact, we can
  explicitly construct this invariant from the data of these joint
  spectra.
\end{theo}

\begin{rema}
  When $m_f = 0$, the system is of toric type, which means that there
  exists a diffeomorphism $G$ from $F(M)$ into its image such that
  $G\circ F$ is the momentum map of an effective $\T^2$-action on
  $M$. In this case, Theorem~\ref{theo:poly_twisting} says that we can
  recover its polygon invariant in compact regions. Since the polygon
  is the only symplectic invariant in this case, this settles the
  inverse spectral problem for systems of toric type. Even this
  seemingly simple result is new; indeed, while in the toric case the
  recovery of the image $F(M)$ was
  sufficient~\cite{charles-pelayo-vungoc,san-alvaro-leonid}, if the
  system is only of toric type, one needs to handle how to
  straighten the deformed integral affine structure, in order to
  recover the diffeomorphism $G$.
\end{rema}

\begin{demo}~\\
  \noindent\textbf{Step 1. Recovering elliptic-elliptic critical values,
    vertical walls and focus-focus critical values.} Thanks to
    \cite{san-alvaro-spectral-limits}, we know how to recover the
    image of the momentum map from the joint spectrum $\Sigma_\h$,
    using the fact that the joint spectrum becomes everywhere dense in
    the image of the momentum map as $\h\to 0$. Another possibility is
    to apply Step 1 in the proof of Theorem 3.1 in
    \cite{san-alvaro-yohann} which uses the fact that Bohr-Sommerfeld
    regular values are also dense as $\h\to 0$.
  
  Therefore, we know where the elliptic-elliptic values and potential
  vertical walls are located. In principle these results only apply to
  less general classes of semiclassical operators such as operators
  with uniformly bounded symbols, but as explained in \cite[Section
  2.4]{san-alvaro-yohann} (see also \cite[Chapter
  10]{dimassi-sjostrand}), we can simply microlocalize in a bounded
  region of phase space containing $S \cap F(M)$ to work with bounded
  symbols; here the properness of $(\hat{J}_\h, \hat{H}_\h)$ is
  crucial.

  Moreover, we can also detect the focus-focus values from the data of
  $\Sigma_\h$ up to $\O(\h^2)$. This is done using Step 3 of the proof
  of Theorem 3.1 in \cite{san-alvaro-yohann}, which only relies on the
  knowledge of a basis of the period lattice over regular values of
  $F$; the idea is to locate the focus-focus values using the
  logarithmic singularity of this basis.  So we know the set
  $(V_1 \cup \ldots \cup V_{m_f}) \cup E \cup W$,
  see~\eqref{equ:Vj}. An alternative proof of Step 1 is mentioned
    in Remark~\ref{rema:DH}.

  \medskip

  \noindent\textbf{Step 2. Recovering a representative of the polygon invariant.}
  Let $(x_{i_1}, y_{i_1}), \ldots, (x_{i_p}, y_{i_p})$ be the
  focus-focus values contained in $S$. For every
  $j \in \{1, \ldots, p \}$, let $r_j > 0$ be such that
  $B((x_{i_j}, y_{i_j}),r_j)$ is contained in the image of an open set
  where Eliasson's normal form of Theorem \ref{theo:eliasson} is
  defined. Let $r = \min_j r_j > 0$. Let $\varepsilon \in (0,r)$, and
  let $\mathcal{U}_{\varepsilon}$ be the union of
  $\varepsilon$-neighborhoods of every $L_{i_j}$ and every element of
  $E \cap S$ and of $W \cap S$ for some well-chosen
  $\varepsilon \in (0,r)$, so as to avoid problems with consecutive
  critical values of $F$ (see Figure
  \ref{fig:choice_compact_labelling}). Let
  $\mathcal{K}(S,\mathcal{U}_{\varepsilon})$ be as in the beginning of
  this section. Of course one does not know $r$ a priori but the rest
  of the proof consists in investigating the limit
  $\varepsilon \to 0$.

  Let $\ell_\h$ be a quantum cartographic map associated with $S$ and
  $\mathcal{U}_{\varepsilon}$, and let $\Phi$ be the corresponding
  cartographic homeomorphism (Lemma \ref{lm:carto} ensures that $\Phi$
  does not depend on $\varepsilon$). By Proposition
  \ref{prop:poly_num}, we can recover the set
  $\Phi(\mathcal{K}(S,\mathcal{U}_{\varepsilon}))$ (note that the
  constants $\nu_\h$ in this proposition are not a problem since the
  position of the polygon $\Phi(F(M))$ in $\RM^2$ does not
  matter). Since the set $\Phi(S \cap F(M))$ is polygonal and we know
  where its vertices should be (since we know the locations of
  elliptic-elliptic and focus-focus values and that the cuts are all
  upwards), and because of our choice of $\mathcal{U}_{\varepsilon}$,
  we can recover this polygonal set from
  $\Phi(\mathcal{K}(S,\mathcal{U}_{\varepsilon}))$ by drawing the
  missing vertices and pieces of edges, see for instance Figures
  \ref{fig:polygon_spinosc_num} and \ref{fig:poly_S2xS2}.

\medskip

  \noindent\textbf{Step 3. Recovering the twisting numbers.}
  By construction, for every $j \in \{1, \ldots, p\}$ the restriction
  of $\Phi$ over some ball intersecting
  $(B((x_{i_j}, y_{i_j}),r) \setminus B((x_{i_j},
  y_{i_j}),\varepsilon)) \cap \{ x \geq x_{i_j} \}$ is an action
  diffeomorphism, so $\Phi \circ F = (J,L_j)$ where the action
  variable $L_j$ is independent of $\varepsilon$. So by Theorem
  \ref{theo:recover-sigma_1}, for every $j \in \{1, \ldots, p\}$, we
  can recover the invariant $\sigma_1((x_{i_j}, y_{i_j}))$ associated
  with $L_j$, by considering $\varepsilon \to 0$ (recall that in this
  theorem, we first let $\h \to 0$ at fixed $c$ close to
  $(x_{i_j}, y_{i_j})$ and then consider the limit
  $c \to (x_{i_j}, y_{i_j})$). This means that we can recover the
  twisting numbers of the focus-focus values contained in $S$, because
  we compare $L_j$ with the privileged action $L_{\text{priv},j}$
  recovered from $\Sigma_\h$ thanks to Proposition
  \ref{prop:priv_label}. Using the representative polygon invariant
  from Step 2, this gives the full twisting index invariant.

  Finally, if $M$ is compact, then $F(M)$ is compact as well, so it
  suffices to take $S$ sufficiently large in order to recover the
  whole polygon associated with $\Phi$ and the corresponding twisting
  numbers, hence the twisting index invariant.
\end{demo}

Let us briefly illustrate the steps of the proof above by working
  out the case of the spin-oscillator, Example
  \ref{ex:spin-osc_quant}. Step 1 is illustrated by
  Figure~\ref{fig:jsp_spinosc}, where we see that the joint
  eigenvalues become dense in the image of the classical momentum
  map. This gives us the position of the elliptic-elliptic critical
  value at $(-1,0)$. The focus-focus value is then obtained by the
  asymptotic behaviour of the period lattice, which is suggested in
  the figure by the accumulation of points around $(1,0)$.

To follow Step 2, we first exclude a neighborhood of the
  elliptic-elliptic critical value and vertical half-strip above the
  focus-focus value, then straighten the lattice points using
  Proposition~\ref{prop:poly_num}, to obtain the representative of the
  polygon invariant depicted in Figure~\ref{fig:polygon_spinosc_num}.

Finally in Step 3 we compute the twisting number associated with
  this representative, here 0, using
  Theorem~\ref{theo:recover-sigma_1}, (more precisely
  Formula~\eqref{equ:sigma}), as illustrated in Figure
  \ref{fig:sigma_spinosc_kvar_err}.

\begin{rema}
  \label{rema:DH} In this remark we indicate another (perhaps
    more satisfactory from an algorithmic point of view) way of
    obtaining Step 1 in the proof of Theorem~\ref{theo:poly_twisting}.

  Locating the elliptic-elliptic values and vertical walls of the
  momentum map can be done by counting the eigenvalues in suitable
  vertical strips. More precisely, let $\delta \in (0,\frac{1}{2})$,
  $c > 0$ and for $x \in J(M)$, set
  $N_\h(x,\delta,c) = \# \Sigma_\h \cap [x - c \h^{\delta}, x + c
  \h^{\delta}]\times \RM$. Then
  \begin{equation} \frac{\h^{2 - \delta}}{2c} N_\h(x,\delta,c)
    \underset{\h \to 0}{\longrightarrow}
    \rho_J(x) \label{eq:DH} \end{equation} where $\rho_J$ is the
  Duistermaat-Heckman function associated with $J$
  \cite{duistermaat-heckman}, using notation from \cite[Section
  5]{san-polytope}. We illustrate this result in Figure
  \ref{fig:DH_S2xS2}. It is standard that $\rho_J$ is continuous and
  piecewise affine and by Theorem 5.3 in the aforementioned paper, a
  change of slope in its graph at $(x_0,\rho_J(x_0))$ indicates the
  presence of one or several elliptic-elliptic or focus-focus values
  in the fiber $J^{-1}(x_0)$. The limit of the top and bottom points
  of this strip gives the potential positions of the elliptic-elliptic
  values, as in Remark~\ref{rema:position-c}.  Of course the
  Duistermaat-Heckman function cannot tell us whether the elliptic
  point lies at the bottom or at the top of the strip (or both), but
  in view of the rest of the proof, it is not an issue. Indeed, we may
  add extra isolated points (corresponding to transversally elliptic
  critical values) to the set $E$. It does not prevent the
  reconstruction of the polygon from the joint spectrum, thanks to the
  data of the surrounding edges. The corners of the polygon will
  finally tell where the true elliptic-elliptic singular values were
  located.  Furthermore, we know that the potential vertical walls can
  only be located at the global minimum or maximum of $J$, and their
  existence would be equivalent to the fact that $\rho_J$ is nonzero
  at these points.
 
  The Duistermaat-Heckman formula also gives an alternative way
    of obtaining the focus-focus critical values. Indeed, since
    focus-focus points are critical points of $J$, from~\eqref{eq:DH},
    we obtain, amongst the (finitely many) interior critical values of
    $J$, the \emph{potential abscissae} of the focus-focus
    values. Then, considering a vertical strip above each of these
    abscissae $x_0$, we may draw the graph of the vertical level
    spacings using Lemma~\ref{lemm:quantum-a}.  The logarithmic
    behavior of the coefficient $a_2(c)$ at the focus-focus value is
    enough to precisely locate that value; more precisely, from
    Proposition~\ref{prop:sigma_smooth} and Equation~\eqref{equ:a1_a2}
    we see that, if $(x_0,y_0)$ is the focus-focus critical value, we
    have
  \[
    a_2(x_0,y) \sim C\ln\abs{y-y_0} \text{ as } y\to y_0.
  \]
  This is illustrated in Figure~\ref{fig:ecarts_50}.

Equation \eqref{eq:DH} is nothing but Weyl's law for $\hat{J}_\h$ in
an interval of size $\h^{\delta}$. Its proof is similar to the usual
case of a fixed interval, see for instance \cite[Theorem
14.11]{zworski}. However one needs to use the fact that if $\chi$ is a
compactly supported smooth function and $P_\h$ is a semiclassical
operator in $S(m)$, $\chi(\h^{-\delta} P_\h)$ is a semiclassical
operator in $S_{\delta}(m)$, and asymptotic estimates for the trace of
a semiclassical operator with symbol in a class $S_{\delta}(m)$, see
for instance \cite[Section 4.4]{zworski} for the definition of these
classes. These results are known for $\h$-pseudodifferential
operators: for instance, the former can be derived by adapting the
arguments in \cite[Section 8]{sjostrand}, and the latter can be
obtained from the usual estimate for the trace of an operator in
$S(m)$ and a rescaling of the semiclassical parameter. These
  results were not proven for Berezin-Toeplitz operators when we wrote
  this manuscript, but they are now available thanks to
  \cite{oltman}. Another important point in the derivation of
\eqref{eq:DH} is that $\rho_J$ is Lipschitz.
\end{rema}

\section{Recovering the Taylor series invariant from the joint
  spectrum}
\label{sec:taylor_series}

In this section, we continue to work in a neighborhood $B$ of a
focus-focus critical value $0\in\RM^2$, for the completely integrable
momentum map $F=(J,H)$. Again, we assume that the corresponding
  critical fiber contains only one critical point; it could be
  interesting to investigate whether the inverse problem can be solved
  in the multiply pinched case, where the Taylor series has to be
  replaced by a more complicated object, see~\cite{pelayo-tang}. 
Here we carry on the analysis of the spectrum of a proper quantum
integrable system $(\hat{J}_\h, \hat{H}_\h)$ associated with the
momentum map $F$ started in Section \ref{sec:lattices} to recover the
Taylor series invariant and the height invariant, both defined in
Section~\ref{sec:tayl-seri-invar}. The study of the height invariant
led us to investigate the behavior of the number of points in an
asymptotic lattice.

Prior to this work, it was unknown whether the joint spectrum of
$(\hat{J}_\h, \hat{H}_\h)$ near $0$ fully determines $F$, up to
symplectic equivalence. An important first step was proven in
\cite{san-alvaro-taylor}: the joint spectrum, restricted to the small
neighborhood $B$, determines the Taylor series invariant. This was an
\emph{uniqueness} statement, whose precise meaning was that whenever
the joint spectra of two such quantum integrable systems coincide up
to $\O(\h^2)$, then their Taylor series invariants coincide. However,
no method of construction of the invariant from the joint spectrum was
given. In this section, we are interested only in the semitoric case,
but the result we show is quite stronger, namely that the Taylor
series invariant can be \emph{constructed} from the joint spectrum of
a single semitoric system near a focus-focus critical value. This
Taylor series invariant determines $F$ near the critical fiber
$F^{-1}(0)$ up to left-composition by a local diffeomorphism. In fact,
we obtain a better result since we also recover the full Taylor series
of the Eliasson diffeomorphism, which completely characterizes $F$ up
to a flat term. Actually, our constructions also allow the recovery of
the Taylor series invariant and the infinite jet of the Eliasson
diffeomorphism for focus-focus singularities in systems that are not
necessarily semitoric.

To our knowledge, constructive statements in inverse spectral theory
are not so common; however, a constructive way to compute the
\emph{linear terms} of the Taylor series invariant was proposed in
\cite{san-alvaro-spin}, under the assumption that the singular
Bohr-Sommerfeld conditions hold; here we want to avoid this assumption
in the context of Berezin-Toeplitz quantization, since the
corresponding Bohr-Sommerfeld conditions have not been proven yet.

\subsection{The height invariant}

As explained in Section~\ref{sec:tayl-seri-invar}, the height
invariant can be considered as the constant term $S_{0,0}$ of the
Taylor series invariant. It has been computed explicitly for some
specific classical systems in
\cite{san-alvaro-spin,alvaro-yohann,ADH-spin,ADH-angular}. In~\cite{san-alvaro-yohann},
it was proven that if two quantum semitoric systems have the same
semiclassical joint spectrum, then they must share the same height
invariant.  In this section, we take another route and obtain a direct
formula for computing this invariant from a single semiclassical joint
spectrum. Since the height invariant has an intrinsic definition in
terms of a symplectic volume, a natural way to recover it from the
joint spectrum is to make use of a suitable Weyl formula. Hence, this
method is quite different from the way the higher order invariants will be
handled in the following sections.
\begin{prop}
  \label{prop:height-invariant}
  The height invariant $S_{0,0}$ associated with the focus-focus
  critical value $c_0=0$ can be explicitly recovered from the joint
  spectrum modulo $\O(\h^2)$ in a vertical strip below $c_0$ by the
  following formula. Let $\delta \in (0,\frac{1}{2})$, $c > 0$ and
  $y \geq 0$, and define
  $N_\h(\delta,c,y) = \# \Sigma_\h \cap [- c \h^{\delta}, c
  \h^{\delta}] \times (-\infty, -y]$. Then
  \begin{equation} S_{0,0} = \lim_{y \to 0} \lim_{\h\to 0} \frac{\h^{2
        - \delta}}{2c}
    N_\h(\delta,c,y).  \label{eq:height_double_limit}\end{equation}
  Furthermore,
  \begin{equation} S_{0,0} = \lim_{\h\to 0} \frac{\h^{2 - \delta}}{2c}
    N_\h(\delta,c,0). \label{eq:height_single_limit}\end{equation}
\end{prop}

\begin{ex}
    \label{ex:height_spin-osc} We consider the quantum Jaynes-Cummings
    system as in Example \ref{ex:spin-osc_quant}. Recall that the
    height invariant of the underlying classical system is
    $S_{0,0} = 1$, see Equation
    \eqref{eq:height_invariant_spinosc}. In Figure
    \ref{fig:height_spinosc}, we recover this height invariant
    numerically from the joint spectrum (see Example
    \ref{ex:spin-osc_jsp}) using Proposition
    \ref{prop:height-invariant},
    Equation~\eqref{eq:height_single_limit}.
    \begin{figure}[H]
      \begin{center}
        \includegraphics[trim=80 10 80
        60,clip,width=0.75\textwidth]{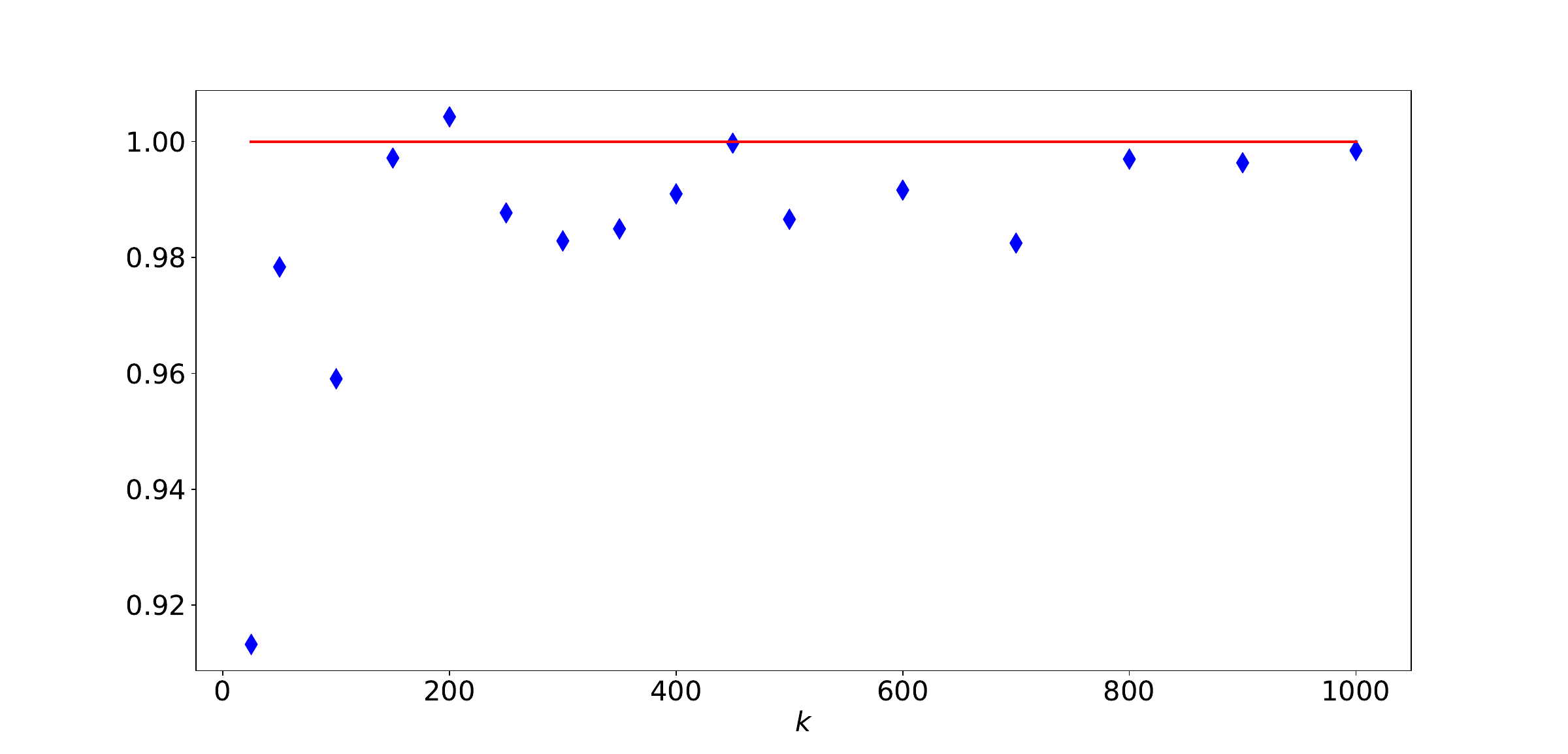}
      \end{center}
      \caption{\small Determination of the height invariant for the
        spin-oscillator using Proposition \ref{prop:height-invariant},
        see Example~\ref{ex:height_spin-osc}. The blue diamonds
        correspond to $\frac{\h^{2 - \delta}}{2c} N_\h(\delta,c,0)$
        for $c=1$, $\delta = 0.4$ and different values of
        $k = \tfrac{1}{\h}$. The solid red line is the theoretical
        value $S_{0,0} = 1$.}
      \label{fig:height_spinosc}
    \end{figure}
  \end{ex}

In order to prove Proposition~\ref{prop:height-invariant} we need to
discuss general results concerning counting functions in asymptotic
lattices or half-lattices.
\begin{lemm}
  \label{lemm:counting_curve}
  Let $(\mathcal{L}_\h,\mathcal{I},B)$ be an asymptotic lattice or
  half-lattice, and let $\Gamma\subset B$ be a compact, smooth curve
  immersed in $B$. Let $\Gamma_\h$ be a thickening of $\Gamma$ of
  width $\O(\h)$. Let $N_\h(\Gamma_\h)$ be the cardinal of
  $\mathcal{L}_\h\cap \Gamma_\h$. Then
  \[
    N_\h(\Gamma_\h) = \O(\tfrac{1}{\h}).
  \]
\end{lemm}

\begin{demo}
  We may cover $\Gamma_\h$ by a number $n_\h$ of balls of radius
  $C\h$, where $C>0$ is large enough and the center of each ball
  belongs to $\Gamma$, in such a way that $n_\h\sim\frac{L}{C\h}$ for
  some $L>0$ ($L$ will be proportional to the length of $\Gamma$). By
  the properties of asymptotic (half-)lattices, the number of points
  in each ball is bounded by a uniform constant independent of $\h$.
\end{demo}

\begin{lemm}\label{lemm:lattice_counting}
  Let $(\mathcal{L}_\h,\mathcal{I},B)$ be an asymptotic lattice or
  half-lattice, and let $\tilde B\Subset B$ be a domain with piecewise
  smooth boundary $\Gamma$.  Let $N_\h(\tilde B)$ be the cardinal of
  $\mathcal{L}_\h\cap \tilde B$. Then
  \[
    N_\h(\tilde B) = \frac{1}{\h^2}\emph{area} (G_0^{-1}(\tilde B\cap
    \underline{\mathcal{L}_\h})) + \O(\tfrac{1}{\h}) ,
  \]
  where $G_0$ is the leading term of an asymptotic chart for
  $\mathcal{L}_\h$, and the set $\underline{\mathcal{L}_\h}$ was
  defined in Lemma~\ref{lemm:restriction_half_lattice_far} (it does
  not depend on $\h$).
\end{lemm}

\begin{demo}
  Let $G_\h$ be an asymptotic chart for $B$. In the case of an
  asymptotic lattice (the case of an asymptotic half-lattice is
  similar, upon replacing $\tilde B$ by
  $\tilde B\cap \underline{\mathcal{L}_\h}$), we have
  \[
    N_\h(\tilde B) = \#\{\lambda_\h \in \tilde B\} \simeq
    \#\{\zeta\in\ZM^2 \ | \ G_\h(\h\zeta)\in \tilde B\},
  \]
  and the approximation $\simeq$ is in general non exact because
  $G_\h(\h\zeta)$ approaches $\lambda_\h$ only up to $\O(\h^\infty)$,
  which is enough for points close to the boundary $\Gamma$ to escape
  $\tilde B$ under this perturbation. However, for any $B'\subset B$,
  let us introduce
  $M_\h(B') := \#\{\zeta\in\ZM^2 \ | \ G_\h(\h\zeta)\in B'\}$; then
  the following inequalities are exact, for $\h$ small enough:
  \[
    M_\h(\tilde B^-_\h) \leq N_\h(\tilde B) \leq M_\h(\tilde B^+_\h),
  \]
  where $\tilde B^-_\h$ is an $\h$-shrinking of $\tilde B$, and
  $\tilde B^+_\h$ an $\h$-enlarging, such that
  $\tilde B^-_\h\subset \tilde B\subset \tilde B^+_\h$ and
  $\Gamma_\h:= \tilde B^+_\h\setminus \tilde B^-_\h$ is an
  $\h$-thickening of $\Gamma$, as in
  Lemma~\ref{lemm:counting_curve}. Applying that lemma yields
  \[
    N_\h(\tilde B) = M_\h(\tilde B^\pm_\h) + \O(\tfrac{1}{\h}) =
    M_\h(\tilde B) + \O(\tfrac{1}{\h}).
  \]
  Similarly, since $G_\h = G_0 + \O(\h)$ uniformly on $\tilde B$, we
  may replace $G_\h$ by $G_0$ in the above estimates:
  \begin{equation}
    N_\h(\tilde B) = M_{0,\h}(\tilde B) + \O(\tfrac{1}{\h}),\label{equ:Nh}
  \end{equation}
  with
  $M_{0,\h}(\tilde B) := \#\{\zeta\in\ZM^2 \ | \ G_0(\h\zeta)\in
  \tilde B\}$.  Finally, we notice that $\h^2 M_{0,\h}(\tilde B)$ is a
  Riemann approximation of the integral
  $\int_{G_0^{-1}(\tilde B)}\dd\zeta = \textup{area}(G_0^{-1}(\tilde
  B))$:
  \[
    \h^2 M_{0,\h}(\tilde B) = \textup{area}(G_0^{-1}(\tilde B)) +
    \O(\h).
  \]
  Together with~\eqref{equ:Nh}, this proves the lemma.
\end{demo}

\begin{demo}[of Proposition~\ref{prop:height-invariant}]
  It follows from the Bohr-Sommerfeld rules (regular, see Theorem
  \ref{theo:BS_reg}, and elliptic, see Theorem
  \ref{theo:BS_ell_trans}) that, away from focus-focus critical
  values, the joint spectrum $\Sigma_\h$ is locally an asymptotic
  lattice or half-lattice. Hence, for any rectangle $R$ containing
  only regular values or transversally elliptic critical values of
  $F$, the number of joint eigenvalues inside $R$ is
  \begin{equation}
    N_\h(R) = \frac{1}{\h^2}\int_{R\cap\underline{\Sigma_\h}}
    \abs{\dd \Phi}\dd c + \O(\tfrac{1}{\h}),
    \label{equ:counting_global}
  \end{equation}
  where the map $\Phi$ was obtained in
  Theorem~\ref{thm:construct_global_labelling}, and
  $\abs{\dd{\Phi}}$ denotes its Jacobian. Indeed, $R$ can be covered by
  asymptotic lattices or half-lattices, and in each one we may find an
  asymptotic chart $G_\h$, such that $\dd \Phi = \dd G_0^{-1}$; hence
  we may apply Lemma~\ref{lemm:lattice_counting} and we see that the
  integrals
  $\int_{G_0^{-1}(\tilde B)}\dd\zeta = \int_{\tilde B}\abs{\dd
    G_0^{-1}} \dd c $ nicely patch together to
  give~\eqref{equ:counting_global}.

  Now, we know from the Bohr-Sommerfeld analysis that $G_0^{-1}$ is
  actually an action diffeomorphism. Therefore, the change of
  coordinates $\zeta = G_0^{-1}(c)$ gives the density
  $\abs{\dd G_0^{-1}} \dd c = \dd \zeta = \frac{1}{(2\pi)^2} \dd \zeta
  \int \dd \theta$, where $(\zeta,\theta)$ are action-angle
  coordinates. Since $F = G_0\circ\zeta$, we have
  $ \int_{\tilde B} \abs{\dd G_0^{-1}}\dd c =
  \frac{1}{(2\pi)^2}\textup{Vol}(F^{-1}(\tilde B))$, where
  $\textup{Vol}$ is the usual symplectic volume in $M$. This gives
  \[
    \int_{R\cap\underline{\Sigma_\h}} \abs{\dd \Phi} \dd c =
    \frac{1}{(2\pi)^2}\textup{Vol}(F^{-1}(R\cap\underline{\Sigma_\h}))
    = \frac{1}{(2\pi)^2}\textup{Vol}(F^{-1}(R)),
  \]
  since $\underline{\Sigma_\h}\subset F(M)$.  Thus,
  Equation~\eqref{equ:counting_global} gives the following ``joint
  Weyl formula'':
  \begin{equation}
    N_\h(R) = \frac{1}{(2\pi\h)^2}\textup{Vol}(F^{-1}(R)) +
    \O(\tfrac{1}{\h}),\label{equ:weyl_conjoint}
  \end{equation}
  Notice that the formula is uniform in $R$, as long as $R$ stays in a
  fixed compact region.
	
  By a simple scaling argument, we may assume without loss of
  generality that the constant $c$ in the proposition is $c=1$.  Let
  $\delta\in(0,\frac{1}{2})$, and let $S_\delta\subset \RM^2$ be a
  vertical strip of width $2\h^\delta$ around the focus-focus value
  $c_0=0$, \emph{i.e.}  $S_\delta=[-\h^\delta,\h^\delta]\times\RM$.
  Let $y\geq 0$, and split $S_\delta$ vertically in three parts,
  $S_\delta^-(y)$, $S_\delta^0(y)$, and $S_\delta^+(y)$, namely:
  \[
    S_\delta^-(y) = [-\h^\delta,\h^\delta]\times(-\infty,-y], \quad
    S_\delta^0(y) = [-\h^\delta,\h^\delta]\times(-y,y), \quad
    S_\delta^+(y) = [-\h^\delta,\h^\delta]\times [y,+\infty).
  \]
  The set $S_\delta^0(y)$ contains the focus-focus value, and the
  joint spectrum near this value is neither an asymptotic lattice nor
  an asymptotic half-lattice. Let $y>0$, so that
  $N_\h(\delta,1,y) =
  N_\h(S_\delta^-(y))$. From~\eqref{equ:weyl_conjoint} we have
  \begin{equation} N_\h(S_\delta^-(y)) =
    \frac{1}{(2\pi\h)^2}\textup{Vol}(F^{-1}(S_\delta^-(y))) +
    \O(\tfrac{1}{\h}). \label{eq:nh_sdelta} \end{equation} Near any point
  $m\in M$ where $\dd J(m)\neq 0$, we can write the symplectic measure
  as
  $\abs{\omega^2}/2 = \abs{\omega_x}\wedge \dd J \wedge \dd \theta$,
  where $\omega_x$ is the natural symplectic form on the local reduced
  manifold $J^{-1}(x)/\ham{J}$, $x=J(m)$, and $\theta$ is the angle
  expressing the time of the Hamiltonian flow of $J$. Hence
  \begin{align*}
    \textup{Vol}(F^{-1}(S_\delta^-(y)))
    & = \int_{-\h^\delta}^{\h^\delta} \dd x \int_{F^{-1}(\{x\}\times(-C,-y])}\abs{\omega_x} \wedge \dd \theta  \\
    & = \int_{-\h^\delta}^{\h^\delta} \dd x \int_{F^{-1}(\{0\}\times(-C,-y])}\abs{\omega_0} \wedge \dd\theta + \O(x) \\
    & = 2\h^\delta \int_{F^{-1}(\{0\}\times(-C,-y])}\abs{\omega_0}\wedge \dd \theta + \O(\h^{2\delta}).
  \end{align*}
  Notice that
  $\int_{F^{-1}(\{0\}\times(-C,-y])}\abs{\omega_0}\wedge \dd \theta =
  2\pi \textup{Vol}_0^-(y)$, where $\textup{Vol}_0^-(y)$ is the volume
  of the sublevel set $H\leq y$ within the reduced symplectic manifold
  $M_0:=J^{-1}(0)/\ham{J}$ (which is actually singular in a set of
  measure zero). This gives
  \begin{equation}
    \frac{\textup{Vol}(F^{-1}(S_\delta^-(y))) }{2\h^\delta} =
    2\pi\textup{Vol}_0^-(y) + \O(\h^{\delta}).
    \label{equ:weyl_y}
  \end{equation}
  We know from the local analysis of focus-focus singularities (see
  for instance~\cite{san-polytope}) that
  \begin{equation} \textup{Vol}_0^\pm(y) = \textup{Vol}_0^\pm(0) +
    \O(y\ln y). \label{eq:continuity_volume} \end{equation} Together
  with \eqref{eq:nh_sdelta}, since the height invariant is precisely
  $S_{0,0}=\frac1{2\pi}\textup{Vol}_0^-(0)$, this gives
  \eqref{eq:height_double_limit}.

  To prove \eqref{eq:height_single_limit}, observe that by simple
  inclusions, we have
  \begin{equation}
    N_\h(S_\delta^-(y)) \leq N_\h(S_\delta^-(0)) \leq N_\h(S_\delta) -
    N_\h(S_\delta^+(y)) .\label{equ:inclusions}
  \end{equation}
  Since $J$ is proper, the vertical extent of joint eigenvalues in the
  strip $S_{\delta}$ is actually bounded; hence in the above formula
  one may replace $S_\delta^-(y)$ by a suitable rectangle
  $ [-\h^\delta,\h^\delta]\times(-C,-y]$, and similarly for
  $S_\delta^+(y)$.

  Of course, an analogous formula holds for
  $S_\delta^+(y)$. Therefore, multiplying
  Equation~\eqref{equ:inclusions} by $\h^{2-\delta}/2$ and taking the
  limit inferior when $\h\to 0$ yields, in view of~\eqref{eq:DH},
  \[
    (2\pi)^{-1} \textup{Vol}_0^-(y) \leq \liminf_{\h \to 0}
    \h^{2-\delta} N_\h(S_\delta^-(0)) /2 \leq
    (2\pi)^{-1}(\textup{Vol}_0 - \textup{Vol}_0^+(y)),
  \]
  where $\textup{Vol}_0$ is the complete volume of $M_0$.

  Using Equation \eqref{eq:continuity_volume} again, and since
  $\textup{Vol}_0 = \textup{Vol}_0^-(0) + \textup{Vol}_0^+(0)$, we
  get, when $y\to 0$,
  \[
    0\leq \liminf_{\h \to 0} \frac{\h^{2-\delta}}{2}
    N_\h(S_\delta^-(0)) - \frac{1}{2\pi}\textup{Vol}_0^-(0) \leq 0.
  \]
  The same holds for the limit superior, which proves the second
  statement of the proposition.
\end{demo}

\begin{rema}
  The ``joint Weyl formula''~\eqref{equ:weyl_conjoint} can be found
  in~\cite{charbonnel} in the pseudodifferential case when $R$
  consists only of regular values. Including elliptic critical values
  (and Berezin-Toeplitz quantization), albeit not surprising, seems to
  be new.
\end{rema}
\begin{rema}
  It would be interesting to obtain the remainder term, or at least
  estimate the convergence speed in this joint Weyl formula. However,
  it is not accessible directly with the results of the present
  article. For instance, in view of the pseudodifferential analysis
  carried out in~\cite{san-focus}, it is expected that the remainder
  $\O(\h^\delta)$ in~\eqref{equ:weyl_y} cannot be uniform as $y\to 0$,
  because of the logarithmic accumulation of joint eigenvalues at the
  origin, as $\h\to 0$.
\end{rema}

\subsection{Linear terms}

The linear and higher order terms in the Taylor series invariant are
obtained from the joint spectrum in slightly different ways.

From Theorem~\ref{theo:recover-sigma_1} and Proposition
\ref{prop:priv_label}, we see that we can recover the first linear
term $[S_{1,0}]$ of the Taylor series invariant from the joint spectrum;
indeed, recall that $\sigma_1^{\text{p}}(0) = S_{1,0}$.  Note that
recovering $[S_{1,0}]$ was already achieved in \cite{alvaro-yohann} but
again under the assumption that the singular Bohr-Sommerfeld rules
hold, which is only a conjecture for the case of Berezin-Toeplitz
operators, as explained above.

In order to recover the term $S_{0,1}$, we proceed similarly to the
way we recovered $S_{1,0}$; as in Lemma~\ref{lemm:def_sigma1_zero}
above, let $B_1\ni x \mapsto (x, \varphi(x))$ be a parametrization of
the radial curve $\gamma_r$ near the origin. Recall that a choice of local action variable $L$ yields the two functions $\tau_1, \tau_2$ defined in Equation \eqref{equ:tau1-tau2}.

\begin{lemm}
  The function $x \mapsto \tau_2(x,\varphi(x)) + \frac{\ln x}{2\pi}$
  defined for $x \in\RM^*_+\cap B_1$ is smooth at $x = 0$, and its
  value at zero is equal to $\sigma_2(0) = S_{0,1}$.
\end{lemm}

\begin{demo}
  We notice, thanks to Definition~\ref{defi:radial-curve}, that the
  function $x \mapsto \tau_2(x,\varphi(x)) + \frac{\ln x}{2\pi}$ is
  the restriction of $\sigma_2$ to the curve $\gamma_r$, and we know
  from Proposition \ref{prop:sigma_smooth} that $\sigma_2$ is smooth
  at the origin.
 
\end{demo}

\begin{lemm}
  In order to compute $\sigma_2(0)$, the radial curve $\gamma_r$
    (see~Definition~\ref{defi:radial-curve}) can be replaced by any
    curve $\gamma$ that is tangent to $\gamma_r$ at the
    origin. In other words, $\sigma_2(0)$ is the limit of
    $\tau_2(c) + \frac{\ln x}{2\pi}$ when $c=(x,y)$ tends to the origin
    along $\gamma$ with $x > 0$.
  \end{lemm}

\begin{demo}
  We argue as in the proof of Lemma \ref{lemm:tangent}.  Let $\gamma$
  be any curve that is tangent to $\gamma_r$ at the origin; it is
  locally the graph of a smooth function $\psi$. We have that
  \[
    x \mapsto \sigma_2(x, \psi(x)) = \tau_2(x, \psi(x)) + \frac{1}{2\pi}
    \Re(\log(x + i f_r(x,\psi(x)))
  \]
  is smooth at $x = 0$, and that
  $\sigma_2(x, \psi(x)) - \sigma_2(x,\varphi(x)) \underset{x \to
    0}{\longrightarrow} 0 $ since $\psi(0) = 0 = \varphi(0)$ and
  $\sigma_2$ is smooth at $(0,0)$. Hence
  \[
    \left(\tau_2(x, \varphi(x)) + \frac{\ln x}{2\pi}\right) -
    \left(\tau_2(x, \psi(x)) + \frac{\ln x}{2\pi}\right) + \frac{\ln
      x}{2\pi} - \frac{1}{2\pi} \Re(\log(x + i f_r(x,\psi(x))))
    \underset{x \to 0^+}{\longrightarrow} 0,
  \]
  so, it suffices to show that
  \[
    \Re(\log(x + i f_r(x,\psi(x)))) -  \ln x  \underset{x \to 0^+}{\longrightarrow}
    0.
  \]
  We can rewrite this quantity as
  \[
    \ln \left| 1 + i \frac{f_r(x,\psi(x))}{x} \right| = \frac{1}{2}
    \ln \left( 1 + \frac{f_r(x,\psi(x))^2}{x^2} \right).
  \]
  But we showed in the proof of Lemma \ref{lemm:tangent} that
  \[
    \frac{f_r(x,\psi(x))}{x} \underset{x \to 0}{\longrightarrow} 0,
  \]
  so the above quantity indeed goes to zero when $x \to 0$.
\end{demo}

\begin{prop}
  \label{theo:recover-sigma_2}
  From the $\h$-family of joint spectra $\Sigma_\h$ of a proper
  quantum semitoric system in a neighborhood of a focus-focus critical
  value $c_0=0$, one can recover, in a constructive way, the
  symplectic invariant $S_{0,1} = \sigma_2(0)$.
\end{prop}

\begin{demo}
  \paragraph{Step 1.} As before, and in view of the previous lemma, we
  consider the curve given by $x \mapsto (x,s(0)x)$. Recall that from
  the joint spectrum, one can recover $s(0)$ as well as
  $\partial_y f_r(0)$, see Lemma \ref{lemm:dxfr_dyfr} and Formula
  \eqref{eq:dyfr_lim}.

  \paragraph{Step 2.} We use once again the above lemmas and Equation
  \eqref{equ:a1_a2} to write
  \[
    \frac{a_2(x,s(0)x)}{\partial_y f_r(x,s(0)x)} + \frac{\ln x}{2\pi}
    \underset{x \to 0^+}{\longrightarrow} \sigma_2(0). \] Since
  $\partial_y f_r(x,s(0)x) = \partial_y f_r(0) + \O(x)$, we obtain
  that
  \[
    \frac{a_2(x,s(0)x)}{\partial_y f_r(x,s(0)x)} + \frac{\ln x}{2\pi}
    = \frac{a_2(x,s(0)x)}{\partial_y f_r(0)} (1 + \O(x)) + \frac{\ln
      x}{2\pi} = \frac{a_2(x,s(0)x)}{\partial_y f_r(0)} + \frac{\ln
      x}{2\pi} + \O(x \ln x), \] where the last equality comes from
  the fact that
  $a_2(x,s(0)x) \sim -\frac{\partial_y f_r(0)}{2\pi} \ln x$ when
  $x \to 0^+$. Hence
  \[
    \frac{a_2(x,s(0)x)}{\partial_y f_r(0)} + \frac{\ln x}{2\pi}
    \underset{x \to 0}{\longrightarrow} \sigma_2(0) \] and all the
  quantities on the left-hand side have been recovered from the
  spectrum by considering good labellings in earlier parts of the paper ($a_2$ in Lemma
  \ref{lemm:quantum-a}, $\partial_y f_r(0)$ and $s(0)$ in Lemma
  \ref{lemm:dxfr_dyfr}).
\end{demo}

\begin{rema}
  This implies, together with Lemma \ref{lemm:quantum-a}, that
  \begin{equation}
    S_{0,1} = \lim_{x \to 0^+} \lim_{\h \to 0} \left(
      \frac{\h}{\partial_y f_r(0)(E_{j,\ell+1} - E_{j,\ell})} +
      \frac{\ln x}{2\pi} \right)
    \label{eq:S01_lim}
  \end{equation}
  where $(\lambda_{j,\ell})_{j,\ell} = (J_{j,l}, E_{j,\ell})_{j,\ell}$
  is a good labelling and, in the above equation, $j,\ell$ are
  $\h$-dependent integers such that the joint eigenvalues
  $\lambda_{j,\ell}$, $\lambda_{j+1,\ell}$ and $\lambda_{j,\ell+1}$
  are well-defined in an $\O(\h)$-neighborhood of $(x,s(0)x)$.
\end{rema}

\begin{ex} \label{ex:S01_spin-osc} We illustrate Proposition
    \ref{theo:recover-sigma_2} by the computation of $S_{0,1}$ for the
    quantum Jaynes-Cummings system defined in Example
    \ref{ex:spin-osc_quant}. Taking into account that
    $\partial_y f_r(0) = 2$, see~\eqref{eq:fr_elia_spinosc},
    Equation~\eqref{eq:S01_lim} yields
    \[
      S_{0,1} = \lim_{x \to 0^+} \lim_{k \to +\infty} \left(
        \frac{1}{2k(E_{0,1} - E_{0,0})} + \frac{\ln x}{2\pi}
      \right)\,.
    \]
    We have seen in Equation \eqref{eq:taylor_series_spinosc} that for
    the classical Jaynes-Cummings system, the theoretical value is
    \[ S_{0,1} = \frac{5 \ln 2}{2\pi}.  \]
    Figure~\ref{fig:S01_spinosc_x01} shows a nice convergence of the
    semiclassical approximation.
    \begin{figure}[H]
      \begin{center}
        \includegraphics[trim=80 10 80
        60,clip,width=0.75\textwidth]{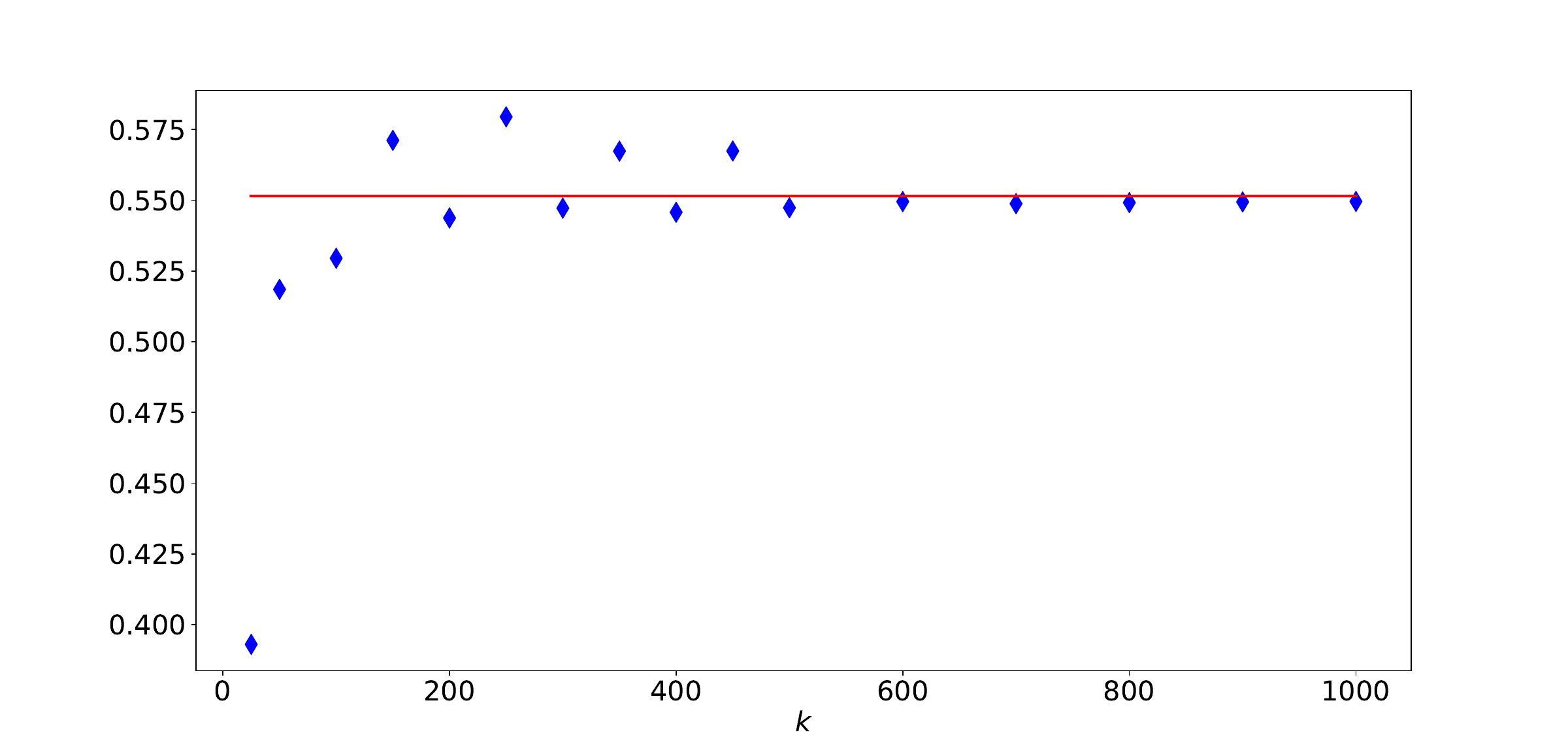}
      \end{center}
      \caption{\small Determination of $S_{0,1}$ for the
        spin-oscillator system, see Example~\ref{ex:S01_spin-osc}. The
        blue diamonds correspond to Formula \eqref{eq:S01_lim}
        evaluated at $(j,\ell) = (0,0)$ and with $x=0.01$, for
        different values of $k$. The red line corresponds to the
        theoretical value $S_{0,1} = \frac{5 \ln 2}{2\pi}$.}
      \label{fig:S01_spinosc_x01}
    \end{figure}
  \end{ex}

\subsection{Higher order terms}

We show in this section how to recover \emph{all} terms of the Taylor
series invariant from the joint spectrum.  The difficulty is that the
Taylor series $S^\infty$ is defined in terms of the normal form
coordinates, \emph{i.e.} in terms of the function $f_r$ (or, more
precisely, of its Taylor series at the origin $[f_r]$), but this
function is also unknown \emph{a priori}. Hence we need to find a
scheme to recover, from the joint eigenvalues, \emph{both} Taylor
series $S^\infty$ and $[f_r]$ simultaneously.

In order to organize the proof, we will treat the coefficients
$S_\alpha$ and all derivatives of $f_r$ at the origin as formal
indeterminates, and use the following notation. Let $\mu$ be an
additional formal parameter.  Let $\mathcal{F}_{\leq n_1,\leq n_2}$ be
the polynomial algebra in the variables $\partial^\beta f_r(0)$ with
$\abs{\beta}\leq n_1$, in the variables $S_\alpha$ with
$\abs{\alpha}\leq n_2$, and in $\mu$.  We will also use the subscript
``$n_j$'' instead of ``$\leq n_j$'' to indicate that only derivatives
of order exactly $n_j$ are concerned.

Let $a_1, a_2$ be the functions defined in Equation \eqref{equ:a} after a choice of action variable $L$.

\begin{prop}
  \label{prop:g_lambda}
  Let $\mu \in \RM$; the function
  \[
    g_{\mu}(x) := a_1(x,\mu x) + \mu a_2(x,\mu x) \qquad \forall x>0
  \]
  admits an asymptotic expansion of the form
  \begin{equation} g_{\mu}(x) \sim \sum_{n \geq 0} x^n ( c_n(\mu) +
    d_n(\mu) \ln x ) \qquad \text{ as } x \to 0^+ .\label{equ:AE_ln}
  \end{equation}
  Moreover, for $n \geq 0$, $d_n(\mu)\in \mathcal{F}_{n+1,0}$, namely:
  \begin{equation}
    d_n(\mu) = -\frac{1}{2\pi n!} \sum_{\ell = 0}^{n+1}
    \binom{n+1}{\ell} \mu^{n+1-\ell} \partial_x^{\ell}
    \partial_y^{n+1-\ell} f_r(0)\label{equ:dn}
  \end{equation}
  and
  $c_n(\mu) \in \mathcal{F}_{\leq n+1, \leq n} \oplus
  \mathcal{F}_{1,n+1} $, namely:
  \begin{equation}\label{equ:cn}
    c_n(\mu)
    = \tilde{c}_n(\mu) + \sum_{\ell = 0}^{n+1} \mu^{n-\ell}
    \Big( \mu (n+1) \partial_y f_r(0) + (n-\ell+1) \partial_x
    f_r(0) \Big) S_{\ell,n+1-\ell}.
  \end{equation}
  Here we slightly abuse notation and use the convention
  $(n-\ell+1) \mu^{n-\ell} = 0$ if $\ell = n+1$ for the sake of
  simplicity.
\end{prop}

\begin{demo}
  Let $\mu \in \RM$. If $F: \RM^2 \to \RM$ is a smooth function, one
  readily checks, for instance by first Taylor expanding in $\mu$, or
  by induction, or by using Fa\`a di Bruno's formula, that the
  coefficient in front of $x^n$ in the Taylor series expansion of
  $x \mapsto F(x,\mu x)$ at zero is
  \[
    \frac{1}{n!} \sum_{\ell=0}^n \binom{n}{\ell} \mu^{n-\ell}
    \partial^{\ell}_x \partial^{n-\ell}_y F(0). \] By Equation
  \eqref{equ:a1_a2},
  $a_2(x,\mu x) = \tau_2(x,\mu x) \partial_y f_r(x,\mu x)$, which
  gives thanks to Proposition \ref{prop:sigma_smooth}
  \[
    a_2(x,\mu x) = \left( \sigma_2(x,\mu x) - \frac{1}{2\pi} \ln x -
      \frac{1}{4\pi} \ln\left( 1 + \left( \frac{f_r(x,\mu x)}{x}
        \right)^2 \right) \right) \partial_y f_r(x,\mu x). \] Since
  $f_r(0) = 0$, the function $x \mapsto \frac{f_r(x,\mu x)}{x}$ is
  smooth at $x=0$; since moreover $\sigma_2$ and $\partial_y f_r$ are
  smooth, this implies that $x \mapsto a_2(x,\mu x)$ has an asymptotic
  expansion of the form \eqref{equ:AE_ln}. Moreover, the coefficient
  of $x^n \ln x$ in this expansion is equal to $-\frac{1}{2\pi}$ times
  the coefficient of $x^n$ in the Taylor series expansion of
  $x \mapsto \partial_y f_r(x,\mu x)$, namely
  \[
    -\frac{1}{2\pi n!} \sum_{\ell=0}^n \binom{n}{\ell} \mu^{n-\ell}
    \partial^{\ell}_x \partial^{n-\ell+1}_y f_r(0).
  \]
  The coefficient of $x^n$ in this expansion is the sum of the
  coefficients of $x^n$ in the respective Taylor series expansions of
  $\sigma_2(x,\mu x) \partial_y f_r(x,\mu x)$ and
  $- \frac{1}{4\pi} \ln\left( 1 + \left( \frac{f_r(x,\mu x)}{x}
    \right)^2 \right) \partial_y f_r(x,\mu x)$. The latter clearly
  lies in $\mathcal{F}_{\leq n+1,0}$. The former is obtained as the
  sum
  \[
    \sum_{k=0}^n [\sigma_2(\cdot,\mu \cdot)]_k [\partial_y
    f_r(\cdot,\mu\cdot)]_{n-k}.
  \]
  Here we denote by $[F]_k$ the coefficient of $x^k$ in the Taylor
  series expansion at $0$ of a function $F \in
  C^{\infty}(\RM,\RM)$. But whenever $k \leq n-1$,
  $[\sigma_2(\cdot,\mu \cdot)]_k [\partial_y f_r(\cdot,\mu
  \cdot)]_{n-k}$ belongs to the algebra generated by the
  $\partial^{\alpha} f_r(0)$ with $|\alpha| \leq n+1$ and the
  $\partial^{\beta} \sigma_2(0)$ with $|\beta| \leq n-1$; the latter
  correspond to the $S_{\gamma}$ with $|\gamma| \leq n$ since, by
  definition of the Taylor series invariant,
  \[
    \partial^{\beta} \sigma_2(0) = \beta_1! (\beta_2 + 1)!  S_{\beta +
      (0,1)}. \] Therefore, for our purpose we need to understand only
  the term corresponding to $k=n$, \emph{i.e.}
  $[\sigma_2(\cdot,\mu \cdot)]_n [\partial_y f_r(\cdot,\mu \cdot)]_0$,
  which equals
  \[
    \frac{\partial_y f_r(0)}{n!} \sum_{\ell = 0}^n \mu^{n-\ell}
    \binom{n}{\ell} \partial_x^{\ell} \partial_y^{n-\ell} \sigma_2(0)
    = \partial_y f_r(0) \sum_{\ell = 0}^n \mu^{n-\ell} (n-\ell+1)
    S_{\ell,n-\ell+1}.
  \]

  Similarly, Equation \eqref{equ:a1_a2} gives
  \[
    a_1(x,\mu x) = \tau_1(x,\mu x) + \tau_2(x,\mu x) \partial_x
    f_r(x,\mu x) \] and Proposition \ref{prop:sigma_smooth} yields
  \[
    \begin{split}  a_1(x,\mu x) & = \sigma_1(x,\mu x) - \frac{1}{2\pi} \arctan\left( \frac{f_r(x,\mu x)}{x} \right) \\
      &+ \left( \sigma_2(x,\mu x) - \frac{1}{2\pi} \ln x -
        \frac{1}{4\pi} \ln\left( 1 + \left( \frac{f_r(x,\mu x)}{x}
          \right)^2 \right) \right) \partial_x f_r(x,\mu
      x). \end{split} \] Similar arguments as above show that
  $x \mapsto a_1(x,\mu x)$ has an asymptotic expansion of the form
  \eqref{equ:AE_ln}, and the coefficient of $x^n \ln x$ in this
  expansion is
  \[
    -\frac{1}{2\pi n!} \sum_{\ell=0}^n \binom{n}{\ell} \mu^{n-\ell}
    \partial^{\ell+1}_x \partial^{n-\ell}_y f_r(0). \] The coefficient
  of $x^n$ in this expansion is the sum of the coefficients of $x^n$
  in the respective Taylor series expansions of $\sigma_1(x,\mu x)$,
  $\sigma_2(x,\mu x) \partial_x f_r(x,\mu x)$,
  $- \frac{1}{2\pi} \arctan\left( \frac{f_r(x,\mu x)}{x} \right)$, and
  $- \frac{1}{4\pi} \ln\left( 1 + \left( \frac{f_r(x,\mu x)}{x}
    \right)^2 \right) \partial_x f_r(x,\mu x)$. The last two belong to
  $\mathcal{F}_{\leq n+1,0}$. Moreover,
  \[
    [\sigma_1(\cdot,\mu \cdot)]_n = \frac{1}{n!} \sum_{\ell = 0}^n
    \mu^{n-\ell} \binom{n}{\ell} \partial_x^{\ell} \partial_y^{n-\ell}
    \sigma_1(0) = \sum_{\ell = 0}^n \mu^{n-\ell} (\ell+1)
    S_{\ell+1,n-\ell},
  \]
  and we can decompose
  \[
    [\sigma_2(\cdot,\mu \cdot) \partial_x f_r(\cdot,\mu \cdot)]_n =
    \sum_{k=0}^{n-1} [\sigma_2(\cdot,\mu \cdot)]_k [\partial_x
    f_r(\cdot,\mu \cdot)]_{n-k} + [\sigma_2(\cdot,\mu \cdot)]_n
    [\partial_x f_r(\cdot,\mu \cdot)]_0
  \]
  where the first term on the right-hand side lies in
  $\mathcal{F}_{n+1, n}$, and the second term reads
  \[
    \frac{\partial_x f_r(0)}{n!} \sum_{\ell = 0}^n \mu^{n-\ell}
    \binom{n}{\ell} \partial_x^{\ell} \partial_y^{n-\ell} \sigma_2(0)
    = \partial_x f_r(0) \sum_{\ell = 0}^n \mu^{n-\ell} (n-\ell+1)
    S_{\ell,n-\ell+1}.
  \]
  Hence the coefficient of $x^n$ in the expansion of
  $a_1(\cdot,\mu \cdot)$ is equal, modulo
  $\mathcal{F}_{\leq n+1, \leq n}$, to
  \[
    \sum_{\ell = 0}^n \mu^{n-\ell} (\ell+1) S_{\ell+1,n-\ell} +
    \partial_x f_r(0) \sum_{\ell = 0}^n \mu^{n-\ell} (n-\ell+1)
    S_{\ell,n-\ell+1}.
  \]

  We deduce from the above computations that
  \[
    \begin{split} d_n(\mu) & = \frac{-1}{2\pi n!} \left(
        \sum_{\ell=0}^n \tbinom{n}{\ell} \mu^{n-\ell}
        \partial^{\ell+1}_x \partial^{n-\ell}_y f_r(0) +
        \sum_{\ell=0}^n \tbinom{n}{\ell} \mu^{n-\ell+1}
        \partial^{\ell}_x \partial^{n-\ell+1}_y f_r(0)
      \right) \\
      & = \frac{-1}{2\pi n!} \left( \sum_{p=1}^{n+1} \tbinom{n}{p-1}
        \mu^{n-p+1} \partial^{p}_x \partial^{n-p+1}_y f_r(0) \right.
      \left. + \sum_{\ell=0}^n \tbinom{n}{\ell} \mu^{n-\ell+1}
        \partial^{\ell}_x \partial^{n-\ell+1}_y f_r(0) \right)
      \\
      & = \frac{-1}{2\pi n!} \left( \partial_x^{n+1} f_r(0) +
        \sum_{\ell=1}^n \tbinom{n+1}{\ell} \mu^{n-\ell+1}
        \partial^{\ell}_x \partial^{n-\ell+1}_y f_r(0) + \mu^{n+1}
        \partial_y^{n+1} f_r(0) \right) \end{split}
  \]
  which yields the desired formula. Furthermore, the above analysis
  shows that $c_n(\mu) = \tilde{c}_n(\mu) + \check{c}_n(\mu)$ where
  $\tilde{c}_n(\mu) \in\mathcal{F}_{\leq n+1, \leq n}$ and
  \[
    \begin{split} \check{c}_n(\mu) & = \sum_{\ell = 0}^n \mu^{n-\ell}
      (\ell+1) S_{\ell+1,n-\ell} + \sum_{\ell = 0}^n \mu^{n-\ell}
      (n-\ell+1) ( \partial_x
      f_r(0)  + \mu \partial_y f_r(0)) S_{\ell,n-\ell+1} \\
      & = \sum_{p = 1}^{n+1} \mu^{n-p+1} p S_{p,n-p+1} + \sum_{\ell =
        0}^n \mu^{n-\ell} (n-\ell+1) ( \partial_x f_r(0) + \mu
      \partial_y f_r(0)) S_{\ell,n-\ell+1} \\
      & = \mu^n \left( \partial_x f_r(0) + \mu  \partial_y f_r(0) \right) (n+1) S_{0,n+1} + (n+1) S_{n+1,0}  \\
      & \ + \sum_{\ell = 1}^n \mu^{n-\ell} \Big( \mu \ell \partial_y
      f_r(0) + (n-\ell+1) \big(\partial_x f_r(0) + \mu \partial_y
      f_r(0)\big) \Big) S_{\ell,n+1-\ell}
    \end{split} \] which yields the desired result.
\end{demo}

\begin{lemm}\label{lemm:vdM}
  Let $n \geq 1$ and let $\mu_0, \ldots, \mu_{n+1} \in \RM$; the
  matrix
  \[
    A_n = \left( \mu_i^{n-j} \left( \mu_i (n+1) \partial_y f_r(0) +
        (n-j+1) \partial_x f_r(0) \right) \right)_{0 \leq i,j \leq
      n+1} \] (again, with the convention that
  $(n-j+1) \mu_i^{n-j} = 0$ if $j = n+1$) has determinant
  \[
    \det(A_n) = (n+1)^{n+2} (\partial_y f_r(0))^{n+2}
    \prod_{i=0}^{n+1} \prod_{j=0}^{i-1} (\mu_i - \mu_j).
  \]
  In particular, if $\mu_0, \ldots, \mu_{n+1}$ are pairwise distinct,
  $A_n$ is invertible.
\end{lemm}

\begin{demo}
  Since $\partial_y f_r(0) \neq 0$, we can factor each column of $A_n$
  by $(n+1) \partial_y f_r(0)$; we obtain that
  $\det(A_n) = (n+1)^{n+2} (\partial_y f_r(0))^{n+2} \det(B_n)$ where
  \[
    B_n = \left( \mu_i^{n-j + 1} + \frac{(n-j+1) \partial_x
        f_r(0)}{(n+1) \partial_y f_r(0)} \mu_i^{n-j} \right)_{0 \leq
      i,j \leq n+1}. \] Now we perform the following determinant
  preserving operations on the columns $C_j$ of $B_n$ by induction:
  \begin{itemize}
  \item replace $C_n$ by
    $\tilde C_n = C_n - \frac{ \partial_x f_r(0)}{(n+1) \partial_y
      f_r(0)} C_{n+1}$,
  \item for $j$ from $n-1$ to $0$, replace $C_j$ by
    $\tilde C_j = C_j - \frac{(n-j+1) \partial_x f_r(0)}{(n+1)
      \partial_y f_r(0)} \tilde{C}_{j+1}$
  \end{itemize}
  to get a new matrix $\tilde B_n$. Then
  \[
    \det(B_n) = \det(\tilde B_n) = \det\left( \left( \mu_i^{n-j + 1}
      \right)_{0 \leq i,j \leq n+1} \right) = \prod_{i=0}^{n+1}
    \prod_{j=0}^{i-1} (\mu_i - \mu_j) \] where the last equality comes
  from the fact that we are computing a Vandermonde determinant.
\end{demo}

\begin{theo}
  \label{theo:full_taylor}
  Given the $\h$-family of joint spectra $\Sigma_\h$ of a proper
  quantum semitoric system in a neighborhood of a focus-focus critical
  value, one can recover, in a constructive way, the complete Taylor
  series invariant, together with the full Taylor expansion of $f_r$,
  hence of the Eliasson diffeomorphism.
\end{theo}

\begin{demo}
  We prove this theorem by induction. By Lemma \ref{lemm:dxfr_dyfr},
  Theorem \ref{theo:recover-sigma_1} and Proposition
  \ref{theo:recover-sigma_2}, we can recover from the joint spectrum the quantities $\partial_x f_r(0)$,
  $\partial_y f_r(0)$, $S_{1,0}$, and $S_{0,1}$ associated with the action variable $L$ coming from a choice of good labelling. So let $n \geq 1$, and assume that we know all
  the derivatives $\partial^{\beta}f_r(0)$ for $|\beta| \leq n$ and
  all the coefficients $S_{\alpha}$ for $|\alpha| \leq n$. Let
  $\mu \in \RM$ and let $g_{\mu}$ be the function defined in the
  statement of Proposition \ref{prop:g_lambda}; since by Lemma
  \ref{lemm:quantum-a}, we can recover $a_1$ and $a_2$ from the joint
  spectrum, we can recover the function $g_{\mu}$. Thanks to the
  induction hypothesis, we can compute the coefficients
  $c_{\ell}(\mu)$ and $d_{\ell}(\mu)$ in the asymptotic expansion
  \eqref{equ:AE_ln} for every $\ell \leq n-1$. Hence we can recover
  $d_n(\mu)$ as the limit
  \[
    d_n(\mu) = \lim_{x \to 0^+} \frac{g_{\mu}(x) - \sum_{\ell=0}^{n-1}
      x^{\ell}(c_{\ell}(\mu) + d_{\ell}(\mu) \ln x)}{x^n \ln x},
  \]
  and henceforth $c_n(\mu)$ as
  \[
    c_n(\mu) = \lim_{x \to 0^+} \frac{g_{\mu}(x) - \sum_{\ell=0}^{n-1}
      x^{\ell}(c_{\ell}(\mu) + d_{\ell}(\mu) \ln x) - d_n(\mu) x^n \ln
      x}{x^n}.
  \]
  Since we know $d_n(\mu)$ for every $\mu$, we can compute
  from~\eqref{equ:dn} all the derivatives $\partial^{\beta} f_r(0)$
  with $|\beta| = n+1$, for instance by taking derivatives with
  respect to $\mu$. Another solution, perhaps preferable from a
  numerical viewpoint, is to invert the linear system
  \[
    D_n \begin{pmatrix} \partial_y^{n+1} f_r(0) \\ \partial_x
      \partial_y^n f_r(0) \\ \vdots \\ \partial_x^{n+1}
      f_r(0) \end{pmatrix} = \begin{pmatrix} d_n(\mu_0) \\ d_n(\mu_1)
      \\ \vdots \\ d_n(\mu_{n+1}) \end{pmatrix} \] where
  $\mu_0, \ldots, \mu_{n+1}$ are pairwise distinct numbers
  and the matrix
  \[
    D_n = \left( \binom{n+1}{j} \mu_i^{n+1-j} \right)_{0 \leq i,j \leq
      n+1} \] is invertible since its determinant is equal to
  \[
    \prod_{j=0} \binom{n+1}{j} \prod_{i=0}^{n+1} \prod_{j=0}^{i-1}
    (\mu_i - \mu_j). \] This in turn implies that we may compute the
  coefficient $\tilde{c}_n(\mu) \in \mathcal{F}_{\leq n+1, \leq n}$
  for every $\mu$. It follows from~\eqref{equ:cn}, with similar
  arguments as above (for instance thanks to Lemma~\ref{lemm:vdM},
  since we obtain a linear system involving the matrix $A_n$ as
  above), that we can recover the coefficients $S_{\beta}$ with
  $|\beta| = n+1$. This concludes the induction step.
\end{demo}

We will not write more explicit formulas for the quadratic terms as
they are already quite involved, but we illustrate their computation in the example below.

\begin{ex} \label{ex:higher_spin-osc} We illustrate part of the
    computation of higher order terms for Eliasson's function and the
    Taylor series invariant for the Jaynes-Cummings system, see
    Example \ref{ex:spin-osc_quant}.  Using the notation of
    Proposition \ref{prop:g_lambda} and the exact values of
    $\partial_x f_r(0)$, $\partial_y f_r(0)$, $S_{1,0}$, $S_{0,1}$,
    $\partial^2_x f_r(0)$, $\partial^2_y f_r(0)$, $S_{2,0}$ and
    $S_{0,2}$ in this precise example (see Equations
    \eqref{eq:taylor_series_spinosc} and \eqref{eq:fr_elia_spinosc}),
    we compute the first coefficients of Equation~\eqref{equ:AE_ln}:
    \[
      d_0(\mu) = -\frac{\mu}{\pi}, \qquad c_0(\mu) = -\frac{1}{2\pi}
      \arctan(2\mu) + \frac{5 \mu \ln 2}{\pi} - \frac{\mu}{\pi} \ln(1
      + 4 \mu^2), \] and
    \[
      d_1(\mu) = -\frac{\mu}{\pi} \partial_x \partial_y f_r(0), \qquad
      c_1(\mu) = 3 \mu S_{1,1} + \mu \partial_x \partial_y f_r(0)
      \left( \frac{5 \ln 2}{\pi} - \frac{1}{2\pi} - \frac{\ln(1 +
          4\mu^2)}{2\pi} \right).
    \]
    This implies that
    \begin{multline*} \partial_x \partial_y f_r(0) = -\frac{\pi}{\mu x
        \ln x} \left( g_{\mu}(x) + \frac{\mu}{\pi} \ln x +
        \frac{1}{2\pi} \arctan(2\mu) - \frac{5 \mu \ln 2}{\pi} +
        \frac{\mu}{\pi} \ln(1 + 4 \mu^2) \right) \\ +
      \O\left(\frac{1}{\ln x}\right)
    \end{multline*}
    and that, in view of~\eqref{equ:cn},
    \begin{equation} S_{1,1} = \frac{1}{3} \left( \frac{c_1(\mu)}{\mu}
        - \partial_x \partial_y f_r(0) \left( \frac{5 \ln 2}{\pi} -
          \frac{1}{2\pi} - \frac{\ln(1 + 4\mu^2)}{2\pi} \right)
      \right) \label{eq:S11_spinosc} \end{equation} where we can
    obtain $c_1(\mu)$ as
    \begin{multline*} c_1(\mu) = \frac{1}{x} \left( g_{\mu}(x) +
        \frac{\mu}{\pi} \ln x + \frac{1}{2\pi} \arctan(2\mu) - \frac{5
          \mu \ln 2}{\pi} + \frac{\mu}{\pi} \ln(1 + 4 \mu^2) +
        \frac{\mu}{\pi} \partial_x \partial_y f_r(0) x \ln x \right)
      \\ + \O(x \ln x). \end{multline*} Hence if we already know all
    the above quantities and simply want to recover
    $\partial_x \partial_y f_r(0)$ and $S_{1,1}$ from the spectrum, we
    can first obtain
    \begin{equation} \partial_x \partial_y f_r(0) = \lim_{x \to 0^+}
      \lim_{\h \to 0} \frac{-\pi \left( \frac{E_{j,\ell} -
            E_{j+1,\ell} + \mu \h}{E_{j,\ell+1} - E_{j,\ell}} +
          \frac{\mu}{\pi} \ln x + \frac{1}{2\pi} \arctan(2\mu) -
          \frac{5 \mu \ln 2}{\pi} + \frac{\mu}{\pi} \ln(1 + 4 \mu^2)
        \right)}{\mu x \ln x} \label{eq:dxdy_spinosc} \end{equation}
    thanks to Lemma \ref{lemm:quantum-a} applied with $c = (x,\mu x)$
    and then use it to recover $c_1(\mu)$ from the joint spectrum, and
    finally $S_{1,1}$ thanks to Formula \eqref{eq:S11_spinosc}. A
    formula for $S_{1,1}$ similar to~\eqref{eq:dxdy_spinosc} can be
    obtained in a similar fashion.

  We recover the derivative $\partial_x \partial_y f_r(0)$ using
  Formula \eqref{eq:dxdy_spinosc} in Figure \ref{fig:dxdyfr_spinosc},
  and we recover the coefficient $S_{1,1}$ of the Taylor series
  invariant using Formula \eqref{eq:S11_spinosc} in Figure
  \ref{fig:S11_spinosc}.
    \begin{figure}[H]
      \begin{center}
        \includegraphics[trim=80 10 80
        60,clip,width=0.75\textwidth]{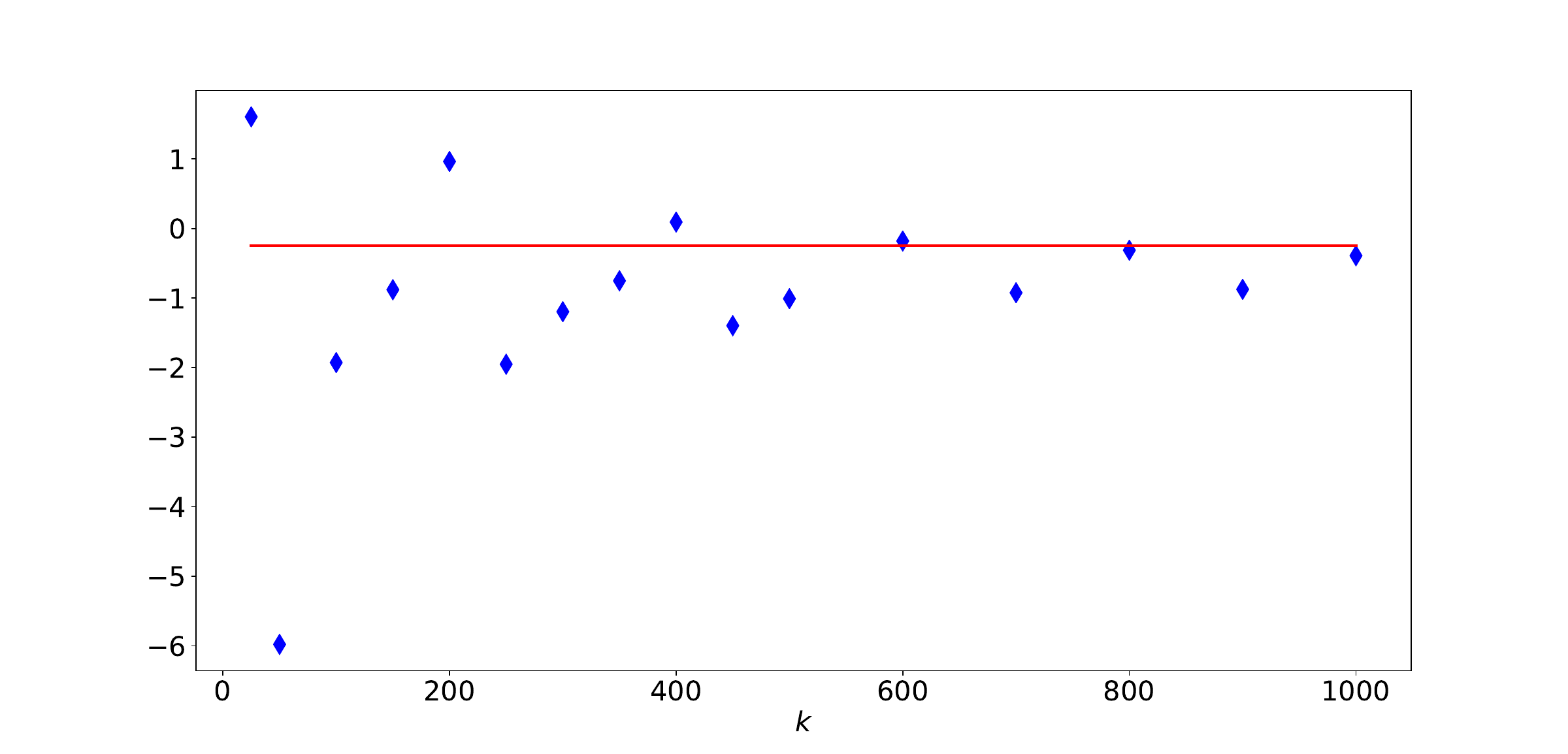}
      \end{center}
      \caption{\small Determination of $\partial_x \partial_y f_r(0)$
        for the spin-oscillator using Formula \eqref{eq:dxdy_spinosc}
        with $x = 0.01$ and $\mu = 1$, for different values of
        $k$. The red line corresponds to the theoretical result
        $\partial_x \partial_y f_r(0) = -\frac{1}{4}$.}
      \label{fig:dxdyfr_spinosc}
    \end{figure}
    \begin{figure}[H]
      \begin{center}
        \includegraphics[trim=80 10 80
        60,clip,width=0.75\textwidth]{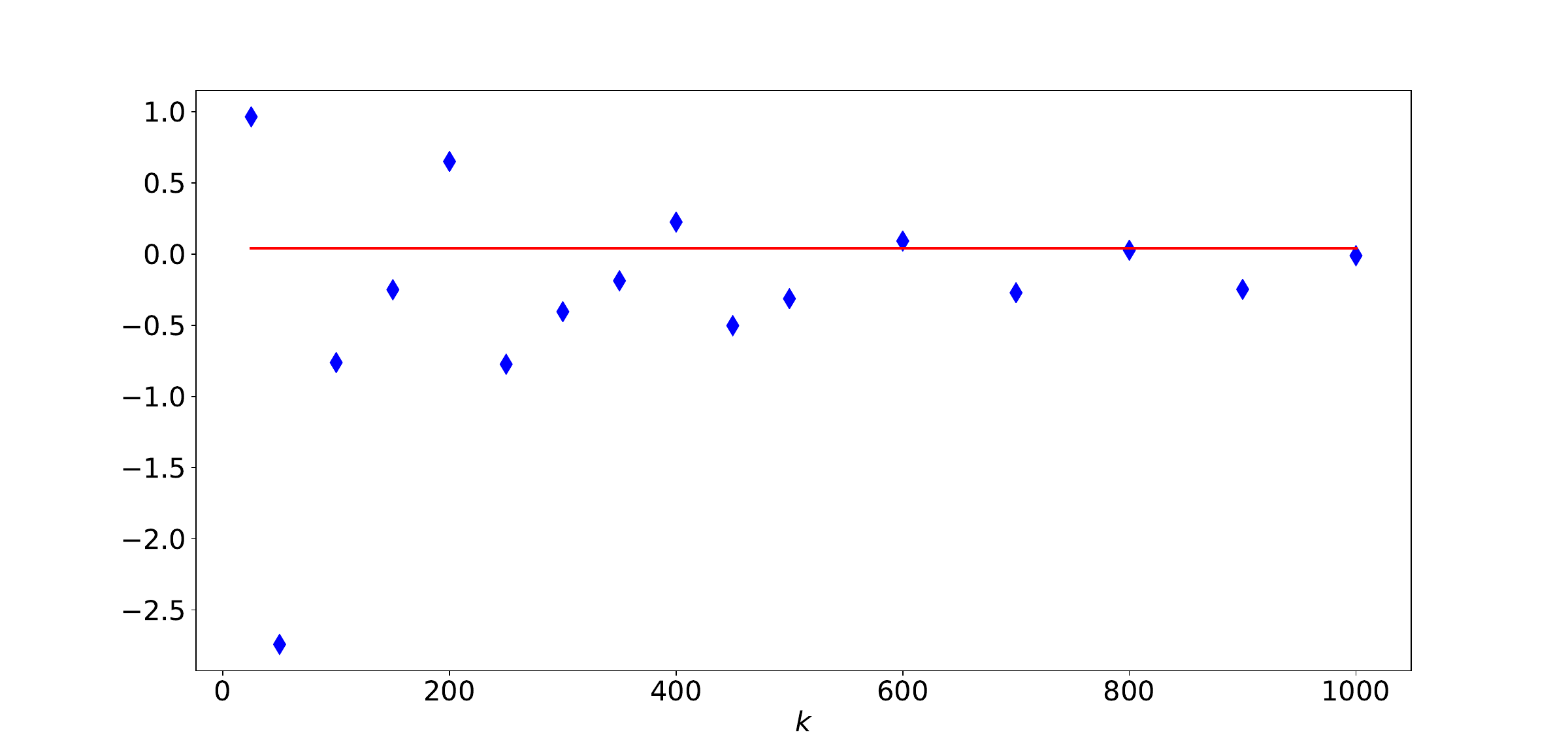}
      \end{center}
      \caption{\small Determination of $S_{1,1}$ for the
        spin-oscillator using Formula \eqref{eq:S11_spinosc} with
        $x = 0.01$ and $\mu = 1$, for different values of $k$. The red
        line corresponds to the theoretical result
        $S_{1,1} = \frac{1}{8\pi}$.}
      \label{fig:S11_spinosc}
    \end{figure}
  \end{ex}

\section{Structure of the joint spectrum near an elliptic-transverse
  singularity}
\label{sect:BScrit}

The goal of this section is to obtain the description of the structure
of the joint spectrum of a two-dimensional proper quantum integrable
system near a transversally elliptic singularity of its classical
counterpart (the local normal form of the momentum map $F$ splits into
one regular block and one elliptic block, see
Theorem~\ref{theo:eliasson}). While this description will contribute
to the proof of the semitoric inverse spectral conjecture
(Section~\ref{sec:asympt-half-latt}), here we don't assume $F$ to be
semitoric, and hence this section, and its main result
Theorem~\ref{theo:BS_ell_trans}, can be read independently of the rest
of the paper. 

In what follows, we endow
$T^*S^1 \times T^*\RM=S^1\times\RM\times\RM\times\RM$ with
coordinates $(x_1, \xi_1, x_2, \xi_2)$ and symplectic form
$\omega_0 = d\xi_1 \wedge dx_1 + d\xi_2 \wedge dx_2$. We have the
following symplectic normal form near a critical fiber.

\begin{lemm}[\cite{dufour-molino}]
  \label{lemm:dufour-molino}
  Let $F = (J,H)$ be an integrable system and let $c = (c_1,c_2)$ be a simple
  transversally elliptic critical value of $F$ with compact fiber
  $F^{-1}(c)$. Then there exist a saturated neighborhood $\mathcal{U}$
  of $F^{-1}(c)$ in $M$, a neighborhood $\mathcal{V}$ of
  $(S^1 \times \{0\}) \times \{(0,0)\}$ in
  $T^* S^1 \times T^*\RM$, a local symplectomorphism
  $\phi:(\mathcal{U},\omega) \to (\mathcal{V}, \omega_0)$ and a local
  diffeomorphism $G_0: (\RM^2,0) \to (\RM^2,c)$ such that
  \[
    (F \circ \phi^{-1})(x_1,\xi_1,x_2,\xi_2) = G_0(\xi_1,
    q(x_2,\xi_2)) \] where
  $q(x_2, \xi_2) = \frac{1}{2} (x_2^2 + \xi_2^2)$.  If moreover
  $(J,H)$ is semitoric, then $\phi$ can be chosen such that
  $J\circ\phi^{-1} - c_1= \xi_1$.
\end{lemm}
The adjective ``simple'' means that the fiber $F^{-1}(c)$ is connected
(this will be the case when we will further assume the system to be
semitoric). This lemma, due to Dufour--Molino~\cite{dufour-molino},
was first shown in the homogeneous setting in~\cite{colinII},
generalized to hyperbolic flows in~\cite{san-cdv}, and extended to all
non-degenerate singularities in~\cite{miranda-zung}.

A corollary of~\ref{lemm:dufour-molino} is the following simpler but
useful local normal form.
\begin{lemm}
  \label{lemm:hyperbolic-local-nf}
  Let $F = (J,H)$ be an integrable system and let $m\in M$ be a simple
  transversally elliptic critical point of $F$. Assume
  $\dd J(m) \neq 0$. Then there exist local symplectic coordinates
  $(x_1,\xi_1,x_2,\xi_2)$ near $m$ in which
  \[
    J - J(m) = \xi_1, \quad H = f(\xi_1, q(x_2,\xi_2))
  \]
  where $f$ is smooth and
  $q(x_2, \xi_2) = \frac{1}{2} (x_2^2 + \xi_2^2)$.
\end{lemm}
\begin{demo}
  First apply Lemma~\ref{lemm:dufour-molino}, so that $J=h(\xi_1,q)$
  for some smooth function $h$. The hypothesis $\dd J(m)\neq 0$
  implies $\partial_1 h \neq 0$; hence by the implicit function
  theorem $\xi_1 = \tilde h(J,q)$ for some smooth function $\tilde h$.
  Writing $z_2:=x_2+i\xi_2$ we define the diffeomorphism:
  \[
    \begin{cases}
      \tilde x_1 = x_1 \partial_1 \tilde h(\xi_1,q(z_2))\\
      \tilde \xi_1 = h(\xi_1,q(z_2))\\
      \tilde z_2 = e^{-ix_1\partial_2 \tilde h(\xi_1,q(z_2))}z_2 .
    \end{cases}
  \]
  Since $\xi_1 = \tilde h(\tilde\xi_1,q)$, we have
  $\dd \xi_1 = \partial_1 \tilde h \dd{\tilde{\xi}_1} +
  \partial_2\tilde h \dd q$, and
  $\tilde x_1 \dd{\tilde{\xi}_1} = x_1 \dd \xi_1 - x_1
  \partial_2\tilde h \dd q$. Writing
  $\tilde{\theta} := \theta + x_1\partial_2 \tilde h$, where $\theta$
  is some determination of the argument of $z_2$, and
  $\tilde q = \frac{1}{2}\abs{\tilde z_2}^2 = q$, we see that
  \begin{equation}
    \tilde x_1 \dd{\tilde{\xi}_1} + \tilde{\theta}\dd{\tilde q}  = x_1
    \dd{\xi_1} + \theta \dd q.\label{equ:exact_symplectic}
  \end{equation}
  Since $(q,\theta)$ is a pair of action-angle variables, we have
  $\dd (\theta \dd q) = \dd \xi_2 \wedge \dd
  x_2$. Hence~\eqref{equ:exact_symplectic} implies that the above map
  $(x_1,\xi_2,z_2)\mapsto(\tilde x_1,\tilde{\xi}_1, \tilde z_2)$ is
  actually a local symplectomorphism.
\end{demo}
\begin{rema}
  Naturally, Lemma~\ref{lemm:hyperbolic-local-nf} can be proven
  directly, without resorting to Lemma~\ref{lemm:dufour-molino}, for
  instance by adapting the method used in~\cite[Theorem 1.5]{san-cdv}.
\end{rema}

The main result of this section is the following.

\begin{theo}
  \label{theo:BS_ell_trans}
  Let $(\hat{J}_\h, \hat{H}_\h)$ be a proper quantum integrable
  system, with momentum map $F = (J,H)$, and let $c$ be a simple
  transversally elliptic critical value of $F$. Then there exists an
  open ball $B \subset \RM^2$ around $c$ in which the joint spectrum
  $\Sigma_\h$ of $(\hat{J}_\h, \hat{H}_\h)$ has the following
  properties:
  \begin{enumerate}
  \item the joint eigenvalues are simple in the sense of
    \cite{charbonnel,san-dauge-hall-rotation}, namely: there exist
    $\h_0 > 0$ such that for every $\h \in \mathcal{I} \cap (0,\h_0]$
    and every $\lambda_\h \in \Sigma_\h \cap B$, the joint spectral
    projector of $(\hat{J}_\h, \hat{H}_\h)$ onto the ball
    $B(\lambda_\h, \h^2)$ has rank $1$,
  \item \label{item:4} there exist a bounded open set
    $U \subset \RM^2$ and a smooth map $G_\h: U \to \RM^2$ with an
    asymptotic expansion $G_\h = G_0 + \h G_1 + \ldots$ in the
    $\Cinf$ topology such that
    $\lambda_\h \in \Sigma_\h \cap B$ if and only if there exist
    $j(\h) \in \ZM$ and $\ell(\h) \in \NM$ such that
    $\lambda_\h = G_\h( \h(j(\h),\ell(\h))) + \O(\h^{\infty})$ where
    the remainder is uniform on $B$. Furthermore, $G_0$ is the same as
    in Lemma \ref{lemm:dufour-molino}.
  \end{enumerate}
  In other words, near $c$, $\Sigma_\h$ is an asymptotic half-lattice in the
  sense of Definition \ref{defi:asymp_half_latt}.
\end{theo}

\begin{rema}
  Given a regular value $\tilde{c}$ of $F$ sufficiently close to $c$,
  $G_0^{-1}$ is an action diffeomorphism as in Section
  \ref{sec:symp_prel}; indeed, away from $0$, $q$ itself defines an
  action variable. Therefore, from Theorem~\ref{theo:BS_ell_trans} we
  recover the description of the spectrum near $\tilde{c}$ as an
  asymptotic lattice (see Theorem \ref{theo:BS_reg}).
\end{rema}

This theorem was initially proved in \cite[Theorem 6.1]{colinII} for
homogeneous pseudodifferential operators. It was stated in
\cite[Théorème 5.2.4]{san-livre} (see also \cite[Theorem
3.38]{san-dauge-hall-rotation}), with a sketch of proof, for
$\h$-pseudodifferential operators. Here, we include both
$\h$-pseudodifferential and Berezin-Toeplitz operators; since we
simply need the explicit description of the principal term in the
asymptotic expansion of the joint eigenvalues, we may treat both cases
at once; differences would appear when looking at subprincipal terms.

\subsection{Semiclassical preliminaries}

Let us collect the tools that will be used throughout the proof,
building on the notions defined in Appendix \ref{sec:semicl-oper},
where semiclassical operators encompass both $\h$-pseudodifferential and
Berezin-Toeplitz operators. The idea is that near a simple
transversally elliptic critical value of $F$, the classical normal
form for $(J,H)$ from Lemma \ref{lemm:hyperbolic-local-nf} can be
quantized to obtain a quantum normal form for the operators
$(\hat{J}_\h, \hat{H}_\h)$, see Proposition~\ref{prop:microlocal_nf}; this
is done using Fourier integral operators and symbolic calculus. Then,
we study the space of microlocal solutions to the joint eigenvalue
equation in the small open set where this normal form is defined
(Lemma~\ref{lemm:sol_microlocale_locale}); this gives the first
Bohr-Sommerfeld conditions associated with the elliptic component
$q$. Then, covering the whole critical fiber with such open sets, we
obtain a flat microlocal bundle, whose cocycle constitutes the
obstruction to the existence of a global solution, and hence to the
existence of a joint eigenfunction of $(\hat{J}_\h, \hat{H}_\h)$. The
final Bohr-Sommerfeld conditions are obtained by writing explicitely
that this cocyle must be a coboundary, using the fact that the
semiclassical invariants of the local normal form are invariant along
the critical set (Lemma~\ref{lemm:independent-f}).

We start with the definition of quantized canonical transformations
which, following the tradition in microlocal analysis, we call Fourier
integral operators. In the following definition, we use the three
cases defined in Section \ref{sec:quantum_semitoric}.

\begin{defi}
  \label{defi:oif}
  Let $m \in M$ and let
  $\phi: (M, \omega, m) \to (\RM^4, \omega_0, 0)$ be a local
  symplectomorphism. A \emph{semiclassical Fourier integral operator}
  $U_\h: \mathcal{H}_\h \to L^2(\RM^2)$ associated with $\phi$ is
  \begin{enumerate}
  \item in case \ref{item:1}, a Fourier integral operator associated
    with $\phi$, in the sense of H\"ormander and Duistermaat
    \cite{hormander,duistermaat-hormander}, but with a semiclassical
    parameter~\cite{duistermaat-oscillatory,guillemin-sternberg-semiclassical};
  \item in cases \ref{item:2} and \ref{item:3}, an operator of the
    form $U_\h = B_k V_k$, with $\h = k^{-1}$, where $V_k$ is a
    Fourier integral operator associated with $\phi$ in the sense of
    Berezin-Toeplitz quantization (see \cite{BG,zelditch,laurent-BS}
    and \cite{yohann-ell} for the case at hand, i.e. Fourier integral
    operators with values in Bargmann spaces) and $B_k$ is the
    semiclassical Bargmann transform, see Appendix
    \ref{sec:semicl-oper}.
  \end{enumerate}
\end{defi}

Like all usual versions of Fourier integral operators, they can be
seen as quantized canonical transformations, which can be precisely
stated by studying their action on semiclassical operators, as
follows.

\begin{theo}[Egorov's theorem]
  \label{theo:egorov}
  Let $U_\h$ be a semiclassical Fourier integral operator associated
  with the symplectomorphism $\phi$. Let $A_\h$ be a semiclassical
  operator with principal symbol $a_0$; then $U_\h A_\h U_\h^*$ is an
  $\h$-pseudodifferential operator with principal symbol
  $a_0 \circ \phi^{-1}$.
\end{theo}

\begin{demo}
  For $\h$-pseudodifferential operators, a proof of this theorem can
  be found in \cite[Section 5.1]{cdv-notes}. For Berezin-Toeplitz
  operators, we first apply the usual Egorov's theorem, see
  \cite[Proposition 13.3]{BG} (for the homogeneous case), and we
  conclude using property \ref{item:B2} of the semiclassical Bargmann
  transform.
\end{demo}

We define microlocal solutions as in Definition \ref{defi:micsol_PDO}
for case \ref{item:1} and Definition \ref{defi:micsol_BTO} for cases
\ref{item:2} and \ref{item:3}. With these definitions, if
$(u_\h)_{\h \in \mathcal{I}}$ is admissible and satisfies
$A_\h u_\h = 0$, then its restriction to any phase space open set
$\mathcal{U}$ is a microlocal solution to
$A_\h u_\h = \O(\h^{\infty})$ over $\mathcal{U}$. One readily checks
that the set of microlocal solutions to the equation
$A_\h u_\h = \O(\h^{\infty})$ over $\mathcal{U}$ is a
$\CM_\h$-module. For more details, see \cite[Section 4.5]{san-focus}
for the $\h$-pseudodifferential case, and \cite[Section 4]{yohann-hyp}
for the Berezin-Toeplitz case. Moreover, if two semiclassical
operators $A_\h$ and $B_\h$ are microlocally equivalent on the open
set $\mathcal{U}$, then $(u_\h)$ is a microlocal solution to
$A_\h u_\h = \O(\h^{\infty})$ on $\mathcal{U}$ if and only if it is a
microlocal solution to $B_\h u_\h = \O(\h^{\infty})$ on
$\mathcal{U}$. Finally, it follows from standard results and from
property \ref{item:B1} of the Bargmann transform that semiclassical
Fourier integral operators behave naturally with respect to microlocal
solutions.

Finally, we will need to use the fact that semiclassical operators are
stable under functional calculus.

\begin{prop}[Joint functional calculus]
  \label{prop:func_calc}
  Let $(A_\h, B_\h)$ be two commuting semiclassical operators of the
  same type \ref{item:1}, \ref{item:2} or \ref{item:3}, with
  respective principal symbols $a_0$ and $b_0$, and let
  $f: \RM^2 \to \RM$ be a smooth compactly supported function. Assume
  that $A_\h^2+ B_\h^2$ either belongs to a bounded symbol class or is
  elliptic at infinity. Then $f(A_\h, B_\h)$ is a semiclassical
  operator with principal symbol $f(a_0, b_0)$.
\end{prop}

\begin{demo}
  For pseudodifferential operators, we refer the reader
  to~\cite{charbonnel} or \cite[Section 8, Theorem
  8.8]{dimassi-sjostrand} for instance. As regards Berezin-Toeplitz
  operators on compact or non-compact manifolds, to our knowledge only
  the case of a single operator can be found in the literature, see
  \cite[Proposition 12]{laurent-BTO}; however, the proof of the latter
  can easily be adapted to the case of several commuting operators
  using Formula $(8.18)$ in \cite{dimassi-sjostrand}.
\end{demo}

\subsection{Microlocal normal form}

The first step towards the proof of Theorem~\ref{theo:BS_ell_trans} is
to obtain a quantum version of the symplectic transformation to a
normal form given by Lemma \ref{lemm:hyperbolic-local-nf}. It could be
also interesting to quantize directly the semi-global normal form of
Lemma~\ref{lemm:dufour-molino}; however, this would require a
semi-global theory of Fourier integral operators, which, for
simplicity, we tried to avoid here. For our local situation, the model
operators constituting the quantum normal form are given by the
following elementary lemma.
\begin{lemm}
  \label{lemm:model_operators}
  Consider the unbounded differential operators $\Xi_\h, Q_\h$ on
  $L^2(\RM^2) \simeq L^2(\RM) \otimes L^2(\RM)$, acting as
  \[
    \Xi_\h = \frac{\h}{i} \deriv{}{x_1}, \qquad Q_\h = \frac{1}{2}
    \left(- \h^2 \frac{\partial^2}{\partial x_2^2} + x_2^2 \right)
  \]
  on compactly supported smooth functions. Then $(\Xi_\h, Q_\h)$ are
  $\h$-pseudodifferential operators with respective principal symbols
  $\xi_1$ and $q$, and extend to commuting self-adjoint operators on
  $L^2(\RM^2)$.
\end{lemm}
Thus, $\Xi_\h$ is just a Fourier oscillator in the variable $x_1$,
while $Q_\h$ is a harmonic oscillator in the variable $x_2$.  Recall
that the eigenvalues of $Q_\h$ acting on $L^2(\RM)$ are simple; more
precisely they are the $\h \left(n + \frac{1}{2} \right)$ for
$n \in \NM$ and the associated eigenspace is generated by
\begin{equation}
  \Psi_{\h,n}(x_2) = \frac{1}{(\pi \h)^{\frac{1}{4}} \sqrt{2^n
      n!}} H_n(x_2) \exp\left(-\frac{x_2^2}{2\h}\right)
  \label{equ:hermite}
\end{equation}
where $H_n$ is the $n$-th Hermite polynomial.

In order to transform the original quantum system
$(\hat J_\h, \hat H_\h)$ into this normal form, we will need to solve
the following system of partial differential equations.
\begin{lemm}[local cohomological equations]
  \label{lemm:cohomological} Let $z_1^0=(x_1^0,\xi_1^0)\in T^*\RM$ and
  let $\Omega\subset T^*\RM \times T^*\RM$ be an open neighborhood of
  $(z_1^0,0)$ of the form $\Omega_1\times\Omega_2$, where
  $\Omega_1=V_{x_1}\times V_{\xi_1}$ is a product of bounded open
  intervals and $\Omega_2=B(0,\epsilon)$ is an open ball at the
  origin.  Let $r,s \in C^{\infty}(\Omega)$ be such that
  $\{ s, \xi_1 \} = \{ r, q \}$. Then there exist
  $\nu \in C^{\infty}(\Omega;\RM)$ and
  $\psi \in
  C^{\infty}(V_{\xi_1}\times\interval[open]{-\epsilon^2/2}{\epsilon^2/2},\RM)$
  such that
  \begin{equation}
    \begin{cases} \{ \xi_1, \nu \} + r = 0, \\[2mm]
      \{ q, \nu \} + s + \psi(\xi_1,q) = 0.
    \end{cases}\label{equ:cohomological}
  \end{equation}
  Moreover, if $r=r_E, s=s_E$ depend smoothly on some additional
  parameter $E\in\RM$, then one can choose solutions $\nu,\psi$ that
  also depend smoothly on $E$.
\end{lemm}
\begin{demo}
  Let us put $\nu=\nu_1+\nu_2$ with
  $\displaystyle\nu_1(x_1,\xi_1,x_2,\xi_2):= -
  \int_{x_1^0}^{x_1}r(t,\xi_1,x_2,\xi_2) \dd t$. Since
  $\{ \xi_1, \nu_1 \} = \deriv{\nu_1}{x_1}$, we see that $\nu_1$
  satisfies the first equation of~\eqref{equ:cohomological}, and
  \[
    \{q,\nu_1\} (x_1,\xi_1,x_2,\xi_2) = - \int_{x_1^0}^{x_1}\{q,r\}
    \dd t = \int_{x_1^0}^{x_1}\{s,\xi_1\} = s(x_1^0,\xi_1,x_2,\xi_2) -
    s(x_1,\xi_1,x_2,\xi_2).
  \]
  Hence $\nu$ is a solution to~\eqref{equ:cohomological} if and only
  if $\nu_2$ satisfies:
  \begin{equation}
    \begin{cases} \{ \xi_1, \nu_2 \} = 0, \\[2mm]
      \{ q, \nu_2 \} + s_0 + \psi(\xi_1,q) = 0,
    \end{cases}\label{equ:cohomological2}
  \end{equation}
  where we define $s_0(\xi_1,x_2,\xi_2) := s(x_1^0,\xi_1,x_2,\xi_2)$.
  Hence we may look for $\nu_2=\nu_2(\xi_1,x_2,\xi_2)$, and the system
  is solved if and only if the last equation
  of~\eqref{equ:cohomological2} holds, where $\xi_1$ can be seen as an
  innocuous parameter. By~\cite[Prop 3.1]{san-miranda}, this is solved
  explicitly by letting $\psi$ be the average of $s_0$ by the
  Hamiltonian $q$-flow $\phy_t$, and
  \[
    \nu_2(\xi_1,x_2,\xi_2) = -\frac{1}{2\pi}\int_0^{2\pi}\left(t
      \phy_t^*s_0 + \psi\right) \dd t.
  \]
  The fact that $\psi$, being invariant under the flow of $q$, must be
  of the form $\psi=\psi(\xi_1,q)$, is classical. Because of the
  explicit formulas, we may directly check that the smooth dependence
  of $r,s$ on an external parameter is transferred to the solutions
  $\nu_1,\nu_2$ and $\psi$.
\end{demo}

To simplify notation, for $A, B, C$ three operators such that $AC$ and
$BC$ are well-defined, we write $(A,B) C := (AC, BC)$, and we adopt
similar notation for left products.

\begin{prop}
  \label{prop:microlocal_nf}
  Let $(\hat{J}_\h, \hat{H}_\h)$ be a proper quantum integrable
  system, with momentum map $F = (J,H)$, and let $c$ be a simple
  transversally elliptic critical value of $F$. Let $m \in F^{-1}(c)$
  and let $\mathcal{U}$ be as in Lemma \ref{lemm:dufour-molino}. Then
  there exist an open set $\mathcal{W} \subset \mathcal{U}$ containing
  $m$, a semiclassical Fourier integral operator
  $U_\h: \mathcal{H}_\h \to L^2(\RM^2) $ and a family of smooth
  functions $L_\h: (\RM^2,c) \to \RM^2$ with an asymptotic expansion
  \[
    L_\h = L_0 + \h L_1 + \cdots \] for the $C^{\infty}$ topology,
  where $L_0$ is a local diffeomorphism, such that
  $U_\h^* U_\h \sim I$ microlocally on $\mathcal{W}$ and
  \[
    U_\h U_\h^* \sim I, \qquad U_\h L_\h(\hat{J}_\h, \hat{H}_\h)
    U_\h^* \sim (\Xi_\h, Q_\h) \] microlocally on
  $\phi(\mathcal{W})$. More precisely, if we assume that
  $\dd J(m)\neq 0$, then there exists a family of smooth functions
  $g_\h\sim g_0+\h g_1 + \cdots:\RM^2\to \RM$ with
  $\partial_y g_0(x,y) \neq 0$, such that, microlocally on
  $\phi(\mathcal{W})$,
  \begin{equation}
    U_\h \hat{J}_\h  U_\h^* \sim \Xi_\h \text{ and }  U_\h \hat{H}_\h  U_\h^* \sim g_\h(\Xi_\h, Q_\h).
    \label{equ:unitary}
  \end{equation}
\end{prop}

Note also that we may (and will) always assume that
$\phi(\mathcal{W})$ is of the form $\Omega_1 \times \Omega_2$ of Lemma
\ref{lemm:cohomological}.

\begin{demo}
  Up to replacing $\hat J_\h$ by a linear combination of
  $\hat J_\h,\hat H_\h$ we may assume that $\dd J(m)\neq 0$, hence we
  can apply the normal form of Lemma~\ref{lemm:hyperbolic-local-nf}.
  Since $\partial_y g_0 \neq 0$, the implicit function theorem implies
  that $U_\h \hat{H_\h} U_\h^* \sim g_\h(\Xi_\h, Q_\h)$ is equivalent
  to $f_\h(\Xi_\h, U_\h\hat{H_\h} U_\h^*) \sim Q_\h$, for some family
  of smooth functions $f_\h \sim f_0 + \h f_1 + \cdots$ such that
  $f_0(\xi_1,g_0(\xi_1,q)) = q$. Hence we want to solve the microlocal
  system:
  \begin{equation}
    \begin{cases}
      U_\h\hat{J}_\h  U_\h^* \sim \Xi_\h  \\
      U_\h f_\h(\hat J_\h, \hat{H_\h}) U_\h^* \sim Q_\h.
    \end{cases}
    \label{equ:system-with-f}
  \end{equation}

  We start by choosing a semiclassical Fourier integral operator
  $U_\h^{(0)}: \mathcal{H}_\h \to L^2(\RM^2)$ associated with the
  symplectomorphism $\phi$ of Lemma~\ref{lemm:hyperbolic-local-nf}
  such that $U_\h^{(0)*} U_\h^{(0)} \sim I$ and
  $U_\h^{(0)} U_\h^{(0)*} \sim I$ microlocally near $m$ and $\phi(m)$
  respectively.

  By Proposition~\ref{prop:func_calc} and Theorem \ref{theo:egorov},
  $U_\h^{(0)} (\hat{J_\h}, f_0(\hat J_\h,\hat{H}_\h)) U_\h^{(0)*}$ is
  a $\h$-pseudodifferential operator with principal symbol
  $ F \circ \phi^{-1} = (\xi_1,q)$, so there exist
  $\h$-pseudodifferential operators $R_\h^{(0)}$ and $S_\h^{(0)}$ such
  that
  \[
    U_\h^{(0)} (\hat{J}_\h, f_0(\hat J_\h,\hat{H}_\h)) U_\h^{(0)*} =
    (\Xi_\h, Q_\h) + \h (R_\h^{(0)}, S_\h^{(0)})
  \]
  microlocally on $\phi(\mathcal{W})$. Let $P_\h$ be a unitary
  $\h$-pseudodifferential operator with principal symbol
  $p_0 = \exp(i\nu_0)$, and let $f_1: \RM^2 \to \RM$ be a smooth
  function. We consider $U_\h^{(1)} = P_\h^* U_\h^{(0)}$ and want to
  determine $\nu_0$ and $f_1$ such that
  \[
    U_\h^{(1)} \left( (\hat{J}_\h, (f_0+\h f_1)(\hat J_\h,\hat{H}_\h))
    \right) U_\h^{(1)*} = (\Xi_\h, Q_\h) + \h^2 (R_\h^{(1)},
    S_\h^{(1)})
  \]
  where $R_\h^{(1)}$ and $S_\h^{(1)}$ are $\h$-pseudodifferential
  operators. A straightforward computation shows that this amounts to
  asking
  \[
    \frac{1}{\h} \left( [\Xi_\h, P_\h], [Q_\h, P_\h] \right) + \left(
      R_\h^{(0)}, \,S_\h^{(0)} + U_\h^{(0)} f_1(\hat{J}_\h, \hat{H}_\h)
      U_\h^{(0)*} \right) P_\h = \h (R_\h^{(1)}, S_\h^{(1)}),
  \]
  which holds if and only if the joint principal symbol of the
  operator on the left-hand side vanishes, in other words if and only
  if
  \[
    \begin{cases} -i \{ \xi_1, p_0 \} + r_0 p_0 = 0, \\[2mm]
      -i \{ q, p_0 \} + \left( s_0 + f_1(\xi_1,g_0(\xi_1,q)) \right)
      p_0 = 0,
    \end{cases}
  \]
  where $r_0, s_0$ are the respective principal symbols of
  $R_\h^{(0)}, S_\h^{(0)}$. This is equivalent (by writing
  $f_1(x,y) = \psi_1(x,f_0(x,y))$) to finding $\nu_0$ and a function
  $\psi_1(\xi_1,q)$ such that
  \begin{equation}
    \begin{cases}
      \{ \xi_1, \nu_0 \} + r_0  = 0, \\[2mm]
      \{ q, \nu_0 \} + s_0 + \psi_1(\xi_1,q) = 0.
    \end{cases} \label{equ:forme_normale}
  \end{equation}
  Let $B_\h = f_0(\hat J_\h,\hat{H}_\h)$; by definition and since
  $[\Xi_\h, Q_\h] = 0$, we have that
  \[
    \begin{split} 0 = [\hat J_\h, B_\h] & \sim [U_\h^{(0)} \hat J_\h U_\h^{(0)*}, U_\h^{(0)} B_\h U_\h^{(0)*}] \\
      & = [\Xi_\h + \h R_\h^{(0)}, Q_\h + \h S_\h^{(0)}] \\
      & = \h \left( [\Xi_\h, S_\h^{(0)}] + [R_\h^{(0)}, Q_\h] \right)
      + \h^2 [ R_\h^{(0)}, S_\h^{(0)}],
    \end{split}
  \]
  which implies that $\{s_0, \xi_1\} = \{r_0, q\}$ on
  $\phi(\mathcal{W})$. Hence we can apply Lemma
  \ref{lemm:cohomological} to obtain $\nu_0, \psi_1$ satisfying
  Equation \eqref{equ:cohomological} on $\phi(\mathcal{W})$.

  Now, let $n \geq 1$ and assume that we have found
  $f_0, \ldots, f_n$, $U_\h^{(n)}$, $R_\h^{(n)}$ and $S_\h^{(n)}$ such
  that $U_\h^{(n)*} U_\h^{(n)} \sim I$,
  $U_\h^{(n)} U_\h^{(n)*} \sim I$ and
  \[
    U_\h^{(n)} \left( \hat{J}_\h, (f_0 + \h f_1 + \cdots + \h^n
      f_n)(\hat J_\h,\hat{H}_\h) \right) U_\h^{(n)*} = (\Xi_\h, Q_\h)
    + \h^{n+1} (R_\h^{(n)}, S_\h^{(n)}).
  \]
  The same argument as above gives $\{\xi_1, s_n\} = \{r_n, q\}$. Let
  $C_\h$ be an $\h$-pseudodifferential operator with principal symbol
  $c_0$ and set $U_\h^{(n+1)} = T_\h^* U_\h^{(n)} $ with
  $T_\h = \exp(i\h^n C_\h)$. We want to solve
  \[
    U_\h^{(n+1)} \left(\hat{J}_\h, (f_0 + \ldots + \h^{n+1} f_{n+1}
      )(\hat{J}_\h, \hat{H}_\h) \right) U_\h^{(n+1)*} = (\Xi_\h, Q_\h)
    + \h^{n+2} (R_\h^{(n+1)}, S_\h^{(n+1)})
  \]
  where $f_{n+1}$ is some smooth function and $R_\h^{(n+1)}$ and
  $S_\h^{(n+1)}$ are $\h$-pseudodifferential operators. This amounts
  to
  \[
    \left( [\Xi_\h, T_\h], [Q_\h, T_\h] \right) + \h^{n+1} \left(
      R_\h^{(n)},\, S_\h^{(n)} + U_\h^{(n)} f_{n+1}(\hat{J}_\h,
      \hat{H}_\h) U_\h^{(n)*} \right) T_\h = \h^{n+2} (V_\h, W_\h)
  \]
  for some $\h$-pseudodifferential operators $V_\h, W_\h$, which is
  true if and only if there exists a smooth function $\psi_{n+1}$ such
  that $c_0$ and $\psi_{n+1}$ satisfy
  \[
    \begin{cases}
      \{ \xi_1, c_0 \} + r_n  = 0, \\[2mm]
      \{ q, c_0 \} + s_n + \psi_{n+1}(\xi_1,q) = 0.
    \end{cases}
  \]
  Here we have used the fact that
  $T_\h = \text{Id} + i \h^n C_\h + \h^{n+1} \tilde{C}_\h$ for some
  $\h$-pseudodifferential operator $\tilde{C}_\h$. This is the same
  system as in Equation \eqref{equ:forme_normale}, and once again we
  can solve it to obtain $c_0$ and $\psi_{n+1}$ using
  Lemma~\ref{lemm:cohomological}.

  Thus by induction, we construct sequences $(f_n)_{n \geq 0}$ and
  $(U_\h^{(n)})_{n \geq 0}$ such that for every $n \geq 0$,
  $U_\h^{(n)*} U_\h^{(n)} \sim I$, $U_\h^{(n)} U_\h^{(n)*} \sim I$ and
  \[
    U_\h^{(n)} (\hat{J}_\h, \, f_0 + \h f_1 + \ldots + \h^n
    f_n)(\hat{J}_\h, \hat{H}_\h) U_\h^{(n)*} = (\Xi_\h, Q_\h) +
    O(\h^{n+1}).
  \]
  From this data, we use Borel's summation theorem to construct $f_\h$
  and $U_\h$ satisfying the desired properties.
\end{demo}

The microlocal solutions to the normal form can be explicitly
described.
\begin{lemm}
  \label{lemm:sol_normal_form}
  Let $\Omega$ and related notation be as in
  Lemma~\ref{lemm:cohomological}. Let the family
  $(\tilde{\nu}_\h, \tilde{\mu}_\h)$ belong to
  $V_{\xi_1}\times (-\epsilon^2/2,\epsilon^2/2)$. There exists a
  microlocal solution to the equation
  \begin{equation}
    \label{eq:mic_sol_fn}
    \left( (\Xi_\h, Q_\h) - (\tilde{\nu}_\h,
      \tilde{\mu}_\h) \right) v_\h = \O(\h^{\infty}) \quad
    \text{ on } \Omega
  \end{equation}
  if and only if
  $\tilde{\mu}_\h \in \h\left( \NM + \frac{1}{2} \right) +
  \O(\h^{\infty})$. If this is the case, the $\CM_\h$-module of these
  microlocal solutions is free of rank $1$.
\end{lemm}
\begin{demo}
  Conjugating by the multiplication operator
  $\exp\left( -\frac{i \tilde{\nu}_\h }{\h} x_1 \right)$, which is a
  unitary microlocal operator, we are reduced to the system:
  \begin{equation}
    \label{eq:mic_sol_reduit}
    \left( (\Xi_\h, Q_\h) - (0,
      \tilde{\mu}_\h) \right) v_\h = \O(\h^{\infty}) \quad
    \text{ on } \Omega - ( \{0\} \times V_{\xi_1} \times \{(0,0)\} ).
  \end{equation}
  (This can be verified by replacing $\tilde{\nu}_\h$ by a constant
  $\xi_1^0$ and checking that all the estimates are locally uniform in
  $\xi_1^0$.) Therefore $v_\h$ does not (microlocally) depend on
  $x_1$, and we are further reduced to the 1D microlocal problem on
  $L^2(\RM_{x_2})$:
  \[
    (Q_\h - \tilde{\mu}_\h)v_\h = \O(\h^{\infty}) \quad \text{on }
    \Omega_2.
  \]
  The conclusion follows then from \cite[Theorem
  4.3.16]{san-livre}. In particular, microlocal solutions
  to~\eqref{eq:mic_sol_fn} have the expected natural form
  \begin{equation}
    v_{n,\h}(x_1, x_2) = c_\h \Psi_{\h,n_\h}(x_2) \exp\left( \frac{i
        \tilde{\nu}_\h }{\h} x_1 \right),
    \label{equ:hermite-sol-mic}
  \end{equation}
  where $\Psi_{\h,n}$ is the Hermite function of~\eqref{equ:hermite},
  $n_\h$ is the integer defined by
  $\tilde\mu_\h = \h(n_\h + \tfrac12)+ \O(\h^\infty)$, and
  $c_\h\in\CM_\h$.
\end{demo}
\subsection{End of the proof}

Using this microlocal normal form, we can now finish proving Theorem
\ref{theo:BS_ell_trans}. As a first step, let $m \in F^{-1}(c)$ and
let $\mathcal{W}$, $U_\h$, $L_\h$ be as in Proposition
\ref{prop:microlocal_nf}. Thanks to this proposition, the family
$(u_\h)_{\h \in \mathcal{I}}$ is a microlocal solution to the system
$(\hat{J}_\h - \nu_\h, \hat{H}_\h - \mu_\h) u_\h = \O(\h^{\infty})$ on
$\mathcal{W}$ if and only if the family
$(v_\h = U_\h^* u_\h)_{\h \in \mathcal{I}}$ is a microlocal solution
to
$\left( (\Xi_\h, Q_\h) - (\nu_\h, f_\h(\nu_\h, \mu_\h)) \right) v_\h =
\O(\h^{\infty})$ on $\phi(\mathcal{W})$. Hence the following is a
direct consequence of Lemma~\ref{lemm:sol_normal_form}.

\begin{lemm}
  \label{lemm:sol_microlocale_locale}
  Let $(\nu_\h, \mu_\h) \in F(\mathcal{W})$. There exists a microlocal
  solution to the equation
  \begin{equation}\label{eq:mic_sol}
    \left( (\hat J_\h, \hat Q_\h) - (\nu_\h,
      {\mu}_\h) \right) v_\h = \O(\h^{\infty}) \text{ on }
    \mathcal{W}
  \end{equation}
  if and only if
  $ f_\h(\nu_\h, \mu_\h) \in \h\left( \NM + \frac{1}{2} \right) +
  \O(\h^{\infty})$. In this case, the $\CM_\h$-module of these
  microlocal solutions is free of rank $1$.
\end{lemm}

The next step is to understand the microlocal solutions
to~\eqref{eq:mic_sol} on the whole $F$-saturated neighborhood
$\mathcal{U}$. For this purpose, we may replace
$(\hat J_\h, \hat H_\h)$ by $G_0^{-1}(\hat J_\h, \hat H_\h)$; this
ensures that, for our new system (which we call
$(\hat J_\h, \hat H_\h)$ again), the semi-global normal form of
Lemma~\ref{lemm:dufour-molino} states that
\[
  F\circ\phi^{-1} = (\xi_1, q). \] In particular, we may apply the
microlocal normal form~\eqref{equ:unitary} (second item of
Proposition~\ref{prop:microlocal_nf}) associated with the restriction
of $\phi$ to a neighborhood of $m$, yielding a Fourier integral
operator $U_\h$ and a function $g_\h$ with $g_0(x,y)=y$.

We shall first need the invariance of the whole semiclassical
expansion of $g_\h$:

\begin{lemm}
  \label{lemm:independent-f}
  Let $B=F(\mathcal{W})$. The function $g_\h:B\to\RM$
  from~\eqref{equ:unitary} is (modulo $\O(\h^\infty)$) independent on
  the choice of the point $m\in F^{-1}(c)$ and of the open set
  $\mathcal{W}$ containing $m$, provided $B$ is fixed and
  $F(\mathcal{W})=B$.
\end{lemm}
\begin{demo}
  Let $\tilde{\mathcal{W}}\ni \tilde m$ be another such open set, and
  assume that $\mathcal{W}\cap
  \tilde{\mathcal{W}}\neq\emptyset$. Then, microlocally on this
  intersection, the composition $P_\h =\tilde{U}_\h U_\h^*$ of the
  corresponding Fourier integral operators $\tilde U_\h$ and $U_\h^*$
  is an $\h$-pseudodifferential operator (because its canonical
  transformation is the identity) and must satisfy:
  \begin{equation}
    P_\h^* \Xi_\h P_\h \sim \Xi_\h \quad \text{ and } \quad P_\h^*
    g_\h(\Xi_\h, Q_\h) P_\h \sim \tilde g_\h(\Xi_\h, Q_\h).
    \label{equ:unicite_g}
  \end{equation}
  Since $P_\h P_\h^* \sim I$, the first of these equalities implies
  that $[P_\h,\deriv{}{x_1}]\sim 0$, \emph{i.e.} the principal symbol
  $p_0$ of $P_\h$ does not depend on $x_1$, microlocally. The second
  equality gives
  $P_\h^* f_\h(\Xi_\h, g_\h(\Xi_\h,Q_\h)) P_\h \sim f_\h(\Xi_\h,
  \tilde g_\h(\Xi_\h, Q_\h))$, \emph{i.e.}
  \[
    P_\h^* Q_\h P_\h \sim a_\h(\Xi_\h,Q_\h)
  \]
  with $a_\h = f_\h^{(2)} \circ \tilde g_\h$ (the function $f_\h$ was
  introduced in~\eqref{equ:system-with-f}). So it suffices to prove
  that $a_\h(\xi_1,q) = q + \O(h^{\infty})$. By looking at the
  principal symbols, the above equality yields
  $p_0 q = p_0 a_0(\xi_1,q)$. Since $P_\h P_\h^* \sim I$, $p_0$ never
  vanishes, and we obtain $a_0(\xi_1,q) = q$. (This conclusion can be
  also directly derived from~\eqref{equ:unicite_g}, which ensures
  $\tilde{g}_0 = g_0$.)

  Hence $a_\h(\Xi_\h,Q_\h) \sim Q_\h + \h T_\h$ where $T_\h$ is an
  $\h$-pseudodifferential operator. Therefore we have
  $Q_\h P_\h \sim P_\h Q_\h + \h P_\h T_\h$, so
  $[Q_\h,P_\h] \sim \h P_\h T_\h$; consequently, the principal symbol
  of $T_\h$ equals $\frac{1}{i p_0} \{ q, p_0 \}$. Since
  $Q_\h = \mathrm{Op}_\h^{W}(q)$, this yields
  $a_1(\xi_1,q) = \frac{1}{i p_0} \{ q, p_0 \} = \{q,\phi_0\}$ where
  $p_0 = \exp(i\phi_0)$. This implies that $a_1(\xi_1,q) = 0$; indeed,
  this comes from integrating the equality
  $a_1(\xi_1,q) = \{q,\phi_0\}$ along the trajectories of the
  Hamiltonian flow of $q$.

  So $P_\h^* Q_\h P_\h \sim Q_\h + \h^2 R_\h$ with $R_\h$ a
  pseudodifferential operator with principal symbol
  $a_2(\xi_1,q)$. Now we write
  $P_\h = \exp(i \h \tilde{P}_\h) P_\h^{(0)}$ with $\tilde{P_\h}$,
  $P_\h^{(0)}$ two pseudodifferential operators such that
  $[Q_\h, P_\h^{(0)}] = \O(\h^3)$ (one can easily achieve this since
  we already know from the previous step that $\{ q, p_0 \} =
  0$). Then
  \[
    \begin{split} \h^2 P_\h R_\h \sim [Q_\h, P_\h] & = [Q_\h, \exp(i \h \tilde{P}_\h)] P_\h^{(0)} + \exp(i \h \tilde{P}_\h) [Q_\h, P_\h^{(0)}] \\
      & = [Q_\h, \exp(i \h \tilde{P}_\h)] P_\h^{(0)} +
      \O(\h^3). \end{split} \] Since
  $\h^{-2} [Q_\h, \exp(i \h \tilde{P}_\h)] P_\h^{(0)}$ is a
  pseudodifferential operator with principal symbol
  $\{q,\tilde{p_0}\} p_0$ where $\tilde{p_0}$ is the principal symbol
  of $\tilde{P}_\h$, this implies that
  $a_2(\xi_1,q) = \{q,\tilde{p_0}\}$ and the same reasoning as above
  yields $a_2(\xi_1,q) = 0$. A straightforward induction yields
  similarly that $a_n(\xi_1,q) = 0$ for every $n \geq 0$, which
  concludes the proof.
\end{demo}

Hence the function $f_\h$ of Lemma~\ref{lemm:sol_microlocale_locale}
does not depend on $\mathcal{W}$ either, and we may now replace
$\hat H_\h$ by $f_\h(\hat J_\h, g_\h(\hat J_\h, \hat H_\h))$, so
that~\eqref{equ:unitary} becomes:
\begin{equation}
  U_\h \hat{J}_\h  U_\h^* \sim \Xi_\h
  \text{ and }  U_\h \hat{H}_\h  U_\h^* \sim Q_\h
  \label{equ:normalized-nf}
\end{equation}
and
\[
  f_\h(x, y) = y + \O(\h^\infty).
\]
We will denote by $(\tilde\nu_\h,\tilde\mu_\h)$ the accordingly
modified joint eigenvalue:
$(\tilde\nu_\h,\tilde\mu_\h) = L_\h(\nu_\h,\mu_\h)$ for some smooth
symbol $L_\h= G_0^{-1} + \O(\h)$.

From Proposition~\ref{prop:microlocal_nf}, and the relative
compactness of $\mathcal{U}$, there exists a finite cover of
$\mathcal{U}$ by open sets $\mathcal{W}_j$, $j=1,\dots, p$ on which
Proposition~\ref{prop:microlocal_nf} applies. Each $\mathcal{W}_j$ is
a neighborhood of a point $m_j\in F^{-1}(c)$. Consider the open set
$\tilde{\mathcal{U}}=F^{-1}(B)$ with
$B=V_J\times\interval[open right]{0}{\frac{\epsilon^2}2}$ for some
$\epsilon>0$ and an open interval $V_J$ containing $J(m)$. By taking
$\epsilon$ and $V_J$ small enough, we may assume that
$B\subset F(\mathcal{W}_j)$.  From now on, we replace $\mathcal{W}_j$
by $\mathcal{W}_j\cap F^{-1}(B)$ and $\mathcal{U}$ by
$\tilde{\mathcal{U}}$. Moreover, if we define $x_1^{(j)}\in S^1$ by
$\phi(m_j) = (x_1^{(j)},J(m),0,0)$ then we may cyclically order these
angles, so that we obtain a cyclic chain of simply connected open sets
$\mathcal{W}_1, \ldots, \mathcal{W}_p$ such that
$\mathcal{W}_j\cap\mathcal{W}_{j+1}$ is connected and non-empty for
all $j$, when the indices are taken modulo $p$. What's more,
$F(\mathcal{W}_j)=F(\mathcal{W}_j\cap\mathcal{W}_{j+1})=B$.  We now
identify $x_1^{(j)}$ with an element of
$\interval[open right]0{2\pi}$, and let
$\phi_j:\mathcal{W}_j\to T^*\RM^2$ be the lift of the restriction of
$\phi$ to $\mathcal{W}_j$ such that
$\phi_j(m_j) = (x_1^{(j)},J(m),0,0) \in \RM^4$. Of course we still
have
\[
  F \circ\phi_j^{-1} = (\xi_1, q).
\]

Therefore, in each $\mathcal{W}_j$ we may apply the microlocal normal
form~\eqref{equ:normalized-nf} {associated with $\phi_j$, yielding a
  Fourier integral operator $U_\h^{(j)}$.  For any integer $n\in\NM$,
  define the ``standard basis''
  $u_{n,\h}^{(j)}:= U_\h^{(j)*} v_{n,\h}$ to be
  Formula~\eqref{equ:hermite-sol-mic} with $c_\h=1$.
  Lemma~\ref{lemm:sol_microlocale_locale} gives constants
  $d_{j}(n,\h)\in\CM_\h$ such that
  \begin{equation}
    u_\h^{(j+1)} = d_j(n,\h) u_\h^{(j)} \quad \text{ on } \mathcal{W}_j\cap\mathcal{W}_{j+1}.
    \label{equ:def-d}
  \end{equation}

  Let us study the structure of these $d_j(n,\h)$ (they can be seen as
  a singular generalization of the ``Bohr-Sommerfeld cocycle''
  of~\cite{san-focus}).

  If $1\leq j \leq p-1$, then
  $\phi_{j}\circ\phi^{-1}_{j+1}=\textup{Id}$ by construction, and
  hence the Fourier integral operator
  $P^{(j)}_\h:=U_\h^{(j)}U_\h^{(j+1)*}$ is actually a semiclassical
  pseudodifferential operator on $L^2(\RM^2)$. As noticed above,
  $[P^{(j)}_\h,\Xi_\h]\sim 0$ and hence the full Weyl symbol of
  $P^{(j)}_\h$ does not depend on $x_1$. In addition, we now have
  $[P^{(j)}_\h, Q_\h]\sim 0$, which says that the Weyl symbol of
  $P^{(j)}_\h$ is a smooth function of $(\xi_1,q)$. Therefore, there
  exists a symbol $a^{(j)}_\h(\xi,q)$ such that
  $P^{(j)}_\h \sim a^{(j)}_\h(\Xi_\h,Q_\h)$. Since $d_j(n,\h)$ is
  defined by
  \begin{equation}
    P^{(j)}_\h v_{n,\h} \sim d_j(n,\h) v_{n,\h}
    \label{equ:def-dj2}
  \end{equation}
  microlocally near a point $(x_1,J(m),0,0)$ with
  $x_1^{j} < x_1 < x_1^{(j+1)}$, we obtain
  from~\eqref{equ:hermite-sol-mic} that
  \begin{equation}
    d_j(n,\h) = a^{(j)}_\h(\tilde \nu_\h, \tilde\mu_\h) +
    \O(\h^\infty),
    \label{eq:d_j}
  \end{equation}
  where $\tilde\mu_\h := \h(n + \frac{1}{2})$. Since $P^{(j)}_\h$ is
  microlocally unitary, this implies that
  $|a^{(j)}_\h(\tilde \nu_\h, \tilde\mu_\h)| = 1 + \O(\h^{\infty})$.

  On the other hand, for $j=p$, on the intersection
  $\mathcal{W}_p\cap \mathcal{W}_1$, the map $\phi_p\circ \phi^{-1}_1$
  is the translation by $(2\pi,0,0,0)$. Hence if we denote by $\tau$
  the operator $\tau(u) = (x_1,x_2)\mapsto u(x_1-2\pi,x_2)$, then the
  composition $ \tau \circ U_\h^{(p)}U_\h^{(1)*}$ is a
  $\h$-pseudodifferential operator $P_\h$ on $L^2(\RM^2)$,
  microlocally in $\phi(\mathcal{W}_p\cap \mathcal{W}_1)$.  It follows
  from~\eqref{equ:def-dj2} that
  \[
    P_\h v_{n,\h} \sim d_p(n,\h) \tau(v_{n,\h}) \quad \text{ on
    }\phi(\mathcal{W}_p\cap \mathcal{W}_1),
  \]
  which implies as before, in view of~\eqref{equ:hermite-sol-mic},
  that
  \begin{equation}
    e^{-\frac{2i\pi\tilde\nu_\h}{\h}}d_p(n,\h) = a^{(p)}_\h(\tilde\nu_\h,\tilde
    \mu_\h) + O(\h^\infty)
    \label{equ:d_p}
  \end{equation}
  for some symbol $a^{(p)}_\h$.

  We may now come back to the eigenvalue problem. If a microlocal
  solution $u_\h$ to~\eqref{eq:mic_sol} on $\mathcal{U}$ exists, then
  its restriction $u_{1,\h}$ to $\mathcal{W}_1$ is a solution on that
  set, and hence, necessarily,
  $\tilde{\mu}_\h \in \h\left( \NM + \frac{1}{2} \right) +
  \O(\h^{\infty})$.  Let $n=n(\h)$ be the integer defined by
  $\tilde{\mu}_\h = \h\left( n + \frac{1}{2} \right) + \O(\h^\infty)$.
  Letting $u_{j,\h}$ be the restriction of $u_\h$ to $\mathcal{W}_j$,
  we get from Lemma~\ref{lemm:sol_microlocale_locale} the existence of
  $c_j(\h)\in\CM_\h$ such that
  \[
    u_{j,\h} \sim c_j(\h) u^{(j)}_{\h},
  \]
  which implies that
  $c_j(\h) u_\h^{(j)} \sim c_{j+1}(\h) u_\h^{(j+1)}$ on
  $\mathcal{W}_j\cap\mathcal{W}_{j+1}$. On the one hand,
  inserting~\eqref{equ:def-d}, we obtain
  $c_j = c_{j+1}d_j + \O(\h^\infty)$, which yields
  \begin{equation}
    d_0 d_1\cdots d_{p-1}  = 1 + \O(\h^\infty).
    \label{equ:cocycle}
  \end{equation}
  On the other hand, using~\eqref{eq:d_j} and~\eqref{equ:d_p} we have
  \[
    d_0 d_1\cdots d_{p-1} = e^{\frac{2i\pi\tilde\nu_\h}{\h} + i
      \sigma_\h(\tilde\nu_\h, \tilde\mu_\h)},
  \]
  where $\sigma_\h$ is a smooth symbol. Therefore, the
  condition~\eqref{equ:cocycle} gives the following
  ``Bohr-Sommerfeld'' rule:
  \[
    \frac{2\pi\tilde\nu_\h}{\h} + \sigma_\h(\tilde\nu_\h, \tilde\mu_\h)
    \in \ZM + \O(\h^\infty).
  \]
  This proves the necessity of item~\ref{item:4} in
  Theorem~\ref{theo:BS_ell_trans}.

  Conversely, if \eqref{equ:cocycle} is satisfied, then one may
  construct a microlocal solution on $\mathcal{U}$ by gluing the
  standard solutions on $\mathcal{W}_j$ by means of a microlocal
  partition of unity. From this, as in \cite[Lemme 2.2.7]{san-livre},
  we obtain a quasimode for the initial spectral problem. But the
  microlocal uniqueness actually gives more: the joint eigenvalues
  must be simple for $\h$ small enough (see \cite[Theorem
  7.1]{san-focus}), and hence coincide module $\O(\h^\infty)$ with the
  microlocal solutions that we have just constructed. This closes the
  proof of the theorem.

  \section{An example: coupled angular momenta}
  \label{sect:coupled_spins}

  Throughout the text, we used the example of the spin-oscillator
    system, on the non-compact manifold $S^2 \times \RM^2$, to
    illustrate our results. More precisely, we defined the classical
    and quantum versions of this system in Examples
    \ref{ex:spin-osc_class} and \ref{ex:spin-osc_quant}, explained how
    to compute its joint spectrum in Example \ref{ex:spin-osc_jsp},
    and used our formulas to recover various symplectic invariants in
    Examples \ref{ex:dfr_spin-osc}, \ref{ex:S10_spin-osc},
    \ref{ex:poly_spin-osc}, \ref{ex:height_spin-osc},
    \ref{ex:S01_spin-osc} and \ref{ex:higher_spin-osc}. In this
    section, we will investigate another example, whose phase space is
    the compact manifold $S^2 \times S^2$, which consists in coupling
    two classical spins. The choice of these two examples was
    motivated by the fact that they are both fundamental in physics,
    that they allow to illustrate both the compact and non-compact
    situation, but also by the explicit computation of their
    symplectic invariants in
    \cite{san-alvaro-spin,alvaro-yohann,ADH-spin,ADH-angular}. Both
    systems have only one focus-focus singularity, and it would be
    interesting to apply our algorithms to compute the invariants
    (especially the polygonal invariant) for a system with two or more
    focus-focus singularities. Such systems are available in
    \cite{HP,joey-yohann,HohMeu} and in the recent \cite{AHP} which includes the computation of the twisting index.

The coupled angular momenta system was introduced
in \cite{sadovski-zhilinski} and consists in coupling two classical
spins in a non-trivial way. More precisely, let $R_2 > R_1 > 0$ and
endow $S^2 \times S^2$ with the symplectic form
$\omega = -(R_1 \omega_{S^2} \oplus R_2 \omega_{S^2})$ and coordinates $(x_1,y_1,z_1), (x_2, y_2, z_2)$. We consider the momentum map:
\[
  F = (J,H), \qquad J = R_1 z_1 + R_2 z_2, \qquad H = (1-t) z_1 + t
  (x_1 x_2 + y_1 y_2 + z_1 z_2) \] depending on a parameter
$t \in [0,1]$. This system is semitoric with exactly one focus-focus
singularity for $t$ chosen in a certain interval depending on $R_1$
and $R_2$, always containing $t=\frac{1}{2}$, see
\cite{alvaro-yohann}.

For quantization purposes, we ask that $R_1$ and $R_2$ are
half-integers. Using the quantization of the sphere described in Example \ref{ex:spin-osc_quant}, we obtain the Hilbert spaces
$\mathcal{H}_k \simeq \CM_{\leq 2kR_1 - 1}[z] \otimes \CM_{\leq 2kR_2
  - 1}[w]$ with inner product
\[
  \langle P_1 \otimes P_2, Q_1 \otimes Q_2 \rangle_k =
  \left(\int_{\CM} \frac{P_1(z) \overline{Q_1(z)}}{(1 + |z|^2)^{2k R_1
        + 1}} |dz \wedge d\bar{z}|\right) \left(\int_{\CM}
    \frac{P_2(w) \overline{Q_2(w)}}{(1 + |w|^2)^{2 k R_2 + 1}} |dw
    \wedge d\bar{w}|\right).  \] Furthermore, $J$ and $H$ are
quantized as the Berezin-Toeplitz operators
\[
  \begin{cases} \hat{J}_k =\frac{1}{2k} \left( (1 + 2k R_1) \hat{Z}_{k R_1} \otimes \mathrm{Id} + (1 + 2k R_2) \mathrm{Id} \otimes \hat{Z}_{k R_2}  \right), \\[2mm]
    \begin{aligned} \hat{H}_k & = \frac{(1-t)}{2k R_1} (1 + 2k R_1)
      \hat{Z}_{k R_1} \otimes \mathrm{Id} \\ & + \frac{t (1 + 2k
        R_1)(1 + 2k R_2)}{4k^2 R_1 R_2} \left( \hat{X}_{k R_1} \otimes
        \hat{X}_{k R_2} + \hat{Y}_{k R_1} \otimes \hat{Y}_{k R_2} +
        \hat{Z}_{k R_1} \otimes \hat{Z}_{k R_2}
      \right). \end{aligned} \end{cases}
\]
with $\hat{X}, \hat{Y}, \hat{Z}$ as in Equation
\eqref{eq:BTO_sphere}. More details can be found in \cite[Section
4]{alvaro-yohann}, including the computation of the joint spectrum of
$(\hat{J}_k, \hat{H}_k)$, which is displayed in Figure
\ref{fig:jsp_coupled_spins}.

\begin{figure}[H]
  \begin{center}
    \includegraphics[trim=60 20 90
    60,clip,scale=0.4]{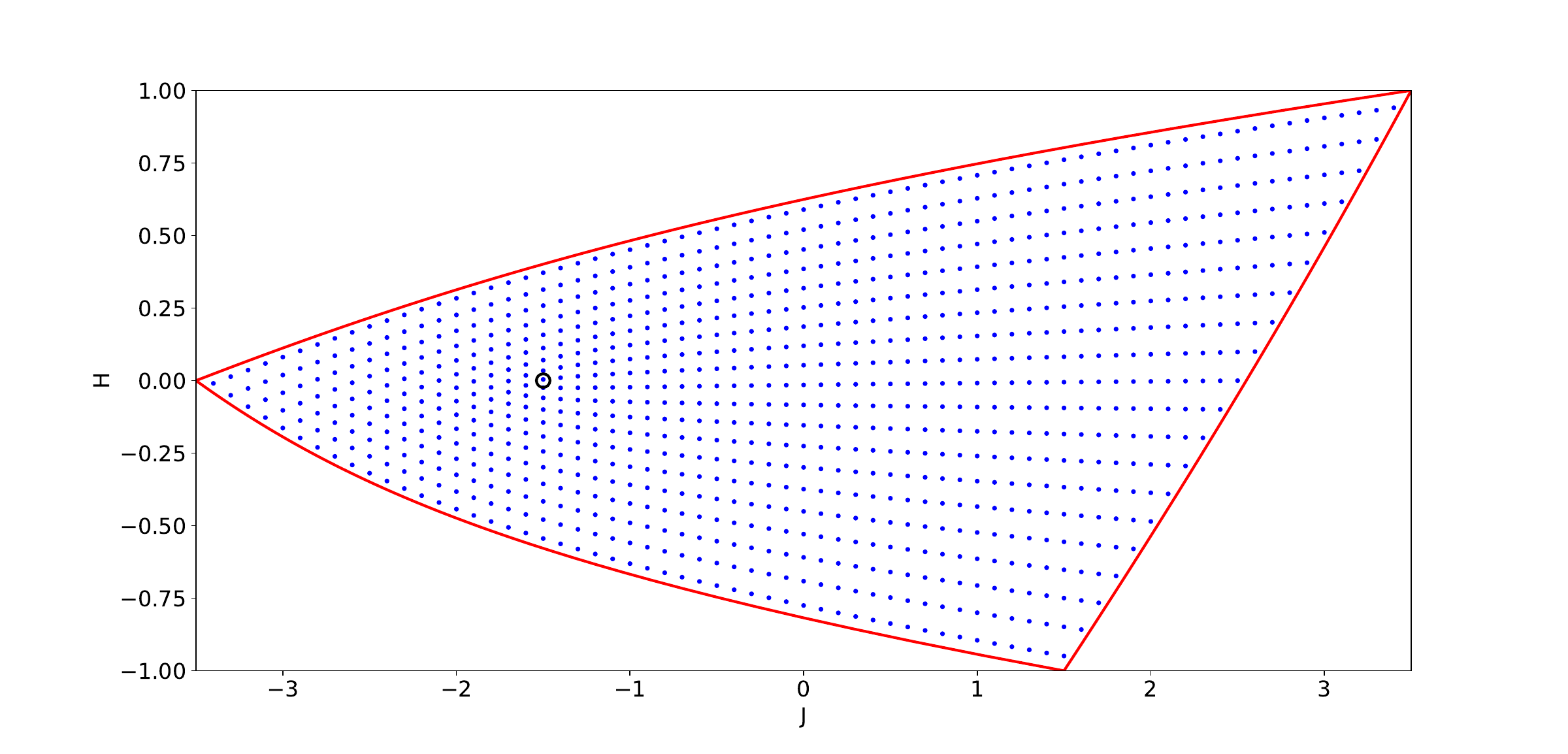}
  \end{center}
  \caption{\small The blue dots are the joint eigenvalues of the quantum
    coupled angular momenta with $R_1 = 1$, $R_2 = \frac{5}{2}$ and
    $t = \frac{1}{2}$ for $k = 10$. The red line corresponds to the
    boundary of the image of the momentum map, and the black circle
    indicates the focus-focus value.}
  \label{fig:jsp_coupled_spins}
\end{figure}

The symplectic invariants were computed in \cite{ADH-angular} for all
values of $R_1, R_2$ and $t$. Here we choose $R_1 = 1$,
$R_2 = \frac{5}{2}$ and $t = \frac{1}{2}$ for our numerical
simulations (part of the invariants were computed for these precises
values of the parameters in \cite{alvaro-yohann}). In this case, the
height invariant is
\begin{equation} S_{0,0} = 2 + \frac{1}{\pi} \left( 3 - 5
    \arctan\left( \frac{3}{4} \right) - 2 \arctan 3
  \right),
  \label{eq:height_spins}
\end{equation}
the Taylor series
invariant reads
\[
  S^\infty = \frac{1}{2\pi} \arctan\left(\frac{13}{9}\right) X +
  \frac{1}{2\pi} \left( \frac{7}{2} \ln 2 + 3 \ln 3 - \frac{3}{2} \ln
    5 \right) Y + \O(2) \] and the first order derivatives of
Eliasson's function are
\begin{equation}
  \partial_x f_r(0) = -\frac{1}{3}, \qquad \partial_y f_r(0) =
  \frac{10}{3}.\label{equ:dx_et_dy_fr_spin-spin}
\end{equation}
Moreover, a representative of the polygonal
invariant with vanishing twisting number and $\epsilon = +1$ is
displayed in Figure \ref{fig:polygon_spins}.

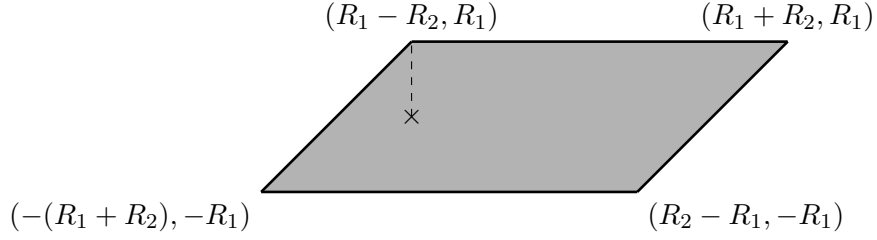
\begin{figure}[H]
  \begin{center}
    \begin{tikzpicture}

      \path[fill, color=gray!40] (-2, -1) -- (0, 1) -- (5, 1) -- (3,
      -1) -- (-2,-1) -- cycle;

      \draw[line width = 1pt] (-2, -1) -- (0, 1); \draw[line width =
      1pt] (-2, -1) -- (3, -1); \draw[line width = 1pt] (0, 1) -- (5,
      1); \draw[line width = 1pt] (3, -1) -- (5, 1);

      \draw [dashed] (0,0) -- (0,1); \draw (0,0) node[] {$\times$};
      \draw (-2,-1) node[below left] {$(-(R_1+R_2),-R_1)$}; \draw
      (0,1) node[above] {$(R_1-R_2,R_1)$}; \draw (3,-1) node[below
      right] {$(R_2 - R_1,-R_1)$}; \draw (5,1) node[above]
      {$(R_1+R_2,R_1)$};
    \end{tikzpicture}
  \end{center}
  \caption{\small A representative of the privileged polygon for the coupled
    angular momenta system.}
  \label{fig:polygon_spins}
\end{figure}

Let us now summarize the various steps necessary to recover these
  invariants from the joint spectrum.

\begin{itemize}
\item We recover the height invariant in Figure~\ref{fig:height_S2xS2}
  using Equation~\eqref{eq:height_single_limit} which gives an
  approximation of the theoretical value~\eqref{eq:height_spins};
\item we recover $\partial_x f_r(0)$ using~\eqref{eq:dxfr_lim} (see
  Figure \ref{fig:dxfr_S2xS2}) and $\partial_y f_r(0)$
  using~\eqref{eq:dyfr_lim} (see Figure \ref{fig:dyfr_S2xS2}), which
  gives~\eqref{equ:dx_et_dy_fr_spin-spin}; hence
  from~\eqref{equ:petit_s}, we obtain $s(0)=\frac{1}{10}$;
\item using this value for $s(0)$, we recover
  $\sigma_1^{\textup{p}}(0)$ in Figure \ref{fig:sigma_S2xS2_kvar} by
  using Formula \eqref{equ:sigma} :
  \begin{equation} \sigma_1(0)= \lim_{x\to 0^+}\lim_{k\to+\infty}
    \frac{E_{0,0} - E_{1,0}}{E_{0,1} - E_{0,0}} + \frac{1}{10k(E_{0,1}
      - E_{0,0})}\, ,
  \end{equation}
  where $E_{0,0}, E_{1,0}$ and $E_{0,1}$ depend on the semiclassical
  parameter $\h = k^{-1}$ and are obtained from the joint eigenvalues
  close to $c=(x,x/10)$ (see~\eqref{equ:semitoric-labelling}) using
  the labelling given by the algorithm of~\cite[Section
  3.5.2]{san-dauge-hall-rotation}, see
  Section~\ref{subsect:asympt_latt}; in this case
  $ \sigma_1(0)\in[0,1[$ and hence by
  Proposition~\ref{prop:relation_inv} we have
  $\sigma_1^{\textup{p}}(0)=\sigma_1(0) = \frac{1}{2\pi}
  \arctan(\frac{13}{9})$;
\item we recover the coefficient $S_{0,1}$ of the Taylor series
  invariant using Proposition~\ref{theo:recover-sigma_2} and
  Equation~\eqref{eq:S01_lim}:
 \begin{equation}
    S_{0,1} = \lim_{x \to 0^+} \lim_{k \to +\infty} \left(
      \frac{3}{10k(E_{0,1} - E_{0,0})} +
      \frac{\ln x}{2\pi} \right)\,,
  \end{equation}
  see Figure \ref{fig:S01_S2xS2_500};
\item finally, we extend the previously obtained labelling to a
  cartographic map (Theorem~\ref{thm:construct_global_labelling}), and
  we recover the associated semitoric polygon thanks to Proposition
  \ref{prop:poly_num}, see Figure \ref{fig:poly_S2xS2}. We know that
  this polygon is the privileged one because the equality
  $\sigma_1^{\textup{p}}(0)=\sigma_1(0)$ stated above shows that the
  corresponding twisting number vanishes.
\end{itemize}

In principle, we could also recover the higher order terms for the
Taylor series invariants as for the example of the
  spin-oscillator system (see Example \ref{ex:higher_spin-osc}), but
in this case the computations are more involved.

In Figure \ref{fig:DH_S2xS2}, we illustrate Remark \ref{rema:DH} by
recovering the Duistermaat-Heckman function of $J$ from the joint
spectrum. We also illustrate the detection of focus-focus values in
Figure \ref{fig:ecarts_50}.

\begin{figure}[H]
  \begin{center}
    \includegraphics[trim=80 10 80
    60,clip,width=0.75\textwidth]{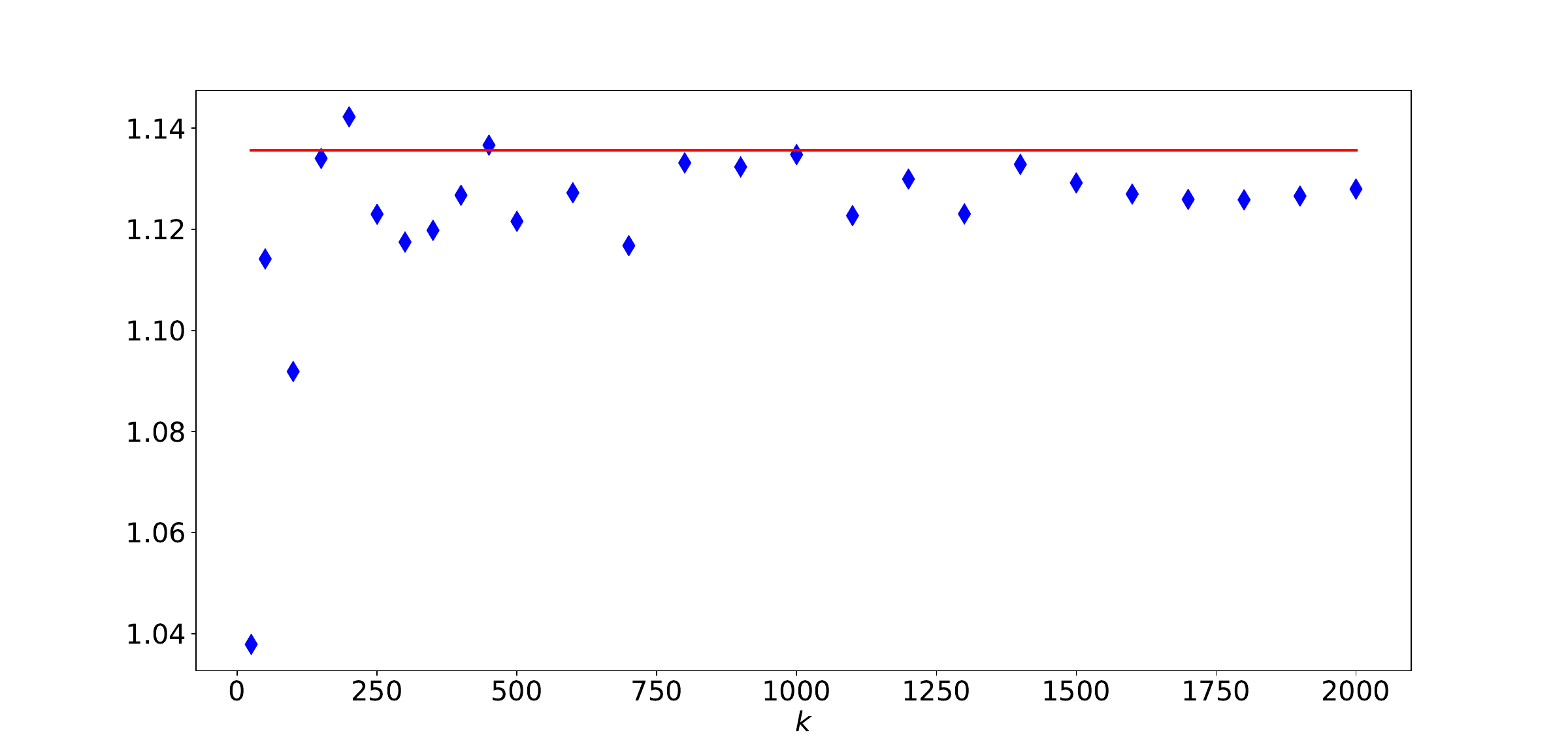}
  \end{center}
  \caption{\small Determination of the height invariant for the
    coupled angular momenta using Proposition
    \ref{prop:height-invariant},
    Equation~\eqref{eq:height_single_limit}. The blue diamonds
    correspond to $\frac{\h^{2 - \delta}}{2c} N_\h(\delta,c,0)$ for
    $c=1$, $\delta = 0.4$ and different values of $k =
    \tfrac{1}{\h}$. The solid red line is the theoretical value given
    in Equation \eqref{eq:height_spins}.}
  \label{fig:height_S2xS2}
\end{figure}

\begin{figure}[H]
  \begin{center}
    \includegraphics[trim=80 10 80
    60,clip,width=0.75\textwidth]{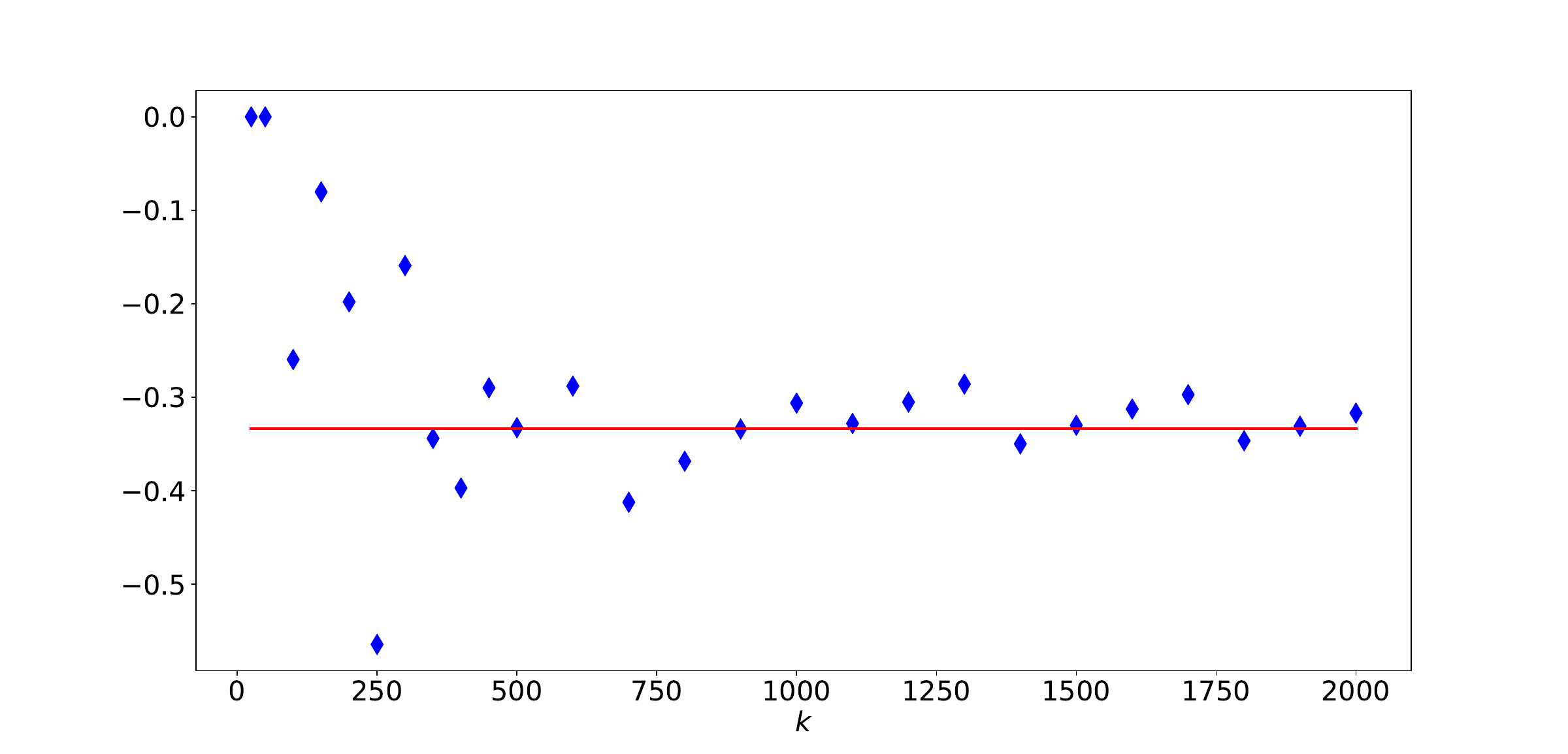}
  \end{center}
  \caption{\small Determination of $\partial_x f_r(0)$ for the coupled
    angular momenta system using Formula \eqref{eq:dxfr_lim} with
    $x = 0.01$, $\mu = 2$ and $(j_1,\ell_1) = (0,0) = (j_2,\ell_2)$,
    for different values of $k$. The red line corresponds to the
    theoretical result $\partial_x f_r(0) = -\frac{1}{3}$.}
  \label{fig:dxfr_S2xS2}
\end{figure}

\begin{figure}[H]
  \begin{center}
    \includegraphics[trim=80 10 80
    60,clip,width=0.75\textwidth]{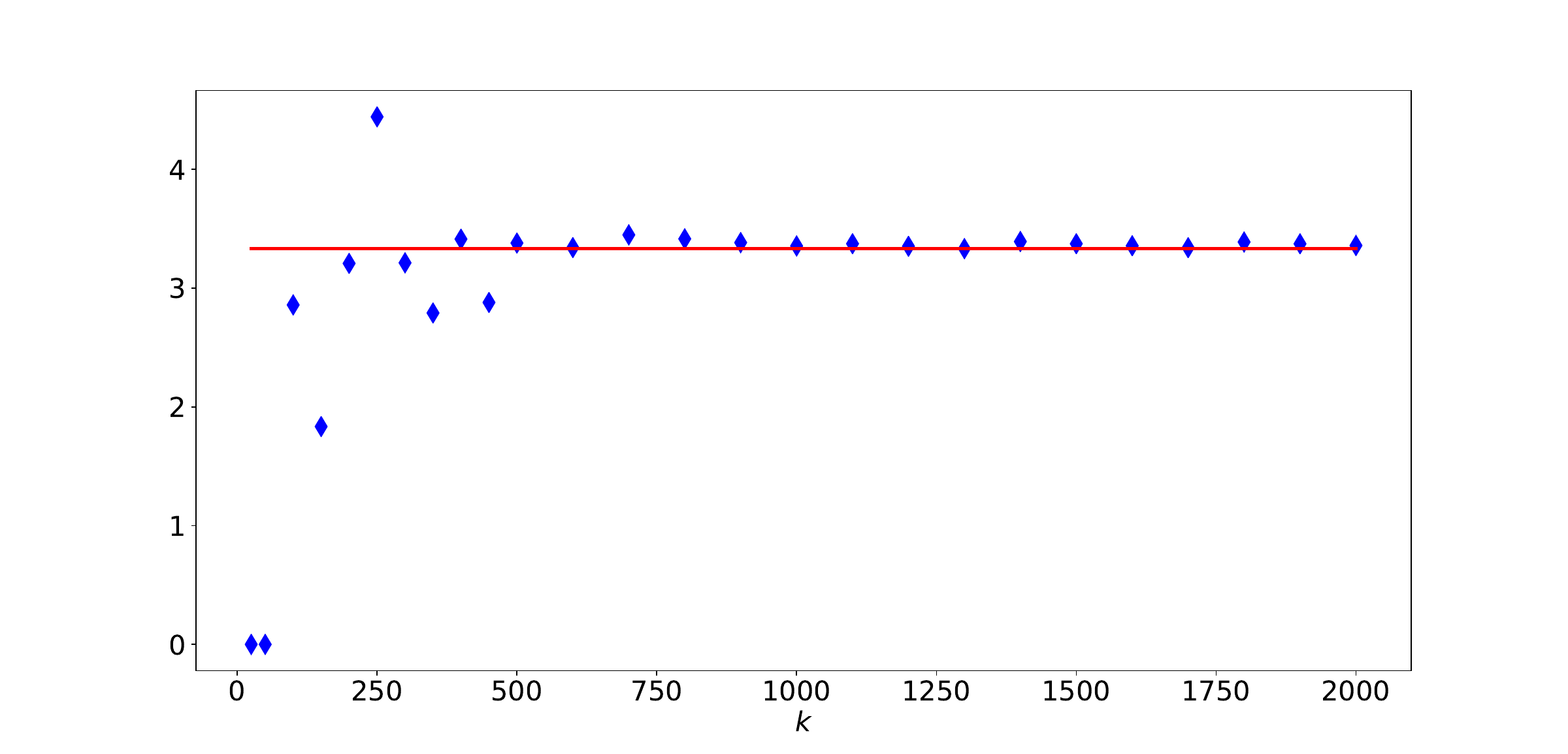}
  \end{center}
  \caption{\small Determination of $\partial_y f_r(0)$ for the coupled
    angular momenta system using Formula \eqref{eq:dyfr_lim} with
    $x = 0.01$, $\mu = 2$ and $(j_1,\ell_1) = (0,0) = (j_2,\ell_2)$,
    for different values of $k$. The red line corresponds to the
    theoretical result $\partial_y f_r(0) = \frac{10}{3}$.}
  \label{fig:dyfr_S2xS2}
\end{figure}

\begin{figure}[H]
  \begin{center}
    \includegraphics[trim=80 10 80
    60,clip,width=0.75\textwidth]{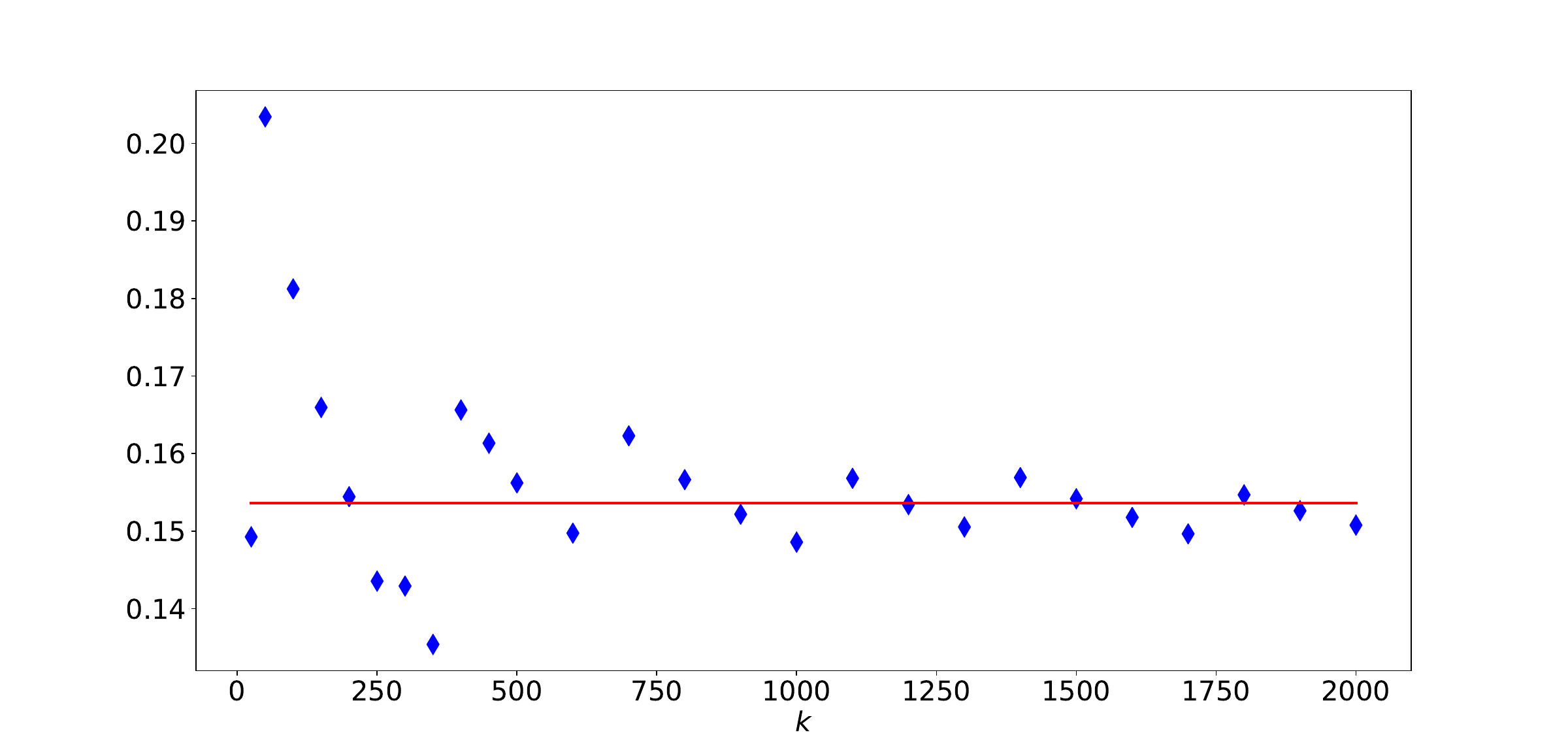}
  \end{center}
  \caption{\small Determination of $\sigma_1^{\textup{p}}(0)$ for the coupled
    angular momenta system. The blue diamonds correspond to Formula
    \eqref{equ:sigma} evaluated at $(j,\ell) = (0,0)$ with $x = 0.01$,
    for different values of $k$. The red line corresponds to
    $\sigma_1^{\textup{p}}(0) = \frac{1}{2\pi}
    \arctan(\frac{13}{9})$.}
  \label{fig:sigma_S2xS2_kvar}
\end{figure}

\begin{figure}[H]
  \begin{center}
    \includegraphics[trim=80 10 80
    60,clip,width=0.75\textwidth]{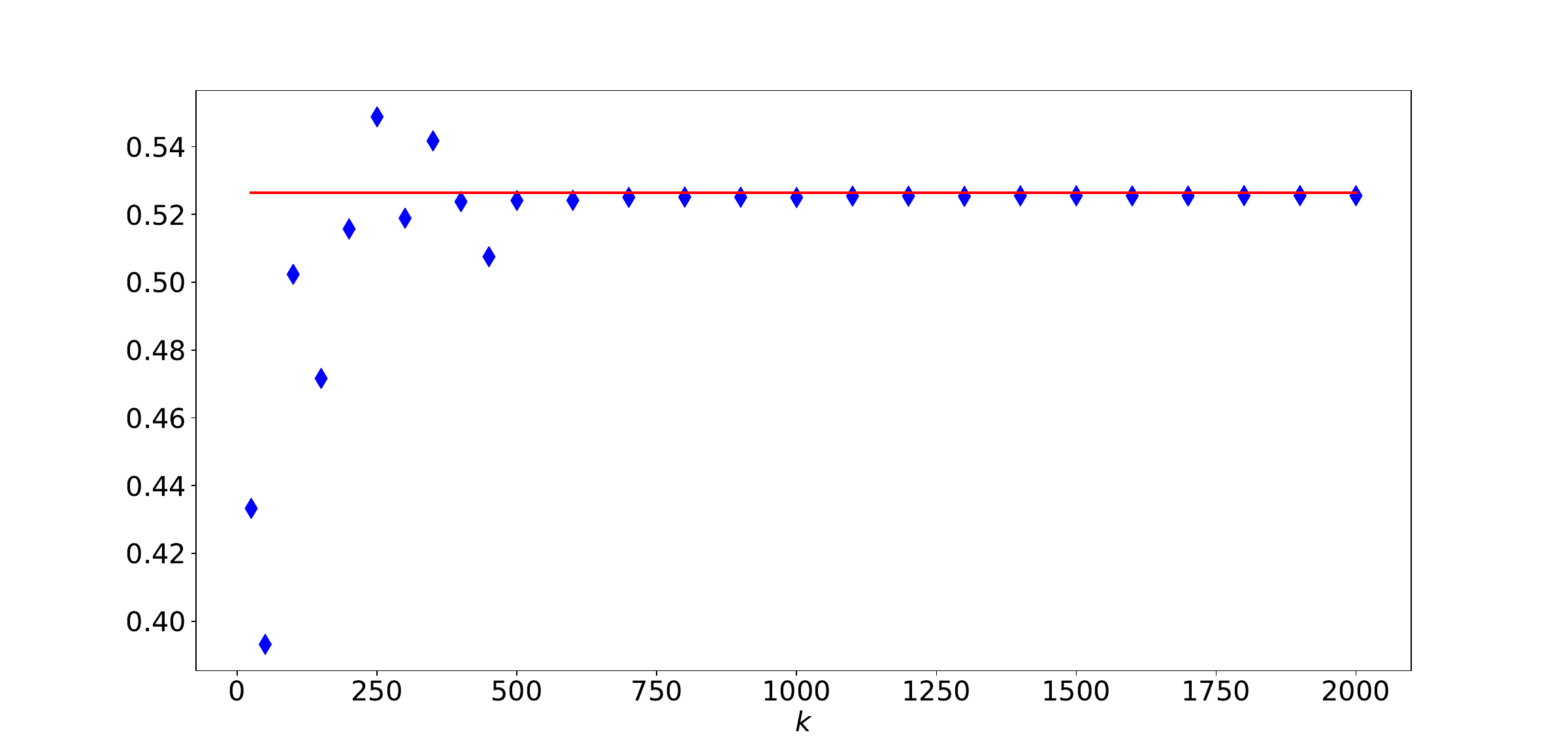}
  \end{center}
  \caption{\small Determination of $S_{0,1}$ for the coupled angular momenta
    system. The blue diamonds correspond to Formula \eqref{eq:S01_lim}
    evaluated at $(j,\ell) = (0,0)$ with $x=0.01$, for different
    values of $k$. The red line corresponds to the theoretical value
    $S_{0,1} = \frac{1}{2\pi} \left( \frac{7}{2} \ln 2 + 3 \ln 3 -
      \frac{3}{2} \ln 5 \right)$.}
  \label{fig:S01_S2xS2_500}
\end{figure}

\begin{figure}[H]
  \begin{center}
    \includegraphics[trim=80 10 80
    60,clip,width=0.75\textwidth]{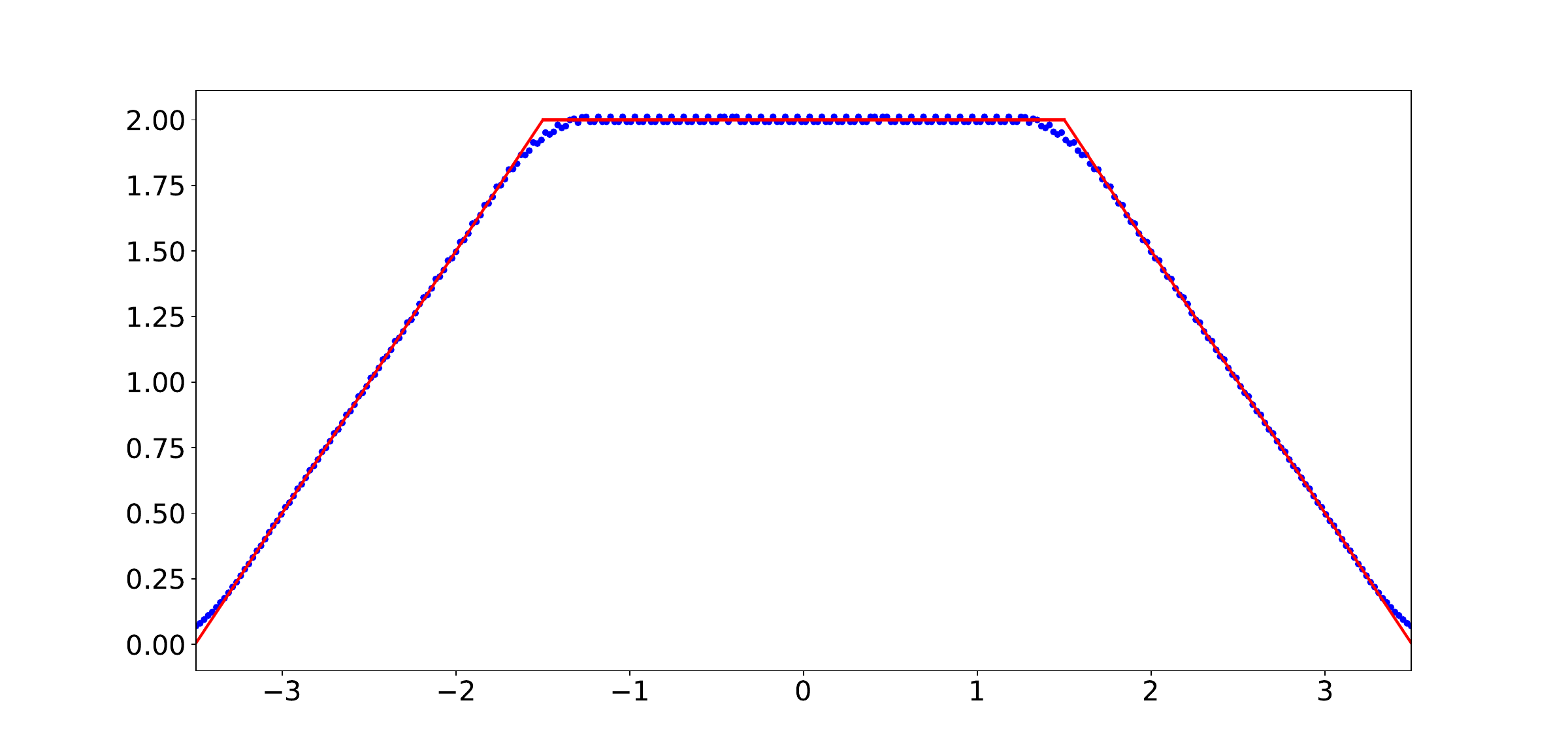}
  \end{center}
  \caption{\small Determination of the Duistermaat-Heckman function
    $\rho_J$ for the coupled angular momenta system using Equation
    \eqref{eq:DH}; the blue dots represent the left hand side of this
    equation, with $k=200$, $\delta = \frac{1}{4}$ and $c=1$. The
    solid red line is the graph of $\rho_J$, which can be computed
    explicitly, for instance from  the polygon in Figure
    \ref{fig:polygon_spins}: indeed, $\rho_J(x)$ is the length of the
    vertical segment obtained by intersecting the polygon with the
    vertical line through $(x,0)$.}
  \label{fig:DH_S2xS2}
\end{figure}

\begin{figure}[H]
  \centering \includegraphics[trim=80 10 80
  60,clip,width=0.75\textwidth]{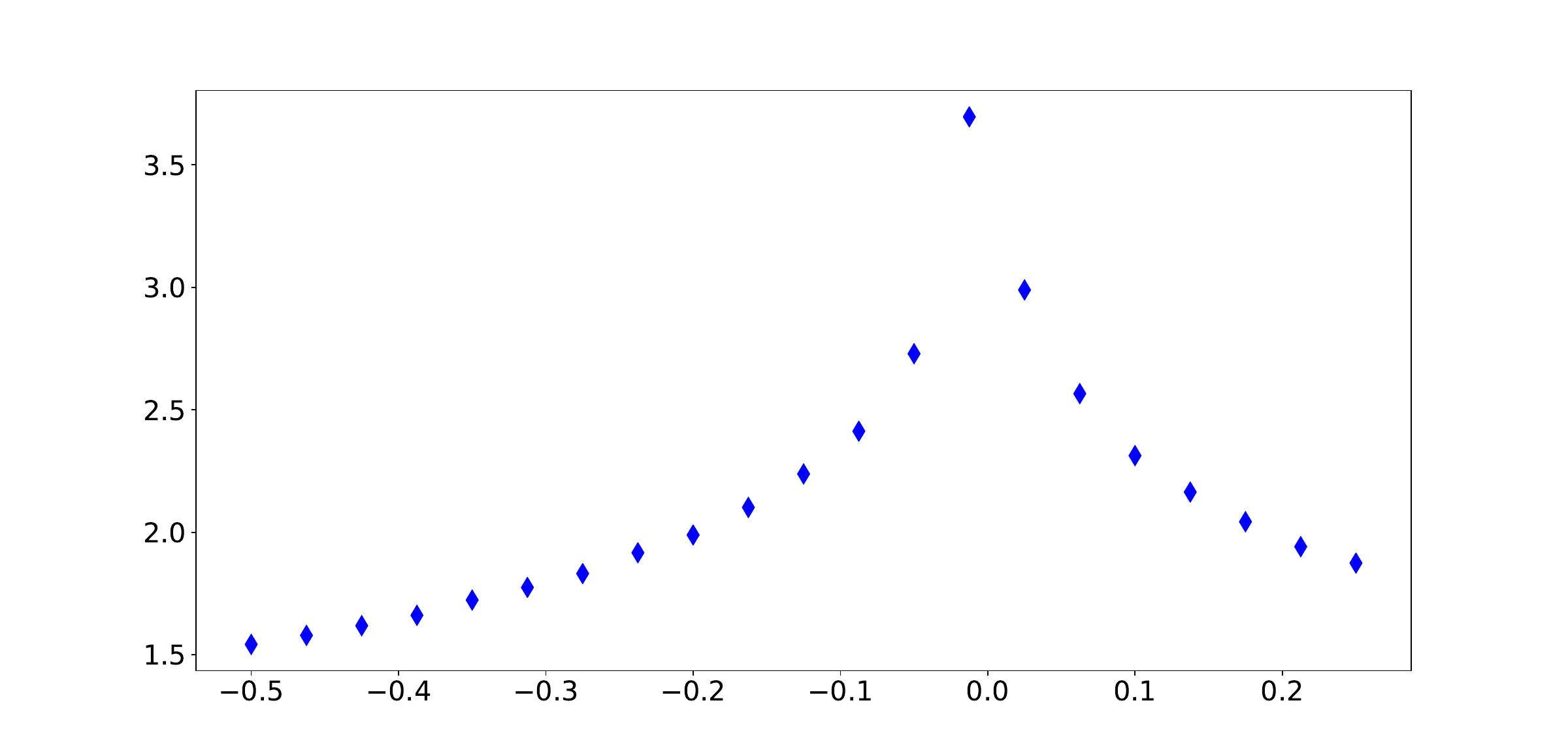}
  \caption{\small Computation of the inverse level spacings
    $\frac{\h}{E_{0,1}-E_{0,0}}$ related to the function $a_2(c)$, see
    Lemma~\ref{lemm:quantum-a}, where $c=(-1.5,y)$ for various values
    of $y$; here $k=50$. Notice the peak indicating the position of
    the focus-focus critical value at $y=0$.}
  \label{fig:ecarts_50}
\end{figure}

\begin{figure}[H]
  \begin{center}
    \includegraphics[trim=80 10 80
    60,clip,width=0.75\textwidth]{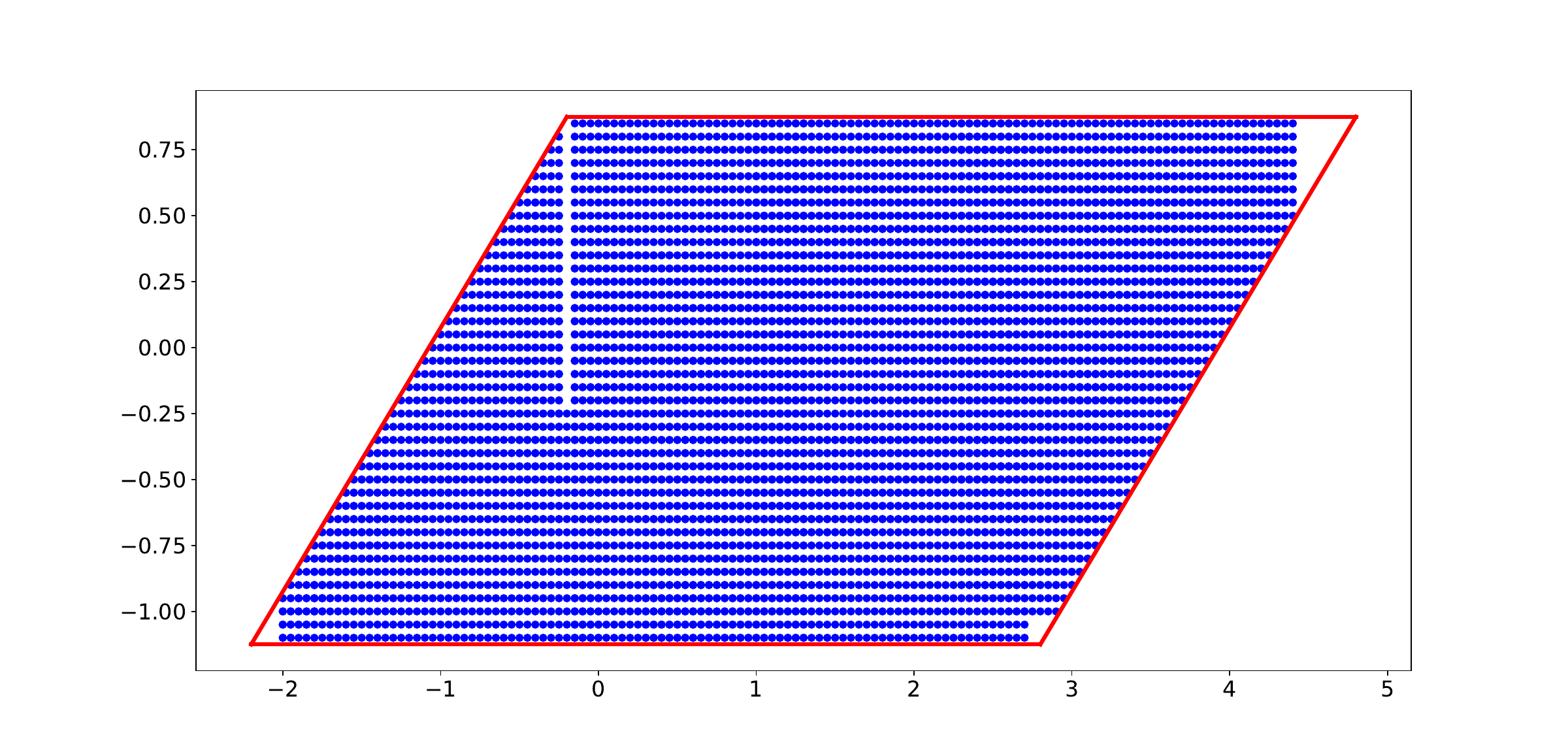}
  \end{center}
  \caption{\small Determination of the privileged polygon for the coupled
    angular momenta system using Proposition \ref{prop:poly_num}; the
    blue dots represent the set
    $\Delta_\h(\mathcal{K}(S,\mathcal{U}))$ where
    $S = \{ (x,y) \in \RM^2 \ | \ -3.3 \leq x \leq 3.1 \}$ for $k=20$,
    while the solid red lines represent a translation of the
    privileged semitoric polygon shown in Figure \ref{fig:polygon_spins}.}
  \label{fig:poly_S2xS2}
\end{figure}

\appendix
\section{Semiclassical operators}
\label{sec:semicl-oper}

In this section, we gather the definition and properties of
semiclassical operators that are used throughout the paper. We allow
$\h$-pseudodifferential operators as well as Berezin-Toeplitz
operators. When quantizing $T^* \RM^d = \CM^d$, these two notions are
related through the Bargmann transform; we also state some useful
properties of this transform.

\subsection{Pseudodifferential operators}

Let $X = \RM^d$ or let $X$ be a compact Riemannian manifold. We
consider $\h$-pseudodifferential operators acting on the (fixed)
Hilbert space $\mathcal{H}_{\h} = L^2(X)$ with usual inner product,
and $\h$ is a continuous parameter taking its values in an interval of
the form $(0,\h_0]$ for some $\h_0 > 0$. In order to define these
operators, we need to separate the two cases.

If $X = \RM^d$ (hence $T^*X \simeq \RM^{2d}$), we say that $A_{\h}$ is
an $\h$-pseudodifferential operator if it is the Weyl quantization of
a symbol $a \in S(m)$ where $m$ is some order function (as in
\cite[Section 4]{zworski}, for instance): $m$ is a measurable function
such that there exist constants $C, N > 0$ so that
\[
  \forall U, V \in \RM^{2d} \qquad 1 \leq m(U) \leq C \langle V - U
  \rangle^N m(V) \] where
$\langle U \rangle = \left( 1 + \| U \|^2 \right)^{\frac{1}{2}}$. Some
usual choices are $m(U) = \langle U \rangle^{\mu}$ or
$m(x,\xi) = \langle \xi \rangle^{\mu}$ for some $\mu \in \RM$. Let
$a = a(\cdot, \h)$ be a family of elements of
$\Cinf(\RM^{2d})$; we say that $a$ belongs to $S(m)$ if
\[
  \forall \alpha \in \NM^{2d} \quad \exists C_{\alpha} > 0 \quad
  \forall \h \in (0,\h_0] \quad \forall (x,\xi) \in \RM^{2d} \qquad
  |\partial^{\alpha} a(x,\xi,\h)| \leq C_{\alpha} m(x,\xi).
\]
We will always assume that $a \in S(m)$ is asymptotic to
$\sum_{j \geq 0} \h^j a_j$, where for every $j \geq 0$,
$a_j \in \Cinf(\RM^{2d})$ is independent of $\h$, in
the sense that
\[
  \forall N \geq 1 \qquad a - \sum_{j=0}^N \h^j a_j \in \h^{N+1}
  S(m).
\]
If it is not identically zero, the function $a_0$ is called the
\emph{principal symbol} of $A_\h$. The Weyl quantization
$A_\h = \mathrm{Op}_\h^{W}(a)$ of $a \in S(m)$ is defined by the
following formula: for every $u \in \mathscr{S}(\RM^d)$ and for every
$x \in \RM^d$,
\[
  (A_\h u)(x) = \frac{1}{(2\pi \h)^d} \int_{\RM^d} \int_{\RM^d} e^{
    \frac{i}{\h} \pscal{x - y}{\xi}} a\left( \frac{x+y}{2}, \xi
  \right) u(y) \dd y \dd\xi. \] By a slight abuse of notation, we say
that $A_\h$ belongs to $S(m)$ when $a$ does.

If $X$ is a compact Riemannian manifold, we always work with the order
function given in local coordinates by
$m(x,\xi) = \langle \xi \rangle^{\mu}$ for some $\mu \in \RM$; we say
that $A_\h$ is an $\h$-pseudodifferential operator in the
Kohn-Nirenberg class $S(m)$ if in local coordinates, after a cut-off
in $x \in X$, $A_\h$ can be written as an $\h$-pseudodifferential
operator with symbol $a \in S(m)$. This does not depend on the choice
of local coordinates. See for instance \cite[Section 14.2]{zworski}
for more details.

We need two notions of ellipticity. Firstly, we say that
$A_\h \in S(m)$ is \emph{elliptic} at $p \in T^*X$ if its principal
symbol does not vanish at $p$. Secondly, we say that $A_\h \in S(m)$
is \emph{elliptic at infinity} in $S(m)$ if there exists $C > 0$ such
that $|a_0| \geq C m$ outside of a compact set.

Additionally, we will need to consider families of elements of
$L^2(X)$ upon which $\h$-pseudodifferential operators act (for
instance, families of eigenvectors of such an operator), and to study
their localization in phase space.

\begin{defi}
  \label{defi:wavefront}
  Let $(u_\h)_{\h \in \mathcal{I}}$ be a sequence of elements of
  $\mathscr{D}'(\RM^d)$, and let $p \in T^*X$. We say that
  \begin{itemize}
  \item $(u_\h)$ is \emph{admissible} if for any
    $\h$-pseudodifferential operator $A_\h$ with compactly supported
    Weyl symbol, there exists $N \in \ZM$ such that
    $\| A_\h u_\h \|_{L^2(\RM^d)} = \O(\h^N)$;
  \item the admissible sequence $(u_\h)$ is \emph{negligible} at $p$
    if there exists an $\h$-pseudodifferential operator $A_\h$,
    elliptic at $p$, such that
    $\| A_\h u_\h \|_{L^2(\RM^d)} = \O(\h^{\infty})$;
  \item $p \notin \mathrm{WF}(u_\h)$ if and only if $(u_\h)$ is
    negligible at $p$. The set $\mathrm{WF}(u_\h) \subset T^*X$ is
    called the \emph{wavefront set} of $(u_\h)$.
  \end{itemize}
\end{defi}

When looking for eigenvalues of a semiclassical operator, it is often
useful to consider functions that solve the eigenvalue equation
``locally in phase space'' in the following sense.

\begin{defi}
  \label{defi:micsol_PDO}
  Let $A_\h$ be an $\h$-pseudodifferential operator. A
  \emph{microlocal solution} to the equation
  $A_\h u_\h = \O(\h^{\infty})$ over the open set $\mathcal{U}$ is an
  admissible family $(u_\h)_{\h \in \mathcal{I}}$ of elements of
  $\mathcal{H}_\h$ such that
  $\mathrm{WF}(A_\h u_\h) \cap \mathcal{U} = \emptyset$.
\end{defi}

By adapting the above definitions, one can compare the action of
operators on a given part of the phase space.

\begin{defi}
  \label{defi:microlocal_equality}
  Let $A_\h, B_\h$ be two $\h$-pseudodifferential operators. We say
  that
  \begin{itemize}
  \item $A_\h$ is \emph{negligible} at $p \in T^*X$ if there exists
    an $\h$-pseudodifferential operator $P_\h$, elliptic at $p$, such
    that $\| P_\h A_\h \| = \O(\h^{\infty})$,
  \item $A_\h$ and $B_\h$ are \emph{microlocally equivalent} at $p$ if
    and only if $A_\h - B_\h$ is negligible at $p$. In this case we
    write $A_\h \sim B_\h$ at $p$,
  \item $A_\h$ and $B_\h$ are \emph{microlocally equivalent} on the
    open set $\mathcal{U} \subset T^*X$ if and only if they are
    microlocally equivalent at every point of $\mathcal{U}$.
  \end{itemize}
\end{defi}

Finally, we use the notation $\CM_\h$ for the ring of constant
symbols, that can be seen as symbols in $S(1)$ on $\{0\}$.

\subsection{Berezin-Toeplitz operators}
\label{subsec:def_BTO}

\paragraph{On a compact phase space.} We now consider a compact
symplectic manifold $(M,\omega)$. In fact, we shall always assume that
$(M,\omega)$ is K\"ahler; this is not really restrictive for the
purpose of the present paper, since:
\begin{itemize}
\item a compact symplectic four-dimensional manifold endowed with
  an effective Hamiltonian $S^1$-action (which is the case if
  there exists a semitoric system on $M$) is automatically K\"ahler by
  \cite[Theorem 7.1]{Karshon} (see case \ref{item:2} in Section
  \ref{sec:quantum_semitoric}),
\item a compact symplectic surface is automatically K\"ahler (see case
  \ref{item:3} in Section \ref{sec:quantum_semitoric}).
\end{itemize}
Furthermore, we will always assume that $M$ is quantizable in the
sense that the cohomology class $\left[ \frac{\omega}{2\pi} \right]$
is integral; this amounts to the existence of a Hermitian and
holomorphic line bundle $(\mathscr{L},h_{\mathscr{L}}) \to M$ whose
Chern connection $\nabla$ has curvature $-i\omega$, called \emph{prequantum
line bundle}.

In this context, we consider Berezin-Toeplitz operators
\cite{berezin,BG,BMS,laurent-BTO,ma-marinescu}, which act on a
sequence of finite-dimensional Hilbert spaces defined as follows. Let
$(\mathscr{K}, h_\mathscr{K}) \to M$ be another Hermitian holomorphic
complex line bundle; for instance one can choose
$\mathscr{K} = \delta$ a half-form bundle (a square root of the canonical bundle) when it exists, to obtain the so-called metaplectic correction. For any
integer $k \geq 1$, $h_{\mathscr{L}}$ and $h_{\mathscr{K}}$ induce a
Hermitian form $h_k$ on $\mathscr{L}^{\otimes k} \otimes \mathscr{K}$,
and we consider the Hilbert space
\[
  \mathcal{H}_k = H^0(M, \mathscr{L}^{\otimes k} \otimes \mathscr{K}),
  \qquad \langle \phi, \psi \rangle_k = \int_M h_k(\phi,\psi) \left|
    \frac{\omega^{\wedge n}}{n!} \right|
\]
of holomorphic sections of the line bundle
$\mathscr{L}^{\otimes k} \otimes \mathscr{K} \to M$. The semiclassical
parameter in this context is $\h = k^{-1}$ and takes only discrete
values. A Berezin-Toeplitz operator is an operator of the form
\[
  T_k = \Pi_k f(\cdot,k) \Pi_k + R_k : \mathcal{H}_k \to \mathcal{H}_k
\]
where
$\Pi_k: L^2(M,\mathscr{L}^{\otimes k} \otimes \mathscr{K}) \to
\mathcal{H}_k$ is the orthogonal projector from the space of square
integrable sections to the space of holomorphic sections of
$\mathscr{L}^{\otimes k} \otimes \mathscr{K} \to M$, $f(\cdot,k)$ is a
sequence of elements of $\Cinf(M)$ with an asymptotic expansion of the
form $f(\cdot,k) = \sum_{\ell \geq 0} k^{-\ell} f_{\ell}$ in the
$\Cinf$-topology and $R_k$ is a sequence of operators whose norm is
$\O(k^{-N})$ for every $N \geq 1$. If not identically zero, the term
$f_0$ in the above asymptotic expansion is called the principal symbol
of $T_k$. When $R_k = 0$, we simply write $T_k(f(\cdot,k))$ for
$ \Pi_k f(\cdot,k) \Pi_k$.

As before, we need to discuss the localization of sequences of
sections in phase space in the semiclassical limit.

\begin{defi}
  \label{defi:microsupport}
  Let $(\psi_k)_{k \geq 1}$ be a sequence such that for each $k$,
  $\psi_k \in \Cinf(M,\mathscr{L}^{\otimes k} \otimes
  \mathscr{K})$, and let $m \in M$. We say that
  \begin{itemize}
  \item $(\psi_k)$ is \emph{admissible} if for every integer
    $\ell \geq 0$, for any vector fields $X_1, \ldots, X_{\ell}$ on
    $M$ and for every compact set $C \subset M$, there exist a
    constant $c > 0$ and an integer $N$ such that
    \[
      \forall p \in C \qquad |\nabla_{X_1} \ldots \nabla_{X_{\ell}}
      \psi_k(p)| \leq c k^N \] (here $|\cdot|$ stands for the
    pointwise norm given by the Hermitian metric on
    $\mathscr{L}^{\otimes k} \otimes \mathscr{K}$),
  \item the admissible sequence $(\psi_k)$ is \emph{negligible} at $m$
    if there exists a neighborhood $\mathcal{V}$ of $m$ such that for
    any integers $\ell, N \geq 0$ and for any vector fields
    $X_1, \ldots, X_{\ell}$ on $M$,
    \[
      \sup_{\mathcal{V}} |\nabla_{X_1} \ldots \nabla_{X_{\ell}}
      \psi_k| = \O(k^{-N}),
    \]
  \item $m \notin \mathrm{MS}(\psi_k)$ if and only if $(\psi_k)$ is
    negligible at $m$. The set $\mathrm{MS}(\psi_k)$ is called the
    \emph{microsupport} of $(\psi_k)$.
  \end{itemize}
\end{defi}
Naturally, the microsupport is the analogue of the wavefront set, see
Section~\ref{sec:bargmann-transform}.  We can then define
negligibility and microlocal equality for Berezin-Toeplitz operators
by applying these definitions to their Schwartz kernels, which are
sections of
$(\mathscr{L}^{\otimes k} \otimes \mathscr{K}) \boxtimes
(\overline{\mathscr{L}}^{\otimes k} \otimes \overline{\mathscr{K}})
\to M \times \overline{M}$. Here $\overline{M}$ is $M$ endowed with
the opposite symplectic and complex structures, and the external
tensor product $\mathscr{L} \boxtimes \mathscr{L}'$ of two line
bundles $\mathscr{L} \to M$ and $\mathscr{L}' \to M'$ is the line
bundle $\pi_1^*\mathscr{L} \otimes \pi_2^*\mathscr{L}'$ where
$\pi_1: M \times M' \to M$ and $\pi_2: M \times M' \to M'$ are the
natural projections.

We also need to define microlocal solutions in this context; however,
there is a subtlety that did not appear in the $\h$-pseudodifferential
case. Indeed, one would like to be able to consider a holomorphic
section $\psi_k$ and to multiply it by a compactly supported smooth
function, but the resulting section will not be holomorphic in
general. This leads to the following definition (see also
\cite[Section 4]{yohann-hyp}).

\begin{defi}
  \label{defi:micsol_BTO}
  Let $T_k$ be a Berezin-Toeplitz operator and let
  $\mathcal{U} \subset M$ be an open set. A \emph{microlocal solution}
  to the equation $T_k u_k = \O(k^{-\infty})$ over $\mathcal{U}$ is an
  admissible sequence $(u_k)_{k \geq 1}$ of elements of
  $\Cinf(\mathcal{U},\mathscr{L}^{\otimes k} \otimes \mathscr{K})$
  such that for every $m \in \mathcal{U}$, there exists a function
  $\chi \in \Cinf(M)$, equal to one near $m$, and compactly supported
  in $\mathcal{U}$, such that
  \[
    \Pi_k(\chi u_k) = u_k + \O(k^{-\infty}), \qquad T_k(\Pi_k(\chi
    u_k)) = \O(k^{-\infty}) \] near $m$.
\end{defi}

\paragraph{On $\CM^d$ or $\CM^d \times M$ with $M$ compact.}

We also need to consider Berezin-Toeplitz operators with symbols
defined on
$(\RM^{2d}, \omega_0 = \dd\xi_1 \wedge \dd x_1 + \ldots + \dd\xi_d \wedge
\dd x_d)$, and in this context we have to to introduce some good symbol
classes as in the previous section. More precisely, we identify
$\RM^{2d}$ with $\CM^d$ using the complex coordinates
$z_j = \frac{1}{\sqrt{2}}(x_j - i \xi_j)$, and endow it with the line
bundle $\mathscr{L}_0 = \CM^d \otimes \CM \to \CM^d$ equipped with its
natural Hermitian form $h$, the connection $\nabla = \dd - i \alpha$
where
\[
  \alpha = \frac{i}{2} \left( z_1 \dd{\bar{z}_1} + \ldots + z_d \dd{\bar{z}_d}
    - \bar{z}_1 \dd z_1 - \ldots - \bar{z}_d \dd z_d \right) \] and the
unique holomorphic structure compatible with both $h$ and $\nabla$. We
consider the quantum spaces
$\mathcal{B}_k(\CM^d) = H^0(\CM^d,\mathscr{L}_0^{\otimes k}) \cap
L^2(\CM^d,\mathscr{L}_0^{\otimes k})$, that is
\begin{multline} \mathcal{B}_k(\CM^d) = \bigg\{f \psi^k \quad | \quad
  f: \CM^d \to \CM \text{ holomorphic},\\ \left. \int_{\CM^d} |f(z)|^2
    \exp(-k \|z\|^2) \ \abs{\dd{z_1} \wedge \dd{\bar{z}_1} \wedge
      \ldots \wedge \dd{z_d} \wedge \dd{\bar{z}_d}} < +\infty
  \right\} \end{multline} for $k \geq 1$, where
$\psi(z) = \exp\left( -\frac{1}{2} \|z\|^2 \right)$ with
$\|z\|^2 = |z_1|^2 + \ldots + |z_d|^2$. These spaces are called
Bargmann spaces and are known to be Hilbert spaces \cite{bargmann}
when equipped with the inner product
\[
  \langle f \psi^k, g \psi^k \rangle_k = \int_{\CM^d}
  f(z)\overline{g(z)} \exp(-k \|z\|^2) \ \abs{\dd{z_1} \wedge
    \dd{\bar{z}_1} \wedge \ldots \wedge \dd{z_d} \wedge
    \dd{\bar{z}_d}}.
\]

The symbol classes that we will consider were discussed in
\cite[Section 3.3]{yohann-ell} and are very similar to the ones used
for $\h$-pseudodifferential operators. Similarly to the previous
section, we set $\langle z \rangle = ( 1 + \|z\|^2)^{\frac{1}{2}}$ and
for $\mu \in \RM$, we consider the weight function
$m(z) = \langle z \rangle^{\mu}$. Then we say that $a(\cdot,k)$
belongs to the symbol class $S(m)$ if
\[
  \forall \alpha, \beta \in \NM^d \quad \exists C_{\alpha,\beta} > 0
  \quad \forall k \geq 1 \quad \forall z \in \CM^d \qquad
  |\partial^{\alpha}_z \partial^{\beta}_{\bar{z}} a(z,k)| \leq
  C_{\alpha,\beta} m(z). \] We will assume as in the previous section
that $a \in S(m)$ is asymptotic to $\sum_{j \geq 0} k^{-j} a_j$, where
for every $j \geq 0$, $a_j \in \Cinf(\CM^d)$ is
independent of $k$, in the sense that
\[
  \forall N \geq 1 \qquad a - \sum_{j=0}^N k^{-j} a_j \in k^{-(N+1)}
  S(m).  \] A Berezin-Toeplitz operator in the class $S(m)$ is an
operator of the form
\[
  A_k = \Pi_k a(\cdot,k) \Pi_k + S_k: \mathcal{B}_k(\CM^d) \to
  \mathcal{B}_k(\CM^d) \] where $\Pi_k$ is the orthogonal projector
$\Pi_k: L^2(\CM^d,\mathscr{L}^{\otimes k}) \to \mathcal{B}_k(\CM^d)$
and $S_k: \mathcal{B}_k(\CM^d) \to \mathcal{B}_k(\CM^d)$ is an
operator whose Schwartz kernel is of the form
\[
  S_k(z,w) = R_k(z,w) \exp(-C k \|z - w\|^2) \] where $C > 0$ does not
depend on $k$ and $R_k$ is a negligible sequence of sections of
$\Cinf(\CM^d \times \CM^d,\mathcal{L}_0^{\otimes k}
\boxtimes \overline{\mathcal{L}_0}^{\otimes k})$. Here the definition
of negligible is the same as in \ref{defi:microsupport}.

The notions of ellipticity and ellipticity at infinity can be defined
as in the previous section. The notions of admissibility,
negligibility and microsupport can be defined as in Definition
\ref{defi:microsupport}, and one can reformulate them in a similar
fashion as Definition \ref{defi:wavefront}; for instance, one can
check that $(\psi_k)$ is negligible at $z_0 \in \CM^d$ if and only if
there exists a Berezin-Toeplitz operator $T_k$, elliptic at $z_0$,
such that $\|T_k \psi_k\|_k = \O(k^{-\infty})$ (see \cite[Lemma
2.7]{yohann-hyp}). Finally, one can define microlocal solutions as in
Definition \ref{defi:micsol_BTO}.

Finally, we will need to handle phase spaces of the form
$\CM^d \times M$ where $M$ is a quantizable compact K\"ahler manifold
(see case \ref{item:3}). In order to do so, we consider the same line
bundle $\mathscr{L}_0 \to \CM^d$ as above and a prequantum line bundle
$\mathscr{L} \to M$ and auxiliary Hermitian line bundle
$\mathscr{K} \to M$. Then the quantum Hilbert spaces are
\[
  \mathcal{H}_k := H^0(\CM^d \times M, \mathscr{L}_0^{k} \boxtimes
  (\mathscr{L}^k \otimes \mathscr{K})) \cap L^2(\CM^d \times M,
  \mathscr{L}_0^{k} \boxtimes \mathscr{L}^k \otimes \mathscr{K})
\]
endowed with the inner product obtained by the same construction as in
the compact case, using the Hermitian metric induced on
$\mathscr{L}_0^{k} \boxtimes (\mathscr{L}^k \otimes \mathscr{K})$ by
those of $\mathscr{L}_0$, $\mathscr{L}$ and $\mathscr{K}$. In fact,
one readily checks that
$\mathcal{H}_k \simeq \mathcal{B}_k(\CM^d) \otimes H^0(N,\mathscr{L}^k
\otimes \mathscr{K})$ as Hilbert spaces. There is no specific
difficulty with this setting: one can work with symbol classes that
are similar to the case of $\CM^d$ in order to handle the lack of
compactness on the first factor. The notions of ellipticity,
ellipticity at infinity, admissibility, negligibility and microsupport
are still well-defined.

\subsection{The Bargmann transform}
\label{sec:bargmann-transform}

The \emph{semiclassical Bargmann transform} is the linear map
$B_k: L^2(\RM^d) \to \mathcal{B}_k(\CM^d)$ given by the following
formula: for every $f \in L^2(\RM^d)$ and for every $z \in \CM^d$,
\[
  (B_k f)(z) = 2^{\frac{d}{4}} \left( \frac{k}{2\pi}
  \right)^{\frac{3d}{4}} \left( \int_{\RM^d} e^{ -\frac{k}{2} \left(
        z^2 + x^2 - 2 \sqrt{2} z \cdot x \right)} f(x) \dd x \right)
  \psi^k(z)
\]
where $z^2 = z_1^2 + \ldots + z_n^2$, $x^2 = x_1^2 + \ldots + x_n^2$
and $z \cdot x = z_1 x_1 + \ldots + z_n x_n $. It is a unitary
operator between those two Hilbert spaces, and has the following
semiclassical properties, see for instance \cite[Sections 13.3 and
13.4]{zworski} for a class of symbols with bounded derivatives, or
\cite[Section 3]{yohann-hyp} for the $d=1$ case (the general case
being completely similar):
\begin{enumerate}[label=(B\arabic*)]
\item \label{item:B1} if $(u_k)_{k \geq 1}$ is an admissible sequence
  of elements of $\mathscr{S}(\RM)$, then
  $(x_1, \ldots, x_d, \xi_1, \ldots, \xi_d) \notin \mathrm{WF}(u_k)$
  if and only if
  $\phi(x_1, \ldots, x_d, \xi_1, \ldots, \xi_d) \notin \mathrm{MS}(B_k
  u_k)$ (in other words, the notions of wavefront set and microsupport
  are equivalent via the semiclassical Bargmann transform),
\item \label{item:B2} if $a(.,k)$ belongs to the class $S(m)$ where
  $m(z) = \langle z \rangle^{\mu}$ for some $\mu \in \RM$, then
  $B_k^* T_k(a(\cdot,k)) B_k$ is a pseudodifferential operator in
  $S(m \circ \phi)$ with principal symbol $a_0 \circ \phi$.
\end{enumerate}
Here $\phi$ is defined as
$\phi(x_1, \ldots, x_d, \xi_1, \ldots, \xi_d) = \frac{1}{\sqrt{2}}(x_1
- i \xi_1, \ldots, x_d - i \xi_d)$. To understand these properties,
one can think of the semiclassical Bargmann transform as a Fourier
integral operator associated with the symplectomorphism
$\phi^{-1}: \CM^d \to \RM^{2d}$.

\bibliographystyle{abbrv}
\bibliography{bibli-utf8}
\end{document}